\documentclass[12pt,a4paper]{JHEP3}
\usepackage{amsmath,epsfig}
\usepackage{amssymb,amsfonts}
\usepackage{latexsym}
\usepackage[latin1]{inputenc}
\usepackage{slashed}
\usepackage{empheq}
\usepackage{tikz}
\usepackage{pgfplots}
\usepackage{cancel}
\usepackage{subcaption}
\usepackage{comment}
\usepackage{subeqnarray}
\usepackage{xcolor}
\usepackage{longtable}
\usepackage{multirow}
\usepackage{epstopdf}
\usepackage{caption,graphicx,color}
\usepackage{amsmath,epsfig}
\usepackage{amssymb,amsfonts}
\usepackage{latexsym}
\usepackage[latin1]{inputenc}
\usepackage{float}
\usepackage{overpic}
\usepackage{slashed}
\usepackage{xcolor}
\usepackage{graphicx}
\usepackage{longtable}

\newcommand{\intd}{\mathrm{d}}
\newcommand{\ex}{\mathrm{e}}

\let\normalcolor\relax

\pgfplotsset{compat=1.16}
\tikzset{
    cross/.pic = {
    \draw[rotate = 45] (-#1,0) -- (#1,0);
    \draw[rotate = 45] (0,-#1) -- (0, #1);
    }
}
\numberwithin{equation}{section}
\let\normalcolor\relax

\relax
\renewcommand{\theequation}{\arabic{section}.\arabic{equation}}

\def\be{\begin{equation}}
	\def\ee{\end{equation}}

\newcommand{\ber}{\begin{eqnarray}}
	\newcommand{\bea}{\begin{eqnarray}}
		\newcommand{\eear}{\end{eqnarray}}
	\newcommand{\eea}{\end{eqnarray}}

\def\bsq{\begin{subequations}}
	\def\esq{\end{subequations}}
\def\hri#1#2{\href{http://arxiv.org/abs/#1}{[ArXiv:#1]#2}}
\def\hre#1#2{\href{http://arxiv.org/abs/#1/#2}{[ArXiv:#1/#2]}}

\def\hrj#1#2{\href{https://doi.org/#1}{#2}}

\DeclareMathAlphabet{\mathmybb}{U}{bbold}{m}{n}
\newcommand{\1}{\mathmybb{1}}

\newbox\pippobox

\def\II{\relax{\rm I\kern-.18em I}}

\def\l{\lambda}
\def\m{\mu}
\def\n{\nu}

\def\g{\gamma}
\def\s{\sigma}
\def\pa{\partial}

\def\sp{\;\;\;,\;\;\;}

\def\f{\varphi}

\def\a{\alpha}
\def\b{\beta}

\def\tr{\ensuremath{\mathrm{Tr}}}
\def\str{\ensuremath{\mathrm{Str}}}

\def\nn{\nonumber}
\def\OO{{\cal O}}
\def\d{\delta}
\definecolor{darkgreen}{rgb}{0.00,0.39,0.00}

\def\TT{\mathcal{T}}

\usetikzlibrary{calc,decorations.markings}
\usetikzlibrary{arrows.meta}

\title{The Tachyon Chern-Simons action with a  generic tachyon field, and baryons in V-QCD}

\author{Jean-Loup Raymond$^a$,  Matti J\"arvinen$^b$, Elias Kiritsis$^{a,c}$,  Francesco Nitti$^{a}$, Edwan Pr\'eau$^d$,
     ~\\	
      $^a$ \href{https://apc.u-paris.fr/}{Universit\'e Paris Cit\' e, CNRS, Astroparticule et Cosmologie}, F-75013 Paris, France.
	~\\

  $^b$ \href{http://www.itp.cas.cn}{Institute of Theoretical Physics, Chinese Academy of Sciences}, Beijing 100190, China
      ~\\

	$^c$ \href{http://hep.physics.uoc.gr}{Crete Center for Theoretical Physics}, Institute for Theoretical and Computational Physics,
	Department of Physics,
	University of Crete, Heraklion, Greece
	~\\
	
	$^d$ \href{https://www.uu.nl/en/research/institute-for-theoretical-physics}{Institute for Theoretical Physics}, Utrecht University, 3584 CC Utrecht, The Netherlands}

\preprint{CCTP-2026-14\\ITCP-2026/14}

\abstract{
Consistency with global flavor anomalies requires the presence of Chern-Simons terms in holographic models of QCD. Such terms  are analyzed in a setup arising in the holographic V-QCD model, where chiral symmetry breaking is implemented through the condensation of a complex scalar field, the tachyon.
Using the superconnection formalism, the Tachyon-Chern-Simons terms are constructed explicitly in the general case, where the tachyon is any complex matrix in flavor space. This general case covers, among other things, backgrounds where different quark flavors have different masses. These new results are used to analyze the structure of the baryon solutions in the presence of nonzero quark masses. Expressions for the baryon number current and the total baryon number are found, and  the baryon number is shown to be equal to the topological instanton number of the baryon solution. The  effective four-dimensional pion action is analyzed and is  shown to  reproduce the chiral Lagrangian, including the Skyrme and Wess-Zumino-Witten terms.
}


\begin{document}

\section{Introduction}

QCD has an intricate anomaly structure. Famously, the nontrivial behavior of the QCD path integral under the axial U$(1)$ transformation leads to the axial anomaly~\cite{Bell:1969ts,Adler:1969gk}. That is, the axial current defined by this U$(1)$ transformation is not conserved. Its conservation law is replaced by a renormalization-group invariant operator identity. In this identity, the nonconservation is characterized in terms of the gluonic operator $\sim  G\wedge G$ where $G$ is the gluon field tensor.
The axial anomaly also has direct phenomenological implications. In particular, it mediates the main decay channel of neutral pions, $\pi^0 \to 2 \gamma$.

However, anomalies in QCD have much more general structure than just that given by the axial U$(1)$ symmetry. Perhaps the most important generalization is given by the global flavor anomalies~\cite{Bardeen:1969md}. To study them, one considers a more general theory, where all the flavor currents of the full chiral symmetry group U$(N_f)_L\times$U$(N_f)_R$ in QCD are coupled to external gauge fields. The addition of the external fields breaks the covariance of the QCD partition function under all global chiral transformations, not just the axial transformation. The resulting non-conservation of the flavor currents is characterized by adding terms in the (non)conservation laws, which are schematically given as $\sim F\wedge F$, where $F$ stands for the field strength tensors of the various external fields.

Considering global flavor anomalies in the context of effective field theory descriptions of QCD, leads to the presence of Wess-Zumino-Witten terms. In a purely pionic four-dimensional description, the relevant term, called the (ungauged) Witten term, can be expressed as a five-dimensional integral of a form, constructed out of the pion fields, which is closed but not (globally) exact~\cite{Witten:1983tw}. Interactions between the pions and vector mesons can then be added by gauging this term~\cite{Witten:1983tw,Gomm:1984at}. The resulting action is uniquely fixed by the anomaly and discrete symmetries of QCD up to boundary terms.

In this article, we shall be interested in the implementation of the global flavor anomalies in dual gauge/gravity descriptions of QCD. It is well known how to do this in simple situations, such as chirally symmetric phases: the bulk gravity description must contain specific Chern-Simons (CS) terms, which are linked to the Witten term of effective field theory~\cite{Witten:1998qj,casero}. In string theory or the top-down context, these terms arise from flavor branes: the brane action contains both a Dirac-Born-Infeld (DBI) sector and a CS sector. As usual in the gauge/gravity dictionary, the chiral symmetry of QCD is promoted  to a gauge symmetry in the dual gravity description, and the dynamics of the gauge fields are described in terms of these DBI and CS terms. The CS sector must exactly reproduce the QCD chiral anomaly on the gravity side.

The CS terms are particularly relevant for the physics of baryons. In gauge/grav\-i\-ty duality, baryons are generically  obtained through solitonic ``instanton'' configurations of the gauge fields~\cite{Witten:1998xy,Kim:2006gp,Hata:2007mb, Lau:2016dxk}. As it turns out, the size of these soliton solutions is stabilized by the CS terms. In their absence, the solitons would tend to shrink to zero size \cite{Hata:2007mb}. Moreover, the CS terms link the topological instanton number of the solitons to the physical baryon charge \cite{Hata:2007mb,jknp}. Finally, they play an important role in the calculation of the moment of inertia that controls the masses of excited baryon states \cite{Hata:2007mb,jknp}.

 Apart from their relevance to baryon physics, the CS terms can drive inhomogeneous instabilities of QCD matter~\cite{Ooguri,Ooguri:2010xs,CruzRojas:2024igr}. While they  are typically irrelevant for homogeneous phases, CS terms do contribute, if one simultaneously turns on baryon number density and magnetic fields, \cite{son,iatrakis,Cai:2024tyv}.

While the CS terms are simple and well known in chirally symmetric configurations, in the presence of chiral symmetry breaking, issues still remain. Despite recent progress in top-down~\cite{SS2} and bottom-up~\cite{jknp} approaches, the complete form  and understanding of the CS terms in  more complex cases, has still been missing. Specific classes of models are five-dimensional dual gravity constructions where chiral symmetry breaking appears through the condensation of a complex scalar field $T^{ij}$, dual to the quark bilinear $\bar \psi^j \psi^i$ (with $i,j$ denoting the flavor indices), which is called the ``tachyon'' because it appears as an open string tachyon in string theory models of D-branes, \cite{AS}. In this setup, the chirally broken CS term is only known in the flavor-independent case, where all quark flavors have the same (zero or nonzero) masses~\cite{jknp}. However,  real QCD is known to have nonzero and strongly flavor-dependent quark masses. Since the presence of the corresponding CS term in the holographic model is required by anomalies, the inability to construct the CS term in the general chirally broken case, therefore presents a direct and concrete obstacle for the construction of a consistent holographic dual for real QCD.

In this article, we remove this obstacle by solving the long-standing problem of constructing the general, tachyon-dependent Chern-Simons term, which we call Tachyon-Chern-Simons (TCS) term. The natural framework for addressing this issue is the V-QCD model~\cite{Jarvinen}. This model is an extension of improved holographic QCD \cite{ihqcd}---a holographic model for pure Yang-Mills theory defined using five-dimensional dilaton gravity---to include a tachyonic flavor sector, which is our focus here. The tachyonic flavor action in V-QCD can be seen as arising from overlapping stacks of $D_4-\overline{D_4}$ branes, with chiral symmetry breaking induced by tachyon condensation~\cite{Bigazzi2005,casero}, following the ideas from Sen~\cite{AS,ta}. However, while solving the CS action for V-QCD is our main goal, our results are also applicable to various other models where chiral symmetry breaking appears through a similar complex scalar field.

Interestingly, a useful tool in the search for the correct TCS terms is the concept of a  superconnection. It  was  first introduced in mathematics, \cite{Quillen}, in order to describe K-theory operations.  It  was subsequently recognized that this formalism  has its natural realization in the theory of D-branes (and  anti D-branes) in string theory, \cite{Witten}. Indeed, some of the calculations of the P-odd action on branes and anti-branes verified that this is the proper formalism to use, \cite{Kennedy,KL,Taka}. Later on, it was shown that this is also the natural formalism from the QFT point of view, \cite{Sugimoto}.

In the framework of holographic QCD,  the construction of the TCS term was already set up in~\cite{casero}, where the supersymmetric D-brane derivation of the tachyonic Wess-Zumino sector~\cite{Kennedy,KL,Taka} was applied to QCD. However, the explicit construction of the relevant TCS term was recognized to be a technically challenging problem, which was only solved in a simple case where the tachyon is proportional to a unit matrix in flavor space in~\cite{casero}. Some progress was obtained in~\cite{jknp}: the TCS action was found in the case where the tachyon is proportional to an arbitrary unitary matrix. This analysis also showed that, if  instead of the D-brane analysis one uses only symmetry and anomaly constraints to restrict the form of the TCS action, only mild modifications of the result are possible.

In this article, we present the construction of the TCS term for a general form of the tachyon matrix field. We use for this, connections to the mathematical machinery of Chern character and superconnections. Specifically, we apply the Quillen formalism,~\cite{Quillen}, to write integral representations for the desired TCS terms. We then argue that a specific choice for the path of integration, which is slightly different from the paths considered in earlier literature~\cite{Quillen,Szabo:2001yd}, produces the TCS action with correct properties. We then use this representation to compute the TCS action explicitly.

Apart from calculating  the TCS term, we also check the implications that the result has for baryonic physics. We derive explicit results for the baryon current and baryon number in our formalism. We show that the baryon number defined through the holographic dictionary in general agrees with the instanton number of the baryon solution in the bulk. Note that the tachyonic flavor-independent TCS term, derived in earlier works~\cite{casero,jknp}, allows one to turn on nonzero quark masses only when all flavors have the same mass. Such flavor-independent solutions have been found earlier explicitly in simpler ``hard-wall'' models~\cite{Gorsky:2012eg,Gorsky2015}. Effects of small flavor-dependent quark masses to the solitons  have also been introduced perturbatively in a top-down approach,~\cite{Hashimoto:2009st,Bartolini:2016jxq,Bartolini:2023oxs}.

This article is organized as follows. In the rest of the introduction, we give a detailed summary of our results and discuss future directions. In section~\ref{sec:VQCD}, we review the holographic model we are using, V-QCD. In section~\ref{sec:TCSQuillen}, we present the basics of the TCS term formalism, its link to the superconnections, and explain how the TCS action can be computed using this formalism. In section~\ref{sec:explicit} we apply the formalism of section~\ref{sec:TCSQuillen} to compute the TCS action as explicitly as possible, and show that it satisfies the required properties. In section~\ref{sec:bar} we switch to baryon physics, deriving an expression for the baryon number, and comparing with the Skyrmion and instanton numbers of the solution. In section~\ref{sec:effective} we check the boundary effective action of the pions at zero and nonzero quark masses, in order to support the analysis of the baryon physics. The appendices contain additional technical details.

\subsection{Summary of results}

\smallbreak

\paragraph{General Tachyon-Chern-Simons terms}

Previous constructions were restricted to tachyon configurations where the tachyon is proportional to a unitary matrix, corresponding essentially to flavor-independent quark masses. We remove this restriction by constructing the action for an arbitrary tachyon matrix. This provides the missing anomaly-consistent topological action for holographic QCD with arbitrary quark masses. From the Wess-Zumino action of string theory, we obtain the five-dimensional integral over the bulk of the TCS forms $\Omega = \Omega_1 + \Omega_3 + \Omega_5 + \dots$,
\be
S = T_4 \int  \textbf{F} \wedge \Omega = \frac{i}{4\pi^2} \int \textbf{F}_0\wedge \Omega_5 + T_4 \int (\textbf{F}_2 \wedge \Omega_3 + \textbf{F}_4 \wedge \Omega_1) . \label{i1}
\ee
where $T_4$ is the $D_4$-brane tension and $\textbf{F} = \textbf{F}_{0,2,4}$ are appropriate bulk Ramond-Ramond forms, \cite{casero}.

We derive explicit expressions for the TCS 1, 3 and 5-forms, $\Omega_1$, $\Omega_3$ and $\Omega_5$. The derivation relies on Quillen's superconnection formalism, together with a homotopy construction, that allows to construct candidate forms that verify descent equations from the generalized Chern character $\chi$,
\be
\intd \Omega = \chi, \qquad \chi(\mathcal{F}) \equiv \str \exp(i\mathcal{F}),\label{i2b}
\ee
where $\mathcal{F}$ is the curvature of the superconnection, and 'Str' denotes the supertrace.

We concentrate on the TCS 5-form $\Omega_5$. We show that it is fixed by the following requirements:
\begin{itemize}
\item It reproduces the superconnection Chern character
\item It reduces to the standard CS form when the tachyon vanishes
\item It has the correct discrete symmetries
\item It reproduces the QCD flavor anomaly
\item It contains no infrared contribution
\end{itemize}

This form naturally separates into three contributions with distinct physical roles:
\begin{itemize}
  \item $\Omega_5^0$, a gauge-invariant bulk contribution that satisfies the descent equation,
  \item $\Omega_5^b$, a contribution that is closed and supported where the tachyon becomes non-invertible,
  \item $\Omega_5^c$, a topological bulk contribution that is closed but not exact, generalizing the Witten WZ term, \cite{Witten:1983tw}.
\end{itemize}
This decomposition arises from the superconnection formalism and separates the gauge-invariant bulk contribution, a localized boundary contribution and the purely topological contribution, making their respective physical roles manifest. The full TCS form is then
\be
\Omega_5 = \Omega_5^0 + \Omega_5^b + \Omega_5^c.\label{i2}
\ee
A similar decomposition exists for $\Omega_3$ and $\Omega_1$. We find explicit formulae for $\Omega_1$ \eqref{eq:Omega1final} and $\Omega_3^0$ \eqref{eq:Omega3final}, and for $\Omega_5^0$ \eqref{eq:Omega5final}, $\Omega_5^b$ \eqref{G4T} and $\Omega_5^c$ \eqref{452}. We obtain explicitly that these formulae reduce to the known formulae of \cite{jknp} when the tachyon is proportional to a unitary matrix.

We find that the integral over the bulk, of the boundary contribution $\Omega_5^b$, can be rewritten as an integral around the submanifold where the tachyon vanishes\footnote{This holds under the hypothesis that the tachyon matrix vanishes wherever it is not invertible. If some but not all of the eigenvalues are zero, then the expression for $\Omega_5^b$ is more complicated, see section \ref{sec:g4omc}.}, where it reduces to the gauged Witten term \cite{Witten:1983tw,Gomm:1984at}.

We also verify that the QCD flavor anomaly receives contributions only from the UV, provided that a regularity condition on the tachyon is enforced in the IR, as it is usually the case in holographic setups \cite{jknp}.

\smallbreak
\paragraph{Baryon at Finite Quark Mass}

The general construction of the TCS form allows one to study baryon configurations with arbitrary quark masses.

In the massless quark case, the baryon number was shown in \cite{jknp} to equal the winding of $\Omega_3^0$ (the gauge-invariant part of the TCS 3-form) at the UV boundary. In the massive quark case, instead, the winding of $\Omega_3^0$ in the UV is trivial.

Before we discuss the baryon number we introduce the notion of a bulk defect.
It is defined as the locus in the bulk where the tachyon becomes non-invertible. This definition explicitly excludes the asymptotic conformal boundary, where the tachyon vanishes.
In the rest of the article, defects refer to this definition.

Bulk defects are classified both by their codimension and by the number of vanishing eigenvalues of the tachyon matrix. Therefore, defects can be thought of as loci where the bulk axial gauge symmetry is restored. Generically, this restoration  is partial (i.e. not all eigenvalues vanish) or complete (i.e. the tachyon matrix vanishes).

In this paper we  show that the baryon number, in the massive quark case, is carried by \textit{bulk point-like defects}\footnote{This is similar to the 't Hooft-Polyakov monopole in three dimensions, \cite{tHooft:1974kcl}.}:
For a point-like defect, we show that the boundary baryon number is given by
\be
N_B = - \frac{1}{4\pi^2}\int_{S^3(r_*)} \Omega_3^0, \label{i3}
\ee
where $S^3(r_*)$ is a sphere of infinitesimal size around the defect, therefore generalizing the result found in simpler models~\cite{Gorsky:2012eg, Gorsky2015}.

If the tachyon is identically zero (i.e. in the chirally symmetric phase), the baryon number is equal to half the difference of the left instanton number and right instanton number. This was shown to be the case also for a tachyon proportional to a unitary matrix and for massless quarks in \cite{jknp}. Here, we extend this result to an arbitrary matrix-valued tachyon field and an arbitrary quark mass: we prove that the baryon number \eqref{i3} is equal to a generalized instanton number, i.e. the second Chern number of the superconnection,
\be
N_B =  - \frac{1}{4\pi^2} \int_{\mathcal{M}} \chi(\mathcal{F})_4 \label{i4},
\ee
where the integrand is the 4-form part of \eqref{i2b}. The obtained baryon number is therefore an integer, and it is gauge-invariant.

For point-like defects where the tachyon vanishes, due to the continuity in the massless quark limit, the baryon number should still be equal to the boundary skyrmion number for the pion matrix. We use this to show that the baryon current is equal to the boundary Skyrme current, up to an improvement term that does not affect the conservation equation. This is  shown in section \ref{sec:bar}.

\smallbreak
\paragraph{Pion Effective Action}

In order to verify that the baryon number is still given by the Skyrme number at nonzero quark mass, we derive the low-energy, four-dimensional pion effective action.\footnote{For earlier holographic analysis of the effective action, see~\cite{Sakai:2004cn,DaRold:2005mxj,Bartolini:2017sxi}.} We show this result for small but finite mass, $m_q\ll \Lambda_{QCD}$.

We recover the expected chiral Lagrangian, including the Skyrme and Wess-Zumino-Witten (WZW) term, with the correct normalization. The Skyrme term arises from the DBI lagrangian, whereas the WZW term  is obtained from the TCS action. In particular, the coefficient of the Wess-Zumino-Witten term is fixed and proportional to the number of colors $N_c$.

We obtain the baryon current from the boundary action as a function of the pion matrix, and we check that it corresponds to the Skyrme current in the small quark mass limit. This fact implies that the improvement term to the Skyrme current, that appears in the baryon current, vanishes in the small quark mass limit.

\subsection{Further directions and open problems}

At this point, it is important to assess the generality of our approach. Supersymmetric D-branes in string theory, when arranged in brane-antibrane configurations have been shown, \cite{Kennedy,KL,Taka}, to have an anomaly form that is an exponential of the Quillen supercurvature. This is mostly the form we are analyzing in this paper, but we also indicate the changes in our formalism if one chooses another function instead of the exponential one.
We do not know from first principles, what is the correct form of the anomaly 6-form on appropriate $D_4$-$\bar D_4$ pairs of 5-dimensional non-critical string theory, that may be appropriate for large-N QCD.
It has been  shown in \cite{jknp}, that when the tachyon has the special form $T=\tau U$, with $\tau$ a real field and $U$ a unitary matrix, the most general TCS term with the symmetries of QCD, but not descending from a superconnection necessarily, has four arbitrary functions of $\tau$. We do not know if this freedom is due to the special form of the tachyon.

Finally, there is the calculation of the generalized anomaly  in a four-dimensional QCD-like theory in \cite{Sugimoto} as a function of the gauge fields and space-time-dependent mass terms, that correspond to the source of the tachyon in the holographic case. Ref. \cite{Sugimoto} has shown, that the 4d anomaly, can be written in terms of a four-form involving the supercurvature.
It is not clear however, that this implies uniquely that the five-dimensional TCS must be written in terms of a superconnection, although this makes it plausible.
These issues will need a further understanding in future work.

Our results open doors for various important applications.
\begin{itemize}
\item As an immediate application, one can use our results when solving for the baryon solution  at nonzero quark masses in V-QCD. However, one can also consider more complicated configurations, such as baryon lattices, or simplified models of nuclear matter (such as the homogeneous approximation of a distribution of baryons modeled in  \cite{Rozali:2007rx,Li:2015uea,Ishii:2019gta,Bartolini:2025sag})  at nonzero quark masses. Our results also allow to consider different quark masses, which makes it possible to investigate for example hyperonic nuclear matter.

\item It has been shown in \cite{Bigazzi:2025zej} that one-flavor baryons could be modelled as extended defects in the Witten-Sakai-Sugimoto model. One could use the TCS action constructed here to study the existence and properties of these defects in the V-QCD model.

\item The TCS action is required to precisely map the extent of inhomogeneous instabilities in holographic QCD, which may be sensitive to the values of the quark masses and the precise form of the TCS action~\cite{Demircik:2024aig}.

\item In holography, the TCS action plays a fundamental role in anomalous transport \cite{Landsteiner:2011cp, Landsteiner:2012kd}: gauge and mixed gauge-gravitational anomalies  are responsible for the transport coefficients which govern the chiral magnetic and chiral vortical effects \cite{Fukushima:2008xe, Son:2009tf, Neiman:2010zi}. Our results can be the starting point  to perform a   systematic  study of anomalous transport associated with magnetic fields and anisotropies  in Improved Holographic QCD and V-QCD,  beyond the existing literature~\cite{Gursoy:2014ela,Gallegos:2018ozs,Gallegos:2024qxo}.  For the chiral vortical effects, our TCS action should be extended to include curvature terms.

\end{itemize}

\section{The V-QCD model}
\label{sec:VQCD}

We start with a quick review of the V-QCD theory~\cite{Jarvinen}. It is a holographic theory that models QCD in $3+1$ dimensions, using a bottom-up approach. The QCD-like theory has $N_c$ colours and $N_f$ flavors. This theory is considered in the so-called Veneziano limit~\cite{VL} of large $N_c$ and $N_f$, keeping the ratio $x=N_f/N_c$ constant,
\be
N_c \to +\infty\, , \quad N_f \to +\infty\, , \quad x \equiv \frac{N_f}{N_c}\, . \label{21}
\ee
In this holographic bottom-up approach, the QCD operators of UV-dimension 4 or smaller are dual to dynamical fields in the bulk, for which a phenomenological action is built based on the  principles of string theory.

\subsection{Field content and dictionary}

The V-QCD model is made of two main building blocks. The first is the modelling of the glue sector of QCD, which is described by the Improved Holographic QCD model (IHQCD), that contains gravity and a dilaton scalar field $\lambda$. The IHQCD dictionary~\cite{ihqcd} is reviewed in Table~\ref{table:ihqcd}.

\begin{table}[htb]
\centering
\begin{tabular}{ |c|c|c|}
 \hline
 \multicolumn{3}{|c|}{IHQCD} \\
 \hline
 Field & QCD operator & Source \\
 \hline
 $\lambda$ & $\tr(G_{\m \n}G^{\m\n})$ & $\lambda_{'t H}$\\
 \hline
 $g_{mn}$ & $T_{\m \n}$ & $g_{\mu \nu}$\\
 \hline
\end{tabular}
\caption{Dictionary for the glue sector of V-QCD, i.e., the IHQCD model, in the Fefferman-Graham gauge.}
\label{table:ihqcd}
\end{table}

In this dictionary, the source for the dilaton field is the 't Hooft coupling $\lambda_{'t H}$, which is linked to the Yang-Mills coupling $g$ by
\be
\lambda_{'t H} \equiv g^2 N_c\,.\label{22}
\ee
The running of the coupling is governed holographically by a potential $V_g(\lambda)$ for the dilaton field $\lambda$ \cite{ihqcd}. It is sourced by $\lambda_{'t H}$ and is dual to the QCD operator
\be
\tr(G_{\mu \nu} G^{\mu \nu})\,,\label{23}
\ee
where $G$ is the (non-abelian) gluon field strength. The bulk metric $g_{mn}$ is sourced by the boundary metric, $g_{\m\n}$, which we take in this article to be the Minkowski metric $\eta_{\m\n}$.\footnote{Our notation here is such that the Greek indices $\mu$, $\nu$, \ldots run over the space-time dimensions while the Latin indices $m$, $n$, \ldots run over all five dimensions.} It is dual to the QCD energy-momentum  tensor $T_{\mu \nu}$. An axion field can also be added to the IHQCD model, dual to the operator $\tr(G\wedge G)$, when a nontrivial $\theta$-angle is considered, \cite{ihqcd,axion}. This field is particularly important for the U$(1)_A$ anomaly and the associated effective action, which has been analyzed in detail in \cite{casero,axion}. In this article, we shall set the axion field to zero as it is not relevant for the five-form TCS term.

\begin{table}[htb]
\centering
\begin{tabular}{ |c|c|c|}
 \hline
 \multicolumn{3}{|c|}{flavor} \\
 \hline
 Field & QCD operator & Source \\
 \hline
 $T^{ij}$ & $\bar{q}^{i} q^{j}$ & $m_q^{ij}$\\
 \hline
 $(A_L)_m^{ij}$ & $\left(J_{L}\right)_{\mu}^{ij} \equiv\bar{q}^{i} (1+\gamma_5)\gamma_\m q^{j}$ & $(A_L^\mathrm{(ext)})_\mu^{ij}$\\
 \hline
  $(A_R)_m^{ij}$ & $\left(J_{R}\right)_{\mu}^{ij} \equiv \bar{q}^{i} (1-\gamma_5)\gamma_\m q^{j}$ & $(A_R^\mathrm{(ext)})_\mu^{ij}$\\
 \hline
\end{tabular}
\caption{Dictionary for the flavor sector of V-QCD in the radial gauge $(A_{L/R})_r=0$}
\label{table:flavor}
\end{table}

We consider solutions of the bulk field equations, in which the bulk metric is of the following form,
\be \label{23-i}
\intd s^2 = e^{2A(r)}\left[\intd r^2 + \eta_{\mu\nu} \intd x^\mu \intd x^\nu\right]\;.
\ee	
This is the most general ansatz with Poincar\'e symmetry in 4 dimensions,
These solutions contain an asymptotically AdS boundary at $r=0$, such that
\be \label{23-ii}
e^A(r) \to \frac{\ell}{r} \;\;\;{\rm as}\;\;\; r\to 0.
\ee

The second building block models the flavor sector and chiral symmetry breaking, from the dynamics of backreacted flavor branes. The corresponding  dictionary is reviewed in Table~\ref{table:flavor}.
In the flavor sector, the chiral symmetry of QCD
\be
\text{U}(N_f)_L\times \text{U}(N_f)_R\, ,\label{24}
\ee
is reproduced in the bulk, by a gauge theory with the same gauge-group structure \eqref{24}. This is regarded as originating from the low-energy gauge theory on a stack of $N_f$ $D_4-\overline{D_4}$ branes in a (five-dimensional, non-critical) string theory, \cite{dissecting}.

The chiral currents of QCD, $J^{\mu}_L$ and $J^{\mu}_R$, are dual to $\text{U}(N_f)_L$ and $\text{U}(N_f)_R$ gauge fields, $A^{m}_L$ and $A^{m}_R$. Such gauge fields are sourced by external boundary gauge field sources  $(A_L^\mathrm{(ext)})$ and $(A_R^\mathrm{(ext)})$, which account for chemical potentials and external field strengths. The conventions for these fields are reported in appendix \ref{sec:convention}.

The quark bilinear operators of QCD are dual to a complex tachyon matrix field $T^{ij}$. It is a complex $N_f \times N_f$ matrix field transforming in a bi-fundamental representation of the gauge group \eqref{24}. A nonzero tachyon in the bulk, implements chiral symmetry breaking, characterized by a nonzero expectation value for the chiral condensate, $\bar q^iq^j$. This symmetry can be spontaneously broken in the absence of quark mass, or explicitly broken if this tachyonic field is sourced by a nonzero quark mass matrix $m_q^{ij}$. While in principle this is a generic complex matrix, a simpler case can be considered, where all quarks have the same mass:
\be
T = \tau U,\label{25}
\ee
where $\tau$ is a scalar function behaving near the AdS boundary $r=0$ as\footnote{In the full model, one must consider more complicated asymptotics with log corrections \cite{Jarvinen}, which we shall consider later.}
\be \label{25-i}
\tau \sim m_q r  + \sigma_q r^3 + \ldots
\ee
Here, $m_q$ is the quark mass and $\sigma_q$ is the magnitude of the chiral condensate. The unitary matrix  $U$,  encodes the moduli of the vacuum expectation value of $\bar{q}^{i} q^{j}$.  Many results presented in this article simplify significantly and become easier to interpret if (\ref{25}) holds. This simplification cannot be applied if one considers quarks of different masses.

\subsection{The action}

The action for the bulk fields is inspired by string theory, but is modified phenomenologically to accommodate  QCD features. For the color sector, the IHQCD action is given by an Einstein-dilaton theory
\be
S_g = M^3 N_c^2 \int \intd^5 x \sqrt{-g} \bigg( R - \frac{4}{3} \frac{(\pa \lambda)^2}{\lambda^2} + V_g(\lambda) \bigg)\,, \label{26}
\ee
where the potential for the glue sector, $V_g(\lambda)$, has asymptotics that are fixed from phenomenological constraints\footnote{Near its maximum, it asymptotes exponentially to a constant, while in the deep IR it is given, to leading order, by the dilaton potential of non-critical string theory in five dimensions, \cite{ihqcd,dissecting}. This IR behavior has a crucial logarithmic correction, as in (\ref{27}), that is responsible for the nontrivial IR asymptotics of YM.}, and $M$ is a constant scale. In the ultraviolet (UV), i.e. near the holographic boundary at $r=0$, this potential approaches a positive constant, while in the infrared (IR), i.e. deep in the bulk, it asymptotes to
\be
V_g \sim V_{g, \mathrm{IR}} \lambda^{4/3} \sqrt{\log(\lambda)}, \quad \lambda \to +\infty\,,\label{27}
\ee
where $V_{g, \mathrm{IR}}$ is a constant. This particular scaling is a critical behavior~\cite{ihqcd} that reproduces many features of the QCD glue sector. In particular, it reproduces Yang-Mills confinement: the Wilson loops can be computed holographically and scale like the area of the loop. Many possible choices give confinement, and in all of them,  the spectrum of glueball excitations~\cite{ihqcd,gkmn} is discrete.
However, the choice in (\ref{27}) is the only one that gives  glueball masses squared, $m_n^2$, that scale linearly asymptotically at large $n$,
\be
m_n^2 \sim  n\label{28} \, .
\ee
 A Gibbons-Hawking UV boundary term is also added,
\be
S_\mathrm{GH} = 2 M^3 N_c^2 \int_\mathrm{UV} \intd^4x \sqrt{- h} K\,, \label{29}
\ee
with $K$ the trace of the extrinsic curvature, and $h$ the pulled-back metric from $g$ to the boundary.

For the flavor sector, the action is again inspired by string theory, more specifically by deforming phenomenologically the action for the stack of $D_4-\overline{D_4}$ space-filling branes, which will be associated to left and right gauge fields respectively~\cite{Bigazzi2005,casero}. This construction leads to the presence of a tachyon field coupled to the gauge fields. The full action in this case is a variant of the Dirac-Born-Infeld action used by Sen~\cite{AS,ta} for $D-\bar D$ systems in string theory.

It is known that in the non-abelian case, this action fails to reconstruct commutators between field strengths that arise in the string theory at sixth order  in derivatives \cite{Tseytlin:1997csa, Sevrin:2001ha}. For our purpose, we shall take an expansion of the DBI action to an order that this action captures correctly. In V-QCD, one uses a phenomenological variant of this action~\cite{Jarvinen,spectrum}
\be
S_\mathrm{DBI} = - \frac{M^3N_c}{2} \int \intd^5x\ \text{Symtr}\bigg[V_f\big(T^\dag T, \lambda) \bigg( \sqrt{- \det(\mathbf{A}_{L, mn})} +\sqrt{- \det(\mathbf{A}_{R, mn})}  \bigg) \bigg]\,,\label{210}
\ee
where the tachyon potential $V_f$ controls the flavor physics, and is taken to also depend on the dilaton field. The ``Symtr'' symbol stands for the symmetrized trace
\be
\text{Symtr}(M_1 M_2 \dots M_n) \equiv \frac{1}{n!}\sum_{\sigma \in \mathfrak{S}_n}\tr(M_{\sigma(1)} M_{\sigma(2)} \dots M_{\sigma(n)})\, , \label{211}
\ee
where $\mathfrak{S}_n$ is the group of permutations.
The fields $\mathbf{A}_{L/R}$ are defined as
\be
\mathbf{A}_{L, mn} \equiv g_{mn} + w(\l) F_{L, mn}  + \frac{\kappa(\l)}{2}\left[(D_mT)^{\dagger}D_nT + (D_nT)^{\dagger}D_mT\right] \, ,\label{212}
\ee
\be
\mathbf{A}_{R, mn} \equiv g_{mn} + w(\l)F_{R, mn}+ \frac{\kappa(\l)}{2}\left[D_mT(D_nT)^{\dagger} + D_nT(D_mT)^{\dagger}\right] \,,\label{213}
\ee
where $\kappa$ and $w$ are functions of $\l$  that control the large mass asymptotics of mesons and thermodynamics properties, and are matched to lattice QCD data.

Recall that $F_{L/R, mn}$ and $D_n T$ are also matrices in flavor space, so that it is understood that the factors $g_{mn}$ in~\eqref{212} and in~\eqref{213} contain a unit matrix in flavor space. Our convention for the field strengths and covariant derivatives is given in appendix \ref{sec:convention}. They are defined as
\be
F_L = \intd A_L - i A_L \wedge A_L\,, \quad F_R = \intd A_R - i A_R \wedge A_R\,,\label{214}
\ee
\be
D T = \intd T + i T A_L - i A_R T\,, \quad DT^\dag = \intd T^\dag - i A_L T^\dag  + i T^\dag A_R\,,\label{215}
\ee
where we already introduced the form notation (such that e.g. $A_L = A_{L, m}\intd x^m$ and $F_L = \frac 12 F_{L, mn}\,\intd x^m\wedge \intd x^n$), which will be useful when writing down the TCS terms below, and $\intd$ is the exterior derivative. The corresponding gauge transformation properties of these fields, parametrized by a group element $(V_L, V_R)$ of \eqref{24} are given as
\be
A_L \to V_L A_L V_L^\dag + i V_L \intd V_L^\dag, \quad A_R \to V_R A_R V_R^\dag + i V_R \intd V_R^\dag,\label{216}
\ee
\be
T \to V_R T V_L^\dag, \quad T^\dag \to V_L T^\dag V_R^\dag.\label{217}
\ee
The UV and IR asymptotics of the V-QCD potentials $V_g$, $V_f$, $\kappa$ and $w$, and of the dynamical fields, are summarized in~\cite{spectrum,JR}, but will not be relevant in the present paper.

In addition to the dynamical action \eqref{210}, the flavor action contains a parity-odd piece. The parity-odd piece contains, among other terms\footnote{There is also a one-form contribution, proportional to the U$(1)_A$ flavor gauge field, that upon dualization mixes with the QCD axion, dual to $G \wedge G$, \cite{casero}, reflecting the U$(1)_A$ anomaly.}, a five-form TCS term. One of the main purposes of this article is to determine this action for a generic tachyon field. This part of the action has been shown to be crucial to baryon physics and anomalies \cite{casero,jknp}.

\section{The Tachyon-Chern-Simons form and Quillen's superconnection formalism}
\label{sec:TCSQuillen}

In this section, we explain the method we use to derive the TCS action in the presence of a generic tachyon field. We first re-express the field content of the flavor sector as a superconnection. Superconnections were first introduced in mathematics, \cite{Quillen}, in order to describe K-theory operations. This formalism was subsequently
recognized to have its natural realization in the theory of D-branes (and their anti D-branes) in string theory, \cite{Witten}. Indeed, some of the calculations of the P-odd action on branes and anti-branes verified that this is the proper formalism to use, \cite{Kennedy,KL,Taka}. Later on, it was shown that this is also the natural formalism from the QFT point of view, \cite{Sugimoto}.

For all these reasons, we first remind the reader below some facts about superconnections, \cite{Quillen}.  We then derive different criteria that the topological action must respect. In particular, we require that this action reproduces the anomaly structure of QCD, both the consistent anomaly, defined as the gauge variation of the topological action, and the covariant anomaly, its gauge invariant counterpart.

Lastly, we discuss a homotopy method that is used to determine the action satisfying these criteria. The proof that the action constructed below satisfies these criteria, is postponed to the next section, where the action is computed explicitly.

\subsection{Quillen's superconnections and the Chern character}

We first return to the field content of the flavor sector, the left and right gauge fields and the tachyon field. These fields can be studied using the formalism of Quillen's superconnections \cite{Quillen}. In this subsection, we review some facts about Quillen superconnections and explain their relevance to the V-QCD model. Mathematical details of their construction and main properties of superconnections are presented in appendix \ref{sec:convention2}.

Superconnections are known to appear naturally in string theory effective descriptions, in the context of tachyon condensation of pairs of $D_p-\overline{D}_p$ branes~\cite{Kennedy,KL,Taka}. In V-QCD, the appearance of the superconnection structure can be viewed as inherited from the underlying string-theory construction~\cite{casero,Jarvinen, dissecting}. The two gauge fields are connection forms coming from $D_4$ and $\overline{D_4}$ branes respectively, while the tachyon is a bi-fundamental scalar (zero-form).

For this discussion, we need to introduce and distinguish different types of degrees. First, we define a $\mathbb{Z}_2$-valued degree as follows. We call fields that transform only under either $\text{U}(N_f)_L$ or $\text{U}(N_f)_R$ ``even'' and fields that transform in a nontrivial mixed representation of $\text{U}(N_f)_L \times \text{U}(N_f)_R$ ``odd'' with respect to the $\mathbb{Z}_2$-grading. A more precise definition of this degree is presented in appendix \ref{sec:convention2}. Using the gauge transformation properties \eqref{216}, \eqref{217}, one can deduce that gauge fields are even, while the tachyon and its hermitian conjugate are odd. In addition to this degree, one can consider the differential form degree of the fields. Lastly, we shall also consider the sum of these two degrees mod $2$. There are, then, three different degrees:
\begin{itemize}
  \item The $\mathbb{Z}_2$ degree defined above, $0$ for even fields (transforming in left or right gauge group representations) and $1$ for odd fields (transforming in mixed representations).

  \item The differential form degree

  \item The total degree ``sdeg'' defined as

\be
\text{sdeg} \equiv \text{form degree} +  \mathbb{Z}_2\text{ degree } [\text{mod }2] \label{31}
\ee
\end{itemize}

The $\mathbb{Z}_2$-degree structure is the one that naturally appears in a supersymmetry context, though in our case there  is no  supersymmetry in the usual sense. We make use of the terminology ``super'' to refer to mathematical objects admitting such a $\mathbb{Z}_2$-grading. In supersymmetry language, the gauge fields are treated as bosons and the tachyon is treated as a fermion.

The field content of the flavor sector can be expressed in a supermatrix,
\be
i \mathcal{A} = i(A + \mathcal{T}) \equiv \begin{pmatrix}
  i A_L & T^\dag \\
  T & i A_R
\end{pmatrix}\,, \quad A\equiv  \begin{pmatrix}
  A_L & 0 \\
  0 & A_R
\end{pmatrix}\,, \quad \mathcal{T} \equiv  \begin{pmatrix}
  0& -iT^\dag \\
  -iT & 0
\end{pmatrix}\,.\label{32}
\ee
which has a graded product law presented in appendix \ref{sec:convention2}. Supermatrices also inherit a notion of degree, from the fields. It is required that all fields in a supermatrix have the same sdeg, which is the degree of the supermatrix. We then introduce the ``supertrace" (Str) as
\be
\str \left(\begin{pmatrix}
 A & B \\
  C &  D
\end{pmatrix}\right) \equiv \tr(A) - \tr(D)\,, \label{33}
\ee
which respects a graded cyclicity property for supermatrices $M$ and $N$,
\be
\str(M N) = (-1)^{\text{sdeg}(M) \text{sdeg}(N)} \str(NM).\label{34}
\ee
Next, we introduce superconnections, which can be written as
\be
D = \intd - i \mathcal{A},\label{35}
\ee
with $\mathcal{A}$ a supermatrix of odd total degree, that satisfies a graded derivation property, defined in \eqref{eq:gradedderiv} in appendix \ref{sec:convention2}. The superconnection \eqref{35} has curvature
\be
i\mathcal{F} \equiv (i D)^2 =  i (\intd  \mathcal{A} - i \mathcal{A}^2) =
\begin{pmatrix}
&iF_L - T^\dagger T & DT^\dagger\\
&DT & iF_R - T T^\dagger
\end{pmatrix},\label{36}
\ee
where we used \eqref{215} and supermatrix graded product rules. The different degrees of the fields and product of fields appearing in \eqref{32} and \eqref{36}, are summarized in Table~\ref{table:degrees}.
We deduce from Table~\ref{table:degrees} and \eqref{32}--\eqref{36} that $\mathcal{A}$ has $\text{sdeg}=1$, as $A_L$, $A_R$ and $T$ all have sdeg$=1$ and $\mathcal{F}$ has $\text{sdeg}= 0$, as all fields in \eqref{36} have sdeg$=0$.

\begin{table}[htb]
\centering
\begin{tabular}{ |c|c|c|c|}
 \hline
 Fields & Form degree & $\mathbb{Z}_2$ degree & Total degree (sdeg) \\
 \hline
 $T, T^\dag$ & $0$ & $1$ & $1$\\
 \hline
 $A_L, A_R$ & $1$ & $0$ & $1$\\
 \hline
 $T^\dag T, T T^\dag$ & $0$ & $0$ & $0$\\
 \hline
 $DT, DT^\dag$ & $1$ & $1$ & $0$\\
 \hline
 $F_L, F_R$ & $2$ & $0$ & $0$\\
 \hline
\end{tabular}
\caption{Degrees of fields appearing in $\mathcal{A}$ and $\mathcal{F}$.}
\label{table:degrees}
\end{table}

We can now state the general form of the topological part in the flavor action. Taking once again input from string theory, we start from the Wess-Zumino action which has been derived from a flat space boundary string field theory approach~\cite{Kennedy,KL,Taka,casero}
\be
S_\mathrm{WZ} =  T_4 \int  C \wedge \str \exp( i \mathcal{F}), \label{37}
\ee
where $T_4$ is the $D_4$ brane tension and we have absorbed a factor of $2\pi \alpha'=1$ into $\mathcal{F}$. In \eqref{37}, $C$ is the formal sum of all Ramond-Ramond forms $C_n$ present in the theory, which can be written
\be
C \equiv \sum_{n} (-i)^{\frac{5-n}{2}}C_n.\label{38}
\ee
The object $\str \exp(i \mathcal{F})$ in \eqref{37} is the generalized Chern character of a superconnection
\be
\chi(\mathcal{F}) \equiv \str(\exp(i \mathcal{F})).  \label{39}
\ee
In \eqref{39}, $\chi$ is also a formal sum of differential forms of all even\footnote{The odd degree terms in $\chi$ vanish by using properties of the supertrace and \eqref{34}.} degrees. This character encodes important topological information, that does not depend on the choice of superconnection \cite{Quillen}. For every superconnection $D$,
\be
\intd \str(D^{2n}) = \str([D,D^{2n}]) = 0,\label{310}
\ee
where in the first equality we used the supercommutator definition
\be
[M, N] \equiv M N - (-1)^{\text{sdeg}(M) \text{sdeg}(N)} NM\, ,\label{310b}
\ee
and that the supertrace vanishes on supercommutators, which can be shown by taking a supertrace of \eqref{310b} and applying \eqref{34}. In the second equality of \eqref{310} we used that $D$ supercommutes with $D^2$. One can sum over $n$ in \eqref{310} to obtain that \eqref{39} is closed
\be
\intd \chi(\mathcal{F}) =  0. \label{311}
\ee
More details can be found in appendix \ref{sec:convention2}.

Equation \eqref{311} implies that locally there exists an $\Omega$ such that
\be
\chi(\mathcal{F}) \equiv \intd \Omega.\label{312}
\ee
The form $\Omega$ in \eqref{312} is the TCS form. $\chi(\mathcal{F})$ contains only even forms, so $\Omega$ can be taken to be a formal sum of differential forms of odd form degree, which we denote $\Omega_{2p+1}$. Note that equation \eqref{312} does not entirely fix $\Omega$. Rather, it is fixed up to a closed form.

We define the Ramond-Ramond field strength $\textbf{F}$ obeying
\be
\textbf{F} \equiv \intd C. \label{313}
\ee
We can integrate by parts \eqref{37} to re-express the action in terms of $\Omega$ as
\be
S_\mathrm{WZ} = T_4 \int  \textbf{F} \wedge \Omega, \label{37b}
\ee
up to neglected boundary terms. $\textbf{F}$ contains, among other terms, a zero-form term $\textbf{F}_0$ proportional to $N_c$, that is responsible for a 5d TCS action. In $4+1$ dimensions, after integrating the $\textbf{F}_0$ term in the Wess-Zumino action we obtain the TCS action at a level $N_c$,
\be
S_{TCS} =  \frac{i N_c}{4\pi^2} \int_{\mathcal{M}}  \Omega_5,\label{314}
\ee
where we denoted by $\mathcal{M}$ the five dimensional manifold (with an asymptotically AdS boundary $\pa \mathcal{M}$) on which our bulk holographic theory lives and on which the TCS action is considered. In the rest of this article, the bulk is denoted $\mathcal{M}$.

The other pieces of the Wess-Zumino action govern different physics. The $\textbf{F}_2$ term is dual to $\textbf{F}_3 = \intd C_2$ which couples magnetically to $D_0$-branes, that are baryon vertices in five dimensions. $\textbf{F}_2$ couples to $\Omega_3$, which is linked to the instanton number through \eqref{312}, consistently with the identification of $D_0$-brane charge and instanton number. The $\textbf{F}_4$ term, which is a dynamical field dual to the axion field strength $\textbf{F}_1$, couples to the one form $\Omega_1$ and is responsible for the U$(1)_A$ anomaly and $\theta$-angle physics \cite{casero, dissecting, axion}.

Importantly, the $\Omega$ form entering \eqref{37b} depends nontrivially on the tachyon, and the explicit form of this action is only known in simpler cases \cite{jknp} like \eqref{25} where $TT^\dag = T^\dag T = \tau^2 I$. The explicit determination of $\Omega$ is a long-standing problem~\cite{casero,jknp} and is one of the main objectives of this article. It is presented in the next sections.

\subsection{Properties of the Tachyon-Chern-Simons action}
\label{sec:criteria}
We now present the generic properties that must be obeyed by the TCS form $\Omega$, as defined in \eqref{312}. We then construct a general formula to compute it, using the superconnection formalism.

We first recall some properties of the standard CS action $\omega$ in the absence of a tachyon field. The results and properties discussed in this section apply to any form degree, but we  show explicit formulae for the 5-form component only as it is the one appearing in \eqref{314}. An explicit formula for $\omega$ valid for all form degrees will be provided later, in \eqref{336}. The standard non-abelian CS action in five dimensions reads
\be
S_{CS} \equiv \frac{i N_c}{4\pi^2} \int \omega =  \frac{N_c}{24\pi^2} \int \intd^5x \left[\tr\bigg(A_L F_L^2 + \frac{i}{2}A_L^3 F_L - \frac{1}{10}A_L^5 \bigg) - (L\leftrightarrow R)\right].\label{315}
\ee
Our first requirement for the TCS form is that it reduces (up to boundary terms) to the CS form $\omega$ when the tachyon field is zero, i.e. when the chiral symmetry of QCD is preserved. We also require that the TCS form $\Omega$, has the same transformation properties as $\omega$ under parity and charge conjugation, reflecting the covariance of QCD under these transformations.

We recall the transformation properties of $\omega$ under the discrete symmetries. Under parity, it transforms as
\be
P_1(\omega) = (-1) \omega, \qquad P(S_{CS}) =(+1) S_{CS},\label{317}
\ee
That is, $S_{CS}$ is $P$-even and $\omega$ is $P_1$-odd, where $P_1$ denotes the action of the parity on the flavor fields (see Appendix~\ref{sec:discsym})  and $P$ is the spatial parity operation $x\to -x$. Under charge conjugation, the action is instead
\be
C(S_{CS}) =(+1) S_{CS}, \qquad C(\omega) = (+1)\omega,\label{318}
\ee
so both $S_{CS}$ and $\omega$ are $C$-even.

In addition, the bulk TCS action has been found in many different holographic QCD models to be related to topological features of the boundary theory. Indeed, $\Omega_5$ is crucial to model holographically the flavor anomalies of QCD~\cite{casero}. In the standard CS context, in order to compute the consistent anomaly, one computes the gauge transformation of $\omega_5$ and requires that it agree with the QCD flavor anomalies, see for instance \cite{jknp},
\be
\delta_\Lambda \omega_5 \equiv \intd I_4^1 = - \frac{1}{6} \intd \tr\bigg[\Lambda_L ((\intd A_L)^2 - \frac{i}{2}\intd(A_L^3)) - (L\leftrightarrow R) \bigg].\label{319}
\ee
Equation \eqref{319} defines the {\em consistent} anomaly 4-form $I_4^1$, where the superscript 1 indicates that it is linear in $\Lambda$. \eqref{319}  guarantees that the consistent anomaly of QCD is reproduced holographically.

Another criterion is that the TCS action does not give rise to contributions from the IR of the theory to observables and in particular to anomalies. We  enforce that the TCS form $\Omega_5$ appearing in \eqref{314} also reproduces the same transformation as $\omega_5$ at the holographic boundary, i.e.
\be
\delta_\Lambda \Omega_5 = \intd I_4^1 = - \frac{1}{6} \intd \tr\bigg[\Lambda_L ((\intd A_L)^2 - \frac{i}{2}\intd(A_L^3)) - (L\leftrightarrow R) \bigg].\label{320}
\ee
It is known that in QCD, adding a counterterm affects the consistent anomaly \cite{Bilal:2008qx}. Equation \eqref{320} is therefore expected to fix the ambiguity in the definition of $\Omega$ in \eqref{312}. We  verify this claim in the next section. In appendix \ref{sec:F}, we show more details about the consistent anomaly and its connection to the covariant anomaly. We verify in particular that the covariant anomaly is well reproduced by $\Omega_5$, and that it is independent of such boundary terms.

The TCS form must not contain an IR boundary term, as these can contribute to the anomaly, which would require fixing IR boundary conditions in order to match QCD anomalies. In the IR, the tachyon field is known to diverge to infinity. One way to guarantee that this criterion is satisfied is then to require that $\Omega$ is suppressed in the large tachyon limit. We show that it is possible to achieve this by choosing\footnote{In string theory, there is no reason to choose anything for this to happen. This is guaranteed by the fact that as the tachyon rolls to its true vacuum, all couplings on the D-brane vanish exponentially, \cite{Kennedy,KL,Taka}. This has been argued by Sen, \cite{AS,ta}, to guarantee that the $D-\bar D$ pair annihilates. It has also been argued that this behavior is akin to confinement of the world-volume gauge fields, \cite{Yi}.}
    an action such that the UV contribution to the anomaly is given in \eqref{320} and the IR contribution to the anomaly is a function of the unitary part of the tachyon $U$ only, which will be set to zero by our regularity conditions\footnote{This IR regularity condition is here equivalent to the fact that space-time derivatives $\pa U$ along the boundary directions vanish in the infrared, such that they do not contribute to the effective action or to the anomaly.}.

To summarize, the criteria that we impose to build the TCS form are:
\begin{enumerate}
  \item \label{cr:character} $\Omega$ must descend from the Chern character,
  \be
  \intd \Omega = \chi(\mathcal{F})\label{321}
  \ee
  \item \label{cr:tachzero} $\Omega$ must reduce to the standard CS action in the absence of tachyons up to boundary terms
  \item $\Omega$ must be $P_1$-odd and $C$-even \label{cr:discretesym}
  \item \label{cr:anomaly} $\Omega$ must reproduce the QCD anomaly on the UV boundary
  \item \label{cr:noIR}$\Omega$ must not contain a boundary term in the infrared
\end{enumerate}

Our strategy is the following. We start by determining a family of TCS forms that satisfy criterion \ref{cr:character} using a homotopy formula adapted to the superconnection context. Then, we show that using the superconnection formalism, the $P_1$-odd part of criterion \ref{cr:discretesym} is automatically satisfied for this family. We then choose a particular representative such that criteria \ref{cr:anomaly} and \ref{cr:noIR} hold. These criteria entirely fix $\Omega$ (up to a gauge choice). We then verify that it is $C$-even, and that criterion \ref{cr:tachzero} holds.

\subsection{A homotopy formula}
In this section, we introduce the tools needed to compute the TCS form. This is a homotopy formula, initially obtained in \cite{Quillen}, as a byproduct of the proof that the Chern classes are independent of the choice of superconnection. We shall first sketch the proof of this fact:

\begin{itemize}
  \item Consider two generic superconnections $D_0$ and $D_1$ on $\mathcal{M}$ (defined below (\ref{314})), and define an arbitrary path $D_s$, parametrized by the parameter  $s\in [0,1]$ between them.

  \item The connection $D_s$ can be extended  to a new superconnection $\bar{D} = \pa_s \intd s + D_s$ on $[0,1] \times \mathcal{M}$, which coincides with $D_s$ at fixed $s$.

  \item $\str(\bar{D}^{2n})$ belongs to some de Rham cohomology class in $[0,1] \times \mathcal{M}$, that restricts to the class of $\str(D_s^{2n})$ for fixed $s$, so the {de Rham} class of $D_s^{2n}$ is {independent} of $s$ (see \cite{Quillen}).

\end{itemize}

Using these objects, one can compute $\intd \str(\bar{D}^{2n})$ in two different ways. On the one hand, we compute the derivative and simplify
\be
\intd \str(\bar{D}^{2n}) = \intd_{\mathcal{M}} \str(D_s^{2n}) + \intd s (\pa_s \str(D_s^{2n}) -  n \intd \str(D_s^{2n-2} \dot{D_s})),\label{322}
\ee
where we defined the exterior derivative restricted to the manifold\footnote{As opposed to the total exterior derivative $\intd$ on $[0,1] \times \mathcal{M}$, which contains also a derivative with respect to $s$.} $\mathcal{M}$ as $\intd_{\mathcal{M}}$, and we indicated with a $\dot{D_s}$ the derivative of $D_s$ with respect to $s$. On the other hand, from \eqref{311} $\str(\bar{D}^{2n})$ is closed, and from the $\intd s$ piece in \eqref{322} we obtain
\be
\pa_s \str(D_s^{2n}) =  n \intd \str(D_s^{2n-2} \dot{D_s}).\label{323}
\ee

Next, note that the Chern character, \eqref{39}, is
\be
\chi(\mathcal{F}_s) = \sum_{n=0}^{\infty} \frac{1}{n!}\str((- D_s^{2})^n)\label{324},
\ee
where $\mathcal{F}_s \equiv i D_s^2$.
Therefore, {using \eqref{323}}, we obtain
\be
\pa_s (\chi(\mathcal{F}_s)) =  \intd \str\left(\sum_{n=1}^{\infty} \frac{1}{(n-1)!}((-1)^n D_s^{2n-2}) \dot{D_s}\right),\label{325}
\ee
which can be simplified as
\be
\pa_s (\chi(\mathcal{F}_s)) = - \intd \str(  \exp(i\mathcal{F}_s) \dot{D_s}).\label{326}
\ee
Equation \eqref{326} is known as a transgression formula. As discussed in~\cite{jknp}, in the context of bottom-up holography, it may be useful to consider more general Chern characters than the exponential expression \eqref{39}. Recall that the string theory result~\eqref{37} was based on a flat space computation\cite{Kennedy,KL,Taka,casero}, so there seems to be no strict requirement for the exponential form of the character in general. If we consider a generalized character given by a generic function $f$ of the curvature, written formally as a series expansion
\be
f(\mathcal{F}) = \sum_{n=0}^{+\infty} f_n (\mathcal{F})^n,\label{327}
\ee
then using \eqref{323}, we obtain
\be
\pa_s (f(\mathcal{F}_s)) = - \intd \str(  f'(\mathcal{F}_s) \dot{D_s}).\label{328}
\ee
In this work, we focus on the case where $f = \exp$. But because the key formula~\eqref{328} holds for any function $f$, much of the analysis carried out in the rest of this article will generalize. See also appendix~\ref{sec:residue} for further comments.

Returning to \eqref{326} and integrating it in $s \in [0,1]$, we obtain
\be
\chi(\mathcal{F}_1) - \chi(\mathcal{F}_0)   = - \intd \int_{0}^{1} \intd s \; \str(  \exp(i \mathcal{F}_s) \dot{D_s}).\label{329}
\ee
For a given path in superconnection space, (\ref{329}) connects the difference of the Chern character, on the two end-points of the path, to the exterior derivative of the integral in the right hand side. As we shall see, choosing the endpoints of the path adequately, \eqref{329} gives an explicit formula to compute $\Omega$ in \eqref{321}.

The left-hand side is path-independent, while the integral on the right hand side does depend on the path. Since the left-hand side of \eqref{329} is path-independent, the path dependence of the integral on the right-hand side can only appear through closed forms, so that it vanishes upon the action of the exterior derivative. In appendix \ref{sec:F} we show that the difference between paths gives an exact form. The dependence on the path in \eqref{329} therefore corresponds to the ambiguity in the definition of the TCS form \eqref{312}. Note that the path dependence does not affect the dynamics of the bulk fields, but instead it is tied to boundary terms and cohomological data.

\subsection{Choice of path and a generic formula}

We shall consider a class of paths, adapted to our objective, defined as follows.
We consider a two-parameter family of superconnections $D_{(a,b)}$, obtained by independently scaling the gauge fields $A_L, A_R$ in~\eqref{35} by a constant $a$, and the tachyon $T$ by a constant $b$, such that
\begin{equation}
\label{Dab} D_{(a,b)} \equiv \intd -i a A - i b \TT \, .
\end{equation}
This defines a plane $(a,b)$ of superconnections. The most general path, described by $s\in[0,1]$, has a superconnection $D_s$ parametrized by two functions $a(s), b(s)$, as
\be
D_s \equiv D_{(a(s), b(s))} \equiv \intd - i \mathcal{A}_s \equiv \intd - i a(s) A - i b(s) \mathcal{T}.\label{331}
\ee
The associated curvature is given by
\begin{align}
i\mathcal{F}(a(s),b(s)) &\equiv \big(i D_s \big)^2 \\
&=  \begin{pmatrix}
ia \intd A_L + a^2 A_L^2  - b^2 T^\dagger T & & & b \intd T^\dagger - i ab A_L T^\dag  + i ab T^\dag A_R\\
b \intd T + i ab T A_L  - i ab A_R T & & & ia \intd A_R + a^2 A_R^2  - b^2 T T^\dag
\end{pmatrix}.
\label{332}
\end{align}

To gain intuition about the role of the tachyon field, we consider briefly the simpler case where $T= \tau U$. We observe  that the $b^2 TT^\dag$ and $b^2 T^\dag T$ terms in \eqref{332} become $b^2 \tau^2  \mathbb{I}_{N_f}$, where $ \mathbb{I}_{N_f}$ is the identity in flavor space, which commutes with everything. As a consequence, there will be an overall $\exp(-b^2 \tau^2)$ factor in the Chern character $\chi$, defined in \eqref{39}. As $b$ increases, for all points of $\mathcal{M}$ where $\tau \neq 0$, $\chi$ vanishes like a Gaussian. At points where $\tau=0$, this exponential factor remains constant equal to $1$. Therefore, as $b$ increases, the support of the Chern character localizes around the points where $\tau = 0$. This motivates us to introduce the locus where $T$ is non-invertible,
\be
\mathcal{Z} \equiv \left\{ x \in \mathcal{M}| \det(T(x))= 0 \right\}.\label{333}
\ee
From now on, $\mathcal{Z}$ will play a central geometric role. In the $\tau U$ case, it is simply the set of points of the manifold where $\tau=0$, and the tachyon vanishes. It is also the set where $U$ is ill-defined and where the phase of the tachyon can wind and carry nontrivial topology.

We now move to the interpretation of the connected components of $\mathcal{Z}$. It includes:
\begin{itemize}
  \item the UV boundary, which was shown in \cite{jknp} to support the topology of the baryon solution in the massless quark case.
  \item bulk defects, which were defined above equation \eqref{i3} to be loci in the bulk where the tachyon matrix becomes non-invertible.
\end{itemize}
Defects of different dimensions can exist in the bulk, with different possible world-volume topologies. We shall return to this discussion when studying the baryon solution with massive quarks in section \ref{sec:bar}.

If there are distinct quark masses, then the $T=\tau U$ decomposition does not apply anymore, and the generic TCS action must be used. The reasoning above equation \eqref{333} still applies, and the Chern character is nonzero only on the submanifold where the tachyon is non-invertible in the $b\to\infty$ limit. We shall denote by  $\sigma_a$ the singular values of the tachyon, see appendix \ref{sec:convention3}\footnote{As the tachyon is a square $N_f\times N_f$ matrix, these are also the common left or right eigenvalues of the tachyon matrix.}. Then, from \eqref{333}, on $\mathcal{Z}$ at least one of the $\sigma_a$ is zero. The physical picture is also slightly different in this case. If not all the singular values vanish on a given defect, this  means that only a subgroup of the chiral symmetry is restored. Depending on how many singular values vanish, the physical interpretation of the solution may change. An analogy can be made between this last point and the case of 't Hooft monopoles \cite{tHooft:1974kcl}, as the breaking of $SU(N)$ to a smaller gauge group occurs. When $N>2$, there are several possible symmetry breaking patterns, each with a different unbroken group. In our case, different ranks of the unbroken group correspond to different numbers of vanishing $\sigma_a$'s at the defect. There are, then, many possible defects, which one can classify by studying the topology of $\mathcal{Z}$ and which eigenvalues vanish on it.

We now come back to the analysis of the plane of superconnections \eqref{Dab}, and the associated paths \eqref{331}. By construction, a path $\gamma$ starting at $s=0$ and ending at $s=1$ defines a form $\Omega^{(\gamma)}$ such that
\be
\chi(\mathcal{F}(a(1),b(1))) - \chi(\mathcal{F}(a(0),b(0))) =  \intd \Omega^{(\gamma)}, \nonumber \ee\be \Omega^{(\gamma)} = - \int_\gamma \intd s\, \str\Big(\pa_s D_s \exp(i \mathcal{F}(a(s),b(s))) \Big),
\label{335}
\ee
where $\mathcal{F}(a(s),b(s))$ is given in \eqref{332} and the equality is meant locally as differential forms.

Note that setting $a=b=0$ leads to the trivial connection $\intd$, whereas setting $a=b=1$ leads to the reference superconnection $\intd - i \mathcal{A}$. If we choose a path with $a(0) = b(0) = 0$ and $a(1) = b(1) = 1$, we see from \eqref{335} that we obtain an action satisfying criterion \ref{cr:character} ($\intd \Omega = \chi$). To compute a TCS form of a given degree, we need all the $(2p+1)$-form terms in (\ref{335}). We denote their sum by $\Omega_{2p+1}^{(\gamma)}$.

Before we move on to the discussion of possible paths, we give a few properties of this construction:

\begin{itemize}

  \item {\em Each path in this space that starts at the origin, (a,b)=(0,0), defines a unique TCS form, and end-points of each path determine the exterior derivative of the TCS action.}\\
This is an immediate consequence of equation (\ref{335}).

\item {\em Actions obtained after integrating over a loop in this plane are exact forms.}\\
Since the starting point and the end point of the path are identical, integrating (\ref{335}) gives closedness of the integral over a loop. Therefore, two paths with identical endpoints differ by a closed form. Furthermore, it is shown in appendix \ref{sec:F} that the loops are also exact forms. Therefore the choice of path only affects boundary terms in the TCS action.

\item {\em Vertical paths at $a=1$ produce gauge-invariant terms.}\\
For vertical paths,
\be
\frac{\partial i\mathcal{A}_s }{\partial s} \propto i \mathcal{T},\label{337}
\ee
which is gauge-covariant, so if $\mathcal{F}_s$ is gauge-covariant too, as is the case at $a=1$, the right hand side of (\ref{335}) is gauge-invariant.

\end{itemize}

Variants of equation \eqref{335} have been used extensively in the Chern-Simons literature \cite{Quillen,Sugimoto,Szabo:2001yd}, with various objectives often related to K-theory or index theorems. Already in the standard CS case, in the absence of a tachyon, an integral formula can be obtained from \eqref{335} by computing it for the path\footnote{Here the subscript ``alt'' refers to the fact that below we consider a different path, which will be our primary choice.} $I_\mathrm{alt}$ with $b(s)=0$, $a(s)=s$,
\be
\omega = \int_0^1 \intd s\, \str(i A \, \exp(i\mathcal{F}(a(s)=s, b(s)=0)) ),\label{336}
\ee
where $A$ was defined in \eqref{32}, and $i\mathcal{F}(a(s)=s, b(s)=0)$ is a specific case of equation \eqref{332}. Equation \eqref{336} can serve as a general definition of $\omega$ for any form degree.

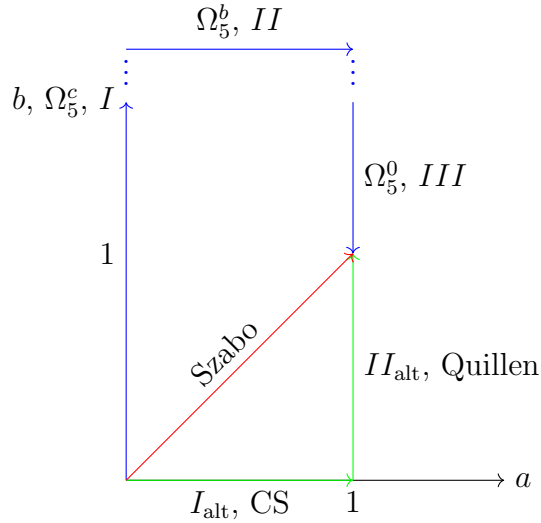
\begin{figure}[htb]
\centering
\begin{tikzpicture}

\draw[->] (0,0) -- (5,0) node[right] {$a$};
\draw[draw=blue, ->] (0,0) -- (0,5) node[left] {$b$, $\Omega_5^c$, $I$};
\node[blue] at (0, 5.5) {$\vdots$};

\draw[draw=green, ->] (0,0) -- (3,0) node[midway, below] {$I_\mathrm{alt}$, CS};

\draw[draw=green, ->] (3,0) -- (3,3) node[midway, right] {$II_\mathrm{alt}$, Quillen};

\draw[draw=red, ->] (0,0) -- (3,3) node[midway, above, sloped] {Szabo};

\draw[draw=blue, ->] (3,5) -- (3,3) node[midway, right] {$\Omega_5^0$, $III$};
\draw[draw=blue, ->] (0,5.7) -- (3,5.7) node[midway, above] {$\Omega_5^b$, $II$};
\node[blue] at (3, 5.5) {$\vdots$};
\node[below] at (3,0) {$1$};
\node[left] at (0,3) {$1$};
\end{tikzpicture}
\caption{Illustration of the parameter space for superconnections \eqref{Dab}. Here ``Szabo'' refers to the path of integration defined in~\cite{Szabo:2001yd} while ``Quillen'' refers to the path defined in~\cite{Quillen}.
 $\Omega_5^c$ is the form described first in the introduction, above  equation (\ref{i2}) that generalizes the Witten WZ term. Forms $\Omega^0_5$ and $\Omega^b_5$ are also defined in the introduction and later on below.
The labels $I_{\text{alt}}$ and $II_{\text{alt}}$ were introduced in the text. The path represented by the blue arrows is our final choice satisfying all criteria.}
\label{plane}
\end{figure}

A major technical difference between the standard CS theory in the absence of tachyon and the one in the presence of tachyons, is that the character in \eqref{336} contains only a finite amount of terms with a given form degree, while in the presence of tachyons, there is {\em an infinity of terms} that we need to resum. The form $\omega$ in \eqref{336}, which is obtained from the horizontal green path in figure \ref{plane}, can be combined with the action obtained from a class of paths introduced by Quillen \cite{Quillen}, the vertical green path in figure \ref{plane}. This second section of the path ($II_\mathrm{alt}$), connects the zero tachyon connection $\intd - i A$ to the full superconnection $\intd - i \mathcal{A}$.

We show the superconnection plane, with the different relevant paths, in figure \ref{plane}. We indicate
\begin{itemize}
  \item \label{path:galt} The path $\gamma_\mathrm{alt} = I_\mathrm{alt} \cup II_\mathrm{alt}$ in green, studied in appendix \ref{sec:gammaalt}.
  \item A diagonal path used in \cite{Szabo:2001yd}, $a(s) = s, b(s) = s$, in red.\label{path:Szabo}
  \item \label{path:g} A path $\gamma = I \cup II \cup III$, in blue, composed of:
  \begin{itemize}
    \item A vertical arrow from $0$ to $+\infty$ at $a=0, b(s)=s$ (path $I$)
    \item A horizontal arrow from $0$ to $1$ at infinity (path $II$)
    \item A vertical arrow from $+\infty$ to $1$, defines of the gauge-invariant TCS action $\Omega_5^0$, with the parametrization used in \cite{Quillen}, $a(s) = 1, b(s) = s$ (path $III$)
  \end{itemize}
\end{itemize}

The vertical path $I$ extends from $b=0$ to infinity at $a=0$. The Chern character, is zero in $a=b=0$, and on $b\to \infty$ it also asymptotes to zero. Therefore, the Chern character is zero on both end-points, and from \eqref{335} it yields a closed form.

To define the part $II$ of the path, we simply consider a horizontal path from $a=0$ to $1$ at a finite $b$, and then take the limit $b\to \infty$, which produces a form that, like the Chern character, localizes around $\mathcal{Z}$, defined in \eqref{333}. We call this piece $\Omega_5^b$.

Lastly, $III$ is the path from $b=\infty$ to $1$ at $a=1$, which defines a gauge-invariant form $\Omega_5^0$.

We now show that the path $\gamma$ generates a generalization of the TCS action found in \cite{jknp} to a generic tachyon field, that satisfies all criteria \ref{cr:character} to \ref{cr:noIR}. For this path, we separate the contribution of each segment, $I (\Omega^c)$, $II (\Omega^b)$ and $III (\Omega^0)$, and we obtain
\be
\Omega =  \Omega^c + \Omega^b + \Omega^0,\nonumber
\ee
\be \Omega \equiv \sum_{p=0}^{\infty} \;\Omega_{2p+1}.\label{eq:Omegagen}
\ee
where the different terms are given by
\be
\Omega^c \equiv \int_{0}^{+\infty} \intd s\, \str \big(i\mathcal{T} e^{i\mathcal{F}(0,s)}\big)\equiv \sum_{p=0}^{\infty} \;\Omega^c_{2p+1},
\label{eq:Omegac}
\ee
\be
\Omega^b \equiv \lim_{b\to +\infty}\int_{0}^{1} \intd s\, \str\big(iA e^{i\mathcal{F}(s,b)}\big)\equiv \sum_{p=0}^{\infty} \;\Omega^b_{2p+1},
\label{eq:Omegab}
\ee
\be
\Omega^0 \equiv \int_{+\infty}^1 \intd s\, \str\big(i\mathcal{T} e^{i\mathcal{F}(1,s)}\big)\equiv \sum_{p=0}^{\infty} \;\Omega^0_{2p+1},
\label{eq:Omega0}
\ee
and where the subscript indicates the form degree. It can be checked that, as anticipated, there are no even-form contributions in these formulae. Depending on the path, the supermatrices either contain only off-diagonal terms, or they contain only diagonal traceless terms.

The supertrace in \eqref{329} provide two terms with opposite signs that are related by $P_1$, so the parity property of the full path $\g$ is guaranteed by the superconnection formalism. We postpone the discussion of the C-parity and of the other criteria to the next section, where we show how this action can be computed explicitly.

\section{Explicit computation of the Tachyon-Chern-Simons action}
\label{sec:explicit}

In this section, we compute explicitly the TCS form $\Omega$, introduced in the previous section, in \eqref{eq:Omegagen}. We first show, that we recover the known results of \cite{jknp}, in the case $T=\tau U$, with $\tau$ a scalar function and $U$ a unitary matrix.

We shall explain how to perform the general computation, starting  with $\Omega^0$, in (\ref{eq:Omega0}). Then, we shall consider the cases of $\Omega^b$ in (\ref{eq:Omegab}) and $\Omega^c$ in (\ref{eq:Omegac}). In the last subsection, we collect the results for $\Omega_3$ and $\Omega_5$ that are defined in \eqref{eq:Omegagen}. The result for $\Omega_1$ is given in appendix \ref{sec:omega1}.

\subsection{The $T=\tau U$ case}
\label{sec:tauU}

Here we assume that the tachyon has the special form $T=\tau U$, where $\tau$ a single function, and $U$ a unitary $N_f\times N_f$ matrix. To compute $\Omega^0$ \eqref{eq:Omega0}, we must first compute the Chern character for $\mathcal{F}(a,b)$, given by \eqref{332}. First, we compute the exponential of $i \mathcal{F}(a,b)$, written as a series expansion
\be
\exp(i \mathcal{F}(a,b)) = \sum_{n=0}^{\infty} \frac{1}{n!} (i \mathcal{F}(a,b))^n \;\;.\label{41}
\ee
Since we are interested in $\Omega_{2p+1}$, defined in \eqref{eq:Omegagen}, we need to select the terms with the appropriate form degrees in \eqref{41}. The field strength $\mathcal{F}(a,b)$ given in \eqref{332}, can be decomposed into three terms of different form degrees as follows,
\be
i\mathcal{F}(a,b) = X(a,b) + Y(a) + Z(b), \label{42}
\ee
with
\be
X(a,b) \equiv \begin{pmatrix}
0 & b \intd T^\dagger - i ab A_L T^\dag  + i ab T^\dag A_R\\
 b \intd T + i ab T A_L  - i ab A_R T &0
\end{pmatrix},\label{43}
\ee
\be
Y(a) \equiv \begin{pmatrix}
ia \intd A_L + a^2 A_L^2& 0\\
 0 & ia \intd A_R + a^2 A_R^2
\end{pmatrix}, \qquad  Z(b) \equiv \begin{pmatrix}
- b^2 T^\dagger T & 0\\
 0& - b^2 T T^\dag
\end{pmatrix}. \label{44}
\ee
The exponential of $\mathcal{F}(a,b)$ in \eqref{41} can then be written as
\be
\exp(i \mathcal{F}(a,b)) = \sum_{n=0}^{\infty} \frac{1}{n!} (X(a,b) + Y(a) + Z(b))^n\label{45}.
\ee

{From now on, to simplify notation, the arguments $(a,b)$ are omitted in formulae whenever $a(s)$ and $b(s)$ are unspecified functions of the path coordinate $s$}. $X$ has form degree one, and $Y$ has form degree two, therefore they appear finitely many times in $\Omega_{2p+1}$. On the other hand, $Z$ is a zero-degree form, so there is no bound on how many times it can appear. Therefore, there are contributions to $\Omega_{2p+1}$ originating from every $(X+Y+Z)^n$, for $n$ sufficiently large. They  come from adding powers of $Z$ in between the $X$ and $Y$ that carry a form degree.

In the $T= \tau U$ case, this problem simplifies significantly, as $Z$ becomes proportional to the identity
\be
Z= -b^2 \tau^2 I.
\label{46}
\ee
This reduces considerably the number of terms that appears at each order $n$ in \eqref{45}. In particular, for $n\geq 2p+1$, this number becomes $n$-independent.
We focus on $\Omega^0_{2p+1}$, expressed from the homotopy formula as in \eqref{eq:Omega0}. As can be seen from this expression, we need the $(2p+1)$-form contribution to the exponential $\exp(i\mathcal{F}(1,s))$. To compute this, one should consider all the possible combinations of terms containing $(2p+1-2k)$ times $X$ and $k$ times $Y$, with $k$ between 0 and $p$. To compute them, we consider the possible ``words'' that one can make with $X$ and $Y$. There are exactly
\be
\begin{pmatrix}
	2p+1-k\\
	k
\end{pmatrix}\label{47}
\ee
possible word arrangements for each $k$. For $\Omega^0_5$, we find that the relevant words are
\begin{align}
&\text{Words}\ (p=2): &  && \nonumber \\
 &XXXXX,& \qquad &(k=0);&\label{48}\\
 &XXXY, \quad XXYX, \quad XYXX, \quad YXXX,& \quad &(k=1);&\label{49}\\
& XYY, \quad YXY, \quad YYX,& \qquad  &(k=2).&\label{410}
\end{align}

The computations of the different words are very similar. We present the computation for $XXXXX$, for illustration. We start by rewriting \eqref{45} as
\be
\exp(i \mathcal{F}(a,b))_{X^5} = \sum_{n=0}^{\infty} \frac{1}{(n+5)!} \begin{pmatrix}
n+5\\
5
\end{pmatrix} X^5 Z^n \label{411},
\ee
where we used (\ref{46}) and the combinatorial factor comes from the expansion of $(X + Z)^{n+5}$. The subscript $X^5$ indicates that we keep the contribution from this particular word. Simplifying  the combinatorial factors in (\ref{411}), and using \eqref{46}, we obtain
\be
\exp(i \mathcal{F}(a,b))_{X^5} = \frac{X^5}{5!} \sum_{n=0}^{\infty} \frac{1}{n!} Z^n  =  \frac{X^5}{5!} \exp(-b^2 \tau^2).
\label{412}
\ee
where we have also used \eqref{46}. The last step is to compute the word's contribution to $\Omega_5^0$, by inserting \eqref{412} into \eqref{eq:Omega0}, which gives
\be
(\Omega_5^0)_{X^5} = \int_{+\infty}^{1} \intd s \exp(-s^2 \tau^2)\;  \str\left(i \mathcal{T} \frac{X(1, s)^5}{5!} \right). \label{413}
\ee

Inserting $\mathcal{T}$ from \eqref{32} and $X(a,b)$ from \eqref{43} into \eqref{413} and computing the integral, after some algebra we obtain
\be
(\Omega_5^0)_{X^5} = \frac{(2+2\tau^2 + \tau^4)}{120} \tr((U^\dag D U )^5)\label{414},
\ee
where we used
\be
D U = \intd U + i U A_L  - i A_R U, \qquad DU^\dag = -U^\dag D U U^\dag. \label{415}
\ee
Note that the contributions in \eqref{414} coming from the $\intd\tau$ terms in $X$ cancel each other.

After computing likewise the contributions from \eqref{49} and \eqref{410}, we obtain that the full $\Omega_5^0$ for $T = \tau U$ is
\be
\Omega_5^0 =
  \frac{(2+2\tau^2 + \tau^4)e^{-\tau^2}}{120} \text{Tr}((DUU^\dagger)^5)\nn
\ee \be
  + \frac{ie^{-\tau^2}(1+\tau^2)}{12}\text{Tr}(U^\dagger DU  F_L (U^\dagger  DU)^2) +\text{Tr}(DUU^\dagger  F_R   (DUU^\dagger)^2 )\nn
\ee \be
-\frac{e^{-\tau^2}}{12} \big[ 2 \text{Tr}(U^\dagger DU  F_L^2)+2 \text{Tr}(DUU^\dagger  F_R^2) + \text{Tr}(DUU^\dagger  F_R U  F_LU^\dagger)+\text{Tr}(DU  F_L U^\dagger  F_R )\big].
\label{416}
\ee
This exactly reproduces the expression $\Omega_5^0$ constructed in \cite{jknp} from symmetry principles. It can be checked that it reduces to the usual CS action when $\tau \to 0$.

We shall now compute $\Omega_5^b$ and $\Omega_5^c$ using their definitions  \eqref{eq:Omegab}, \eqref{eq:Omegac}. We start with
$\Omega_5^c$, which is easier to handle. According to \eqref{332}, the superconnection curvature on the path of~\eqref{eq:Omegac} is given by
\be
i\mathcal{F}(0, s) =  X(0, s) + Z(s) = is \intd \mathcal{T}+ s^2 \mathcal{T}^2 = \begin{pmatrix}
  -s^2 T^\dag T & s\intd T^\dag \\
  s\intd T& -s^2 TT^\dag
\end{pmatrix}\label{417}.
\ee
Therefore, the only word that contributes in the calculation of the exponential is $X^{5}$.
Similarly to \eqref{413}, we can then write
\be
\Omega_5^c =  \int_{0}^{+\infty} \intd s \exp(-s^2 \tau^2)  \str\left(i \mathcal{T} \frac{X(0,s)^5}{5!} \right) \label{418}.
\ee
Computing the supertrace, after some algebra, \eqref{418} reproduces the WZW term, \cite{wit-bar}
\be
\Omega_5^{c} = -\frac{1}{60} \tr[(U^\dag \intd U)^5]\;. \label{419}
\ee
Therefore, beyond  the $T=\tau U$ case, we expect that this path generalizes the WZW term to a generic tachyon. Note also  that \eqref{419} is indeed a closed form, but is not exact. Given that the chiral lagrangian has fields that parametrize the coset space of chiral symmetry breaking $(\text{SU}(N_f)_L \times \text{SU}(N_f)_R)/\text{SU}(N_f)_V$, {$\Omega_5^c$} is classified the cohomology group of this coset \cite{wit-bar},
\be
{\rm H}^5 ((\text{SU}(N_f)_L \times \text{SU}(N_f)_R)/\text{SU}(N_f)_V, \mathbb{Z}).\label{420}
\ee

We shall now compute $\Omega_5^b$, which generalizes $\intd G_4$ in \cite{jknp}, namely the gauged WZW term, \cite{Witten:1983tw,Gomm:1984at}.
There is already an $iA$ factor in \eqref{eq:Omegab}, hence, to compute $\Omega_5^b$, we need the $4$-form contribution to
$\exp(i\mathcal{F}(s, b))$ at fixed $b$.

There are five words to compute: $YY$, $YXX$, $XYX$, $XXY$ and $XXXX$. For these words, the limit in $b$ behaves differently. For the term $YY$, we find that
\be
(\Omega_5^b)_{Y^2} = e^{-b^2 \tau^2} \omega_5,\label{421}
\ee
where $\omega_5$ was given in \eqref{315}. That is, this word is proportional to the standard CS form $\omega_5$, which is independent of $\tau$. Then, in the large $b$ limit, the behavior of this term is that of the exponential. For this, we need to distinguish points in  the set $\mathcal{Z}$, defined in \eqref{333}, i.e. the set of points where $\tau$ vanishes, and points that belong to  $\mathcal{M} - \mathcal{Z}$ where $\tau \neq 0$.  For points in $\mathcal{Z}$, \eqref{421} is independent of $b$, and is equal to the standard CS, $\omega_5$,  in (\ref{315}). For points in $\mathcal{M} - \mathcal{Z}$, however, \eqref{421} goes to zero like a Gaussian. We obtain
\be
e^{-b^2 \tau^2}\; \omega_5 \xrightarrow[b\to +\infty]{}\theta_{\mathcal{Z}}\; \omega_5  \label{422},
\ee
{where $\theta_{\mathcal{Z}}$ is a Heaviside function on $\mathcal{Z}$, which is equal to $1$ on $\mathcal{Z}$ and $0$ on $M-\mathcal{Z}$.}
In the chirally broken phase (still for $T=\tau U$), $\mathcal{Z}$ is a measure zero set. Therefore this term can be ignored in the action.

The words with $X$ letters behave quite differently as $b$ is taken to infinity.
They contain terms with exterior products of different numbers  of $\intd \tau$ differentials, arising from expanding the $\intd T$ terms in $X$ given in (\ref{43}).
The fate of these terms depends on the number of $\intd \tau$ factors that they contain (either zero or one, since $\intd \tau^2 = 0$). The terms with no $\intd \tau$ come with an additional even power of $b\tau$, such that  they vanish in the large $b$ limit, like
\be
e^{-b^2 \tau^2} (b\tau)^{2m} \xrightarrow[b\to +\infty]{} 0 \sp m\in \mathbb{N}^*,\label{423}
\ee
both on $\mathcal{Z}$ and $\mathcal{M} - \mathcal{Z}$.

Only terms with exactly one $\intd\tau$ factor survive, and they are proportional to
\be
e^{-b^2 \tau^2}b^{2m} \tau^{2m-1} \intd \tau.\label{424}
\ee
This differential form converges in the distributional sense to a Dirac $\delta$-distribution,
\be
\lim_{b\to \infty}e^{-b^2\tau^2} b^{2m} \tau^{2m-1} \intd \tau \propto \delta(\tau) \intd\tau.\label{425}
\ee
Note that this implies that $\Omega_5^b$ is a form defined on $\mathcal{Z}$.

Collecting the contributions from different words, we find the same contribution as in \cite{jknp},
\be
\Omega_5^b = \delta(\tau) \intd \tau \wedge  G_4,\label{426}
\ee
where
\be
 \label{G4final}
 24\, G_4 =\Big\{ 2\Big[ \tr(A_L\wedge F_L\, \wedge U^\dagger DU)+ \tr(A_L\,\wedge U^\dagger DU\wedge F_L)\Big]+
\ee
\be \nonumber
+\Big[\tr(A_L\,\wedge U^\dagger DU\, \wedge U^\dagger F_R\,U)+\tr(A_L\,\wedge U^\dagger F_R U \wedge U^\dagger DU)\Big]+
\ee
\be \nonumber
+i\Big[ \tr(A_L\wedge A_L\, \wedge U^\dagger A_R U \wedge U^\dagger DU)- \tr(A_L\wedge A_L\, \wedge U^\dagger DU\, \wedge U^\dagger A_R\,U)\Big]+
\ee
\be \nonumber
+i\Big[ \tr(A_L\wedge F_L\,\wedge U^\dagger A_R\,U)+ \tr(A_L\, \wedge U^\dagger A_R\,U\wedge F_L)\Big]+
\ee
\be \nonumber
+2 i \tr(A_L\wedge A_L\wedge A_L\,\wedge U^\dagger DU) -2 \tr(A_L\wedge A_L\wedge A_L\, \wedge U^\dagger A_R\,U)
\ee
\be \nonumber
 +2\tr(A_L\, \wedge U^\dagger DU\, \wedge U^\dagger DU\, \wedge U^\dagger A_R\,U) -\tr(A_L\, \wedge U^\dagger DU\wedge A_L\, \wedge U^\dagger DU)+
\ee
\be \nonumber
-2 i \tr(A_L\, \wedge U^\dagger A_R\,U\wedge A_L\, \wedge U^\dagger DU)
-2 i\tr(A_L\, \wedge U^\dagger DU\, \wedge U^\dagger DU\, \wedge U^\dagger  DU)\Big\}+
\ee
\be \nonumber
+\ \Big(L\leftrightarrow R\Big) +\tr(A_L\, \wedge U^\dagger A_R\,U\wedge A_L\, \wedge U^\dagger A_R\,U) \ .
\ee

Interestingly, \eqref{426} can be inverted to yield a path formula for $G_4$, which will be useful for our discussion of the baryon number below. For this purpose, we introduce the following operator,
\be
P_\mathcal{Z} (B)= \int_{0}^{+\infty} \intd \tau \frac{\pa B}{\pa \intd \tau} \label{eq:pz},
\ee
where $B$ is a generic differential form over $\mathcal{M}$.
Operation with $P_\mathcal{Z}$ breaks into two steps. First, the derivative with respect to $\intd \tau$ in \eqref{eq:pz} selects the terms with a single $\intd \tau$ and removes the instance of $\intd \tau$. Then, according to the previous discussion, in the large $b$ limit, these differential forms asymptote to a fraction of a Dirac $\delta$-distribution. The ``trick'' is that the integral in $\tau$ does not depend on the value of $b$. Therefore, we can compute the integral at finite $b$ and multiply by $\delta(\tau)$ to compute the limit in large $b$, $\Omega_5^b$. We therefore have:
\be
G_4  = P_\mathcal{Z}( \Omega_5^b).\label{429-2}
\ee
$\Omega_5^b$ is known as the gauged Wess-Zumino-Witten term\footnote{More specifically, $G_4$ and the Witten term differ by the exterior derivative of a 3-form,
see the Appendix in \cite{jknp}.}, \cite{wit-bar-2}, that was found by gauging $\Omega_5^c$. Note that unlike the bottom-up calculations, \cite{jknp}, the IR contributions from $G_4$ are automatically excluded in the superconnection calculation.

Lastly, in \cite{jknp}, $\Omega_5^b$ was written as $\Omega_5^b = \intd G_4$. Instead, here we obtain it as
\be
\Omega_5^b = \intd_\tau G_4, \qquad \intd_\tau(.) \equiv \delta(\tau) \intd \tau \wedge (.).\label{428}
\ee
{$\intd_\tau$ anticommutes with the exterior derivative $\intd$, and $\intd_\tau^2= 0$. We then define a notion of $\intd_\tau$-closed and $\intd_\tau$-exact forms, which will be used in sections \ref{sec:explicit} and \ref{sec:bar}. Integrals over $\mathcal{M}$ of $\intd_\tau$-exact forms, like $\Omega_5^b$ then become integrals over $\mathcal{Z}$,
\be
\int_{\mathcal{M}} \intd_\tau G_4 = \int_{\mathcal{Z}} G_4,
\ee
This reformulation makes the localization on $\mathcal{Z}$ structurally analogous to ordinary boundary terms in de Rham theory. Unlike the bottom-up computation of \cite{jknp}, where it is enforced that the IR contribution from $G_4$ vanishes, in the superconnection formalism the vanishing of the IR contribution from $\Omega_5^b$ is automatic.

\subsection{The general method for $\Omega^0$}
We shall now study the generic tachyon case. We can introduce a first decomposition of the tachyon field that generalizes the $\tau U$ ansatz. It is the matrix polar decomposition
\be
T= HU \label{429},
\ee
where $H$ is a hermitian matrix, and $U$ is a unitary matrix. In the $\tau U$ case, $Z(b)$ as defined in \eqref{44}, is proportional to the identity, as is clear from  \eqref{46}. It therefore commutes with every supermatrix. With the more general decomposition \eqref{429},  this  is not anymore the case.

All the terms appearing in products, are still given in the form of ``words" composed of the two ``letters" $X = i[D, \mathcal{T}]$ and $Y = iF$, with any powers of $Z = \mathcal{T}^2$ in between them. Because nothing is assumed to commute, the instances of $Z$ in between other letters can no longer be factored out as in the $\tau U$ case. We need to compute first the exponential of $i\mathcal{F}$.

The words to compute are the same as before, and we again consider the $X^5$ term as an example. We can already write a generalization of \eqref{411},
\be
\exp(i \mathcal{F}(a,b))_{X^5} = \sum_{n=0}^{+\infty} \frac{1}{(n+5)!} \sum_{i_1 + ... + i_6=n} (Z^{i_1} X Z^{i_2}X Z^{i_3}X Z^{i_4}X Z^{i_5}X Z^{i_6}) \label{432}.
\ee
It is possible to find a residue formula for $\Omega^0$, obtained in Appendix \ref{sec:residue}, but this is not the case for $\Omega^b$ and $\Omega^c$. It is also possible to make some progress by re-expressing $T$ in terms of its singular value decomposition (SVD). We shall use in the following this other possible decomposition, guaranteed by the singular values' theorem in linear algebra. The decomposition reads
\be
T= V_R \Sigma V_L^\dagger, \label{430}
\ee
where $V_R$ and $V_L$ are unitary, but $\Sigma$ is now diagonal and non-negative. The SVD simultaneously diagonalizes $TT^\dag$ and $T^\dag T$, and therefore also $Z$ in \eqref{432}. {$\Sigma$ contains the positive square-roots of the eigenvalues of both $T^{\dagger}T$ and $T T^{\dagger}$, $\sigma_a$, in the diagonal} and we can order them in ascending order. Note that from \eqref{430} there is a gauge transformation \eqref{217} that transforms the tachyon to $\Sigma$. Equations \eqref{429} and \eqref{430} are related by
\be
H = V_R \Sigma V_R^\dag, \qquad U = V_R V_L^\dagger. \label{431}
\ee
Some facts about the decomposition of the tachyon are given in Appendix \ref{sec:convention3}. The strategy is to go to this singular value decomposition, which diagonalizes $Z$, and allows the exponential to be reorganized into functions of singular values. We write
\be
Z = b^2 \mathcal{T}^2 = - b^2 \begin{pmatrix}
 T^\dagger T& 0 \\
 0 & TT^\dagger
\end{pmatrix}= - b^2 \begin{pmatrix}
V_L \Sigma^2 V_L^\dagger& 0 \\
 0 & V_R \Sigma^2 V_R^\dagger
\end{pmatrix} \equiv -b^2 V \Sigma^2 V^\dag, \nonumber \ee \be V =\begin{pmatrix}
V_L & 0 \\
 0 & V_R
\end{pmatrix} \label{433}.
\ee
Note that $\Sigma^2$ contains the eigenvalues of $TT^\dag$ and $T^\dag T$. This can be used to rewrite \eqref{432} as
\be
\exp(i \mathcal{F}(a,b))_{X^5} = \sum_{n=0}^{+\infty} \frac{(-b^2)^n}{(n+5)!} \sum_{i_1 + ... + i_6=n} (\Sigma^{2 i_1} \tilde{X} \Sigma^{2 i_2} \tilde{X}  \Sigma^{2 i_3} \tilde{X} \Sigma^{2 i_4} \tilde{X} \Sigma^{2 i_5} \tilde{X} \Sigma^{2 i_6}), \label{434}
\ee
where we have defined
\be
\tilde{X} \equiv V^\dag X V.\label{435}
\ee
Since $X$ is gauge-covariant, $\tilde{X}$ is $X$ expressed in the diagonal tachyon gauge. Next, we write \eqref{434} in flavor indices, and note $\Sigma_{a_1 a_2} = \delta_{a_1 a_2} \sigma_{a_1}$. We obtain
\be
(\exp(i \mathcal{F}(a,b))_{a_1 a_6})_{X^5} =  \sum_{n=0}^{+\infty} \frac{(-b^2)^n}{(n+5)!}  \sum_{a_2, ... a_5} (\tilde{X}_{a_1 a_2} \tilde{X}_{a_2 a_3} \tilde{X}_{a_3 a_4} \tilde{X}_{a_4 a_5} \tilde{X}_{a_5 a_6})\times \nn \ee\be \times h_{n}(\sigma_{a_1}, ..., \sigma_{a_5}, \sigma_{a_6})\nn \ee\be
h_{n}(\sigma_{a_1}, ..., \sigma_{a_{m}}, \sigma_{a_{m+1}}) \equiv \sum_{i_1 + ... + i_{m+1}=n}  \sigma_{a_1}^{2i_1} \sigma_{a_2}^{2i_2} \dots \sigma_{a_{m+1}}^{2i_{m+1}}, \label{436}
\ee
where $h_n$ is the fundamental symmetric polynomial of degree $n$ in the squared singular values of the tachyon. A crucial property of \eqref{436} is that it significantly simplifies when different eigenvalues are equal. If we take all the $\sigma$ to be the same, as in the $T=\tau U$ case, then we obtain
\be
h_{n}(\sigma_{a_1}, ..., \sigma_{a_{m}}, \sigma_{a_{m+1}})  = \begin{pmatrix}
  n+m\\m
\end{pmatrix} \tau^{2n},\label{437}
\ee
which is precisely the factor appearing \eqref{411} with $m=5$. This object can be computed exactly also when eigenvalues are not equal, and then the series in $n$ can be resummed exactly too. We define
\be
\mathcal{L}(b(s)) \equiv \sum_{n=0}^{+\infty} \frac{(-b(s)^2)^n}{(n+m)!}  \sum_{i_1 + ... + i_{m+1}=n} \sigma_{a_1}^{2i_1} \sigma_{a_2}^{2i_2} \dots \sigma_{a_{m+1}}^{2i_{m+1}}\label{438},
\ee
{where $\mathcal{L}$ implicitly bears $m+1$ indices $a_i$, which are omitted to simplify the notations. In \eqref{438}, we defined $m \equiv 2p+1-k$, on which $\mathcal{L}$ also implicitly depends.}

The $\mathcal{L}$ operators encode the combinatorics of inserting arbitrary powers of $Z$ in the words considered in the previous section. The relevant $\mathcal{L}$ operators for each word and their exact values are summarized in appendix \ref{app:Ldef}. We obtain from \eqref{436} and \eqref{438} with $m=5$
\be
(\exp(i \mathcal{F}(a,b))_{a_1 a_6})_{X^5} =  \sum_{a_2, ... a_5} \mathcal{L}( b)  (\tilde{X}_{a_1 a_2} \tilde{X}_{a_2 a_3} \tilde{X}_{a_3 a_4} \tilde{X}_{a_4 a_5} \tilde{X}_{a_5 a_6}).\label{439}
\ee
where the left-hand side was defined in (\ref{411}). Similar formulae are valid for the other words. Note that, in virtue of gauge invariance of the formula,  the $\tilde{X}$ can be replaced with $X$, defined in \eqref{43}.

At this stage, one may worry about the fact that we expressed these operators as functions of the eigenvalues of the tachyon. We would like to have an explicit expression, expressed in terms of $T$ instead. Thankfully, the eigenvalues $\sigma_a$ can be re-expressed as functions of the tachyon field using Vi\`ete's relations between the coefficients of a polynomial and its roots, which are in this case the tachyon eigenvalues $\sigma_a^2$.

This a priori depends on the number of different eigenvalues of the tachyon. The eigenvalues $\sigma^2$ are roots of the characteristic polynomial of $TT^\dag$, and for small enough number of different eigenvalues, there exists formulae to re-express them as functions of the coefficients. For more than four eigenvalues, the $\sigma_a$ can still be connected to the coefficients, but inverting them suffers from the usual Galois theory obstruction, and should be treated differently, for instance by working in the gauge where $TT^\dag$ and $T^\dag T$ are diagonal, at the cost of losing the usual radial gauge $A_r=0$ used in holography. We shall instead consider smaller values of $N_f$. We  now show the results for $N_f=2$ and $N_f=3$. The $N_f=4$ case can be done using the Ferrari solution of the quartic polynomial equation.

For the $SU(2)$ flavor group, the characteristic polynomial is given by
\be
\mathcal{X}^2 - \tr(TT^\dag) \mathcal{X} + \det(TT^\dag)\label{441},
\ee
therefore the two eigenvalues can be rewritten
\be
\label{442}
x_\pm = \frac{1}{2}\tr(TT^\dag) \pm \frac{1}{2}\sqrt{\tr(TT^\dag)^2 - 4 \det(TT^\dag)}.
\ee

The $SU(3)$ analogue can be done using Cardano's method. For this, recall that the characteristic polynomial for a $3$ by $3$ matrix $M$ is
\be
\mathcal{X}^3 - \tr(TT^\dag)\mathcal{X}^2 + \frac{\tr(TT^\dag)^2 - \tr((TT^\dag)^2)}{2}\mathcal{X} - \det(TT^\dag)\label{443}.
\ee
After some algebra, we obtain the following result for the three different eigenvalues $x_a$ ($a=1,2,3$)
\begin{small}
\be
x_a =   \frac{\tr(TT^\dag)}{3} + 2\sqrt{\frac{\tr((TT^\dag)^2)}{6} - \frac{\tr(TT^\dag)^2}{18}} \cos\bigg(\frac{2a\pi}{3} + \nonumber
\ee
\be
 \frac{1}{3}\arccos\bigg( \frac{\frac{\tr(TT^\dag)^3}{18} - \frac{\tr(TT^\dag) \tr((TT^\dag)^2)}{2} - 3\det(TT^\dag)}{\frac{\tr(TT^\dag)^2}{3} - \tr((TT^\dag)^2) } \sqrt{\frac{3}{\frac{\tr((TT^\dag)^2)}{2} - \frac{\tr(TT^\dag)^2}{6}}}\bigg)\bigg)
\label{eq:cardan}.
\ee
\end{small}
where above we used the Cardano formula adapted to three real roots.

The full formula for $\Omega^0_5$ in \eqref{eq:Omega0} is obtained by substituting (\ref{442}) or (\ref{eq:cardan}) for the eigenvalues into (\ref{438}).
{Then, we substitute (\ref{438}) into (\ref{439})  and integrate on the path.
Since $i \mathcal{T}$ in \eqref{eq:Omega0} does not depend on $s$, we can integrate \eqref{439} directly, and define an integrated weight as}
\be
\mathcal{L}_{a_1, ..., a_{2p+2-k}} \equiv \int_{+\infty}^1 \intd s \; i^k \; s^{2p+1-2k} (\mathcal{L}(b(s)))
\label{genericL}.
\ee
{Recall that for $\Omega_5^0$, we have $p=2$ and that $k$ is the number of $Y$ letters in the word.} We shall define one more object to deal with these weighted sums, which is the \emph{$\mathcal{L}$-weighted trace}:
\be
\tr_{\mathcal{L}}(X_1, ..., X_n, X_{n+1}) \equiv \sum_{a_1, ..., a_{n+1}} (\mathcal{L})_{a_1 ... a_n a_{n+1}} (X_1)_{a_1 a_2} ...(X_n)_{a_n a_{n+1}} (X_{n+1})_{a_{n+1} a_1},
\label{440}
\ee

We emphasize that the commas in $\tr_{\mathcal{L}}$ are {\em not} usual matrix products.
Contributions from other words (than $X^5$) can also be rewritten in terms of such weighted traces, although with a different weight $\mathcal{L}$ for each word.

With the definition \eqref{440}, the contribution to $\Omega_5^0$ from the word $XXXXX$ can now be expressed as
\be
(\Omega^0_5 )_{X^5} = - \tr_{\mathcal{L}}\big(T^\dagger DT, DT^\dagger, DT, DT^\dagger, DT, \mathbb{I} \big) -\big( L \leftrightarrow R,T \leftrightarrow T^\dag \big). \label{449}
\ee
The contributions from other words can be computed along the same lines, from which we obtain closed forms for both $\Omega_3^0$ and $\Omega_5^0$
\be
\Omega^0_3 = - \tr_{\mathcal{L}}\big(T^\dagger DT, F_{L}, \mathbb{I}\big) + \tr_{\mathcal{L}}\big(F_L, DT^\dag T , \mathbb{I}\big)  \nonumber
\ee
\be  +\tr_{\mathcal{L}}\big(T^\dagger DT, DT^\dag, DT , \mathbb{I}\big) - \big( L \leftrightarrow R,T \leftrightarrow T^\dag \big) ,
\label{eq:Omega3final}
\ee

\be
\Omega^0_5 =  -\tr_{\mathcal{L}}\big(T^\dagger DT, DT^\dagger, DT, DT^\dagger, DT, \mathbb{I} \big) +  \tr_{\mathcal{L}}\big(T^\dagger DT, DT^\dagger, DT, F_L, \mathbb{I} \big) \nonumber
\ee
\be
 +  \tr_{\mathcal{L}}\big(T^\dag DT, DT^\dag, F_R, DT, \mathbb{I} \big)   +  \tr_{\mathcal{L}}\big( T^\dag DT, F_L ,DT^\dag, DT , \mathbb{I} \big) \nonumber
\ee
\be
  - \tr_{\mathcal{L}}\big(F_L, DT^\dag, DT, DT^\dag T , \mathbb{I} \big)
-\tr_{\mathcal{L}}\big(T^\dagger DT, F_L, F_L , \mathbb{I} \big) \nonumber
\ee
\be
  -\tr_{\mathcal{L}}\big(T^\dag F_R, DT, F_L, \mathbb{I} \big)  + \tr_{\mathcal{L}}\big(F_L, F_L , DT^\dag T, \mathbb{I} \big)  -  \big( L \leftrightarrow R,T \leftrightarrow T^\dag \big).
\label{eq:Omega5final}
\ee
$\Omega_1$ can also be expressed in a similar form, which is written in appendix \ref{sec:omega1}.

\subsection{General method for $\Omega^b$ and $\Omega^c$}
\label{sec:g4omc}
We now determine the closed contributions $\Omega^b$ and $\Omega^c$, defined in \eqref{eq:Omegab} and \eqref{eq:Omegac}. Unlike $\Omega^0$, these terms are not determined by the descent equations. Their role is respectively to account for localized contributions on $\mathcal{Z}$, defined in \eqref{333}, and for the appearance of the Witten-Wess-Zumino term, see equation \eqref{452}.

To compute $\Omega^b$ and $\Omega^c$ there is one major difference. Unlike $\Omega^0$, the forms $\Omega^b$ and $\Omega^c$ are not gauge-invariant. Therefore going back from $\tilde{X}$ to $X$ as we did earlier is no longer possible. To obtain the result, one must write the generic field $T$ in terms of the tachyon field in the diagonal gauge, $\Sigma$, and then  re-express $V_L$, $V_R$ and $\Sigma$ in terms of $T$. We  introduce the following definitions:
\be
\widetilde{\intd T} \equiv V_R^\dag \intd T V_L = \intd \Sigma - (\intd V_R^\dag V_R) \Sigma - \Sigma (V_L^\dag \intd V_L) \label{450},
\ee
\be
\widetilde{\intd T^\dag} \equiv V_L^\dag \intd T^\dag V_R = \intd \Sigma  +  \Sigma  (\intd V_R^\dag V_R) + (V_L^\dag \intd V_L)\Sigma\label{451}.
\ee
The main difference with the previous case of $\Omega^0$ is that now the tilded fields are not equal to the fields in the diagonal gauge. In particular, $\widetilde{\intd T}$ is {\em not} exact. We can perform the same procedure as in the previous section, and the result is
\be
\Omega_5^{c} = \tr_{\mathcal{L}}(\Sigma \widetilde{\intd T},  \widetilde{\intd T^\dag},  \widetilde{\intd T},  \widetilde{\intd T^\dag},  \widetilde{\intd T}, \mathbb{I})- (T \leftrightarrow T^\dag) \label{452}.
\ee

We then {substitute} \eqref{450} and \eqref{451} in \eqref{452}. The resulting expression is a generalization of the Witten term, $\tr(U \intd U^\dag)^5$, valid for a generic tachyon. In particular, it can be shown that the terms that do not include any $\intd\Sigma$, are precisely the same terms as the ones that appear when expanding $\tr(U \intd U^\dag)^5$ with $U = V_R V_L^\dag$. Moreover, since on $\mathcal{M} - \mathcal{Z}$ the tachyon matrix can be continuously deformed to its unitary part\footnote{To see this, consider \eqref{430}, and replace $\Sigma$ with $\Sigma_s$, the diagonal matrix whose coefficients are $\sigma^a / ((1-s) + s\sigma^a)$. It is a continuous path in $GL_n(\mathbb{C})$ as the eigenvalues of $\Sigma_s$ cannot vanish. Moreover, one has $\Sigma_0 = \Sigma$ and $\Sigma_1 = \mathbb{I}$. Then at $s=1$ one has $T=U$.},
$\Omega_5^c$ in the generic tachyon case lies in the same cohomology class \eqref{420} as the Witten term $\tr(U^\dag \intd U)^5$.

Finally, to compute $\Omega^b$, we may follow the same steps and expand the words, but now choosing at least one letter to be $\intd T$. The main difference with the $\tau U$ case is that there can be a priori terms with different $\intd \sigma$'s (of the form $\intd \s_a\wedge \intd\s_b$) that are not zero.

For the same reason as in the $T=\tau U$ case, the word $YY$ contains no $\intd \s$, and it vanishes on $\mathcal{M} - \mathcal{Z}$, due to the $b\to\infty$ limit involved in the definition of $\Omega^b$, \eqref{eq:Omegab}. In order to understand the behavior of the terms with several $\intd T$, we study the term $XXXX$. Computing the contribution of this word to $\Omega_5^b$, we obtain
\be
(\Omega_5^b)_{X^4} =
\sum_{a_0\dots a_4} (\mathcal{L}(b))_{a_0\dots a_4} b^4  (i\tilde{A}_L)_{a_0 a_1} \int_{0}^{1} \intd s \widetilde{X^\dag_s}_{a_1 a_2} \widetilde{X_s}_{a_2 a_3} \widetilde{X^\dag_s}_{a_3 a_4} \widetilde{X_s}_{a_4 a_0}  - (L \leftrightarrow R).\label{455}
\ee
We denoted
\be
\tilde{A}_L = V_L^\dag A_L V_L,\label{456}
\ee
\be
\widetilde{X_s} = V_R^\dag (\intd T + i s T A_L - is  A_R T) V_L = \Sigma \intd V_L^\dag V_L  + \intd \Sigma + V_R^\dag \intd  V_R \Sigma  + i s \Sigma  \widetilde{A_L} - is  \widetilde{A_R} \Sigma ,\label{457}
\ee
\be
\widetilde{X^\dag_s} = V_L^\dag (\intd T^\dag + i s T^\dag A_R - i s A_L T^\dag) V_R =  V_L^\dag \intd V_L \Sigma + \intd \Sigma + \Sigma \intd V_R^\dag V_R  + i s \Sigma  \widetilde{A_R} - i s \widetilde{A_L} \Sigma\label{458}.
\ee
The integral \eqref{455} can be done after expanding each term according to \eqref{457}-\eqref{458}.
{Note that $b$ is taken to be fixed on this path, so $\mathcal{L}(b)$ can be taken out of the integral over the path coordinate $s$.} We are only interested in the limit of the $\mathcal{L}(b)$ operators as $b$ goes to infinity. In the limit of large $b$, {the $\mathcal{L}(b)$ operators asymptote  to zero like a Gaussian $e^{-b^2 TT^\dag}$ on $\mathcal{M}-\mathcal{Z}$, such that} the integral  localizes around $\mathcal{Z}$ again, where $\mathcal{Z}$ is defined in \eqref{333}.

We start from a simpler case where the tachyon vanishes over the whole $\mathcal{Z}$, and we also assume that $\mathcal{Z}$ is a codimension $1$ manifold. At every point on $\mathcal{Z}$ there is a single direction orthogonal to $\mathcal{Z}$. Moreover, because here we assumed $T=0$ on $\mathcal{Z}$, the $\intd \sigma_a$ terms are all orthogonal to $\mathcal{Z}$. Therefore they are all aligned, and their wedge products vanish. Hence, the contribution to $\Omega_5^b$ from terms with more than one $\intd T$ vanishes in this case.

For practical applications, we shall need to relax the hypothesis that the codimension of $\mathcal{Z}$ is $1$ and consider higher codimension defects. We still assume for now that $T=0$ on $\mathcal{Z}$. It is then possible to reduce the general case to the codimension $1$ case. To see this, we consider a small neighborhood of spatial extension $\epsilon$ around the defect. This neighborhood is excised out of the manifold, and we consider configurations with $T=0$ on the newly created bounding surface of codimension $1$. The discussion above can be applied to this surface, and one retrieves the physical solution by sending $\epsilon \to 0$.

For instance, in the presence of a point-like defect, one excises a small ball of radius $\epsilon$ around it. At finite $\epsilon$, we take the tachyon to vanish on the sphere of radius $\epsilon$, that is a codimension $1$ submanifold, so terms with more than one $\intd T$ in $\Omega_5^b$ can be ignored, and taking $\epsilon$ to zero, we obtain the physical solution with a point-like defect.

Therefore, when $T=0$ on the defect, terms with more than one $\intd \Sigma$ are zero as in the $T=\tau U$ case for any defect topology. Only terms with a single $\intd \Sigma$ survive, and $\intd \Sigma$ can be commuted to the left of expressions by using the cyclicity of the weighted traces. As a result, we obtain that for defects where the tachyon vanishes identically, {$ \Omega_5^b$}  is given by
\begin{equation}
\Omega_5^b = \delta(T) \; \intd T \wedge G_4(T). \label{O5bT0}
\end{equation}
The expression for $G_4(T)$ is the same as the one derived in \eqref{G4final} for the $T=\tau U$ case, which we rewrite here denoting the unitary matrix $U$ in the polar decomposition $T=HU$ \eqref{429}, which is given by $U = V_R V_L^\dag = (TT^\dag)^{-1/2}T$.
\be
 \label{G4T}
 24\, G_4 =\Big\{ 2\Big[ \tr(A_L\wedge F_L\, \wedge U^\dagger DU)+ \tr(A_L\,\wedge U^\dagger DU\wedge F_L)\Big]+
\ee
\be \nonumber
+\Big[\tr(A_L\,\wedge U^\dagger DU\, \wedge U^\dagger F_R\,U)+\tr(A_L\,\wedge U^\dagger F_R U \wedge U^\dagger DU)\Big]+
\ee
\be \nonumber
+i\Big[ \tr(A_L\wedge A_L\, \wedge U^\dagger A_R U \wedge U^\dagger DU)- \tr(A_L\wedge A_L\, \wedge U^\dagger DU\, \wedge U^\dagger A_R\,U)\Big]+
\ee
\be \nonumber
+i\Big[ \tr(A_L\wedge F_L\,\wedge U^\dagger A_R\,U)+ \tr(A_L\, \wedge U^\dagger A_R\,U\wedge F_L)\Big]+
\ee
\be \nonumber
+2 i \tr(A_L\wedge A_L\wedge A_L\,\wedge U^\dagger DU) -2 \tr(A_L\wedge A_L\wedge A_L\, \wedge U^\dagger A_R\,U)
\ee
\be \nonumber
 +2\tr(A_L\, \wedge U^\dagger DU\, \wedge U^\dagger DU\, \wedge U^\dagger A_R\,U) -\tr(A_L\, \wedge U^\dagger DU\wedge A_L\, \wedge U^\dagger DU)+
\ee
\be \nonumber
-2 i \tr(A_L\, \wedge U^\dagger A_R\,U\wedge A_L\, \wedge U^\dagger DU)
-2 i\tr(A_L\, \wedge U^\dagger DU\, \wedge U^\dagger DU\, \wedge U^\dagger  DU)\Big\}+
\ee
\be \nonumber
+\ \Big(L\leftrightarrow R\Big) +\tr(A_L\, \wedge U^\dagger A_R\,U\wedge A_L\, \wedge U^\dagger A_R\,U) \ ,
\ee
where above $U = (TT^\dag)^{-1/2}T$.

Note that in this simpler case, where all the eigenvalues vanish on the defect, all of the terms that do not vanish in the $b\to\infty$ limit, are simple traces (instead of ${\cal L}$-traces).

In the rest of this section, we generalize the study of $\Omega_5^b$ to the case where the tachyon is not identically zero on $\mathcal{Z}$, i.e. we allow for some of the eigenvalues of the tachyon to remain nonzero on this manifold. We shall then obtain more complicated structures that are limits of ${\cal L}$-traces. An example can be seen in (\ref{eq:suma1}).

When some, but not all eigenvalues $\sigma_a$ vanish on $\mathcal{Z}$, the $\intd \sigma_a$ are no longer required to be orthogonal to $\mathcal{Z}$ and therefore $\intd \sigma_a \wedge \intd \sigma_b$ terms do not automatically vanish. As a consequence, one needs to keep track also of terms which have more than one $\intd T$ as they can a priori survive the $b\to \infty$ limit. These terms may be particularly relevant when considering defects with nonzero tachyon eigenvalues, and intersections of such defects. An example with a defect with $\sigma_1 = 0$ (in green,  in figure \ref{fig:defectinter}) intersecting another defect with $\sigma_2 = 0$ (in blue, in figure \ref{fig:defectinter}) is shown in figure \ref{fig:defectinter}. In this example, one may expect terms coming from $\Omega_5^b$ of the form
\be
\intd \sigma_1 \wedge \intd \sigma_2 \wedge G_3, \label{def:G3}
\ee
that are supported on the intersection of the two defects.

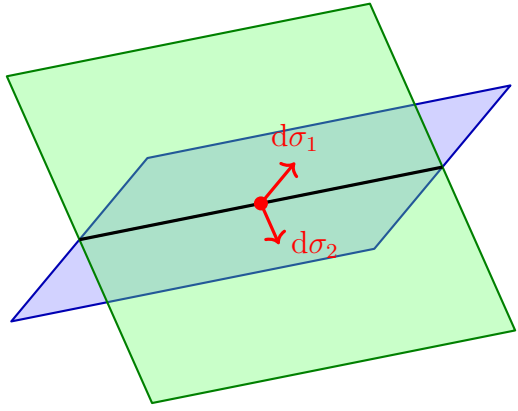
\begin{figure}[htb]
\centering
\begin{tikzpicture}[
    scale=1.2,
    line join=round
]

\coordinate (A) at (0,0);
\coordinate (B) at (4,0.8);
\coordinate (C) at (5.5,2.6);
\coordinate (D) at (1.5,1.8);

\filldraw[
    fill=blue!50,
    fill opacity=0.35,
    draw=blue!70!black,
    thick
]
(A)--(B)--(C)--(D)--cycle;

\coordinate (P) at ($(A)!0.5!(C)$);

\coordinate (L1) at ($(P)+(-2.0,-0.4)$);
\coordinate (L2) at ($(P)+(2,0.4)$);

\draw[dashed, thick] (L1)--(L2);

\coordinate (E) at ($(L1)+(0.8,-1.8)$);
\coordinate (F) at ($(L2)+(0.8,-1.8)$);
\coordinate (G) at ($(L2)+(-0.8,1.8)$);
\coordinate (H) at ($(L1)+(-0.8,1.8)$);

\filldraw[
    fill=green!60,
    fill opacity=0.35,
    draw=green!50!black,
    thick
]
(E)--(F)--(G)--(H)--cycle;

\draw[very thick, black] (L1)--(L2);

\draw[->, very thick, red]
(P) -- ++(0.4/2,-0.9/2)
node[right] {$\intd \sigma_2$};

\draw[->, very thick, red]
(P) -- ++(0.75/2, 0.9/2)
node[above] {$\intd \sigma_1$};

\fill[red] (P) circle (2.2pt);

\end{tikzpicture}

\caption{Illustration of a defect configuration which requires keeping terms with more than one $\intd T$ in $\Omega_5^b$. The green plane has $\sigma_1 = 0$ and the blue plane has $\sigma_2 = 0$. Their intersection is shown in black.}
\label{fig:defectinter}
\end{figure}

We shall not explicitly compute all of these terms, but they can be straightforwardly extracted by taking the limit $b\to \infty$ in \eqref{455} and in analogous formulae for the other words.

We shall now restrict the study to the structure of the terms with a single $\intd \s_a$ orthogonal to $\mathcal{Z}$ and study the large $b$ limit of the $\mathcal{L}$-traces. All the different words contain many different terms, all of which have to be checked for the different $\mathcal{L}$ operators possible for this word, where $\mathcal{L}$ was defined in \eqref{438}. Many of these operators are identical, and many of them vanish in the limit $b\to \infty$. Below we summarize some results:
\begin{itemize}
  \item $\mathcal{L}$ operators with all identical eigenvalues behave exactly as in the $\tau U$ case.
  \item For operators with two different eigenvalues, we need to check eight cases: $(\mathcal{L}_b) b^4 \sigma_a^3 \intd \sigma_a$, $(\mathcal{L}_b) b^4  \sigma_a^2 \sigma_b \intd \sigma_a$, $(\mathcal{L}_b) b^4  \sigma_a \sigma_b^2 \intd \sigma_a$, $(\mathcal{L}_b) b^4  \sigma_b^3 \intd \sigma_a$, and the same terms with $\intd \sigma_a$ replaced by $\intd \sigma_b$. However, some of these cases do not appear in the computation because of the symmetries of the word. A subset of these have divergent integrals, but the divergent ones do not arise in the computation. The rest have zero integral, except the first, $(\mathcal{L}_b) b^4 \sigma_a^3 \intd \sigma_a$. This expression becomes (a fraction of) a Dirac $\delta$-distribution as $b\to\infty$.
  \item All the operators with two different eigenvalue indices or more, behave the same way. The only terms that survive in the limit of $b\to\infty$, become  $\delta$-distributions. They all involve limits of the 1-form
  \be
  (\mathcal{L}_b) b^4 \sigma_a^3 \intd \sigma_a.\label{459}
  \ee
\end{itemize}
The previous discussion applies to all types of words\footnote{An example of term in $\Omega_5^b$ which contains all the cases listed above is the following one,
\be
\sum_{a_0\dots a_4} (\mathcal{L}_b)_{a_0 a_1 a_1 a_3 a_4} b^4 (i \widetilde{A}_L)_{a_0 a_1}
\intd \sigma_{a_1}  (\intd V_R V_R^\dag)_{a_1 a_3}\sigma_{a_3}  (V_L^\dag \intd V_L)_{a_3 a_4} \sigma_{a_4} (\intd V_R V_R^\dag)_{a_4 a_0} \sigma_{a_0}.
\label{abo}\ee}.

We now treat an explicit example, corresponding to a contribution to $\Omega_5^b$ which contains three times the Maurer-Cartan form $\alpha_L\equiv V_L^\dag \intd V_L$\footnote{There are four such contributions, corresponding to choosing for which of the $\tilde{X}$ we select $\intd \Sigma$ in \eqref{455}. The four terms behave in a similar way, so we may focus on \eqref{461} without loss of generality.},
\be
(\Omega_5^b)_{\alpha_L^3}\equiv
\sum_{a_0\dots a_4} (\mathcal{L}_b)_{a_0 a_1 a_1 a_3 a_4} b^4 \times \nonumber \ee\be \times (i \widetilde{A}_L)_{a_0 a_1}
\intd \sigma_{a_1} \sigma_{a_1} (\intd V_L^\dag V_L)_{a_1 a_3}  (V_L^\dag \intd V_L)_{a_3 a_4} \sigma_{a_4}^2 (\intd V_L^\dag V_L)_{a_4 a_0}.\label{461}
\ee
Using the limits of the $\mathcal{L}$ operators, all the terms vanish unless $\sigma_{a_1} = \sigma_{a_4}$, which is true if the eigenvalues are equal or if $a_1 = a_4$. In the first case there is no constraint on $a_1, a_4$ and all the terms in \eqref{461} survive. In the second case, using the constraint $a_1 = a_4$ we obtain
\be
(\Omega_5^b)_{\alpha_L^3}=
\sum_{a_0, a_1, a_3}  - \delta(\sigma_{a_1})\intd \sigma_{a_1} \wedge  (i \widetilde{A}_L)_{a_0 a_1} (\intd V_L^\dag V_L)_{a_1 a_3}  (V_L^\dag \intd V_L)_{a_3 a_1} (\intd V_L^\dag V_L)_{a_1 a_0},\label{462}
\ee
up to a constant multiplicative factor. Recognizing the sum over $a_0, a_3$ as a matrix product, we obtain
\be
(\Omega_5^b)_{\alpha_L^3}=
\sum_{a_1}  \delta(\sigma_{a_1})\intd \sigma_{a_1} \wedge  (\intd V_L^\dag V_L i \widetilde{A}_L)_{a_1 a_1} (\intd V_L^\dag \intd V_L)_{a_1 a_1}.
\label{eq:suma1}
\ee
Note that this expression is {\em not} a trace. A posteriori, we can see that the result strongly depends on how many eigenvalues of $\Sigma$ are set to zero on $\mathcal{Z}$. Indeed, if they are all zero, then the requirement $a_1 = a_4$ is not needed as $\sigma_{a_1} = \sigma_{a_4}$ is satisfied automatically. In this case, after taking the limit, we are left with a trace
\be
(\Omega_5^b)_{\alpha_L^3}=
 \delta(T)\intd T \wedge  \tr( \intd V_L^\dag V_L i \widetilde{A}_L \intd V_L^\dag \intd V_L).\label{464}
\ee
This term can be re-expressed in terms of $A_L$ as
\be
(\Omega_5^b)_{\alpha_L^3}=
 - \delta(T)\intd T \wedge  i \tr(A_L (\intd V_L V_L^\dag)^3),\label{465}
\ee
which is one of the terms appearing in the second to last term in \eqref{G4final}. It has the correct normalization once the overall coefficient is restored.

The generic terms in $\Omega_5^b$ are similar to \eqref{eq:suma1}, but their structure can be more complicated. Indeed, mixing different contributions in \eqref{457} and \eqref{458}, where $\Sigma$ is on the left and on the right of Maurer-Cartan forms for $V_L$ and $V_R$, there can be terms where the requirements on the indices are even more restrictive. For instance, \eqref{abo} requires $\sigma_{a_1} = \sigma_{a_3} = \sigma_{a_4} = \sigma_{a_0}$ to not vanish in the large $b$ limit, and if all the eigenvalues of $\Sigma$ are different, it gives
\be
\sum_{a_1}  \delta(\sigma_{a_1})\intd \sigma_{a_1} \wedge  (V_R^\dag \intd V_R)_{a_1 a_1} (V_L^\dag \intd V_L)_{a_1 a_1} (V_R^\dag \intd V_R)_{a_1 a_1}  (i \widetilde{A}_L)_{a_1 a_1}.
\label{eq:suma12}
\ee

Summarizing the results regarding $\Omega^b$, we obtained that for cases where the tachyon is zero on $\mathcal{Z}$, the expression in \cite{jknp} for $G_4$, reproduced  in \eqref{G4final}, still holds in the generic tachyon case. However, when it is not assumed that the tachyon is zero on $\mathcal{Z}$, but instead that only some of the eigenvalues $\sigma_a$ vanish there, we found that different terms like \eqref{eq:suma1}, \eqref{eq:suma12} can survive the large $b$ limit, and one needs to consider such non-trace terms. Lastly, if one considers different defect topologies, for instance defect intersections like figure \ref{fig:defectinter}, terms with several $\intd \sigma^a$ can arise from this construction, giving rise to possible corner terms in the TCS action.

\subsection{Properties and descent equations\label{properties}}

In this section, we show that the TCS form we computed from the path $\gamma$ (blue in figure \ref{plane}), respects the criteria defined {in section \ref{sec:criteria}}. As we already discussed, $\Omega$ descends from a Chern character by construction, so criterion \ref{cr:character} is satisfied by construction.

Criterion \ref{cr:tachzero}, i.e., that in the chirally symmetric phase the TCS action \eqref{314} reduces to the standard CS action \eqref{315}, is verified in appendix \ref{sec:limitT0}. It is argued that, possibly up to gauge-invariant boundary terms,
\be
\lim_{T \to 0}  \Omega_5 = \omega_5.\label{469a}
\ee

The parity part of criterion \ref{cr:discretesym} is also automatic. We now check that the full from  we obtained is $C$-even. For this, recall (see Appendix \ref{sec:discsym}) that the action of  charge conjugation is given by
\be
C(A_L, A_R, T, T^\dag) \equiv (-A_R^t, -A_L^t, T^t, T^{\dag,t}),
\label{eq:cdef}
\ee
where $t$ denotes the matrix transpose. Consequently, we obtain
\be
C(DT, DT^\dag) = (DT^t, DT^{\dag, t}).\label{469}
\ee
Therefore, the action of the charge conjugation on the letter $X$ in \eqref{43} is just to transpose it, while the action on $Y$ in \eqref{44} is to transpose and exchange $L$ and $R$. Then, the action of $C$ on the TCS action is to first exchange the order of the letters in the words, then to exchange the left and right traces in the supertrace and change the sign of every gauge field term. Both of these operations leave the overall sign unchanged because of the cyclicity of the trace and the grading property of the supertrace \eqref{34}. For instance, for the word $XXXY$ contributing to $\Omega_5^0$, the left contribution is sent to
\be
XXXY_L \to -YXXX_R\label{470}
\ee
Therefore, the TCS action is necessarily C-even, when it is written from a path formula as we wrote, and criterion \ref{cr:discretesym} is also automatically satisfied.

Next, we check the anomaly criterion \ref{cr:anomaly}. Generalizing the argument of \cite{Szabo:2001yd} to the character $\chi$, defined in \eqref{39}, we can write down the anomaly as a boundary form. To do so, we introduce the {nilpotent} BRST operator $\delta$. It is  defined as the operator that acts on fields, by performing an infinitesimal gauge transformation associated to \eqref{216}, \eqref{217}. Then, we define
\be
\Lambda \equiv V^\dag \delta V = \begin{pmatrix}
  \Lambda_L & 0\\
  0 & \Lambda_R
\end{pmatrix}, \qquad  V = \begin{pmatrix}
  V_1 & 0\\
  0 & V_2
\end{pmatrix}\label{471-b}
\ee
where $V_1 \in \text{U}(N_f)_L$ and $V_2 \in \text{U}(N_f)_R$. We choose conventions for $\Lambda$ such that $\delta$ and $\intd$ anticommute, and define the operator
\be
\mathcal{D} \equiv \intd + \delta, \qquad \mathcal{D}^2 = 0.\label{472}
\ee

Using a similar argument as the one used to derive \eqref{326} and replacing $\intd$ with $\mathcal{D}$, we can derive a similar transgression formula that computes the anomaly. This is done by constructing paths in superconnection space that go from $\Lambda$ to $\mathcal{A}+ \Lambda$ with this new operator $\mathcal{D}$. As already noted in \cite{Szabo:2001yd}, these paths construct the anomaly forms $I$ (see equations \eqref{319}-\eqref{320}), which depend on the path, such that the correct descent equations are satisfied,
\be
\delta \Omega_5 =  \intd I_4^1\;.\label{471}
\ee
Here, the index $1$ in $I_4^1$ indicates that $I$ contains a single power of $\Lambda$. Performing a similar generalization of the paths as we did for the construction of $\Omega$, and considering paths along which the tachyon and the gauge fields run at different rates, $a(s)$ and $b(s)$, we consider the superconnections
\be
\mathcal{D}_s \equiv \mathcal{D} - i a(s) A - i b(s) \mathcal{T} + \Lambda \equiv \mathcal{D} - i\mathcal{A}_s + \Lambda. \label{eq:dsano}
\ee
Its curvature $\mathcal{F}_s = \mathcal{D}_s^2$ can be substantially simplified and becomes
\be
i \mathcal{F}_{s} = i \mathcal{F}(a,b) + (1-a(s))\intd \Lambda, \label{475}
\ee
where $\mathcal{F}(a,b)$ was given in \eqref{332}.

We then introduce the lifted exterior derivative and Chern character following \cite{MSZ},
\be
\bar{\intd} \equiv \intd  + \delta + \intd s \pa_s ,  \quad \bar \chi \equiv \chi + \intd s  \chi_s  + \chi^{1}, \label{eq1}
\ee
where $\chi$ and $\chi_s$ are appropriate forms that are of order zero in $\Lambda$.
$\chi^1$ is a form of order one in $\Lambda$, which we indicate with the superscript $1$ in the following.
 We don't need to consider terms with more than one $\Lambda$ for our purposes, \cite{MSZ}. We derive the descent equations, using the closure of the extended Chern form on the principal fibre bundle,  $\bar{\intd} \bar \chi=0$, which reads
\be
\bar{\intd} \bar \chi = \intd\chi  + (\delta \chi + \intd \chi^{1}) + \intd s(\pa_s \chi - \intd \chi_s) +  \intd s(\pa_s \chi^{1} - \delta \chi_s )  = 0. \label{eq2}
\ee
We obtain
\be
\intd \chi=0\;, \qquad  \pa_s \chi^{1} - \delta \chi_s=0\;,
\label{eq3a}\ee

\be
\pa_s \chi = \intd \chi_s\;, \qquad  \delta \chi + \intd \chi^{1} = 0\;. \label{eq3}
\ee
The  first equation in (\ref{eq3}) is the descent equation that leads to \eqref{326}. Moreover, integrating this same equation over a path $\mathcal{C}$, we obtain (using equation \eqref{335}),
\begin{equation}
\label{eq4c}  \int_{\mathcal{C}} \intd s  \, \chi_s = \Omega^{(\mathcal{C})}.
\end{equation}
Using that the Chern character $\chi$ is gauge invariant, $\delta \chi = 0$, the second equation in \eqref{eq3} implies that $d\chi^1=0$ and therefore, locally $\chi^1$ is the exterior derivative of a $4$-form.

The second equation of (\ref{eq3a})  can be integrated over a path and combined with \eqref{eq4c}, to obtain
\begin{equation}
\label{eq4E} \delta  \Omega_5^{(\mathcal{C})} = \chi_5^{1}(s=1) - \chi_5^{1}(s=0),
\end{equation}
where we projected to the 5-form component. Substituting the descent equation \eqref{471} into (\ref{eq4E}), we obtain
\be
\intd I_4^1 = \chi_{5}^{1}(s=1) -   \chi_{5}^{1}(s=0).\label{eq4f}
\ee
Note that the right hand side of equation \eqref{eq4f} is the analogue of \eqref{329}, but for the generalized superconnection \eqref{eq:dsano} together with  selecting $5$-form terms with a single $\Lambda$. The derivation leading to \eqref{329} does not rely on a particular form of the superconnection, and can then be applied to the superconnection \eqref{eq:dsano}. To find $I_4^1$, we then apply \eqref{329} with $D_s$ substituted with $\mathcal{D}_s$, and select the terms of order $1$ in $\Lambda$. We obtain an integral formula, analogous to \eqref{335} for the anomaly, which takes the form
\be
I^{1}_4 = - \int_\mathcal{C} \intd s\, \str\Big( \pa_s \mathcal{D}_s \exp(i \mathcal{F}_s) \Big)_4^1.
\label{475b}
\ee
In \eqref{eq:dsano}, only $\mathcal{A}_s$ depends on $s$. Therefore, substituting \eqref{475} into \eqref{475b} yields
\be
I_4^1 = \int_{0}^1 \intd s\, \str\left(i \frac{\partial \mathcal{A}_s}{\partial s} \exp(i\mathcal{F}(a,b) + (1-a(s))\intd \Lambda)\right)_4^1\;,\label{476}
\ee
which can be integrated by parts to obtain
\be
 I_4^1  =  \int_{0}^{1}\intd s\, (1-a(s))\str\left(\Lambda \wedge \intd \left(i \frac{\partial\mathcal{A}_s}{\partial s} \wedge e^{i\mathcal{F}(a,b)}\right)\right)_4^1.
\label{eq:anompath}
\ee
Observe that $I_4^1$ vanishes for the paths at $a(s)=1$, consistently with the fact that the actions obtained from these paths are gauge-invariant.

We now apply this result to the alternative path $\gamma_\mathrm{alt}$, and to the path $\gamma$ that was used to compute $\Omega$, both of which are defined in figure \ref{plane}. Treating first $\g_{\text{alt}}$, we note that the anomaly from $II_\mathrm{alt}$ vanishes as it has $a(s)=1$. The anomaly then originates from the path $b=0, a=s$. For the alternative path, the  anomaly form $I_{4}^{1} (\gamma_\mathrm{alt}) $ is precisely the QCD anomaly arising from $\omega_5$ in \eqref{319},
\be
 I_{4}^{1} (\gamma_\mathrm{alt})  \equiv  \int_{0}^{1}\intd s\, (1-s)\str(\Lambda \wedge \intd (i A \wedge (is \intd A + s^2 A^2))) = -\frac{1}{6} \str(\Lambda \wedge \intd( A\intd A  - \frac{i}{2} A^3))\label{478}.
\ee
Equation \eqref{478} shows that $\gamma_\mathrm{alt}$ satisfies criterion \ref{cr:anomaly}, but it contains an a priori nontrivial contribution from the infrared, requiring specific boundary conditions in the IR. This arises because $\gamma_\mathrm{alt}$ does not satisfy criterion \ref{cr:noIR} (see appendix \ref{sec:gammaalt}).

We now return to the main path of interest, $\gamma$. For the portion $I$ of the path that generates $\Omega_5^c$ (with $a=0, b=s$), and in the case $T=\tau U$, we can check that we obtain the anomaly from the WZW-term
\begin{equation}
I_4^1 (I) \equiv \int_{0}^{+\infty} \intd s \str(\Lambda \wedge \intd (i \mathcal{T} \wedge e^{i\mathcal{F}(0, s)})) \nonumber \ee\be = -\frac{1}{12}\tr(\Lambda_L \wedge \intd (U^\dag \intd U)^3) - (L\leftrightarrow R, U\leftrightarrow U^\dag).\label{477}
\end{equation}

For the portion $II$ which generates $\Omega^b$, a subtlety arises due to the localization of this action on $\mathcal{Z}$, where $\mathcal{Z}$ is defined in \eqref{333}, that is tied to $\intd_\tau$, as defined in \eqref{428}. Similarly to \eqref{428}, in the generic tachyon case, one can define
\be
\intd_{\sigma_a}\equiv \delta(\sigma_a) \intd \sigma_a \wedge (-.),\label{480}
\ee
which defines a set of $N_f + 1$ different operators, $\{\intd,(\intd \sigma_a)_{1\leq a\leq N_f}\}$. The $\intd_{\sigma_a}$'s all anticommute with each other and with $\intd$, and they all square to zero. Equation \eqref{eq:anompath} misses contributions from $\intd_{\sigma_a}$-exact terms and must be generalized in order to compute correctly the anomaly in the presence of these other operators.

To retrieve the gauge variation of $\Omega_5^b$, we rely instead on the explicit expression \eqref{G4T}, which is valid for defects with $T=0$. The generalization of the descent equations to the $N_f + 1$ operators $\{\intd,(\intd \sigma_a)_{1\leq a\leq N_f}\}$, which we do not derive here, could be used to derive a generic formula without relying on this assumption. It was shown in \cite{jknp} that in the case where the explicit expression~\eqref{G4T} is valid, the variation of $\Omega_5^c$ and $\Omega_5^b$ reproduces the QCD anomaly \eqref{320}, and then $\Omega$ satisfies also \ref{cr:anomaly}.

Note that the path $\gamma$, where the portion $II$ is sent to infinity, is such that the anomaly localizes on $\mathcal{Z}$, since $\Omega^b$ itself localizes on $\mathcal{Z}$. As we shall see in the next section, this also has topological implications. Note that $\intd_{\sigma_a}$-exact terms affect only the anomaly on $\mathcal{Z}$. In the IR, the tachyon asymptotes to infinity, so the anomaly contribution from $\Omega_5^b$ is zero there. Hence, because $\Omega_5^0$ is gauge-invariant and $\Omega_5^b$ contributes only on $\mathcal{Z}$, only $\Omega_5^c$ can produce an IR contribution to the anomaly. This contribution depends only on $U$ as can be seen from \eqref{477}. Therefore, there is no anomaly contribution from the infrared unless $\intd U$ is nonzero there. The IR contribution to the anomaly from $\Omega_5^c$, which is given by $I_4^1$ in equation \eqref{477}, vanishes due to the IR regularity condition on $U$.

\section{The massive quark baryon}
\label{sec:bar}

In this section, we use the action $\Omega$,{ defined in \eqref{eq:Omegagen} and determined in the previous section} to address the problem of constructing baryonic solutions in the holographic theory of V-QCD with massive quarks. When quarks are massive, the near-boundary expansion of $T$ reads
\be
T =  (m_q r (-\log(r \Lambda_{UV}))^{-\rho} + \sigma_q r^3 \mathcal{U}(-\log(r \Lambda_{UV}))^{\rho})\left(1 + \mathcal{O}(\frac{1}{\log(r \Lambda_{UV})})\label{51}\right).
\ee
In \eqref{51} the quark mass $m_q$ and chiral condensate $\sigma_q$ are generic hermitian matrices, which were defined in \eqref{25-i}. $\mathcal{U}$ is the unitary pion matrix, and $\rho$ and $\Lambda_{UV}$ are two parameters of the V-QCD model.  If $m_q$ and $\sigma_q$ are proportional to the identity matrix, i.e. if the quark masses are all equal, then the tachyon can be written $T=\tau U$ and using the generic TCS action is not necessary. However, if $m_q$ or $\sigma_q$ are not proportional to the identity, then $T$ cannot be written $T= \tau U$ and the generic TCS action is needed.

\subsection{Boundary conditions and definition}

Consider the action in the bulk $\mathcal{M}$, composed of the TCS action
\be
S_{TCS} = \frac{i}{4\pi^2} \int_{\mathcal{M}} (\Omega_5^0 + \Omega_5^c + \Omega_5^b),
\label{eq:Scs}
\ee
defined in \eqref{eq:Omegac}, \eqref{eq:Omegab} and  \eqref{eq:Omega0}, and the DBI action, \eqref{210}. We separate the $\text{SU}(N_f)$ part and the $\text{U}(1)_V$ field $\Phi$ as $A_L = \Phi + L$, $A_R = \Phi + R$.

In \cite{jknp}, where the mass of the quarks was set to zero, the baryon number was carried by the skyrmion winding of $U$ in (\ref{429})  at the boundary.
On the other hand, from \eqref{51}, we deduce that in the presence of quark masses, $U$ must asymptote to the identity matrix at the boundary\footnote{We recall that $U$ is defined in general as the unitary part in the polar decomposition of the tachyon \eqref{429}}. Therefore, in order to have a solution with nontrivial baryon number, the topology of the solution in the massive quark case cannot be the same as for massless quarks.

We show below that a solution carrying baryon number, can be obtained with massive quarks, by considering a (single) point-like defect in the bulk, at a point $(r=r_*, x=0)$, where $T=0$. This was considered in \cite{Gorsky:2012eg, Gorsky2015} in the simpler case of the hard-wall holographic QCD. Note that the location of the defect {belongs to} $\mathcal{Z}$ as defined in \eqref{333}. We  show that this defect can carry a nonzero instanton number, which in turn equals the baryon number in the boundary theory. {The defect describes a single baryon, as we verify by computing explicitly the baryon number.}

As explained in the previous section, the unitary part of the tachyon $U$ is not well-defined on $\mathcal{Z}$, defined in \eqref{333}, and in particular it can wind around $r_*$. To study the topology of the solution, one should excise a small ball around $r_*$ and study the field configuration on the small sphere $S^3(r_*)$ of infinitesimal radius. {Field configurations that wind around $S^3(r_*)$ can then contribute to the baryon current.} We now show that if the fields around the defect have appropriate winding, this results in a nonzero boundary baryon number.

In the following, we work in the $A_r=0$ gauge, which is the reference gauge for the holographic dictionary. However, after we develop our tools and formulae, we can translate the baryon number in any gauge, provided the gauge transformations do not change the topology of the bulk configuration.

We are interested in the value of the baryon current density, defined as
\be
j^M \equiv \frac{\delta S_{\text{on shell}}}{\delta \Phi_M(x, r=0)},\label{53}
\ee
where $S_{\text{on shell}}$ is the on-shell action and $\Phi$ the U$(1)_V$ gauge field. The variation of the action is
\be
\delta S = \int_{\mathcal{M}} \bigg( \frac{\partial L}{\partial \Phi_M} \bigg) \delta \Phi_M +  \bigg( \frac{\partial L}{\partial \partial_N \Phi_M} \bigg) \delta \partial_N \Phi_M ,\label{54}
\ee
where we denoted by $L$ the total Lagrangian corresponding to the DBI and TCS actions. Integrating by parts the previous expression, we obtain for the on-shell action variation
\be
\delta S_{\text{on shell}} = \int_{\partial \mathcal{M}} \bigg( n_N \frac{\partial L}{\partial \partial_N \Phi_M}  \bigg)\delta \Phi_M,
\label{c6}
\ee
with $n_N$ a normal vector to the boundary $\partial \mathcal{M}$. There are three pieces in $\partial \mathcal{M}$, that we decompose as\footnote{We follow~\cite{jknp} for the notation of the components of $\partial \mathcal{M}$.}
\be
\partial \mathcal{M} \equiv (B_3 \cup D_3 \cup S^3(r_*)) \times \mathbb{R}_t,\label{520}
\ee
where $\mathbb{R}_t$ is the time line. $B_3$ is the UV boundary at $r=0$, $D_3$ is spatial infinity for $r>0$, and $S^3(r_*)$ is the small sphere around the excised defect. This leads to a priori three contributions to the baryon current. Our boundary conditions will be such that the contributions to the baryon number from spatial infinity ($D_3$) vanish. We can therefore rewrite the variation of the on-shell action as
\be
\delta S_{\text{on shell}} =  \int_{B_3\times \mathbb{R}_t} \intd^4x \bigg(- \frac{\partial L }{\partial \partial_r \Phi_M}
 \bigg) \delta \Phi_M(x, r=0)+ \nn \ee\be
  + \int_{S^3(r_*) \times \mathbb{R}_t} \intd^4x' \bigg( n_N \frac{\partial L }{\partial \partial_N \Phi_M}
   \bigg) \delta \Phi_M(x', r=r_*).
\label{Sonshell}
\ee
The variation of $\Phi$ at $r_*$ is a priori challenging to express as a function of the variation of the source $\Phi(r=0)$. The relation can in principle be written in term of the kernel\footnote{Note that $K$ is the boundary-to-bulk propagator evaluated at $r_*$} $K(x,x')$ as
\be
\delta \Phi_P(x', r=r_*) = \int_{B_3\times \mathbb{R}_t} \intd^4x \;K(x', x,r_*)\; \delta \Phi_P(x, r=0).\label{57}
\ee
Substituting \eqref{57} into (\ref{Sonshell}), we obtain an expression for the boundary baryon current in terms of the kernel $K$
\be
j^M = \bigg(- \frac{\partial L }{\partial \partial_r \Phi_M}  \bigg)(x) + \int_{B_3 \times \mathbb{R}_t} \intd^4x' K(x', x) \bigg( n_N \frac{\partial L }{\partial \partial_N \Phi_M}  \bigg)(x').
\label{JM}
\ee
Note that the position of the singularity $r_*$ and the exact computation of the kernel $K(x,x')$ are dynamical problems, which would require solving the equations of motion.

However, as we shall show, it is not necessary to know the kernel $K$ in order to compute the integrated baryon charge $N_B$. Indeed, the equations of motion are such that $\Phi$ only appears with derivatives, so that from a given solution to the equations of motion $\bar{\Phi}$, we may construct another solution\footnote{Our IR boundary condition for $\Phi$ is $\partial_r \Phi = 0$, which is still satisfied after a constant shift of $\Phi$.\cite{jknp}} $\Phi$ by performing a constant variation $\delta \Phi$ everywhere in the bulk,
\be
\Phi(x, r) = \bar{\Phi}(x, r) + \delta \Phi.\label{59}
\ee
where $\delta \Phi$ is independent of coordinates. This represents a variation of the baryon chemical potential $\mu_B$, instead of a generic variation of $\Phi(x, r)$
\be
\delta \Phi(x, r) = \delta \mu_B.\label{510}
\ee
We substitute \eqref{510} in \eqref{57} to find
\be
\int_{B_3\times \mathbb{R}_t }\intd^4x K(x,x') = 1 \label{510b}.
\ee
Then, substituting \eqref{510b} into \eqref{JM} we obtain the following expression for the baryon number,
\be
N_B =  \int_{B_3} \intd^3x \bigg(- \frac{\partial L}{\partial \partial_r \Phi_0} \bigg)  + \int_{S^3(r_*)} \intd^3x' \bigg( n_N \frac{\partial L}{\partial \partial_N \Phi_0} \bigg).
\label{eq:NB1}
\ee

Note that it is also possible to compute the integrated charge $N_B$ by substituting \eqref{510} directly into \eqref{54} with $\delta \partial_N \Phi_M = 0$, without integrating by parts and using the equations of motion explicitly. This method gives
\be
N_B = \frac{\partial S_\text{on shell}}{\partial \mu_B}.
\label{eq:baryonnumber}
\ee
Because the DBI Lagrangian, $\Omega_5^c$ and $\Omega_5^0$ do not contain $\Phi$, but only $\pa \Phi$ terms, the transformation \eqref{59} leaves them invariant and therefore only $G_4$ contributes in \eqref{eq:baryonnumber}. However, to obtain the expression \eqref{eq:NB1}, we used the equations of motion explicitly and we shall therefore consider possible contributions from the DBI and from $\Omega_5^c$ and $\Omega_5^0$. Both expressions \eqref{eq:NB1} and \eqref{eq:baryonnumber} can be used to compute $N_B$ and it can be checked that they give the same result.

\subsection{The baryon number}

We start from expression \eqref{eq:NB1}, which can be rewritten as
\be
N_B = \int_{\partial \mathcal{M}} \frac{\partial L}{\partial \intd \Phi}, \label{513}
\ee
where the integral is performed on the spatial part of $\partial \mathcal{M}$ only (ie. at fixed time).
This expression may receive contributions both from the TCS and DBI actions.
We start by discussing the TCS contribution and postpone the analysis of the DBI contribution, which requires results that are  introduced later in this section, and then shown in appendix \ref{sec:instanton}. It will be shown that the DBI contribution to $N_B$ vanishes.

For the TCS terms, note that $\Omega^c_5$ cannot contribute, as it does not depend on $\intd \Phi$. We consider the contribution to \eqref{513} from the TCS action, computed on a given path $\bar{\gamma}$, i.e. $\partial \Omega^{(\bar{\gamma})}/\partial \intd \Phi$, where $\Omega^{(\bar{\gamma})}$ was given in \eqref{335}. We obtain
\be
\frac{\partial \Omega_5}{\partial \intd \Phi} = \int_{0}^{1} \intd s\, \bigg[ a'(s) \str\left(i A \frac{\partial}{\partial \intd \Phi} e^{i\mathcal{F}(a,b)}\right) + b'(s) \str\left(i \mathcal{T} \frac{\partial}{\partial \intd \Phi} e^{i\mathcal{F}(a,b)}\right) \bigg]_\text{3-form}.\label{514}
\ee
This may rewritten in a simpler form. Indeed, the variation with respect to $\intd \Phi$ of the Chern character density can be computed from the expansion in words, \eqref{48}-\eqref{410}. The derivation $\partial /\partial \intd \Phi$ removes one $Y$ from each words and adds a factor $i a(s)$. The newly obtained words are then simply the words in $\Omega_3$. In other words, we may write
\be
\frac{\partial \Omega_5}{\partial \intd \Phi} = \int_{0}^{1} \intd s\, \  i a(s)\bigg[ a'(s) \str(i A e^{i\mathcal{F}(a,b)}) + b'(s) \str(i \mathcal{T} e^{i\mathcal{F}(a,b)}) \bigg]_\text{3-form}.\label{515}
\ee
From \eqref{331} and \eqref{335} this is
\be
\int_0^1\intd s a(s) \Omega_{3,s},  \quad
\Omega_3 \equiv \int_{0}^{1} \intd s\, \Omega_{3,s},\label{516-ii}
\ee
where $\Omega_{3,s}$ denotes the density of the TCS three-form, with $s$ being here the path coordinate on the full path that defines $\Omega$. We have, after restoring the normalization factor $i/4\pi^2$ of the TCS action,
\be
N_B = -\frac{1}{4\pi^2} \int_{\partial \mathcal{M}} \int_{0}^{1} \intd s\, a(s) \Omega_{3,s} + \int_{\partial \mathcal{M}} \frac{\partial L_{DBI}}{\pa \intd \Phi},\label{516}
\ee

In \eqref{516}} $L_{DBI}$ is the DBI Lagrangian \eqref{210}. We now compute the first term in \eqref{516}, that is the baryon number from the TCS action. Using \eqref{516-ii}, we obtain
\be
N_B = \frac{i}{48\pi^2}\int_{\mathcal{Z}} (-4i \tr(A_L F_L) + \tr(A_L^3) + 4i \tr(A_R F_R)  - \tr(A_R^3)) - \frac{1}{4\pi^2}\int_{\partial \mathcal{M}} \Omega_3^0 \nn\ee\be + \int_{\partial \mathcal{M}} \frac{\partial L_{DBI}}{\pa \intd \Phi}
\label{NB2}
\ee
with $\Omega_3^0$ expressed in \eqref{eq:Omega3final}. Recall that $\partial \mathcal{M}$ can be decomposed as in \eqref{520}, while for the point-like defect considered here $\mathcal{Z}$ can be decomposed as
\be
 \mathcal{Z} = (B_3 \cup S^3(r_*)) \times \mathbb{R}_t.\label{eq:decompZ}
\ee
We shall now analyze where the various terms in \eqref{NB2} have support:
\begin{itemize}
  \item The $B_3$ contributions of all three integrals in \eqref{NB2} vanish using the UV boundary conditions, i.e. the absence of sources for the gauge fields $A_L = A_R=  0$, and $U= \mathbb{I}$.

  \item The $D_3$ contributions coming from terms containing $F_L$ and $F_R$ vanish. Indeed, requiring finiteness of the DBI contribution to the energy imposes that $F_L$ and $F_R$ vanish at spatial infinity. $A_L$ and $A_R$ are then pure gauge, and we  show later that $\tr(A_L^3 - A_R^3)$ does not contribute either due to the radial gauge condition $A_{L,r} = A_{R,r} = 0$. Finiteness of the energy requires that the contribution from $U$ in $\Omega_3^0$ in $D_3$ also vanishes.
  \item Using the results of appendix \ref{sec:instanton}, we find that the contribution of the DBI term at the defect vanishes.

\end{itemize}

One can worry about possible IR contributions to \eqref{NB2}. Notice that in the integration over $\mathcal{Z}$, defined in \eqref{333}, there is no IR term as the tachyon diverges to infinity there. The asymptotics of the tachyon also imply that the DBI contribution from the IR vanishes. The only possible IR contributions to the baryon number then come from the integral of $\Omega_3^0$. We  consider boundary conditions such that the IR contribution from $\Omega_3^0$ vanishes as it was done in \cite{jknp}, i.e. it is assumed that there is no winding number in the IR.

Therefore, the only possible contributions to the baryon number come from the first and second integrals in $S^3(r_*)$. It is argued in appendix \ref{sec:instanton} that in the presence of a TCS term and a bi-fundamental scalar field action, the finiteness of the action implies that $F_L$ and $F_R$ vanish at $r_*$. We therefore obtain
\be
N_B = \frac{i}{48\pi^2} \int_{S^3(r_*)}  \bigg[ \tr(A_L^3) - \tr(A_R^3)  \bigg] - \frac{1}{4\pi^2}\int_{S^3(r_*)}\Omega_3^0  \label{eq:NB3},
\ee
where $\Omega_3^0 $ is given by \eqref{eq:Omega3final}.

Expression \eqref{eq:NB3} can also be derived without using the equations of motion. Indeed, starting again from \eqref{eq:baryonnumber}, the only term in the action that depends on $\Phi$ but not $\intd \Phi$ is\footnote{Recall that here we consider variations $\delta \Phi$ that are independent of $x$ and $r$, \eqref{510}.} $G_4$, which then leads to:
\be
N_B = \int_{\mathcal{Z}} \frac{\pa G_4}{\pa \Phi}.\label{519}
\ee
The integrand can then be computed directly from the explicit expression \eqref{G4final}. After some algebra, we obtain the same result as \eqref{eq:NB3} up to a total derivative term, which does not contribute around the defect where $T = 0$, as the sphere $S^3(r_*)$ surrounding it has no boundary, and vanishes elsewhere. Equation \eqref{eq:NB3} is the direct generalization of the integrated current found in \cite{jknp} {in equation (6.1)}. The {gauge field contribution $\tr(A_L^3 - A_R^3)$} comes from the standard CS action $\omega_5$, which multiplied by $\theta_{\mathcal{Z}}$ {in \eqref{422}}.

We shall now show that the first integral in \eqref{eq:NB3} vanishes in the radial gauge. Because $F_{L}, F_R$ have only zero components on $S^3(r_*)$, $A = A_L, A_R$ can be rewritten in a pure gauge form (with $V\equiv V_L, V_R$) as
\be
A = i V \intd V^\dag + \mathcal{O}(\xi).
\label{eq:puregauge}
\ee
This implies that around the defect, the integrals of $\tr(A^3)$ are integrals of a Maurer-Cartan form $V \intd V^\dag$ cubed,
\be
\frac{i}{48\pi^2} \int_{S^3(r_*)}  \bigg[ \tr(A_L^3) - \tr(A_R^3)  \bigg] = \frac{1}{48\pi^2} \int_{S^3(r_*)}  \bigg[ \tr((V_L \intd V_L^\dag)^3) - \tr((V_R \intd V_R^\dag)^3) \bigg] \label{521-ii},
\ee
which is a closed but non-exact form on $S^3$ (and at spatial infinity). Therefore, these integrals are quantized by cohomology
\be
\frac{i}{24\pi^2}\int \tr(A_{L/R}^3) \in {\rm H}^3(S^3; \mathbb{Z}) = \mathbb{Z}.\label{522}
\ee
Therefore, the integrals of $\tr(A_L^3)$ and $\tr(A_R^3)$ in \eqref{eq:NB3}, \eqref{521-ii} are half-integers. Moreover, in the radial gauge $(A_L)_r = (A_R)_r = 0$, the integers in  \eqref{522} vanish. This can be shown as follows. In virtue of equation (\ref{eq:puregauge}), the radial  gauge fixing implies

\be
\partial_r V_L={\cal O}(\xi)\sp  \partial_r V_R = {\cal O}(\xi)\;.\label{523}
\ee

Therefore, $V_L$ and $V_R$ do not depend on the holographic coordinate $r$. {Consequently, $V_L$ and $V_R$ are equal in two points facing each other on the $S^3$.} This implies that for the sphere of radius $\xi = \epsilon$ around $r_*$, $V_L$ and $V_R$ are determined by a map from a closed ball of radius $\epsilon$ into $SU(2)$. Because balls are contractible, {they have trivial nonzero homotopy groups,} and therefore $V_L$ and $V_R$ cannot wind around $r_*$. Therefore, in the radial gauge we obtain

\be
\int_{S^3(r_*)} \tr(A_{L}^3) = \int_{S^3(r_*)} \tr(A_{R}^3) = 0\label{524}
\ee
By a similar argument, the same contribution vanishes on $D_3$.

Combining \eqref{eq:NB3} and \eqref{524} we obtain
\be
N_B = - \frac{1}{4\pi^2}\int_{S^3(r_*)}\Omega_3^0  \label{eq:NB5},
\ee
The term $\Omega_3^0$ in \eqref{eq:NB5} is the gauge-invariant TCS 3-form {defined in \eqref{eq:Omega0}}. In equation \eqref{eq:NB3}, only the first term can introduce a nonzero gauge variation (under large gauge transformations), but these transformations are forbidden as it can be shown that they introduce a nonzero anomaly from the defect. We conclude that $N_B$ is indeed gauge-invariant.

We shall show later that $N_B$ is also quantized by topology, as it is the second Chern number of the (suitably compactified) bulk.

We consider as above the small sphere $S^3(r_*)$ around the defect, and expand the ${\cal L}$-traces in $\Omega_3^0$ in the size of the sphere in \eqref{eq:Omega3final}. Then, we use once more the fact that in the small size limit, the integrals involving $F_L$ and $F_R$ vanish, as derived in appendix \ref{sec:instanton}. Finally, using the fact that the windings of $A_L $ and $A_R$ are zero in the radial gauge as shown in equation \eqref{524}, equation \eqref{eq:NB5} becomes
\be
N_B = -\frac{1}{4\pi^2} \int_{S^3(r_*)} \Omega_3^0 = -\frac{1}{24\pi^2} \int_{S^3(r_*)}   \tr((U^\dag \intd U)^3)\;.
\label{eq:NBfinal}
\ee
Equation (\ref{eq:NBfinal})  shows that the baryon number is equal to a ``skyrmion winding number" at $r_*$ in this setup.

\subsection{The connection with the baryon number of the Skyrme model}
We would like to connect the baryon number defined and calculated in the previous subsection, to the skyrmion number defined from the pion matrix $\mathcal{U}$, \cite{Witten:1983tw,wit-bar-2}. The pion matrix is related to the vev of the $\bar q q$ bilinear and therefore,  to the tachyon near-boundary expansion in \eqref{51}.

For the baryon solution, the spatial UV boundary is to be thought of as a sphere $S^3$ with the sphere $S^2$ at infinity being identified to a point\footnote{This is because finiteness of the tachyon kinetic term imposes that $\mathcal{U}$ has the same value on the $S^2$ at infinity. Therefore, the field $\mathcal{U}$ is described by a map from the compactified UV, $S^3$, into the gauge group.}.

We are considering solutions such that the boundary expansion of $T$ is continuous when $m_q\to 0$. In the massless case, \eqref{51} implies $U= \mathcal{U}$, where $U$ is the unitary part of the tachyon.
In the massive case, $U=1$ at the  UV boundary,  but $U$ is now winding around $r_*$.
Since the winding numbers must be integers, for the winding-number-one case, continuity in the massless limit implies
\be
-\frac{1}{24\pi^2} \int_{B_3} \tr((\mathcal{U}^\dag \intd \mathcal{U})^3) = -\frac{1}{24\pi^2} \int_{S^3(r_*)} \tr((U^\dag \intd U)^3)  = 1,
\label{eq:equalwinding}
\ee
{where $B_3$ and $S^3(r_*)$ are defined in \eqref{520}.}

This equality shows that the baryon number is indeed equal to the skyrmion number, as defined from the vev of the tachyon. We shall show, that it implies that the local skyrmion current can be expressed as the pullback of the baryon current in $r_*$ up to an improvement term.

We consider a {smooth function $\Psi$} of degree $\deg(\Psi) = 1$, \footnote{A continuous map $\Psi$ from $S^3$ to $S^3$ induces an endomorphism in the de Rham homology group $H_3(S^3) = \mathbb{Z}$. By acting on an element of $H_3(S^3)$ associated to the integer $m$, it gives an element of the same group associated to the integer $n \times m$, where $n \in \mathbb{Z}$ is the degree of $\Psi$.}
\be
\Psi : S^3_{UV} \equiv \mathbb{R}^3_{UV} \cup \{\infty\} \to  S^3_{r_*},\label{526}
\ee
(the stereographic projection is an example), then one has that the pullback of $U(x', r_*)$ by $\Psi$, which we denote
\be
V \equiv \Psi^*(U),\label{527}
\ee
also has winding one in the UV boundary, because
\be
n_V = n_U \times \deg(\Psi) = n_U,
\label{eq:windpullback}
\ee
where $n_V$ and $n_U$ are respectively the winding numbers of $V$ and $U$. Therefore, the winding numbers of $V$ and $\mathcal{U}$ are equal. Since the compactification of the UV boundary is topologically an $S^3$, both $V$ and $\mathcal{U}$ are associated to the same element in $\pi_3(S^3) = \mathbb{Z}$. Since $V$ and $\mathcal{U}$ are homotopic, there exists a homotopy map $H(., s)$ continuous in $s$ satisfying
\be
H(., 0) = V,\qquad H(., 1) = \mathcal{U}.\label{529}
\ee

This homotopy alone is however not enough to express pointwise $\mathcal{U}$ as a function of $U$ {on $S^3(r_*)$}. We want to connect the baryon current in $r_*$,
\be
J_U = * \tr((U^\dag \intd U)^3),
\label{eq:JB2}
\ee
to the Skyrme current,
\be
J_\mathcal{U} = *_\pa \tr((\mathcal{U}^\dag \intd \mathcal{U})^3).
\label{eq:JSkyrme}
\ee
In \eqref{eq:JB2} the $*$ denotes the Hodge dual with respect to the bulk metric pulled-back to the defect $S^3(r_*) \times \mathbb{R}_t$, while in \eqref{eq:JSkyrme} $*_\pa$ is the Hodge dual with respect to the boundary. For this purpose, we define the winding number density
\be
w_U \equiv \tr((U^\dag \intd U)^3).\label{531}
\ee
{Using properties of the pullback,} we obtain,
\be
w_V \equiv \tr((V^\dag \intd V)^3) = \tr((\Psi^*(U^\dag) \intd \Psi^*(U))^3) = \Psi^*w_U
\label{eq:winddensity}
\ee
We act with $\intd$ on \eqref{eq:winddensity},
\be
\intd w_V = \intd \Psi^*w_U = \Psi^* \intd w_U = \Psi^* \tr((U^\dag \intd U)^4) = \Psi^* 0 = 0.\label{533}
\ee
by graded cyclicity of the trace. Moreover, we can act with $\intd$ on $w_\mathcal{U}$ to find the same result,
\be
\intd w_\mathcal{U} = \tr((\mathcal{U}^\dag \intd \mathcal{U})^4) = 0.\label{534}
\ee
Now, this means that both $ w_\mathcal{U}$ and $ w_V$ are closed forms, living on the compactified $S^3_{UV}$. Moreover, thanks to \eqref{eq:equalwinding} and \eqref{eq:windpullback}, we obtain
\be
n_\mathcal{U} = n_U = n_V\label{535}.
\ee
We are therefore comparing the winding number of two three-forms on $S^3$, which takes its value in
\be
\pi_3(S^3) = \mathbb{Z}.\label{536}
\ee

The Hurewicz theorem implies the following isomorphism\footnote{{In \eqref{537}, $\pi_3$ is the third homotopy group. ${\rm H}_3(S^3)$ is the third homology group of $S^3$. ${\rm H}^3(S^3, \mathbb{Z})$ is the third cohomology group of $S^3$, describing the closed differential forms with integer coefficients modulo the exact ones on the manifold.}}, see for instance \cite{hatcher}
\be
\pi_3(S^3) \cong H_3(S^3) \cong H^3(S^3, \mathbb{Z}) = \mathbb{Z}\label{537}.
\ee
The second isomorphism comes from the fact that the homology groups of spheres are torsion-free.

$\mathcal{U}$ and $V$ are in the same homotopy class, so $w_\mathcal{U}$ and $w_V$ are in the same cohomology class. By definition of the cohomology group we obtain
\be
w_\mathcal{U} = w_V + \intd \alpha_2,\label{538}
\ee
where $\alpha_2$ is a given two-form which encodes the point-wise mismatch of $V$ and $\mathcal{U}$.
 We can use these to rewrite the currents as
\be
J_\mathcal{U} = *w_\mathcal{U} =  \Psi^*J_U + *_\partial \intd \alpha_2 \equiv J_V + *_\partial \intd \alpha_2,
\label{eq:localcurr}
\ee
where we denoted $ *_\partial $ the Hodge star relative to the boundary coordinates. It is clear from \eqref{eq:localcurr} that the pulled-back current from $S^3_*$ to the UV and the Skyrme current differ by a form which has no impact on the conservation equation,
\be
\intd *_\partial J_\mathcal{U} = \intd *_\partial \Psi^*J_U  + \intd *_\partial *_\partial \intd \alpha_2 = \intd *_\partial \Psi^*J_U = \intd *_\partial J_V.\label{540}
\ee
because $*_\partial^2$ is a sign and $\intd^2=0$. Therefore, we are free to use $J_V$ as the baryon current, up to a local counterterm that does not contribute to the conservation equation.

\subsection{The topological bulk current and instanton number}

In this subsection, we derive the generalization of the instanton number in the presence of general tachyon field. We also  identify the bulk instanton number with the baryon number (\ref{eq:NBfinal}) we found. {Ignoring the overall normalization factor, the instanton number density at zero tachyon is}
\be
\chi_4(T=0) = -\frac{1}{2} (F_L^2 - F_R^2)\label{541}.
\ee
The natural generalization of the instanton number density was found by Quillen in \cite{Quillen}, and is the 4-form contribution to the Chern character \eqref{39},
\be
\chi_4 = \str(e^{i \mathcal{F}})\big|_{4-form}.\label{542}
\ee
Note that this quantity is gauge-invariant and closed, \eqref{311}. We already have the tools to write down the formula for $\chi_4$ for a generic tachyon field. {We perform the computation of $\exp(i\mathcal{F})$ and then follow the same steps as when computing $\exp(i\mathcal{F}(a,b))$ in the previous section. Instead of $\mathcal{F}(a(s),b(s))$, here we only need $\mathcal{F}(1,1)$, so we obtain traces weighted by $\mathcal{L}(1)$, {defined in \eqref{438}}}
\be
\chi_4 = \tr_{\mathcal{L}(1)}(DT^\dag, DT, DT^\dag, DT, \mathbb{I}) - \tr_{\mathcal{L}(1)}(DT, DT^\dag, DT, DT^\dag, \mathbb{I}) \nonumber
\ee
\be
-i \tr_{\mathcal{L}(1)}(DT^\dag, DT, F_L, \mathbb{I}) + i \tr_{\mathcal{L}(1)}(DT , DT^\dag, F_R, \mathbb{I})\nonumber
\ee
\be
+i \tr_{\mathcal{L}(1)}(DT^\dag, F_R, DT, \mathbb{I}) - i \tr_{\mathcal{L}(1)}(DT ,F_L, DT^\dag, \mathbb{I})  \nonumber
\ee
\be
-i \tr_{\mathcal{L}(1)}(F_L, DT^\dag, DT,  \mathbb{I}) + i \tr_{\mathcal{L}( 1)}(F_R, DT^\dag, DT,  \mathbb{I})\nonumber
\ee
\be
- \tr_{\mathcal{L}(1)}(F_L, F_L, \mathbb{I}) + i \tr_{\mathcal{L}(1)}(F_R, F_R, \mathbb{I}).
\label{eq:instdensity}
\ee
We then use Stokes' theorem with $\chi_4 = \intd \Omega_3^0$ to re-express the generalized instanton number in terms of the integral of $\Omega_3^0$
\be
N_I =- \frac{1}{4\pi^2}\int_{\mathcal{M}} \chi_4 = -\frac{1}{4\pi^2} \int_{\partial \mathcal{M}} \Omega_3^0 = N_B.\label{544}
\ee
We have already written down the TCS 3-form in this case in equation (\ref{eq:Omega3final}). {$N_I$ is quantized to be an integer for topological reasons}, \cite{Quillen}. We conclude that for a generic tachyon matrix, the baryon number is equal to the generalized instanton number. The instanton number is quantized to be an integer, and therefore so is the baryon number. The bulk topological current associated to $N_I$ is simply
\be
j = *\chi_4. \label{545}
\ee

\section{The boundary effective action}
\label{sec:effective}

In this section, we derive the boundary effective action for the pions with and without quark mass. We  show that the action matches the chiral Lagrangian with a Skyrme term and a Wess-Zumino-Witten term. More details about the computations in this section can be found in appendix \ref{sec:effectiveappendix}. Note that we do not discuss the effect of the $U(1)$ axial anomaly here, as it is irrelevant for our purposes.

In holography, the pions are identified as  the lightest excitations arising from normalizable  pseudo-scalar\footnote{With respect to boundary Poincar\'e symmetry}  fluctuations of the bulk fields in the flavor sector.  As in regular QCD, the lightest modes are associated to the spontaneously broken (approximate) boundary global  $\text{SU}(N_f)$ axial symmetry.

 We decompose the tachyon field as
\be
T=H \, U
\ee
where $H$ is Hermitian and $U$ is unitary.  The pions  are part of  the perturbations of  the unitary matrix $U(x^\mu, r)$, that mix with  the longitudinal part of the bulk axial gauge fields. The fluctuations of $H$ only contribute heavier modes, so in order to study the pion sector, we can leave $H$ unperturbed.

To simplify the study, we consider the following ansatz\footnote{Without assuming the ansatz \eqref{f73} the computations in this section are more complex but conceptually similar.}, in which $H$ is proportional to the identity matrix:
\be
T(x^\mu, r) = \tau(r) U(x^\m, r), \qquad U \equiv \exp\left[ i \theta^a(x^\m, r) T_a\right].
\label{f73}
\ee
where and $\tau(r)$ is an $x^\mu$-independent scalar function, determined by the V-QCD background.
This ansatz  requires that $m_q$ and $\sigma_q$ are proportional to the unit matrix in \eqref{51}.

We  now consider $x^\mu$-dependent  fluctuations in $U$ and the gauge fields $A_{L}$, $A_R$ over the V-QCD background characterized by a fixed  bulk  metric,  dilaton and tachyon modulus  $\tau(r)$.

The pion effective action is obtained by finding  the bulk perturbations corresponding to the lightest normalizable pseudo-scalar modes  and evaluating the bulk action on-shell. For this, we
 consider the same expansion of the DBI action \eqref{210} to quadratic order in field strengths $F_{L,R}$ as in \cite{jknp}, and re-express it using the ansatz \eqref{f73} as
\be
S_{\mathrm{DBI}} =  - M^3 N_c \int \intd^5x V_f(\lambda, \tau)\sqrt{- \det \tilde{g}}\bigg(\frac{1}{2} + \frac{\kappa \tau^2}{2}\tilde{g}^{mn} \tr(D_m U^\dag D_n U) \nonumber \ee \be + \frac{w^2 }{8} \tilde{g}^{mp}\tilde{g}^{nq} \tr(F_{L, mn} F_{L, pq} + F_{R, mn} F_{R, pq})\bigg)\label{f74}.
\ee
The effective metric $\tilde{g}$ is defined as
\be
\tilde{g}^{mn} \equiv (g_{mn} + \kappa \partial_m \tau \partial_m \tau)^{-1}, \label{f75}
\ee
which is  diagonal since $\tau $ is only a function of $r$. The functions that characterize the background, $\kappa$, $w$, $\tau$, $\lambda$ and the metric are functions of $r$ only.

We rewrite the left and right gauge fields as a function of vector and axial fields,
\be
A \equiv \frac{A_L - A_R}{2}, \quad V \equiv \frac{A_L + A_R}{2}. \label{f75-ii}
\ee
Since the pion modes mix only with the axial sector, the vectorial gauge field $V$ in \eqref{f75-ii} will be set to zero for this analysis. The DBI action for gauge fields and $U$ becomes
\be
S_{\mathrm{DBI}}  =  -\frac{1}{2}  \int \intd^5x \bigg[ 2 B_1(r) \tr(D_r U^\dag D_r U)  +2 B_2(r) \eta^{\m\n}\tr(D_\mu U^\dag D_\nu U)  \nonumber \ee\be + B_3(r) \left[(F_{\mu r}^{A,a})^2 -  ([A_r, A_\m]^a)^2\right] + B_4(r) \left[ (F_{\mu \nu}^{A,a})^2 - ([A_\m, A_\n]^a)^2\right]\bigg],
\label{f76}
\ee
where $F_{m n}^{A,a} = \pa_m A_n^a - \pa_n A_m^a$ and we defined the background-dependent functions $B_i$ as
\be
B_1 \equiv M^3 N_c V_f \sqrt{- \tilde{g}} \frac{\kappa \tau^2}{2} \tilde{g}^{rr}, \quad
B_2 \equiv M^3 N_c V_f \sqrt{- \tilde{g}} \frac{\kappa \tau^2}{2} g^{xx},
\label{f78}\ee
\be
B_3 \equiv M^3 N_c V_f \sqrt{- \tilde{g}} \frac{w^2}{2} \tilde{g}^{rr} g^{xx},\quad
B_4 \equiv  M^3 N_c V_f \sqrt{- \tilde{g}} \frac{w^2}{4}  (g^{xx})^2.
\label{f80}\ee
We first identify the pion effective action to quadratic order and leave the higher order terms for later. To do so, we compute the action \eqref{f76} to quadratic order in the fields, from which we deduce the linearized equations of motion. Working at linear order is enough to obtain the kinetic term and mass spectrum for the boundary excitations, which are identified with bulk normalizable modes.

At the linearized level, the non-abelian couplings and the TCS terms do not contribute to the equations of motion and the DBI action is equivalent to $N_f^2$ copies of an  abelian theory (which is studied in appendix \ref{sec:effectiveappendix1}).

Since all  flavors obey the same equations and are decoupled,  we can drop the flavor indices $a$. The action for the pseudo-scalar and axial vector  sector is then
\be
S  =-\frac{1}{2} \int \intd^4x \intd r\,  \Big[ B_1(r)  (\partial_r \theta +2A_r )^2 + B_2(r) ( \partial_\mu \theta  + 2 A_\mu)^2 + B_3(r) (F^{A}_{r\m})^2  \nonumber \ee \be  + B_4(r)(F^{A}_{\mu \nu})^2 \Big] \label{f87},
\ee
{where $\theta(r,x)$ was defined in \eqref{f73} and $A(r,x)$  in \eqref{f75-ii}.}

In  the  radial gauge $A_r=0$, the equations of motion
 resulting from \eqref{f87} are
\be
-\partial_r(B_3 \partial_r A_\mu) - 2B_4 \eta^{\a \n} \pa_\a F^A_{\n \m} + 2 B_2 (\partial_\mu \theta + 2 A_\mu) = 0,
\label{f88}
\ee
\be
\partial_r(B_1 \partial_r \theta) +  B_2 \eta^{\mu \nu} \partial_\mu   (\partial_\nu \theta + 2 A_\nu) = 0.
\label{f89}
\ee
together with the constraint equation,
\be
B_3 \eta^{\mu \nu} \partial_\mu \partial_r A_\nu + 2 B_1 \partial_r  \theta = 0\label{f86}.
\ee
We shall now solve these equations of motion and find normalizable modes for $m_q = 0$ and $m_q \neq 0$ in order to derive the pion effective action to quadratic order. This is the purpose of the next two subsections. In the third subsection, we derive the non-linear terms in the effective action, including the Skyrme term and the Wess-Zumino-Witten term. Lastly, we study the boundary baryon current associated to this effective action.

\subsection{Quadratic effective action at $m_q= 0$}
\label{sec:61}

We begin with the computation of the effective action in the massless quark case. For vanishing quark mass, it is expected that the lightest mode corresponding to the Goldstone boson is massless. We shall confirm the existence of such a massless mode.

Working in Fourier space, we consider null 4-momentum $k_\mu$, that we parametrize  as
\be
k_\mu = k \hat{k}_\mu \label{f92},
\ee
where $k$ is the energy,  and the spatial 3-vector in $\hat{k}_\mu$ has unit norm.

We   expand the vector $A_\mu(k)$ on a basis adapted to the null momentum vector:
\be
A_\mu(k,r) \equiv A(k,r) \hat{k}_\mu + \bar{A}(k,r) \hat{\bar{k}}_\mu + A^i(k,r) \hat{k}_{i\mu}\,, \label{f89b}
\ee
where the  basis  vectors are chosen to satisfy:
\be
\eta^{\mu \nu} \hat{k}_\mu \hat{\bar{k}}_\nu = 1, \quad  \eta^{\mu \nu} \hat{k}_{i\mu} \hat{k}_{j \nu} = \delta^{ij}, \qquad \eta^{\mu \nu} \hat{k}_{i\mu} \hat{k}_{\nu} = \eta^{\mu \nu} \hat{k}_{i\mu} \hat{\bar{k}}_\nu = \eta^{\mu \nu} \hat{\bar{k}}_\mu \hat{\bar{k}}_\nu =  \eta^{\mu \nu}\hat{k}_\mu \hat{k}_\nu = 0\label{f93}.
\ee
For example, for the  spatial momentum  along the $z$ axis, we have explicitly:
\be
\hat{k}_\mu \equiv (1,0,0,1), \quad \hat{\bar{k}}_\mu \equiv \frac{1}{2}(1,0,0,-1)\label{f90},
\ee
\be
\hat{k}_{1, \mu} \equiv (0,1,0,0), \quad \hat{k}_{2,\mu} \equiv (0,0,1,0)\label{f91}.
\ee

 The equations of motion \eqref{f86}, \eqref{f88} and \eqref{f89} can then be written in the basis \eqref{f89b}. We find that the transverse sector decouples from the longitudinal sector, for which we obtain
\be
 \partial_r(B_3 \partial_r \bar{A}) - 4 B_2 \bar{A} = 0, \label{f95}
\ee
\be
2 B_1 \partial_r \theta - i k B_3 \partial_r (\bar{A}) = 0,\label{f97}
\ee
\be
\partial_r(B_3 \partial_r A) +2B_4 k^2 \bar{A} - 2 B_2 ( 2 A - ik\theta ) = 0.\label{f104}
\ee
We now solve \eqref{f95}-\eqref{f97}  to determine if a normalizable massless mode exists for the fields $\bar{A}$ and $\theta$. Below, we  show that a normalizable massless mode can be found with the ansatz
\be
\theta(k,r) \equiv 2\pi(k)\theta_0(r), \qquad A_\mu(k,r) = -i k_\mu \pi(k)\xi(r).\label{f94}
\ee
This is the ansatz that was used in \cite{casero, Sakai:2004cn}. Inserting (\ref{f94}) in the action \eqref{f87} and using the radial field equations (\ref{f95}-\ref{f104}), one arrives at the boundary action
\be
S = -\frac{1}{2} \int \intd^4x \left[ \bigg( \int \intd r\,4 B_1 (\pa_r \theta_0)^2  \bigg) \pi^2 + \bigg(\int\intd r\,B_3 (\pa_r \xi)^2 + 4B_2 (\xi + \theta_0)^2   \bigg) (\pa_\mu \pi)^2 \right]\label{f95-ii}.
\ee
Therefore, we have  a  normalizable mode   if the two radial integrals in the above expression are finite: they correspond to the pion kinetic term and the pion mass term, with
\be
m_\pi^2 = \int_0^{+\infty}\intd r\, 4 B_1(r) (\pa_r \theta_0)^2. \label{f102}
\ee

To show that a massless normalizable mode exists,  we work perturbatively in the UV. For this, we use the UV expansion of $B_1, B_2, B_3$, presented in appendix \ref{sec:d21}, which we recall here,
\be r\to 0 : \qquad B_1 \sim B_{1, UV} r^3\log(r \Lambda_{UV})^{2\rho}, \quad B_2 \sim B_{2, UV} r^3\log(r \Lambda_{UV})^{2\rho}, \nonumber \ee\be B_3 \sim B_{3, UV} r^{-1}, \quad B_4 \sim B_{4, UV} r^{-1} \label{f102-ii},
\ee
where $\Lambda_{UV}$ is a UV scale,  $B_{i, UV}$ are constants and $\rho$ is a model-dependent parameter whose value has   no  effect on the present discussion.

Due to \eqref{f102-ii} the second term in \eqref{f95} can be neglected  to leading order in the near-boundary expansion.  Therefore, to leading order the solution to \eqref{f95} for $\bar{A}$ is
\be \label{f102-b}
\bar{A} = \bar{A}_0 + \bar{A}_1 r^2 + \mathcal{O}(r^2 \log(r)).
\ee
where  $\bar{A}_0$ and  $\bar{A}_1$ are two integration constants controlling the independent solutions for $\bar{A}$. Each of them  sources  an  independent solution to \eqref{f97} for $\pa_r \theta_0$. After some algebra, we obtain that the leading solution  for $\bar{A}$ corresponds to  $\pa_r \theta_0 \sim r^{-3}$ in the UV, which makes \eqref{f102} UV-divergent. This solution is therefore  non-normalizable.  The subleading  solution  in \eqref{f102-b}, proportional to  $\bar{A}_1$, leads to an expression for $\pa_r \theta_0$ which is  normalizable  in the UV, however it can be checked that this  is IR-divergent\footnote{This mechanism can be observed directly in the U$(1)$ case in AdS that is covered in appendix~\ref{sec:effectiveappendix1}, where an analytical solution can be found.}. Therefore, both integration constants in (\ref{f102-b}) must be set to zero for a normalizable mode, and  using \eqref{f97} we obtain:
\be
\bar{A} = 0 \quad \implies \quad \partial_r \theta_0 = 0, \label{f103}
\ee
i.e. $\theta_0$ is a constant.  By \eqref{f102}, if the  mode under consideration  is normalizable, then it  is massless.

To check normalizability,  we also need to evaluate the second term in (\ref{f95-ii}), for which we need the radial function $\xi(r)$. Inserting (\ref{f94}) in (\ref{f104})  and using the facts  that $\bar{A}=0$ and that $\theta_0$ is constant, we can rewrite (\ref{f104})  as:
\be
 \partial_r(B_3 \partial_r \tilde{A}) - 4 B_2 \tilde{A} = 0, \quad \tilde{A}(k, r) \equiv A - ik\theta_0 . \label{f106}
\ee

The 4-vector $\tilde{A}_\mu$ associated with $\tilde{A}$ is obtained by multiplying with the basis element $\hat{k}_\mu$. Using \eqref{f94} and performing the inverse Fourier transform, $\tilde{A}_\mu(k,r)$ becomes
\be
\tilde{A}_\mu(x,r) = A_\mu(x,r) +  \frac{1}{2}\pa_\m \theta(x,r) = \pa_\mu \pi (x)\tilde{\xi}(r)  ,\label{f108}
\ee
where we defined:
\be
\tilde{\xi}(r) \equiv \xi(r)+ \theta_0. \label{f108-ii}
\ee
With this definition, the  coefficient of the kinetic term in \eqref{f95-ii} can be rewritten as:
\be
\int_0^\infty\intd r\, (B_3 (\pa_r \tilde{\xi})^2 + 4B_2 \tilde{\xi}^2) = - (B_3 \tilde{\xi} \pa_r \tilde{\xi})(r=0).
\label{f109}
\ee
To obtain the right hand side of \eqref{f109} we integrated by parts and used the  field equations to reduce the integral to two  boundary terms, of which only the one in $r=0$ is non-vanishing\footnote{That the IR contribution vanishes can be seen by solving asymptotically \eqref{f104} with the IR behavior of the $B_i$'s, given in equations \eqref{f83A} and \eqref{f84A}.}. From \eqref{f102-ii}, the UV scaling of $B_3$ is $B_3\sim B_{3,UV}r^{-1}$. Solving perturbatively \eqref{f106} in the UV and using \eqref{f102-ii}, we deduce
\be
\tilde{\xi} = c_{\xi, 0} + c_{\xi, 1} r^2 + \dots,\label{f109b}
\ee
where $c_{\xi, i}$ are constants and the dots denote subleading terms in the UV. Using \eqref{f102-ii} and \eqref{f109b}, we deduce that the UV boundary term in \eqref{f109} is finite. Therefore, we can canonically normalize the kinetic term of the pions by choosing the normalization of $\tilde{\xi}$ such that
\be
\int_0^\infty\intd r\, (B_3 (\pa_r \tilde{\xi})^2 + 4B_2 \tilde{\xi}^2) = - (B_3 \tilde{\xi} \pa_r \tilde{\xi})(r=0) = 1.
\label{f109-ii}
\ee

To complete our analysis, we now show how the normalization choice \eqref{f109-ii}, fixes the (constant) value of $\theta_0$ in terms of the pion decay constant $f_\pi$. We use the holographic definition of $f_\pi$ established in \cite{casero}, expressed in terms of the UV asymptotics of the transverse modes at zero momentum
\be
f_\pi^2 = - B_3 (\upsilon \pa_r \upsilon)(r=0, p^2 = 0), \label{f110}
\ee
where we have defined $\upsilon$ the wavefunction of a transverse mode, for instance $A^1$, defined in \eqref{f89b}, \eqref{f91}, which satisfies the equation of motion
\be
\pa_r (B_3 \pa_r \upsilon) - 4 B_2  \upsilon =0, \qquad \upsilon(0) = 1.\label{f111}
\ee
Equation \eqref{f111} admits a single normalizable mode. Using that \eqref{f106} and \eqref{f111} are the same equation, $\upsilon$ and $\tilde{\xi}$ should differ by an overall constant multiplicative factor. After some algebra presented in appendix \ref{sec:d22} we obtain
\be
\upsilon = \tilde{\xi} f_\pi. \label{f115}
\ee
Evaluating \eqref{f115} in the UV $(r=0)$, where $\upsilon = 1$ and $\xi = 0$, then leads to
\be
\theta_0 =  \frac{1}{f_\pi}.\label{f116}
\ee

\subsection{Quadratic effective action at $m_q \neq 0$}
We now consider a nonzero but small quark mass and derive the effective pion action. As before, we take the flavor indices to be implicit. We first study the linearized equations of motion. With nonzero quark mass, the pions are massive, so we can split the fields into longitudinal and transverse fields.
\be
A_\m \equiv \pa_\m \varphi + A^{\perp}_\m, \quad \partial^\mu A_\mu^{\perp} =0, \quad F^{A}_{\mu \nu} \equiv \partial_\mu A_\nu - \partial_\nu A_\mu. \label{f124}
\ee
We expand the longitudinal fields as a series in separable orthogonal modes, which will be justified a posteriori,
\be
\varphi = - \sum_n \varphi_n(r)  \alpha_n(x), \label{f125}
\ee
\be
\theta =  2 \sum_n \theta_n(r)  \alpha_n(x). \label{f126}
\ee
We make the ansatz that the same modes $\alpha_n(x)$ appear in \eqref{f125} and \eqref{f126}. Inserting these expansions into \eqref{f88} gives the equation of motion for the radial modes
\be
\partial_r(B_3 \partial_r \varphi_n) + 4 B_2 (\theta_n - \varphi_n) = 0.
\label{f127}
\ee
We then divide \eqref{f127} by $4B_2$, take a derivative with respect to $r$, multiply by $\alpha_n$ and use \eqref{f89}. This gives
\be
B_1 \partial_r \bigg(\frac{1}{4 B_2} \partial_r (B_3\pa_r \varphi_n)\bigg) \alpha_n - B_1 \pa_r \varphi_n \alpha_n   = - \frac{1}{4}B_3 \pa_r \varphi_n \Box \alpha_n \label{f131A}.
\ee
In order to find a solution to equation \eqref{f131A}, we introduce a Sturm-Liouville equation for $f_n \equiv  B_3\pa_r \varphi_n$,
\be
- B_1 \left( \partial_r \left(\frac{1}{B_2} \pa_r f_n\right) - \frac{4}{B_3} f_n \right)=  m_n^2 f_n,\label{f132}
\ee
where we denoted by $m_n^2$ the eigenvalue of the Sturm-Liouville operator. Substituting $f_n = B_3 \pa_r \varphi_n$ into \eqref{f132} leads to
\be
B_3 m_n^2 \partial_r \varphi_n = 4  B_1 \partial_r \theta_n. \label{f129}
\ee
Substituting \eqref{f129} into \eqref{f127}, we obtain
\be
\partial_r(B_1 \partial_r \theta_n) + m_n^2 B_2 (\theta_n - \varphi_n) = 0.
\label{f128}
\ee

In appendix \ref{sec:d23} we justify  the fact that the modes $f_n$ are orthonormalizable and derive the implications for $\theta_n$ and $\varphi_n$. Inserting the ansatz \eqref{f125}-\eqref{f126} in the action \eqref{f87}, we obtain
\be
S = -\frac{1}{2} \sum_{m,n} \int \intd^4x \left[ \bigg( \int\intd r\,4 B_1 \pa_r \theta_n \pa_r \theta_m  \bigg) \a_n \a_m \right.\nonumber\ee\be \left. + \bigg(\int \intd r\, B_3 \pa_r \varphi_n \pa_r \varphi_m + 4B_2 (\theta_n - \varphi_n)(\theta_m - \varphi_m)   \bigg) \pa_\mu \a_n \pa_\mu \a_m \right]\label{f131}
\ee
We choose the normalization of the modes, such that the kinetic term is canonically normalized,
\be
\int_0^{\infty} \intd r\, \left[ B_3 \partial_r \varphi_{n}\partial_r \varphi_{m}+
  4 B_2 (\theta_{n} - \varphi_{n}) (\theta_{m} - \varphi_{m})\right]  = \delta_{mn} \, .  \label{f137}
\ee
It is shown in appendix \ref{sec:d23} that \eqref{f137} also fixes the value of the radial integral appearing in front of the kinetic term,
\be
\int_0^{\infty} \intd r\, \left[4 B_1 \partial_r \theta_n \partial_r \theta_m \right] = m_n^2 \delta_{mn}.\label{f134}
\ee
The effective quadratic action $S_{(2)}$ from \eqref{f87} then takes the form
\be
S_{(2)} =  -  \frac{1}{2} \int \intd^4x \sum_n [(\partial_\mu \alpha_n)^2 + m_n^2 (\alpha_n)^2].
\label{f138}
\ee
The lightest mode in this tower is identified to be the pion field, $\alpha_0 = \pi$.
The effective action for the pions to quadratic level can be extracted from \eqref{f138},
\be
S_{\pi, (2)} =  - \frac{1}{2}\int \intd^4x \bigg( (\partial_\mu \pi^a)^2 + m_\pi^2  (\pi^a)^2 \bigg)\label{f174}.
\ee

\subsection{Non-linear terms in the pion effective action}
Up to this point, we have restricted ourselves to the quadratic action in the pion fields in order to identify the spectrum and normalization of the low-energy modes. We now derive the nonlinear pion effective action. Our goal is to recover the Skyrme model together with the Wess-Zumino-Witten term directly from the holographic action. In order to retrieve the full effective action, in principle we need to solve the non-linear equations of motion.
Although this is a hard problem, we  show that it can be solved perturbatively with the bulk fields expanded as
\begin{equation}
\label{exp} \theta = \varepsilon\theta^{(0)} + \varepsilon^2\theta^{(1)} + \dots \sp A = \varepsilon A^{(0)} + \varepsilon^2 A^{(1)} + \dots \,.
\end{equation}
This expansion is valid as long as the higher order terms remain smaller for all values of the bulk coordinate $r$.

In this subsection we work in the massive quark case. The leading order fields $\theta^{(0)},  A^{(0)}$ are solutions of the linearized problem we solved in the previous subsection. Therefore, the non-linear terms in the effective action can be computed by substituting the leading bulk fields inside \eqref{f76}. We first derive the non-linear effective action, and we shall justify a posteriori the perturbative expansion in the pions.

We substitute the solution to the linearized equations of motion in \eqref{f76} without assuming that the pions of different flavors commute. The term proportional to $B_4$ in the DBI action gives the Skyrme term
\be
S_{Skyrme} = - \int \intd^4x  \frac{1}{32 g^2 f_\pi^4} ([\partial_\mu \pi, \partial_\nu \pi]^a)^2\label{f175},
\ee
where $g$ is determined by
\be
\int_{0}^{\infty}\intd r\, \varphi_0^4(r) B_4(r) \equiv  \frac{1}{32 g^2 f_\pi^4}\label{f176}.
\ee

Substituting the ansatz \eqref{f124}-\eqref{f126} into the TCS action gives the Witten-Wess-Zumino term,
\be
S_{TCS} = \frac{2 N_c}{15\pi^2} \int \intd^4x \epsilon^{\mu \nu \rho \sigma} \tr( \pi \partial_\mu \pi \partial_\nu \pi\partial_\rho \pi\partial_\sigma \pi) \int \intd r\, \pa_r (\theta_0^5),\label{f177}
\ee
where the $r$ integral is a total derivative.

We shall now determine the UV and IR values of $\theta_0$ in order to compute the radial integral in \eqref{f177}. $\theta_0$ can be expressed in terms of the field $\varphi_0$ by integrating \eqref{f129},
\be
\theta_0(r) = \frac{m_\pi^2}{4} \int_0^r \intd r'\, \frac{\varphi_0' B_3}{B_1}  + \theta_0(0). \label{f139}
\ee
The leading near-boundary expansions for the two fields $\theta_0$ and $\varphi_0$ that solve \eqref{f127} and \eqref{f129} will play an important part in our calculation. These expansions are given by\footnote{We checked in particular that no logarithmic terms appear in these expansions.}
\be
\theta_0 = \theta_0(0)  + c_{\theta, 1}r^2 + \OO(r^4) \label{f142},
\ee
\be
\varphi_0 =  \varphi_0(0)  +  c_{\varphi, 1}r^2 + \OO(r^4) \label{f143},
\ee
where our choice of boundary conditions is such that the external field strengths vanish in the UV, so that we can fix the gauge and take the gauge field source $\varphi_0(0)$ to vanish. The normalization condition \eqref{f137} further enforces
\be
\theta_0(0) = \varphi_0(0) = 0.\label{f144}
\ee

We now compute $\theta_0(r)$ in \eqref{f139}. The corresponding integral cannot be done analytically as the full form of the function $B_1$ and $B_3$, or their ratio, is not known exactly in the V-QCD model, and therefore $\varphi_0'$ is also not known exactly. However, we argue in the following, that in the small mass limit, this integral is dominated by the near-boundary regime $(r\to0)$, where the integrand can be expressed analytically in terms of boundary data. In particular, from \eqref{f143}, we obtain that, at leading order in the $r\to0$ limit, $\varphi_0' \sim 2c_{\varphi, 1} r + \OO(r^3)$. The UV asymptotics of the ratio $r B_3/B_1$ are them given by
\be
\frac{r B_3}{B_1} = \frac{w^2(0) v^3}{2 \kappa(0) \ell^2 (m_q v + \sigma v^3)^2}\left(1 + \mathcal{O}\left(\frac{1}{\log(v \Lambda_{UV})}\right)\right)\label{f146b},
\ee
where we defined
\be
v \equiv r (-\log(r \Lambda_{UV}))^{-\rho},
\ee
and used the UV expansion \eqref{51} inside \eqref{f78}-\eqref{f80} and \eqref{f143}.

In the IR, after some algebra, we obtain that the ratio $B_3/B_1$ asymptotes to zero as a power law, using the IR scalings of the V-QCD background (see Appendix \ref{sec:d21}). We checked numerically that in the small $m_q$ limit, the region where $B_3/B_1$ has significant support is located near the UV, and therefore, the integral can be estimated using the UV expansions of the integrand. Extremizing the denominator in \eqref{f146b}, which is proportional to $m_q v + \sigma v^3$, we can  estimate that the region where the integrand in (\ref{f139}) has significant support is centerered around the value:
\be
v \approx \sqrt{\frac{m_q}{\sigma}}.\label{f148}
\ee

As a consistency check, this estimate is indeed small in the small $m_q$ regime, further justifying the use of the near-boundary expansion of the fields. Consistently, if we replace $r$ in the small $r$ expansions by the value in (\ref{f148}) (up to logarithmic corrections) and expand in small $m_q$, then we find that all higher order terms in $r$ in \eqref{f146b} are at higher order in $m_q$, such that the expansion \eqref{f146b} is consistent as a small $m_q$ expansion:
\be
\frac{r B_3}{B_1} \cong  \frac{w^2 g^{xx} r}{2 \kappa \tau^2} \sqrt{\sigma/m_q}  \cong \frac{w^2(0) }{2 \kappa(0)  m_q^2 \ell^2} \bigg(1 + \mathcal{O}\left(\frac{1}{\log(m_q)}\right)\bigg).\label{f149}
\ee
where in the first equality we used the definitions (\ref{f102-ii}).
We conclude that it is consistent, in a small mass expansion, to consider the UV expansion of the functions inside the integral (\ref{f139}).

Next, we show that under this approximation, $m_\pi^2 r B_3/ B_1$ approaches a Dirac $\delta$-distribution in the $m_q \to 0$ limit. It can be shown, see \eqref{f160A}, that we can start from \eqref{f146b} and consider instead the UV expansion of $m_q r B_3/ B_1$ to leading order,
\be
f(v) \equiv \frac{m_q w^2(0) v^3}{2 \ell^2 \kappa(0) (m_q v + \sigma v^3)^2}.\label{f150}
\ee
$f(v)$ admits a maximum $f_{\text{max}}$ at a location $v_{\text{max}}$, that is given by
\begin{equation}
\label{f150b} f_{\text{max}} = \frac{9w^2(0)}{16 \ell^2 \kappa(0) \sqrt{3m_q\s}} \quad,\quad v_{\text{max}} = \sqrt{\frac{m_q}{3\s}} \, .
\end{equation}
In particular, the maximum diverges in the limit of zero quark mass, while its location approaches the boundary. On the other hand, $f(v)$ vanishes linearly in the limit $m_q\to 0$, for any $r>0$. This indicates that $f(v)$ behaves as a Dirac $\delta$-function in this limit. The precise limit of $f(v)$ can be computed from its integral
\be
\int_{0}^{+\infty} \intd v\, f(v) =  \frac{m_q w^2(0)}{2 \ell^2 \kappa(0)}\int_{0}^{+\infty} \intd v\, \frac{ v^3}{(m_q v + \sigma v^3)^2} =\frac{w^2(0)}{4 \ell^2 \kappa(0) \sigma},\label{f151}
\ee
from which we infer that
\be
f(v) \xrightarrow[m_q \to 0]{} \frac{w^2(0)}{4 \ell^2 \kappa(0) \sigma} \delta(v) \label{f152}.
\ee
We conclude that, starting from (\ref{f139}),
\be
\theta_0(r) = \frac{1}{4} \int_0^r \intd r\, f(r) \frac{m_\pi^2 \varphi_0'}{m_q r} = \frac{1}{4}\frac{w^2(0)}{4 \ell^2 \kappa(0) \sigma}  \frac{16 \sigma \kappa(0) \ell^2}{w^2(0) f_\pi} + \mathcal{O}(m_q) \nonumber \ee\be = \frac{1}{f_\pi} + \mathcal{O}(m_q),\label{f161}
\ee
In appendix \ref{sec:d22}, in the absence of logs in the tachyon expansion \eqref{51}, we derive an estimate for small but nonzero $m_q$, \eqref{f160A-ii},
\be
\theta_0(r) = \frac{1}{f_\pi} \frac{\sigma r^2}{m_q + \sigma r^2} + \dots,\label{f162-ii}
\ee
where the $\dots$ indicate subleading terms in the small quark mass limit. From \eqref{f161}\footnote{Indeed, the second equality in \eqref{f161} is valid for any $r>0$, and for $r=0$ we have $\theta_0(0)=0$ from \eqref{f144}.} and from \eqref{f162-ii}, we obtain the same massless limit of $\theta_0$,
\be
\theta_0(r) = \Theta(r) \frac{1}{f_\pi},\label{f162}
\ee
where we defined the Heaviside function
\be
 \Theta(r) = \left\{ \begin{matrix}
 0 \text{ if } r\leq 0\\
 1 \text{ if } r> 0
 \end{matrix} \right.
\ee

One may worry about the discontinuity of $\theta_0$ in the massless case. However, since in the massless case $B_1$ vanishes in the UV, equation \eqref{f97} is automatically satisfied in $r=0$ for $\bar{A} =0$ and any value of $\theta_0(0)$. In particular, in the massless quark case the equations of motion \eqref{f95}-\eqref{f97} are solved for $\pa_r\theta_0 \propto \delta(r), \bar{A}=0$, as long as $\theta_0(r>0) = \frac{1}{f_\pi}$, which is consistent with the pointwise small quark mass limit result in \eqref{f162}.

Substituting \eqref{f162} into \eqref{f177} gives the expected normalization of the Wess-Zumino-Witten term. Using \eqref{f162}, \eqref{f174}, \eqref{f175}, \eqref{f177}, we obtain
\be
S =  - \frac{1}{2} \int \intd^4x \bigg( (\partial_\mu \pi^a)^2 + m_\pi^2 (\pi^a)^2+ \frac{1}{16 g^2 f_\pi^4} ([\partial_\mu \pi, \partial_\nu \pi]^a)^2 \bigg)  \nonumber \ee \be  +  \frac{2 N_c}{15\pi^2 f_\pi^5} \int \intd^4x \epsilon^{\m \n \rho \sigma} \tr( \pi \partial_\mu \pi \partial_\nu \pi\partial_\rho \pi\partial_\sigma \pi)\label{f179}.
\ee
\smallbreak

We solved the linearized problem and substituted the modes we obtained from the quadratic action in the full action. The result \eqref{f179} obtained using this approach, is only legitimate if it can be shown that the non-linear terms that appear in the full equations of motion, can be considered perturbatively in $\epsilon$, see \eqref{exp}.

For the small field expansion to be consistent, the non-linear corrections must remain smaller than the leading linear piece in the asymptotic UV and IR region. We checked that, in the equations of motion, all the terms coming from the non-linear contributions are subleading in $r$ in the UV with respect to the linearized equations. Therefore, the perturbative approach is correct in the UV.

In the IR, it has been argued in \cite{jknp,Ishii:2019gta} that the TCS contributions must vanish faster than the contributions from the DBI. After some algebra it can be shown that the nonlinear corrections to the equations of motion are then also subleading in the IR.

Another possible issue with our calculation, is that non-linear corrections to the pion wave function may in principle induce additional terms in \eqref{f179} of order-5 in the pion, which could in principle change the coefficient of the WZW term. However, such contributions have to vanish due to the linearized variational problem. This indeed implies that the corrections to the quadratic action $S_2$, due to arbitrary small variations $\delta\theta,\delta A$ of the linearized solutions $\theta^{(0)},A^{(0)}$ (that do not change the boundary conditions), start at quadratic order in the variations
\begin{equation}
\label{lvp} S_2[\theta^{(0)}+\d\theta,A^{(0)}+
\d A] = S_2[\theta^{(0)},A^{(0)}] + \OO(\d\theta^2,\d\theta\d A, \d A^2).
\end{equation}
In particular, this applies to non-linear corrections to the fields as in \eqref{exp}, with $\delta\theta = \varepsilon^2 \theta^{(1)}, \delta A = \varepsilon^2 A^{(1)}$. Since it can be checked that the non-linearities $\theta^{(1)},A^{(1)}$ are of order 3 and higher in the pion field $\pi(x)$, the corresponding corrections to the on-shell action \eqref{f179} therefore start at order 6 in $\pi$.

\subsection{The boundary baryon current and the bulk baryon current}
Since \eqref{f179} is the Skyrme model with an additional Wess-Zumino-Witten term, we know what is the (topological) baryon current for this action, which is given by \cite{wit-bar}
\be
J^\mu_B = - \frac{1}{24\pi^2} \epsilon^{\mu \nu \rho \sigma} (\mathcal{U}^\dag \partial_\nu \mathcal{U})(\mathcal{U}^\dag \partial_\rho \mathcal{U})(\mathcal{U}^\dag \partial_\sigma \mathcal{U}) = \frac{i}{3\pi^2 f_\pi^3} \epsilon^{\mu \nu \rho \sigma} \partial_\nu \pi \partial_\rho \pi \partial_\sigma \pi,\label{f191}
\ee
where we used $\mathcal{U} = \exp(2 i \pi(x)/f_\pi)$. This effective baryon current that we obtained is indeed, up to an improvement term, the same baryon current $J_{\mathcal{U}}$ that appeared from the bulk computation in \eqref{eq:localcurr}. From this expression, the current $J_B$ can also be re-expressed as the pulled-back baryon current from the singularity,
\be
J_B  = J_{\mathcal{U}} = \Psi^*(J_U) + *_\partial \intd \alpha_2,\label{f193}
\ee
where $J_U$ was defined in \eqref{eq:JB2}, and the improvement term $*_\partial \intd \alpha_2$ is irrelevant to this discussion, but is required to asymptote to zero in the small $m_q$ limit. The pulled-back current in $r_*$ can then be connected to $\tr((U^\dag \intd U)^3)$, then linked to the TCS 3-form as before
\be
J_U(r_*) = - \frac{1}{4\pi^2} *_{r_*} \Omega_3^0,\label{f194}
\ee
where we denoted by $*_{r_*}$ the Hodge dual on $S^3(r_*) \times \mathbb{R}_t$. $\Omega_3^0$ can be re-expressed in terms of the bulk instanton density using
\be
\intd \Omega_3^0 = \chi_4,\label{f195}
\ee
and therefore we obtain a bulk current that is directly related to the baryon current we computed from the effective action,
\be
j = *\chi_4\label{f196}
\ee
This bulk current is indeed what was obtained in \eqref{545}. In particular both these currents are conserved topologically as their charge is the instanton number.

\section*{Acknowledgements}\label{ACKNOWL}
\addcontentsline{toc}{section}{Acknowledgements}

We thank F. Bigazzi, A. Cotrone, C. Ecker, N. Jokela, A. Paredes,  C. Rosen for useful discussions.
	
	This work was partially supported by  the H.F.R.I. call ``Basic research Financing" (Horizontal support of all Sciences)
	under the National Recovery and Resilience Plan ``Greece 2.0" funded by the European Union -NextGenerationEU
	(H.F.R.I. Project Number: 15384), by the In2p3 grant ``Extreme Dynamics", the ANR grant ``XtremeHolo"
	(ANR project n.284452), by the H.F.R.I. Project Number: 23770  of the H.F.R.I call
	``3rd Call for H.F.R.I.'s Research Projects to Support Faculty Members \& Researchers", the ERC starting grant 101078061 SINGinGR, under the European Union's Horizon Europe program for research and innovation"
	and the UoC grant number 12030.
	
	EP has received funding from the European Union's Horizon 2024 research and innovation program under the Marie Sklodowska-Curie grant agreement No 101210184.

\newpage

\appendix
	
\begin{appendix}
\renewcommand{\theequation}{\thesection.\arabic{equation}}
\addcontentsline{toc}{section}{Appendices}
\section*{APPENDIX}

\section{Conventions and definitions}
In this appendix we review our conventions and known facts about fields and superconnections. In the first part, we define our conventions for the gauge fields, $A_L$ and $A_R$, and the tachyon field $T$. Then, we review the construction and some useful properties of superconnections. The third section recalls the definition of the discrete symmetries used in criterion \ref{cr:discretesym}.

\subsection{Conventions for gauge fields and the tachyon}
\label{sec:convention}
The gauge fields $A_L$ and $A_R$ are respectively $\text{U}(N_f)_L$ and $\text{U}(N_f)_R$ gauge fields, and $T$ is a bi-fundamental scalar field. On one hand, we take the $\text{SU}(N_f)$ generators, $\lambda^a$, $a=1, ..., N_f^2-1$, to be normalized in the following way for both left and right fields
\be
\tr(\lambda^a \lambda^b) = \frac{1}{2}\delta^{ab}, \qquad (\lambda^a)^\dagger = \lambda^a\, .\label{a1}
\ee
On the other hand, we take the U$(1)$ generator $\lambda^0 = \mathbb{I}$. For both $\text{U}(N_f)_L$ and $\text{U}(N_f)_R$ the gauge fields can be rewritten
\be
A_L = A_L^A \lambda^A = A_L^{\text{U}(1)} \lambda^0 +  A_L^a \lambda^a\, , \quad A_R = A_R^A \lambda^A = A_R^{\text{U}(1)} \lambda^0 +  A_R^a \lambda^a\,,\label{a2}
\ee
with $A = 0, \dots, N_f^2 - 1$. We use form notation.  The field strengths associated to $A_L$ and $A_R$ are then
\be
D_L = \intd - i A_L, \qquad i F_L = (i D_L)^2 = i (\intd A_L- i A_L\wedge A_L)\, ,\label{a3}
\ee
\be
D_R = \intd - i A_R, \qquad i F_R = (i D_R)^2 = i (\intd A_R- i A_R\wedge A_R)\, .
\label{a4}
\ee
The covariant derivatives of the field strengths $F_{L,R}$ are zero by the Bianchi identity
\be
D_L F_L= D_R F_R = 0\, .\label{a5}
\ee
Under a generic gauge transformation with group elements $V_L \in \text{U}(N_f)_L$ and $V_R\in \text{U}(N_f)_R$, the gauge fields and their field strengths transform as
\be
A_{L} \to V_{L}A_{L}V_{L}^\dagger + i V_{L} \intd V_L^\dag\,, \qquad A_{R} \to V_{R}A_{R}V_{R}^\dagger + i V_{R} \intd V_R^\dag \,, \label{a6}
\ee
\be
F_{L} \to V_{L}F_{L}V_{L}^\dagger\,,\qquad F_{R} \to V_{R}F_{R}V_{R}^\dagger\,. \label{a7}
\ee
The tachyon field is bi-fundamental, i.e. it transforms as
\be
T \to  V_R T V_{L}^\dagger\,, \qquad  T^\dagger \to V_{L} T^\dagger V_R^\dag\,.\label{a8}
\ee
The tachyon field's covariant derivative is then given in terms of the gauge fields by
\be
DT = \intd T + i T A_L - i A_R T\,, \qquad DT^\dagger = \intd T^\dagger - i A_L T^\dagger + i T^\dagger A_R\,.\label{a9}
\ee
and as usual transforms homogeneously under gauge transformations.

From this covariant derivative, one can define two products that transform in the adjoint representation under left and right transformations
\be
T^\dagger DT \to V_{L} \left(T^\dagger DT\right) V_L^\dagger\,,\qquad  T DT^\dagger \to V_{R} \left(T DT^\dagger \right)V_R^\dagger \,.
\label{a10}
\ee

\subsection{The superconnection formalism}
\label{sec:convention2}
In order to define the superconnections, recall that the two gauge fields are defined as connection forms on the two different $\text{U}(N_f)$-bundles\footnote{In the string theory context they are associated with stacks  $D_4$ and $\overline{D_4}$ branes respectively in five non-compact dimensions.}, which we denote by
\be
E_L, \quad E_R. \label{a21}
\ee
The direct sum of the two bundles,
\be
E = E_L \oplus E_R,  \label{a22}
\ee
defines a ``super" bundle, i.e. a bundle that admits a $\mathbb{Z}_2$-grading structure. The grading distinguishes between elements associated with $E_L$  and those associated with $E_R$. It is different from the degrees defined in Table \ref{table:degrees}. We define $E_L$ as the ``even" part, and $E_R$ as the ``odd" part, which corresponds to a grading given by $\epsilon$,
\be
\epsilon = \begin{pmatrix}
  +1 & 0\\
  0& -1
\end{pmatrix} \label{a23}.
\ee
We define three different notions of degree, that appear in section \ref{sec:TCSQuillen}. First, $\epsilon$ can be used to define more precisely the notion of $\mathbb{Z}_2$ degree for fields. We call fields that commute with $\epsilon$ even (i.e. of $\mathbb{Z}_2$ degree $0$), and fields that anticommute with $\epsilon$ odd (i.e. of $\mathbb{Z}_2$ degree $1$). Second, we use the differential form degree of the fields. Third, to define the notion of supermatrix, we  need the total degree. This last degree is defined as the sum of the $\mathbb{Z}_2$-degree and the differential form degree, and is defined mod $2$.

To summarize, the three different degrees are:
\begin{itemize}
  \item The $\mathbb{Z}_2$ degree defined above, $0$ for fields which commute with $\epsilon$ and $1$ for fields which anti-commute with $\epsilon$
  \item The differential form degree
  \item The total degree ``sdeg'' defined as
\be
\text{sdeg} \equiv \text{form degree} +  \mathbb{Z}_2\text{ degree } [\text{mod }2] \label{31a}
\ee
\end{itemize}

The values of these three different degrees for the flavor fields appearing in this work are shown in table \ref{table:degreesAppendix}.

\begin{table}[htb]
\centering
\begin{tabular}{ |c|c|c|c|}
 \hline
 Fields & Form degree & $\mathbb{Z}_2$ degree & Total degree (sdeg) \\
 \hline
 $T, T^\dag$ & $0$ & $1$ & $1$\\
 \hline
 $A_L, A_R$ & $1$ & $0$ & $1$\\
 \hline
 $T^\dag T, T T^\dag$ & $0$ & $0$ & $0$\\
 \hline
 $DT, DT^\dag$ & $1$ & $1$ & $0$\\
 \hline
 $F_L, F_R$ & $2$ & $0$ & $0$\\
 \hline
\end{tabular}
\caption{The different degrees of each of the flavor fields.}
\label{table:degreesAppendix}
\end{table}

We can define a ``supermatrix'', as a  $2N_f \times 2N_f$ matrix of differential forms, organized in $N_f \times N_f$ blocks, such that all blocks have the same total degree. Block diagonal fields are even fields, block antidiagonal fields are odd fields with respect to the $\epsilon$ grading. The product of two supermatrices is graded by the differential form degree, and is given by
\be
M . M' = \begin{pmatrix}
A & B\\
C & D
\end{pmatrix} .
\begin{pmatrix}
A' & B'\\
C' & D'
\end{pmatrix}
=
\begin{pmatrix}
AA' + (-1)^{deg(C')}BC' & & &AB' + (-1)^{deg(D')}BD'\\
DC' + (-1)^{deg(A')}CA' & & & DD' + (-1)^{deg(B')}CB'
\end{pmatrix}. \label{a13}
\ee
Here, ``$deg$'' refers to the differential form degree of the fields. The product between the blocks of the supermatrix are regular matrix wedge products. We define the supertrace from $\epsilon$ as
\be
\str(.) = \tr(\epsilon .), \quad \str(M) = \str\begin{pmatrix}
A & B\\
C & D
\end{pmatrix} = \tr(A) - \tr(D) \label{a24}.
\ee
Notably, \eqref{a24} has a graded cyclicity property
\be
\str(M N) = (-1)^{\text{sdeg}(M) \text{sdeg}(N)} \str(N M),\label{a25}
\ee
where $\text{sdeg}$ is the total degree of $M$.

The field content of the holographic flavor sector,  can be rewritten in the supermatrix language. $T$ and $T^\dag$ are odd with respect to the $\mathbb{Z}_2$ degree, while $A_L$ and $A_R$ are even. Both have total degree $\mathrm{sdeg=1}$, so that we can define the odd supermatrix
\be
i\mathcal{A} =
\begin{pmatrix}
iA_L & T^\dagger\\
T & iA_R
\end{pmatrix} \label{a17}.
\ee
Such an object can be used to define a superconnection operator $D$ as
\be
D \equiv \intd - i \mathcal{A}. \label{a18-ii}
\ee
$D$ satisfies a graded derivation property, i.e. for a given differential form $\alpha$ of form degree $\deg(\a)$, and a given supermatrix of differential forms $M$, it satisfies
\be
D(\alpha M) = (\intd \a) M  + (-1)^{\deg(\a)} \a DM. \label{eq:gradedderiv}
\ee
The curvature of a superconnection is defined similarly to standard connections
\be
i \mathcal{F} \equiv (iD)^2 \label{a18}.
\ee

We now compute \eqref{a18} explicitly. For convenience, we  introduce the diagonal and anti-diagonal supermatrices
\be
\mathcal{A} = A + \mathcal{T}, \quad A = \begin{pmatrix}
A_L & 0\\
0 & A_R
\end{pmatrix}, \quad D_A \equiv \intd - i A, \quad
\mathcal{T} = -i \begin{pmatrix}
0 & T^\dagger\\
T & 0
\end{pmatrix}.\label{a20-ii}
\ee
Then, expanding \eqref{a18} in terms of these objects we obtain
\be
i \mathcal{F} = i(\intd A - i A^2) + i(\intd \mathcal{T} - i [A, \mathcal{T}]) + \mathcal{T}^2,  \label{a20}
\ee
where we denoted with brackets the supercommutator, i.e. the graded commutator taken with respect to the supermatrix product \eqref{a13}. For generic supermatrices $M$ and $N$, the graded commutator is defined as
\be
[M, N] = MN - (-1)^{\text{sdeg}(M)\text{sdeg}(N)} NM.\label{a22-ii}
\ee
We now analyze more closely each term in \eqref{a20}. We first identify the curvature of $A$, defined as
\be
iF \equiv (i D_A)^2 =i(\intd A - i A^2) = \begin{pmatrix}
	i F_L & 0\\
	0 & i F_R
\end{pmatrix}.\label{a23-ii}
\ee
The supercommutator term in \eqref{a20} can then be simplified as
\be
i(\intd \mathcal{T} - i [A, \mathcal{T}])=i\left[D_A, \mathcal{T} \right]  = \begin{pmatrix}
0 &  DT^\dagger\\
DT & 0
\end{pmatrix},\label{a24-ii}
\ee
whereas the last term $\mathcal{T}^2$ reads
\be
\mathcal{T}^2 = - \begin{pmatrix}
T^\dagger T & 0\\
0& TT^\dag
\end{pmatrix}\label{a25-ii}.
\ee
As a result, we can finally rewrite \eqref{a20} as
\be
i\mathcal{F} = i F + i\left[D_A, \mathcal{T} \right] + \mathcal{T}^2 =
\begin{pmatrix}
iF_L - T^\dagger T & DT^\dagger\\
DT & iF_R - T T^\dagger
\end{pmatrix}.\label{a26-ii}
\ee
The formulae above give explicit superconnection formulae for $X, Y, Z$ in \eqref{42}.

In the rest of this appendix, we review a useful lemma used in \eqref{310} and proven in \cite{Quillen}, which shows that the Chern character, defined as in \eqref{39}
\be
\chi(\mathcal{F}) = \str \exp(i\mathcal{F}),\label{a27-ii}
\ee
is closed. To do so, we relate its exterior derivative to $\str([D, D^{2n}])$ terms, and then show that it vanishes. We compute the supertrace of a {supercommutator} with $D$
\be
\str([D, M]) = \str(\intd M - i [\mathcal{A} , M]). \label{strdm}
\ee
Moreover, using the definition of the supertrace, we find that for generic supermatrices, we obtain
\be
\str([M, N]) = \str(MN) - (-1)^{\text{sdeg}(M)\text{sdeg}(N)} \str(NM),\label{a29}
\ee
and using \eqref{a25}, we find
\be
\str([M, N]) = \str(MN) - \str(MN) =0\;.\label{a30}
\ee
We conclude that the supertrace vanishes on supercommutators. Using this result in \eqref{strdm} and using linearity,
\be
\str(\intd M - i [\mathcal{A} , M]) = \intd\str(M) - i \str([\mathcal{A} , M])= \intd \str(M),\label{a31}
\ee
and we obtain
\be
 \intd \str(M) = \str([D, M]).\label{a32}
\ee
Applying it to $M= (iD)^{2n}$ and using \eqref{a18}, we can show that
\be
\intd \str( (i\mathcal{F})^n ) =  (-1)^n \str([D, D^{2n}]) = 0,\label{a33}
\ee
as $D$ commutes with $D^2$.

\subsection{Discrete symmetries}
\label{sec:discsym}
We recall here the definition of charge conjugation and parity symmetries on the bulk fields, \cite{casero}.

\paragraph{Parity:} the parity transformation factorizes as
\be
P = P_1 . P_2,\label{a34}
\ee
where $P_2$ reverses the space coordinates
\be
P_2(x_1, x_2, x_3) = (-x_1, -x_2, -x_3),\label{a35-ii}
\ee
and $P_1$ acts on the flavor fields by exchanging left and right,
\be
P_1(A_L, A_R, T, T^\dag) = (A_R, A_L, T^\dag, T)\label{a36}.
\ee

\paragraph{Charge conjugation:} this transformation acts on flavor fields as:
\be
C(A_L, A_R, T, T^\dag) = (-A_R^t, -A_L^t, T^t, T^{\dag,t}),
\label{eq:cparity}
\ee
where $t$ denotes the matrix transpose.

\section{The near-boundary expansion  of the tachyon field}
\label{sec:convention3}

In this section, we recall generic statements about the tachyon matrix that are of use in this article.
First, we shall recall the singular value decomposition (SVD) of a matrix. A linear algebra theorem states that for a generic matrix, we may write
\be
T = V_R \Sigma V_L^\dag,
\label{eq:svd}
\ee
where $V_R, V_L$ are two unitary matrices, and $\Sigma$ is a diagonal matrix with real non-negative coefficients. Although this decomposition holds for any rectangular complex matrix, in our case, $T$, $V_R$, $\Sigma$ and $V_L$ are all square $N_f \times N_f$ matrices. This decomposition does not assume that the matrix is invertible or diagonalizable.

Note that this decomposition implies the existence of a gauge where the tachyon is diagonal, with non-negative eigenvalues everywhere. Applying equation (\ref{a8}) with group elements $(V_L^\dag, V_R^\dag)$, we reach this gauge --- that we call the diagonal tachyon gauge --- where the following holds
\be
\bar{T} = \bar{T}^\dag = \Sigma\label{a43-ii}.
\ee
The notation $\bar{T}$ refers to the tachyon field evaluated in the diagonal gauge. Applying this gauge condition fixes the gauge up to a reordering of the eigenvalues of $\Sigma$.

It is also useful to consider the polar decomposition of a complex square matrix,
\be
T = H U, \label{a35}
\ee
with $H$ Hermitian and $U$ unitary. The relation between the polar and the SVD decompositions is given by
\be
H = V_R \Sigma V_R^\dag,\label{a44}
\ee
and
\be
U = V_R V_L^\dag,\label{a45}
\ee
which are indeed respectively Hermitian and unitary. This formulation is especially useful in the massless quark case where $H = \tau I$, with $I$ the identity matrix, and then $U$ can be identified with the pion matrix, \cite{jknp}.

$H$ and $U$ can be expressed as explicit functions of $T$ and $T^\dag$. For this, we first compute $TT^\dag$,

\be
T T^\dag = H^2  = V_R \Sigma^2 V_R^\dag.\label{a46}
\ee
Then, note that $\Sigma^2$ only has positive or zero eigenvalues, as it is the square of $\Sigma$. Defining $(TT^\dag)^{1/2}$ as
\be
(TT^\dag)^{1/2} \equiv  V_R \Sigma V_R^\dag,\label{a46-ii}
\ee
$H$ can then be written as
\be
H  = V_R \Sigma V_R^\dag = (TT^\dag)^{1/2}\label{a47}.
\ee
On $\mathcal{M} - \mathcal{Z}$, defined in \eqref{333}, $\Sigma$ has only strictly positive eigenvalues, and then we can define $\Sigma^{-1}$. Then \eqref{a46-ii} is invertible, and inserting \eqref{a47} into $T = HU$ leads to
\be
U  = (TT^\dag)^{-1/2} T.
\label{eq:defU}
\ee

\subsection{Expansion of the tachyon field}
We shall now work out a boundary expansion of the tachyon and connect it to the fields $V_R$, $V_L$, $\Sigma$, $H$ and $U$ introduced in the decompositions \eqref{eq:svd} and \eqref{a35}. In V-QCD the near-boundary tachyon expansion is given by \cite{Jarvinen}
\be
T = (m_q r (-\log(r \Lambda_{UV}))^{-\rho} + \sigma_q \mathcal{U} r^3(-\log(r \Lambda_{UV}))^{\rho})\left(1 + \mathcal{O}(\frac{1}{\log(r \Lambda_{UV})})\label{f140b}\right).
\ee
In this expression, the (bare) quark mass matrix is denoted $m_q$, and $\sigma_q \mathcal{U}$ is proportional to the vev of the quark mass bilinears. We wish to relate them to the tachyon unitary part $U$ defined in \eqref{a35}. The logarithms in this expression are irrelevant to this identification, and therefore we drop them in this appendix. In addition, we take $m_q$ and $\sigma_q$ to be proportional to the identity matrix. This provides the simplified expansion
\be
T  = m_q r\mathbb{I} + \sigma_q r^3 \mathcal{U} + \mathcal{O}(r^5),
\label{eq:tacexp}
\ee
where we denoted $\sigma_q$ the chiral condensate and $\mathcal{U}$ the pion matrix on the boundary. We can reconstruct \eqref{eq:tacexp} by starting with the following ansatz for the expansions of $H$ and $U$,
\be
H  = m_q r \mathbb{I} + r^3 \mathcal{H} + \mathcal{O}(r^5)\label{a43},
\ee
\be
U  = \mathbb{I} + r^2 \mathcal{A} +  \mathcal{O}(r^4).
\label{a26}
\ee
Given \eqref{a43}, the expansion \eqref{a26} is unique when the tachyon is invertible, that is, on $\mathcal{M}- \mathcal{Z}$, where $\mathcal{Z}$ is defined in \eqref{333}.
$U$ is unitary, therefore it is the exponential of an anti-hermitian matrix using Lie theory. Because $m_q$ is real and proportional to the identity, $T$ asymptotes to $m_q \mathbb{I}$ on the boundary, therefore $U$ asymptotes to the identity matrix on the boundary. This implies that $\mathcal{A}$ is necessarily an anti-hermitian matrix. We compute the product of \eqref{a43} and \eqref{a26} to obtain
\be
T  = m_qr \mathbb{I}+  \big(\mathcal{H} + m_q  \mathcal{A}\big)r^3 + \mathcal{O}(r^5) \label{a27}.
\ee
It matches with \eqref{eq:tacexp} if and only if
\be
\mathcal{H} + m_q \mathcal{A} = \sigma_q  \mathcal{U} \label{a28}.
\ee
Therefore, we see that $ m_q \mathcal{A}$ and $\mathcal{H}$ are the hermitian and anti-hermitian part of the pion matrix (times $\sigma_q$). Inverting this relation, we obtain
\be
m_q\mathcal{A}  = \frac{\sigma_q\mathcal{U} - \sigma_q \mathcal{U}^\dag}{2},\label{a56}
\ee
\be
\mathcal{H}  = \frac{\sigma_q\mathcal{U} + \sigma_q\mathcal{U}^\dag}{2}\label{a57}.
\ee
We can also express $\mathcal{U}$ as a function of the boundary values of the tachyon as
\be
 \sigma_q\mathcal{U} = \frac{1}{3!}\partial_r^3 (H)\big|_{bdry} + \frac{m_q}{2!} \partial_r^2 (U)\big|_{bdry}\label{a58},
\ee
\be
 \sigma_q\mathcal{U} = \frac{1}{3!}\partial_r^3 ((TT^\dagger)^{1/2})\big|_{bdry} +  \frac{m_q}{2!} \partial_r^2 ((TT^\dagger)^{-1/2} T)\big|_{bdry}\label{a59}.
\ee
where in the second equation we used the relations \eqref{a47}-\eqref{eq:defU} between $H,U$ and $T$. In the massless case, $T$ asymptotes to $\sigma_q \mathcal{U}$ as can be seen from \eqref{eq:tacexp}. This implies that $\mathcal{U}$ and $U$ are equal at the boundary.

We shall now study the transformation properties of $\mathcal{U}$. For this purpose, we recall the gauge transformation law of the tachyon under two independent unitary elements $V_L$ and $V_R$ from \eqref{a8}
\be
T \to V_R T V_L^\dag, \qquad T^\dag \to V_L T^\dag V_R^\dag. \label{a51}
\ee
Acting with this general gauge transformation on the expansion (\ref{eq:tacexp}), we obtain immediately that to avoid changing the source, the bulk gauge transformation must reduce to a vectorial gauge transformation $V_L = V_R \equiv V$ on the boundary. Note that this is only true if the mass matrix is proportional to the unit matrix, as we assumed previously. Under a vectorial gauge transformation in the full bulk that asymptotes to $V$ on the boundary, we obtain from \eqref{eq:tacexp},
\be
T  = m_q r\mathbb{I} + \sigma_q r^3 V \mathcal{U} V^\dag + \mathcal{O}(r^5).\label{a61}
\ee
Note that, choosing the appropriate $V$, we can diagonalize $\mathcal{U}$. This is not the most general transformation that we can apply, as the transformation needs not be vectorial in the bulk. The most general allowed transformation is instead
\be
V_L  = V (\mathbb{I} + r^2 \mathcal{A}_L + \mathcal{O}(r^4))\label{a62},
\ee
\be
V_R  = V (\mathbb{I} + r^2 \mathcal{A}_R + \mathcal{O}(r^4)),\label{a63}
\ee
where $A_L$ and $A_R$ are antihermitian matrices. Now, the tachyon expansion becomes
\be
T  = m_q r\mathbb{I} + \sigma_q r^3 V \left(\mathcal{U} + \frac{m_q}{\sigma_q} (\mathcal{A}_R - \mathcal{A}_L) \right) V^\dag + \mathcal{O}(r^5),\label{a64}
\ee
where $\frac{m_q}{\sigma_q} (\mathcal{A}_R - \mathcal{A}_L)  \equiv \mathfrak{A}$ is an antihermitian matrix. Therefore, by performing an axial transformation in the bulk we can add to $\mathcal{U}$ a general anti-hermitian matrix. In particular we can choose $\mathfrak{A}$ to cancel $m \mathcal{A}$ in \eqref{a27}. We obtain that there exists a gauge transformation such that
\be
T  = m_q r\mathbb{I} + \sigma_q r^3 V \mathcal{H} V^\dag + \mathcal{O}(r^5),\label{a65}
\ee
and we can again choose $V$ to make $\mathcal{H}$ diagonal. Under the gauge transformation with parameters $V_L$, $V_R$ in \eqref{a62}, \eqref{a63}, the gauge fields transform as in \eqref{a6},
\be
A_L \to V_{L}A_{L}V_{L}^\dagger + i V_{L} \intd V_L^\dag\,, \qquad A_{R} \to V_{R}A_{R}V_{R}^\dagger + i V_{R} \intd V_R^\dag,\label{a65-ii}
\ee
such that the near-boundary expansion of gauge fields,
\be
A_L = A_L^{(0)}(x)  + r^2 (A_L^{(1)}(x)  + \log(r) \tilde{A}_L^{(1)}(x)) + \mathcal{O}(r^4 \log(r))\label{a65-iii},
\ee
becomes, under \eqref{a65-ii}
\be
A_L = V A_L^{(0)}(x) V^\dag + i V\intd V^\dag - 2i r \intd r V \mathcal{A}_L V^\dag + r^2 (V A_L^{(1)}(x)V^\dag  + \log(r) V_L \tilde{A}_L^{(1)}(x)V_L^\dag \nonumber \ee
\be + (V \mathcal{A}_L A_L^{0} - A_L^{0} \mathcal{A}_L V^\dag) + i(V \mathcal{A}_L) \intd V^\dag - V \intd (\mathcal{A}_L V^\dag) ) + \mathcal{O}(r^3)\label{a65-iv},
\ee

We shall now focus on the transformations of $U$. Using \eqref{a46}, we obtain that $(TT^\dag)^{-1/2}$ transforms under a gauge transformation in the following way
\be
(TT^\dag)^{-1/2} \to V_R (TT^\dag)^{-1/2} V_R^\dag\label{a67},
\ee
which implies the following gauge transformation laws for $U$ and $U^\dag$
\be
U \to V_R U V_L^\dag,  \qquad U^\dag \to V_L U^\dag V_R^\dag.\label{a68}
\ee
Note that this is the same transformation law as the tachyon itself, i.e. $U$ is bi-fundamental, and its covariant derivative is therefore
\be
DU = \intd U + i U A_L - i A_R U, \qquad DU^\dagger = \intd U^\dagger - i A_L U^\dagger + i U^\dagger A_R.\label{a69}
\ee
Note that, using $UU^\dag = I$ and these expressions, the following relation holds
\be
DU^\dag = - U^\dag DU U^\dag\label{a70}.
\ee

\section{Details of the explicit computation of the Tachyon-Chern-Simons action}
\label{sec:details}

In this appendix we collect additional computations related to the TCS action. In \ref{sec:gammaalt} we derive the action from the path $\gamma_\mathrm{alt}=I_{\rm alt}\cup II_{\rm alt}$ (see \ref{path:galt}) in figure \ref{plane}. In \ref{sec:omega1} we show the computation of $\Omega_1$ with the path $\gamma$ (see \ref{path:g}), which is considerably simpler than $\Omega_3$ and $\Omega_5$. In \ref{sec:residue} we show a residue formula that can be obtained for $\Omega^0$. In the rest of the appendix, we list the expressions for the $\mathcal{L}$ operators appearing in $\Omega^0$ and the equations of motion for this piece.

\subsection{The alternative path: $\gamma_\mathrm{alt}$}
\label{sec:gammaalt}
In this subsection, we compute the TCS $5$-form, from an alternative path $\gamma_\mathrm{alt}$, defined in figure \ref{plane}. We show that it does not produce the correct QCD anomaly in the IR. Using the path formula \eqref{335}, we obtain the analogue of \eqref{eq:Omegagen}
\be
\Omega_5^{\gamma_\mathrm{alt}} = \int_{0}^{1}\intd s\, \str(i A e^{i \mathcal{F}(s,0)}) + \int_{0}^{1}\intd s\, \str(i \mathcal{T} e^{i \mathcal{F}(1,s)}), \label{b111}
\ee
where $\mathcal{F}(a,b)$ is defined in \eqref{332}, $A$ and $\mathcal{T}$ in \eqref{32}. The horizontal path $I_\mathrm{alt}$, defined in figure \ref{plane}, is a path at zero tachyon, turning on the gauge fields progressively. This is precisely the path taken to compute the CS action in the absence of tachyon. This path defines $\omega_5$, as can be seen from \eqref{336}. This path is at $b=0$, so $Z(b) = 0$ in \eqref{42}. Therefore, the exponential \eqref{45} contains only $X$ and $Y$, which both have nonzero form degree. The CS five-form from this path then takes contributions from finitely many terms in the series \eqref{45}. Then computing the supertrace gives $\omega_5$, defined in \eqref{315}:
\be
\omega_5 = -\frac{i}{6} \tr\left(AF^2 + \frac{i}{2}A^3 F - \frac{1}{10}A^5\right)\label{b2}.
\ee
On the vertical path $II_\mathrm{alt}$, , defined in figure \ref{plane}, we use \eqref{332}, with $a=1$, $b=s$, with $s$ integrated from $0$ to $1$. We define the action given by this path as,
\be
Z^0 \equiv \int_{0}^{1}\intd s\, \str(i \mathcal{T} e^{i \mathcal{F}(1,s)}).\label{z0}
\ee
This piece of the action vanishes when the tachyon field is set to zero everywhere. Indeed, setting $T$ to zero, projects the plane in figure \ref{plane} to the horizontal axis. We obtain that \eqref{b111} can be rewritten as
\be
\Omega_5^{\gamma_\mathrm{alt}}  = \omega_5  + Z^0_5,\label{b18-ii}
\ee
with $Z^0_5$ the 5-form contribution in \eqref{z0}. Because vertical paths at $a=1$ are gauge-invariant, the variation of $\Omega_5^{\gamma_\mathrm{alt}}$ under gauge transformations is the variation of $\omega_5$. Therefore, the QCD anomaly is necessarily correctly reproduced in the UV. However, with this path, the anomaly also receives a contribution from the infrared.

We now write $Z^0_5$, the $5$-form contribution to \eqref{z0}, for $T=\tau U$. We obtain
\be
Z_5^0 = \frac{-2 + e^{-\tau^2}(2+2\tau^2 + \tau^4)}{120}\tr[(U^\dag DU)^5]
\nonumber \ee
\be-\frac{i}{12}(1-e^{-\tau^2}(1+\tau^2)) \left[\tr[(U^\dag DU)^3 F_L] -\tr[(U DU^\dag)^3 F_R]   \right]
\nonumber \ee
\be
+\frac{-1 + e^{-\tau^2}}{12}\left[-2\tr[U^\dag DU F_L^2]  +2\tr[U DU^\dag F_R^2] + \tr[UF_L DU^\dag F_R - U^\dag F_R DU F_L] \right]\label{b5}.
\ee
In the limit of $\tau \to \infty$, $Z^0$ does not vanish, but instead asymptotes to a closed gauge-invariant form, $-\Omega_5^0 (T \to 0, U)$ that depends on the gauge fields and on $U$. Equation \eqref{b5} can then be rewritten
\be
Z_5^0 = \Omega_5^0(T) - \Omega_5^0(T\to 0, U).\label{b6}
\ee
In \eqref{b6}, $\Omega_5^0$ is the $5$-form contribution to \eqref{eq:Omega0}.

Remarkably, in the $T= \tau U$ case, it can be checked that
\be
\omega_5 - \Omega_5^0(T\to 0, U)= \intd G_4 + \Omega_5^c,
\ee
where $G_4$ can be found in \eqref{G4final} and $\Omega_5^c$ was defined in \eqref{eq:Omegac}. Then, \eqref{b18-ii} can be rewritten as
\be
\Omega_5^{\gamma_\mathrm{alt}}  = \Omega_5^0+ \intd G_4 + \Omega_5^c,\label{b7}
\ee
Note that the only difference, in this case, with the action computed in \eqref{eq:Omegagen} is that in the action the $G_4$ form is integrated on the full boundary, instead of $\mathcal{Z}$, defined in \eqref{333}.

Therefore, the full action contains an IR term and violates criterion \ref{cr:noIR}. Moreover, the variation of this action in the IR is given by the variation of the standard CS forms \eqref{b2}. Therefore, there could be an anomaly contribution from the IR, illustrating how this seemingly more natural path is not appropriate for our purposes.

\subsection{The case of $\Omega_1$}
\label{sec:omega1}
In this subsection, we compute $\Omega_1$, the one-form component of \eqref{eq:Omegagen}. We write the series expansion of the exponential
\be \label{eq:exponential}
\ex^{i\mathcal{F}(a,b)} = \sum_{n=0}^{+\infty} \frac{1}{n!} (i \mathcal{F}(a,b))^n.
\ee
and expand $i \mathcal{F}(a,b)$ as in \eqref{42}-\eqref{45}
\be
i \mathcal{F}(a,b) = X(a,b) + Y(a) + Z(b)\label{b9},
\ee
where $\mathcal{F}(a,b)$ is defined in \eqref{332}.

We start with $\Omega_1^0$, the one-form component of \eqref{eq:Omega0}, for a generic tachyon field, generalizing the expression found in \cite{casero}. For this purpose, we shall identify all the possible terms in the sum that can contribute to degree-one forms. Since $Y(a)$, defined in \eqref{44}, contributes to order two and we want a one-form, we shall consider only the terms with $X(a,b) = i[D, \mathcal{T}](a,b)$ (also defined in \eqref{43}), appearing only once in the products. There are still contributions from every order $n$ in \eqref{eq:exponential}, coming from powers of $Z(b)$ in \eqref{44}.

 Applying this procedure to $\Omega^0$, setting $a=1, b=s$, we obtain,
\be
\ex^{i\mathcal{F}_{1,s}}\Big|_{\text{1-form}} = \sum_{n=0}^{+\infty} \frac{1}{(n+1)!} \sum_{i+j=n} (s^2 \mathcal{T}^2)^i is [D, \mathcal{T}] (s^2 \mathcal{T}^2)^j.
\label{expleibniz}
\ee
In this case, we can simply use the linearity and cyclicity of the supertrace with respect to the super-product, and rewrite this sum
\be
\str \left( i \mathcal{T} e^{i\mathcal{F}_{1,s}}\right)\Big|_{\text{1-form}} = \sum_{n=0}^{+\infty} \frac{1}{(n+1)!}  \sum_{i+j=n} \str\left( (s^2 \mathcal{T}^2)^i is [D, \mathcal{T}] (s^2 \mathcal{T}^2)^{j} i \mathcal{T} \right)\label{b11}.
\ee
We use the graded cyclicity of the supertrace \eqref{34} to obtain
\be
\str \left( i \mathcal{T} \ex^{i\mathcal{F}_{1,s}}  \right)\Big|_{\text{1-form}} = \sum_{n=0}^{+\infty} \frac{1}{(n+1)!} \sum_{i+j=n} s^{2n}\str \left( \mathcal{T}^{2n}  i \mathcal{T}  i s [D, \mathcal{T}]\right)\label{b12}.
\ee
The sum over $i+j$ contains $n+1$ equal terms, so it gives a multiplicative $(n+1)$ factor. The sum over $n$ in \eqref{b12} can be computed as
\be
\str \left( i \mathcal{T}  \ex^{i\mathcal{F}_{1,s}}\right)\Big|_{\text{1-form}} =  \str \left(\ex^{(s \mathcal{T})^2} i \mathcal{T} is [D, \mathcal{T}]\right)
\label{leibniz}.
\ee
From (\ref{leibniz}) we compute on the right hand side
\be
i\mathcal{T} is[D, \mathcal{T}]= -s \begin{pmatrix}
T^\dagger DT & 0\\
0 & T DT^\dagger
\end{pmatrix}.
\label{eq:TDT}
\ee
We obtain for the supertrace
\be
\str \left(\ex^{i\mathcal{F}_{1,s}} i \mathcal{T} \right)\Big|_{\text{1-form}} =-
 \tr \left( \ex^{-s^2 T^\dagger T} s T^{\dagger}DT \right) + \tr \left( \ex^{-s^2 TT^\dagger } s TDT^{\dagger} \right)\label{b15}.
\ee
It follows that the value of $\Omega^0_1$ is
\be
\Omega^0_1 = \int_{+\infty}^{1}\str \left(\ex^{i\mathcal{F}} i \mathcal{T} \right)\Big|_{\text{1-form}} = \frac{1}{2} \tr \left( e^{-T^\dagger T} (T^\dagger T)^{-1} T^\dagger DT - e^{-TT^\dagger } (T T^{\dagger})^{-1} T DT^\dagger  \right)=\label{b16}
\ee
\be
 = \frac{1}{2} \tr \left( e^{-T^\dagger T} T^{-1} DT - e^{-TT^\dagger } (T^{\dagger})^{-1} DT^\dagger  \right).\label{b17}
\ee
(\ref{b17})  was obtained by relying on the cyclicity of the trace, which in the case of $\Omega^0_1$ is sufficient in order to resum the exponential.
However, it is impossible to generalize this argument to a general value of $p$.

The other two contributions are obtained by the same method, from \eqref{eq:Omegac} and \eqref{eq:Omegab}. After some  algebra, we obtain
\be
\Omega_1^c = -\frac{1}{2} \tr(T^{-1} \intd T - T^{\dag, -1}\intd T^\dag),\label{b18}
\ee

and for the last piece we obtain
\be
\Omega_1^b = \lim_{b\to +\infty} \int_{0}^{1}\intd s \tr(iA_L \ex^{-b^2 T^\dag T} - iA_R \ex^{-b^2 T T^\dag}).\label{b19}
\ee
The integration over $s$ is trivial. After diagonalizing $TT^\dag$, the exponentials can be rewritten $V_L \ex^{-b^2 \Sigma^2} V_L^\dag$, where $V_L$ is the matrix in \eqref{eq:svd}. Taking the limit $b\to +\infty$ then gives a vanishing result for nonzero eigenvalues of $\Sigma$. For vanishing eigenvalues the limit is $1$. We define the indicator function
\be
\1 (x) \equiv \begin{pmatrix}
  &1 \text{ if } x=0\\
  &0 \text{ if } x\neq 0
\end{pmatrix}\label{b19-ii}
\ee
Then, the limit of the exponential appearing in \eqref{b19} becomes
\be
 \lim_{b\to +\infty} \ex^{-b^2 T^\dag T} = V_L \begin{pmatrix} \1(\sigma_1) & & \\ & \ddots & \\ & & \1(\sigma_{N_f})\end{pmatrix} V_L^\dag. \label{b19-iii}
\ee

In the simpler case where all eigenvalues vanish on $\mathcal{Z}$, defined in \eqref{333}, \eqref{b19-iii} becomes simply the unit matrix on $\mathcal{Z}$, and then from \eqref{b19} we obtain
\be
\Omega_1^b = \tr(i (A_L - A_R))\theta_{\mathcal{Z}}(x).\label{b20}
\ee
Note that, even without assuming that all eigenvalues vanish on $\mathcal{Z}
$, the limit \eqref{b19-iii} is nonzero on a subset of measure zero of $\mathcal{M}$, therefore it can be safely ignored for equations of motion.

The final result for the full TCS 1-form is
\be
\Omega_1 = \frac{1}{2} \tr \left( e^{-T^\dagger T} T^{-1} DT - e^{-TT^\dagger } T^{\dagger,-1} DT^\dagger  \right)
-\frac{1}{2} \tr(T^{-1} \intd T - T^{\dagger,-1} \intd T^\dag) \nonumber \ee \be + \tr(i (A_L - A_R))\theta_{\mathcal{Z}}(x)
\label{eq:Omega1final}.
\ee

\subsection{A residue formula for $\Omega^0$}
\label{sec:residue}

In this appendix, we derive a residue formula for  $\Omega^0$.
To this end, we  start from~\eqref{eq:Omega0} and apply a resummation technique in order to obtain a more explicit expression for the three-form component, $\Omega^0_3$. We also give schematically the result for $\Omega^0_5$. As we shall see, $\Omega^b$ and $\Omega^c$ both present obstacles to applying this method. Moreover, the derivation assumes that all eigenvalues of $TT^\dagger$ are positive. If some of the eigenvalues are zero, a special treatment is needed.

We first introduce a shorthand notation for the components in $\mathcal{F}(1,s)$ in~\eqref{eq:Omega0}
\be
\mathcal{F}(1,s) = -i(Y(1)+X(1,s)+Z(s)) \equiv \mathcal{F}^{(2)}+ s\, \mathcal{F}^{(1)} + s^2\,\mathcal{F}^{(0)}.\label{b22}
\ee
where the upper index refers to the degree of the form. Explicitly,
\be
 \mathcal{F}^{(2)} = F = \begin{pmatrix}
	F_L & 0\\
	0 & F_R
\end{pmatrix} \ , \, \quad  \mathcal{F}^{(1)} =  \begin{pmatrix}
	0 & -iDT^\dagger\\
	-iDT & 0
\end{pmatrix} \ ,\label{b23}
\ee
\be
\mathcal{F}^{(0)} = - i \mathcal{T}^2 = \begin{pmatrix}
	iT^\dagger T & 0\\
	0 & iTT^\dagger
\end{pmatrix}.\label{b24}
\ee
Inserting this in the exponential~\eqref{eq:exponential}, after some combinatorics, and extracting the three-form piece, we obtain
\be \label{eq:exp3formsum}
  e^{i\mathcal{F}_s}\Big|_\mathrm{3-form} = \sum_{k_1 \ldots k_3=0}^{\infty} \frac{i^{2+\sum_ik_i}}{(2+\sum_ik_i)!}(s^2\mathcal{F}^{(0)})^{k_1}s\mathcal{F}^{(1)}(s^2\mathcal{F}^{(0)})^{k_2}\mathcal{F}^{(2)}(s^2\mathcal{F}^{(0)})^{k_3} + \left(1\leftrightarrow 2\right) \nonumber
\ee
\be
  +\sum_{k_1 \ldots k_4=0}^{\infty} \frac{i^{3+\sum_ik_i}}{(3+\sum_ik_i)!}(s^2\mathcal{F}^{(0)})^{k_1}s\mathcal{F}^{(1)}(s^2\mathcal{F}^{(0)})^{k_2}s\mathcal{F}^{(1)}(s^2\mathcal{F}^{(0)})^{k_3}t\mathcal{F}^{(1)}(s^2\mathcal{F}^{(0)})^{k_4}.\label{b25}
\ee
In order to resum, we use the integral representation
\be
\frac{1}{n!} = \frac{1}{2\pi i}\int_{\mathcal{C}} du\, \frac{e^u}{u^{n+1}}\label{b26}
\ee
for all the factorials appearing in~\eqref{eq:exp3formsum}. Since $n$ is an integer, the contour $\mathcal{C}$ can be taken to encircle the origin counterclockwise.

\begin{figure}[htb]
\centering
\begin{tikzpicture}
\def\bigradius{3}
\def\littleradius{0.5}

\draw [help lines,->] (-1.25*\bigradius, 0) -- (1.25*\bigradius,0);
\draw [help lines,->] (0, -1.25*\bigradius) -- (0, 1.25*\bigradius);

\path (-1,0) pic {cross=4pt};
\path (-3,0) pic {cross=4pt};
\node[draw] at (3, 3) (a) {$u$};

\node at (-4.5, 0) (b) {$\dots$};

\draw[decoration={markings, mark=at position 0.25 with {\arrow{>}}},
        postaction={decorate}] (-1,0) circle (0.5);
        \draw[decoration={markings, mark=at position 0.25 with {\arrow{>}}},
        postaction={decorate}] (-3,0) circle (0.5);

\end{tikzpicture}
\caption{Contour $\mathcal{\tilde C}$ in equation \eqref{b28}. The contour encircles counterclockwise the poles at $u = -\sigma_a^2$ for every tachyon eigenvalue $\sigma_a$.}
\label{fig:contour}
\end{figure}
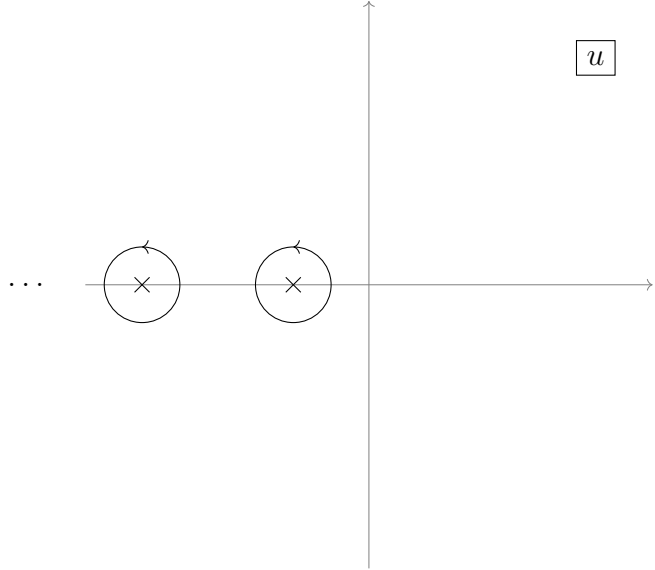

After rearranging and doing the sums over $k_i$, we find
\be \label{eq:exp3form}
  e^{i\mathcal{F}_s}\Big|_\mathrm{3-form} = \frac{1}{2\pi i}\int_{\mathcal{\bar C}} du \,e^{u}\Bigg[\frac{1}{u-is^2\mathcal{F}^{(0)}}is\mathcal{F}^{(1)}\frac{1}{u-is^2\mathcal{F}^{(0)}}i\mathcal{F}^{(2)}\frac{1}{u-is^2\mathcal{F}^{(0)}}+ \left(1\leftrightarrow 2\right) +\nonumber
\ee
\be
  +\frac{1}{u-is^2\mathcal{F}^{(0)}}it\mathcal{F}^{(1)}\frac{1}{u-is^2\mathcal{F}^{(0)}}is\mathcal{F}^{(1)}\frac{1}{u-is^2\mathcal{F}^{(0)}}is\mathcal{F}^{(1)}\frac{1}{u-is^2\mathcal{F}^{(0)}}\Bigg] \ .\label{b27}
\ee
Here the sums over $k_i$ are power series of $s^2/u$ which only converge within some origin-centered disk on the $s^2/u$-plane, whose radius is determined by the first singularity of the matrix inverse $1/[1-i(s^2/u) \mathcal{F}^{(0)}]$. On the $u$-plane, the convergent region is found at large $|u|$. Consequently, in order to remain in the region where the integrand is well defined, we first need to modify the contour of integration to $\mathcal{\bar C}$, which circles counterclockwise all singularities of the expression. These singularities correspond to zero eigenvalues of the tachyon field. At this point, it is enough to take the contour $\mathcal{\bar C}$ to be a single circle with a large enough radius so that the sums leading to~\eqref{b27} are convergent. Note that $\mathcal{F}^{(0)}$ is block diagonal, and for block diagonal matrices, the supermatrix product is equal to the regular matrix product, so the inverse matrices in~\eqref{eq:exp3form} can be defined as usual.

Inserting~\eqref{b27} in~\eqref{eq:Omega0}, for $\Omega^0$ we have always $s>1$. Since the singularities are actually related to the eigenvalues of $-TT^\dagger$ or  $-T^\dagger T$, and we assume that all eigenvalues are nonzero, they occur at negative values of $u$. Therefore, we can further modify the contour of $u$-integration such that $\mathrm{Re}\, u<0$ along it. The final contour for   $\tilde{\mathcal{L}}$ is shown in figure \ref{fig:contour}. This choice of contour allows us to rescale $u \mapsto u s^2$, such that the $s$-integral still remains convergent. We obtain
\be \label{eq:Omega3first}
   \Omega^0_3 = - \frac{i}{2\pi i} \int_1^\infty ds \int_{\mathcal{\tilde C}} du \,\frac{e^{us^2}}{s^3}\ \str\Bigg\{\Bigg[\frac{1}{u-i\mathcal{F}^{(0)}}i\mathcal{F}^{(1)}\frac{1}{u-i\mathcal{F}^{(0)}}i\mathcal{F}^{(2)}\frac{1}{u-i\mathcal{F}^{(0)}}+ \left(1\leftrightarrow 2\right) \nonumber
\ee
\be
  +\frac{1}{u-i\mathcal{F}^{(0)}}i\mathcal{F}^{(1)}\frac{1}{u-i\mathcal{F}^{(0)}}i\mathcal{F}^{(1)}\frac{1}{u-i\mathcal{F}^{(0)}}i\mathcal{F}^{(1)}\frac{1}{u-i\mathcal{F}^{(0)}}\Bigg]\mathcal{T}\Bigg\}.\label{b28}
\ee

Finally, we integrate over $s$, and express the integral over $u$ in terms of residues. The result is
\be
   \Omega^0_3 =-\frac{i}{2}\sum_{\mathrm{Res}\, u}\left(e^u-u\, \mathrm{Ei}(u)\right)\ \str\Bigg\{\Bigg[\frac{1}{u-i\mathcal{F}^{(0)}}i\mathcal{F}^{(1)}\frac{1}{u-i\mathcal{F}^{(0)}}i\mathcal{F}^{(2)}\frac{1}{u-i\mathcal{F}^{(0)}} + \left(1\leftrightarrow 2\right) \nonumber
\ee
\be
  +\left(\frac{1}{u-i\mathcal{F}^{(0)}}i\mathcal{F}^{(1)}\right)^3\frac{1}{u-i\mathcal{F}^{(0)}} \Bigg]\mathcal{T}\Bigg\}\label{b29}
\ee
where the residues are located at the eigenvalues of $i\mathcal{F}^{(0)}$. Note that the branch choice of $\mathrm{Ei}(u)$, defined as
\be
\mathrm{Ei}(-u) = -\int_{u}^{+\infty} \frac{e^{-s}}{s} ds \label{b49},
\ee
should be such that it agrees with the $s$-integral in~\eqref{eq:Omega3first}, which is analytic when $\mathrm{Re}\,u<0$, where the integration contour lies. The branch cut of the function can be taken to be e.g. on the positive $u$-axis, which is different from the usual principal branch of this function. Since the branch cut does not enter the region encircled by the integration contour, it does not contribute to the result.

The generalization to $\Omega^0_5$ is straightforward, given schematically as
\be
   \Omega^0_5 =-\frac{i}{4}\sum_{\mathrm{Res}\, u}\left(e^u(1+u)-u^2\, \mathrm{Ei}(u)\right)\times  \nonumber
\ee
\be
   \times\ \str\Bigg\{\Bigg[\frac{1}{u-i\mathcal{F}^{(0)}}i\mathcal{F}^{(1)}\left(\frac{1}{u-i\mathcal{F}^{(0)}}i\mathcal{F}^{(2)}\right)^2\frac{1}{u-i\mathcal{F}^{(0)}}+\mathrm{other\ words} \Bigg]\mathcal{T}\Bigg\}\label{b30},
\ee
where ``other words'' indicates the words defined in section \ref{sec:explicit}. The term we have written down corresponds to the word $XYY$ in~\eqref{410}, and we should also add the contributions from all other words in~\eqref{48}--\eqref{410}.

In \eqref{b30}, the poles come from the terms $\frac{1}{u-i\mathcal{F}^{(0)}}$. Each residue then corresponds to a choice of eigenvalue of $i \mathcal{F}^{(0)}$ for each term $\frac{1}{u-i\mathcal{F}^{(0)}}$ in the product, while keeping the first and the last eigenvalue equal (because of the trace). As an example, one residue is associated to picking the first eigenvalue of $\mathcal{F}^{(0)}$ for every $\frac{1}{u-i\mathcal{F}^{(0)}}$ in the product. Then, the order of each pole corresponds to the number of times this eigenvalue appears in the given term. The result obtained by computing each residue is precisely the same as the result obtained using the formula for $\mathcal{L}$ obtained in section~\ref{sec:explicit}.

More generally, if one is interested in $\str(f(i\mathcal{F}))$ instead, one should  expand:
\be
f(i\mathcal{F}_s)  = \sum_{n}^{+\infty} f_n (i\mathcal{F}_s)^n\label{b31}
\ee
and insert the integral representation:
\be
f_n = \frac{1}{2\pi i}\int_{\mathcal{C}} du\, \frac{f(u)}{u^{n+1}}.\label{b32}
\ee
After expanding in words and rearranging the sums, we obtain the same expression as (\ref{eq:Omega3first}), with $f$ instead of the exponential. Then the residue formula obtained is:
\be
   \Omega^0_{(2p+1)} = - i \sum_{\mathrm{Res}\, u} \left( \int_{1}^\infty ds   \frac{f(us^2)}{s^{2p+1}}\right) \times  \nonumber
\ee
\be
   \times\ \str\Bigg\{\Bigg[\left(\frac{1}{u-i\mathcal{F}^{(0)}}i \mathcal{F}^{(1)} \right)^{2p+1}\frac{1}{u-i\mathcal{F}^{(0)}} +\mathrm{other\ words} \Bigg]\mathcal{T}\Bigg\}\label{b33}
\ee
Computing these residues, we obtain deformed versions of the $\mathcal{L}$ operators defined in \eqref{438} (see also appendix~\ref{app:Ldef}).
They are generically given by replacing the factorials in the definition of $\mathcal{L}$ by the coefficients in the series expansion of $f$,
\be
(\mathcal{L})_{a_1, ..., a_{m}, a_{m+1}} =  \sum_{n=0}^{+\infty}  (-1)^{n} b^{2n} f_{n+m} h_{n}(x_{a_1}, ..., x_{a_{m}}, x_{a_{m+1}}),\label{b34}
\ee
\be
\mathcal{L}_{a_1, ..., a_{2p+2-k}} =- \int_{0}^{1} \intd s i^k b(s)^{2p+1-2k} (\mathcal{L})_{a_1, ..., a_{2p+2-k}}.\label{b35}
\ee

\subsection{$\mathcal{L}$ operators for $\Omega_5^0$}\label{app:Ldef}
In this subsection, we show more details for the construction of the $\mathcal{L}$ operators in the simpler case of $\Omega_3^0$. We then list the $\mathcal{L}$ operators obtained for $\Omega_5^0$.

We illustrate the procedure by computing {the contribution of} the word $XY$, which is defined in \eqref{42}-\eqref{45} and we call the sum $(\Omega_3^0)_{XY}$. $(\Omega_3^0)_{XY}$ appears in $\Omega^0_3$, the $3$-form component of \eqref{eq:Omega0}. According to section \ref{sec:explicit}, the sum to calculate is
\be
(\Omega_3^0)_{XY} =  \sum_{n=0}^{+\infty} \frac{1}{(n+2)!} \sum_{i + j + k=n} \str((s^2 \mathcal{T}^2)^i is[D, \mathcal{T}] (s^2 \mathcal{T}^2)^j (iF)   (s^2 \mathcal{T}^2)^k  i\mathcal{T})\label{b36}.
\ee
It can be checked that this sum is gauge-invariant, independently of the other terms. We use (\ref{eq:TDT}), and after computing the supertrace, diagonalize $T^\dag T$ and $TT^\dag$ and rewrite them as $V_L \Sigma^2 V_L^\dag$ and $V_R \Sigma^2 V_R^\dag$ respectively, as defined in \eqref{eq:svd}. Note that we are not performing any gauge fixing here. Rather, we rewrite the tachyon field as a function of $V_L$ and $V_R$ in \eqref{eq:svd}. Crucially, this procedure preserves the gauge invariance of $\Omega^0$. We obtain from \eqref{b36}
\be
(\Omega_3^0)_{XY} =  \sum_{n=0}^{+\infty}  \frac{s^{2n+1}(-1)^n i}{(n+2)!} \bigg[   \nonumber
\ee
\be \sum_{i+j+k=n}\big[-\tr(U_L ( \Sigma^2)^i U_L^\dagger T^\dagger DT U_L ( \Sigma^2)^j U_L^\dagger F_L U_L  ( \Sigma^2)^{k} U_L^\dagger ) \nonumber
\ee
\be
  +\tr(U_R ( \Sigma^2)^i U_R^\dagger T DT^\dagger U_R ( \Sigma^2)^j U_R^\dagger F_R U_R  ( \Sigma^2)^{k} U_R^\dagger )\big] \bigg]\label{b37}.
\ee

We shall focus on the first trace, as the two others are related by $T \leftrightarrow T^\dagger$ and $L\leftrightarrow R $. Next, we use the cyclicity of the trace and rewrite
\be
\tilde{F_{L}} = U_{L}^\dagger F_{L} U_{L}, \quad \tilde{DT} = U_{R}^\dagger DT U_L, \quad \tilde{T^\dag} = U^\dagger_L T^\dagger U_R\label{b38},
\ee
\be
\tilde{F_{R}} = U_{R}^\dagger F_{R} U_{R}, \quad \tilde{DT^\dagger} = U_{L}^\dagger DT^\dagger U_R, \quad \tilde{T} = U^\dagger_R T U_L\label{b39}.
\ee
We shall denote by $x_a$ the diagonal entries of $\Sigma^2$, defined in \eqref{eq:svd}. We write the expressions in matrix flavor indices $a, b, c\dots$. After some algebra we obtain:
\be
(\Omega_3^0)_{XY} = \sum_{n=0}^{+\infty}  \frac{s^{2n+1}(-1)^n i}{(n+2)!} \sum_{a, b}  (-(\tilde{T^\dagger DT})_{ab} (\tilde{F_{L}})_{ba}+(\tilde{TDT^\dag})_{ab} (\tilde{F_{R}})_{ba} ) \times \nonumber \ee\be \times \sum_{i+j+k=n} x_a^i x_b^j x_c^k \delta_{ac}\label{b40}.
\ee
where {\it we do not use} the Einstein summation convention for flavor indices. The last  sum on the right of equation (\ref{b40}) can be rewritten in terms of the complete homogeneous symmetric polynomials of degree $n$:
\be
h_{n}(x_a, x_b, x_c) \equiv \sum_{i+j+k=n} x_a^i x_b^j x_c^k\label{b41}.
\ee
This last polynomial can be computed, explicitly, depending on the number of distinct  $x_a$, $x_b$ or $x_c$,
\be
h_{n}(x_a, x_b, x_c) =
\left\{
    \begin{array}{ll}
        \frac{x_a^{n+2}}{(x_a - x_b)(x_a-x_c)} + \frac{x_b^{n+2}}{(x_b - x_a)(x_b-x_c)}  + \frac{x_c^{n+2}}{(x_c - x_a)(x_c-x_b)}  & \mbox{if } x_a \neq x_b \neq x_c \neq x_a\\
        \frac{(n+2)(n+1)}{2!}x_a^{n} & \mbox{if } x_a = x_b = x_c
    \end{array}
\right.\label{b42}.
\ee
For equal parameters, for instance $x_a=x_b \neq x_c$, $h_n$ becomes a derivative of a symmetric polynomial of higher degree, but with less variables
\be
h_{n}(x_a, x_b, x_c) =
        \frac{d}{d x_a} h_{n+1}(x_a, x_b)\quad  \mbox{if }\quad  x_a = x_b \neq x_c\label{b43}.
\ee
In addition, one must pay attention to the fact that when $k$ parameters are equal, there is an additional combinatorial $\frac{1}{(k-1)!}$ factor. This can be interepreted in terms of the residue formula given in the previous subsection as the order of a pole increasing.

We now define the $\mathcal{L}$ operator (that depends on $s$)
\be
(\mathcal{L}_s)_{abc} = \sum_{n=0}^{+\infty}  \frac{s^{2n}(-1)^n}{(n+2)!} h_{n}(x_a, x_b, x_c)\label{b44}.
\ee
The trace enforces that the only $\mathcal{L}_s$ objects we need have equal $x_a$ and $x_c$ because of the $\delta_{ac}$ appearing in \eqref{b40}. We write this with indices $aba$, and we can rewrite $(\mathcal{L}_s)_{aba}$ as:
\be
(\mathcal{L}_s)_{aba}  \equiv
\left\{
    \begin{array}{ll}
        \frac{d}{d x_a} \left[ \frac{e^{-s^2 x_a}}{s^4 (x_a-x_b)} + \frac{e^{-s^2 x_b}}{s^4 (x_b-x_a)}   \right] & \mbox{if } x_a \neq x_b\\
         \frac{1}{2} e^{-s^2 x_a} & \mbox{if } x_a = x_b
    \end{array}
\right. \label{b45}.
\ee
Using this definition for $(\Omega_3^0)_{XY}$ we obtain:
\be
(\Omega_3^0)_{XY} = is \sum_{a, b} (\mathcal{L}_s)_{aba} \left[-(\tilde{T^\dagger DT})_{ab} (\tilde{F_{L}})_{ba} + (\tilde{T DT^\dagger})_{ab} (\tilde{F_{R}})_{ba} \right].
\label{sum:XY}
\ee

It can be shown that this result is gauge-invariant. Indeed, $(\mathcal{L}_s)$ is a gauge-invariant object as the eigenvalues of $TT^\dag$ entering its definition are independent of the gauge. Then, by an appropriate gauge transformation, we can go from the tilde fields \eqref{b38}-\eqref{b39} to the original fields in \eqref{sum:XY}, with the result
\be
(\Omega_3^0)_{XY} = is \sum_{a, b} (\mathcal{L}_s)_{aba} \left[-(T^\dagger DT)_{ab} (F_{L})_{ba} + (T DT^\dagger)_{ab} (F_{R})_{ba} \right].
\label{sum:XY2}
\ee

Now, note that $\mathcal{L}_s$ is the only $s$-dependent function in $(\Omega_3^0)_{XY}$. We can integrate it with respect to $s$ from $1$ to $\infty$, to obtain
\be
(\mathcal{L})_{aba} \equiv -\int_{1}^{+\infty} is (\mathcal{L}_s)_{aba} ds =
\left\{
    \begin{array}{ll}
        -i\frac{e^{-x_b}- e^{-x_a} + x_b (\mathrm{Ei}(-x_b)-\mathrm{Ei}(-x_a))}{2(x_a - x_b)^2} & \mbox{if } x_a \neq x_b\\
        -i\frac{e^{-x_a}}{4x_a} & \mbox{if } x_a = x_b,
    \end{array}
\right.
\label{eq:defLab}
\ee
with $\mathrm{Ei}(u)$ the exponential integral function defined in \eqref{b49}. Note that the operators $\mathcal{L}_s$ depend on the degree of the symmetric polynomials appearing in the summations like \eqref{b40} (see \eqref{b41} for the definition of this degree). The degree is $2p+2-k$, that is the number of letters in the word plus one. Here $k$ is the number of occurences of the letter $Y$ in the word. However, $\mathcal{L}$ depends explicitly on $k$ as can be seen from \eqref{genericL}. We  indicate this dependence, when there is an ambiguity. For instance, the $XYY$ word with $k=2$, which appears in $\Omega_5$ and the $XXX$ word with $k=0$ in $\Omega_3$ both have equal $\mathcal{L}(b(s))$ operators with four indices. However, due to \eqref{genericL} their integrated $\mathcal{L}$ operators are different as they depend on $k$. For the other words, we show the result simply for the $\mathcal{L}$ operators. They are in general defined by
\be
(\mathcal{L}_s)_{a_1, ..., a_{m}, a_{m+1}} =  \sum_{n=0}^{+\infty}  \frac{(-1)^{n} s^{2n}}{(n+m)!}h_{n}(x_{a_1}, ..., x_{a_{m}}, x_{a_{m+1}})\label{b50}.
\ee
The integrated $\mathcal{L}$ on the path from $\infty$ to $1$ that arises in
$\Omega^0$ (see \eqref{eq:Omega0}) is
\be
\mathcal{L}_{a_1, ..., a_{2p+2-k}} = \int_{+\infty}^1 ds i^k s^{2p+1-2k} (\mathcal{L}_s)_{a_1, ..., a_{2p+2-k}},
\label{genericL2}
\ee
with $k$ the number of $Y$'s in the words, and $2p+1$ the form order of the TCS form.

The value of the $\mathcal{L}$ operators in \eqref{genericL2} does not depend on the order of the indices. We first show the result for the $\mathcal{L}$ operators with $1$ and $2$ different indices, then for $3$ different indices. For the word $XXX$ contributing to $\Omega_3^0$ we obtain

\be
(\mathcal{L})^{(k=0)}_{aaaa}  = -\frac{e^{-x_a} (1 + x_a)}{12 x_a^2}
\label{Laaa}
\ee
\be
(\mathcal{L})^{(k=0)}_{aaba} =(\mathcal{L})^{(k=0)}_{abaa} =  \frac{e^{-x_a - x_b}}{4\,x_a\,(x_a - x_b)^3}
\left[
-2 e^{x_a}x_a
+ e^{x_b}\left(
x_a(2+x_a) - 2x_a x_b + x_b^2 \right.\right. \nonumber
\ee
\be
  \left.\left.
- 2 e^{x_a}x_a x_b\big(\operatorname{Ei}(-x_b)-\operatorname{Ei}(-x_a)\big)
\right)
\right].
\label{Laab}
\ee
\be
 (\mathcal{L})^{(k=0)}_{abba}  = -\frac{1}{(x_a - x_b)^3}
\left[
e^{-x_a}
- e^{-x_b}
+ \tfrac{1}{2}(x_a+x_b)\big(\operatorname{Ei}(-x_a)-\operatorname{Ei}(-x_b)\big)
\right].
\label{Labb}
\ee
We shall now cover the case of $\Omega_5^0$. For the terms $XYY$, $YXY$, $YYX$ we obtain:
\be
\mathcal{L}^{(k=2)}_{aaaa} = \frac{e^{-x_a}}{12 x_a}\label{b55}.
\ee
\be
\mathcal{L}^{(k=2)}_{aaba} =
\frac{e^{-x_a - x_b}}{4 (x_a - x_b)^3}
\Big[
- e^{x_a}(-1 + x_b)
+ e^{x_b}\Big(
-1 - x_a + 2x_b  \nonumber
\ee
\be
+ e^{x_a}x_b^2\big(\operatorname{Ei}(-x_a) - \operatorname{Ei}(-x_b)\big)
\Big)
\Big].\label{b56}
\ee
\be
\mathcal{L}^{(k=2)}_{abba} =  \frac{e^{-x_a - x_b}}{2 (x_a - x_b)^3}
\Big[
e^{x_a}(-1 + x_a)
+ e^{x_b}\Big(
1 - x_b   \\
 - e^{x_a} x_a x_b (\operatorname{Ei}(-x_a) - \operatorname{Ei}(-x_b))\label{b57}.
\Big)
\Big]
\ee
All terms with the same number of $a$'s and $b$'s are equal. For the terms $XXXY$, $XXYX$, $XYXX$ and $YXXX$ we obtain:
\be
\mathcal{L}_{aaaaa} =-i\frac{( e^{-x_a} (1 + x_a))}{48 x_a^2}\label{b58}.
\ee
\be
\mathcal{L}_{aaaba} =
\frac{i\,e^{-x_a - x_b}}{12\,x_a\,(x_a - x_b)^4}
\big[
3 e^{x_a} x_a (-1 + x_b)
+ e^{x_b}\Big(
x_a^3 - 3x_a^2(-1 + x_b)  \nonumber
\ee
\be
    + 3x_a(-1 + x_b)^2 - x_b^3
- 3 e^{x_a} x_a x_b^2\big(\operatorname{Ei}(-x_a) - \operatorname{Ei}(-x_b)\big)
\Big)
\big]\label{b59}.
\ee
\be
\mathcal{L}_{aabba} =  \frac{i\,e^{-x_a - x_b}}{4 (x_a - x_b)^4}
\bigg[
- e^{x_a}(-3 + 2x_a + x_b)
+ e^{x_b}\bigg(
-3 - x_a + 4x_b  \nonumber
\ee
\be
+ e^{x_a} x_b (2x_a + x_b)\big(\operatorname{Ei}(-x_a) - \operatorname{Ei}(-x_b)\big)
\bigg)
\bigg]\label{b60}.
\ee
The $\mathcal{L}$ operators for the word $XXXXX$ appearing in $\Omega^0_5$ are:
\be
\mathcal{L}_{aaaaaa} = -\frac{e^{-x_a} (2 + 2x_a + x_a^2)}{240 x_a^3}\label{b61}.
\ee
\be
\mathcal{L}_{aaaaba} =  \frac{e^{-x_a - x_b}}{48\,x_a^2\,(x_a - x_b)^5}
\bigg[
12 e^{x_a} x_a^2 (-1 + x_b)
+ e^{x_b}\bigg(
x_a^2 (2 + x_a)(6 + x_a(3 + x_a))  \nonumber
\ee
\be
 - 4 x_a^2 (6 + x_a(4 + x_a)) x_b
+ 6 x_a^2 (3 + x_a) x_b^2
- 4 x_a (2 + x_a) x_b^3
+ (1 + x_a) x_b^4 \nonumber
\ee
\be
 - 12 e^{x_a} x_a^2 x_b^2 \big(\operatorname{Ei}(-x_a) - \operatorname{Ei}(-x_b)\big)
\bigg)
\bigg]\label{b62}.
\ee
\be
\mathcal{L}_{aaabba} =  \frac{e^{-x_a - x_b}}{12\,x_a\,(x_a - x_b)^5}
\bigg[
-6 e^{x_a} x_a (-2 + x_a + x_b)
+ e^{x_b}\bigg(
 -x_a (12 + x_a (6 + x_a))
  \nonumber
\ee
\be
  + 3 x_a (6 + x_a) x_b
 - 3 x_a x_b^2
 + x_b^3
 + 6 e^{x_a} x_a x_b (x_a + x_b)
   \big(\operatorname{Ei}(-x_a) - \operatorname{Ei}(-x_b)\big)
\bigg)
\bigg]\label{b63}.
\ee
\be
\mathcal{L}_{aabbba} =  \frac{e^{-x_a - x_b}}{4 (x_a - x_b)^5}
\bigg[
6 e^{x_a}(x_a - 1)
+ e^{x_b}\bigg(
6(1 - x_b) \nonumber
\ee
\be
 - e^{x_a}(x_a^2 + 4 x_a x_b + x_b^2)
  \big(\operatorname{Ei}(-x_a) - \operatorname{Ei}(-x_b)\big)
\bigg)
\bigg]\label{b64}.
\ee

The values of the other $\mathcal{L}$ operators that appear in $\Omega_5$ with at most two different eigenvalues can be deduced from \eqref{b55}-\eqref{b64} using that the $\mathcal{L}$ operators are invariant under exchanging indices. When $N_f>2$ there can be more than two different eigenvalues, and therefore one should compute the $\mathcal{L}$ with $N_f$ different indices. For instance, for $N_f = 3$, one should compute $\mathcal{L}_{aaabca}$, $\mathcal{L}_{aabbca}$ and $\mathcal{L}_{abbcca}$, $\mathcal{L}_{aabca}$, $\mathcal{L}_{abbca}$ and $\mathcal{L}_{abca}$. All the other operators may be included by using the symmetry in letters from these ones. We obtain:

\be
\mathcal{L}^{(k=2)}_{abca} = -\frac{1}{4 (x_a-x_b)^2 (x_a-x_c)^2 (x_b-x_c)}
\Bigg[
- e^{-x_b}(-1+x_b)(x_a-x_c)^2 \nonumber
\ee
\be
+ e^{-x_c}(-1+x_c)(x_a-x_b)^2\nonumber
\ee
\be
+ e^{-x_a}(x_b-x_c)\big(x_b+x_c-2 x_b x_c + x_a(-2+x_b+x_c)\big)\nonumber
\ee
\be
- x_a(x_b-x_c)\big(-2 x_b x_c + x_a(x_b+x_c)\big)\operatorname{Ei}(-x_a)\nonumber
\ee
\be
+ x_b^{2}(x_a-x_c)^2 \operatorname{Ei}(-x_b)
- x_c^{2}(x_a-x_b)^2 \operatorname{Ei}(-x_c)
\Bigg]\label{b65}.
\ee

\be
\mathcal{L}^{(k=1)}_{aabca} = \frac{i}{4 (x_a-x_b)^3 (x_a-x_c)^3 (x_b-x_c)}
\,e^{-(x_a+x_b+x_c)}
\Bigg[
e^{x_a+x_b} (x_a-x_b)^3 (-1+x_c)
\nonumber
\ee
\be
+ e^{x_c}\Big(
e^{x_a}(-1+x_b)(-x_a+x_c)^3
+ e^{x_b}(-x_b+x_c)\big(
x_a^3 + x_b^2(1-2x_c)
\nonumber
\ee
\be
+ x_b(1-2x_c)x_c + x_c^2
-3x_a^2(-1+x_b+x_c)
+ x_a\big(x_b^2+(-3+x_c)x_c+x_b(-3+7x_c)\big)
\big)
\Big)
\nonumber
\ee
\be
+ e^{x_a+x_b}\Big(
-(x_b-x_c)\big(-3x_a^2 x_b x_c + x_b^2 x_c^2 + x_a^3(x_b+x_c)\big)\operatorname{Ei}(-x_a)
\nonumber
\ee
\be
- x_b^2(-x_a+x_c)^3 \operatorname{Ei}(-x_b)
+ x_c^2(x_a-x_b)^3 \operatorname{Ei}(-x_c)
\Big)
\Bigg]\label{b66}.
\ee

\be
\mathcal{L}^{(k=1)}_{abbca} = -\frac{i}{4 (x_a-x_b)^3 (x_a-x_c)^2 (x_b-x_c)^2}
\Bigg[
e^{-x_c}(x_a-x_b)^3
- 3 e^{-x_a} x_a (x_b-x_c)^2\nonumber
\ee
\be
+ e^{-x_a} x_a^2 (x_b-x_c)^2
+ e^{-x_a} x_b (x_b-x_c)^2
+ e^{-x_a} x_a x_b (x_b-x_c)^2
\nonumber
\ee
\be
- e^{-x_c} (x_a-x_b)^3 x_c
+ 2 e^{-x_a} (x_b-x_c)^2 x_c
- 2 e^{-x_a} x_b (x_b-x_c)^2 x_c
\nonumber
\ee
\be
- e^{-x_b} (x_a-x_c)^2\!\left[(-3+x_b)x_b + x_a(1+x_b-2 x_c) + 2 x_c\right]
\nonumber
\ee
\be
- x_a^3 (x_b-x_c)^2 \operatorname{Ei}(-x_a)
- x_a^2 x_b (x_b-x_c)^2 \operatorname{Ei}(-x_a)
\nonumber
\ee
\be
+ 2 x_a x_b (x_b-x_c)^2 x_c \operatorname{Ei}(-x_a)
+ x_a x_b^2 (x_a-x_c)^2 \operatorname{Ei}(-x_b)
\nonumber
\ee
\be
+ x_b^3 (x_a-x_c)^2 \operatorname{Ei}(-x_b)
- 2 x_a x_b (x_a-x_c)^2 x_c \operatorname{Ei}(-x_b)
\nonumber
\ee
\be
+ (x_a-x_b)^3 x_c^2 \operatorname{Ei}(-x_c)
\Bigg]\label{b67}.
\ee

\be
\mathcal{L}^{(k=0)}_{aaabca} =
-\frac{1}{24 x_a (x_a-x_b)^4 (x_a-x_c)^4 (x_b-x_c)}
\,\operatorname{Ei}\!\big(-(x_a+x_b+x_c)\big)\nonumber
\ee
\be
\Bigg[
-6 e^{x_a+x_c} x_a (-1+x_b) (x_a-x_c)^4
+ 6 e^{x_a+x_b} x_a (x_a-x_b)^4 (-1+x_c)
\nonumber
\ee
\be
+ 2 e^{x_b+x_c}\Big(
x_a x_b\!\big[-x_a^3 (12+x_a(6+x_a))
+ 3 x_a^2 (6+x_a(6+x_a)) x_b\nonumber
\ee
\be
- 3 x_a (2+x_a)^2 x_b^2
+ (3+x_a(3+x_a)) x_b^3\big]
\nonumber
\ee
\be
+ x_a\!\left[x_a^3 (12+x_a(6+x_a))
- 6 x_a^2 (6+x_a) x_b^2
+ 8 x_a (3+x_a) x_b^3
- 3 (2+x_a) x_b^4\right] x_c
\nonumber
\ee
\be
- 3 x_a\!\left[x_a^2 (6+x_a(6+x_a))
- 2 x_a^2 (6+x_a) x_b
+ 2 x_a x_b^3
- x_b^4\right] x_c^2
\nonumber
\ee
\be
+ \left[3 x_a^2 (2+x_a)^2
- 8 x_a^2 (3+x_a) x_b
+ 6 x_a^2 x_b^2
- x_b^4\right] x_c^3
\nonumber
\ee
\be
+ \left[-x_a^3 + 3 x_a^2 (-1+x_b)
- 3 x_a (-1+x_b)^2 + x_b^3\right] x_c^4
\Big)
\nonumber
\ee
\be
- 3 e^{x_a} x_a (x_b-x_c) (x_a^2-x_b x_c)
\big(-4 x_a x_b x_c + x_a^2 (x_b+x_c) + x_b x_c (x_b+x_c)\big)
\,\operatorname{Ei}(-x_a)
\nonumber
\ee
\be
+ 3 e^{x_a} x_a x_b^2 (x_a-x_c)^4 \operatorname{Ei}(-x_b)
- 3 e^{x_a} x_a (x_a-x_b)^4 x_c^2 \operatorname{Ei}(-x_c)
\Bigg]\label{b68}.
\ee

\be
\mathcal{L}^{(k=0)}_{aabbca} = \frac{1}{4 (x_a-x_b)^4 (x_a-x_c)^3 (x_b-x_c)^2}
\Bigg[
e^{-x_c} (x_a-x_b)^4 (-1+x_c)
\nonumber
\ee
\be
+ e^{-x_b} (x_a-x_c)^3 \big(2(-2+x_b)x_b + x_a(1+x_b-2x_c) - (-3+x_b)x_c\big)
\nonumber
\ee
\be
+ e^{-x_a} (x_b-x_c)^2 \Big(
x_b^2 - 2(-1+x_b)x_b x_c + (3-4x_b)x_c^2
- 2 x_a^2(-3+2x_b+x_c)
\nonumber
\ee
\be
+ x_a\big((-4+x_b)x_b + 2(-4+5x_b)x_c + x_c^2\big)
\Big)
\nonumber
\ee
\be
+ (x_b-x_c)^2 \big(x_a^3(x_a+2x_b) - 6x_a^2 x_b x_c + x_b(2x_a+x_b)x_c^2\big)\operatorname{Ei}(-x_a)
\nonumber
\ee
\be
- x_b (x_a-x_c)^3 \big(x_b(x_a+2x_b) - (2x_a+x_b)x_c\big)\operatorname{Ei}(-x_b)
\nonumber
\ee
\be
- x_c^2 (x_a-x_b)^4 \operatorname{Ei}(-x_c)
\Bigg]\label{b69}.
\ee

\be
\mathcal{L}^{(k=0)}_{abbcca} =
-\frac{1}{2 (x_a-x_b)^3 (x_a-x_c)^3 (x_b-x_c)^3}
\Bigg[
- e^{-x_c} x_a (x_a-x_b)^3\nonumber
\ee
\be
+ e^{-x_c} x_a (x_a-x_b)^3 x_b
+ e^{-x_c} x_b (-x_a+x_b)^3
\nonumber
\ee
\be
+ 2 e^{-x_a} x_a (x_b-x_c)^3
+ 2 e^{-x_c} (x_a-x_b)^3 x_c
+ e^{-x_a} x_b (x_b-x_c)^3 x_c
\nonumber
\ee
\be
- e^{-x_c} (x_a-x_b)^3 x_c^2
+ e^{-x_a} x_a^2 (-x_b+x_c)^3
+ e^{-x_a} x_b (-x_b+x_c)^3
\nonumber
\ee
\be
+ e^{-x_a} x_c (-x_b+x_c)^3
+ e^{-x_b} (x_a-x_c)^3 \big(x_a+(-2+x_b)x_b+x_c-x_a x_c\big)
\nonumber
\ee
\be
- x_a x_b (x_b-x_c)^3 x_c \operatorname{Ei}(-x_a)
- x_a^3 (-x_b+x_c)^3 \operatorname{Ei}(-x_a)
\nonumber
\ee
\be
- x_b^3 (x_a-x_c)^3 \operatorname{Ei}(-x_b)
- x_a x_b x_c (-x_a+x_c)^3 \operatorname{Ei}(-x_b)
\nonumber
\ee
\be
- x_a (x_a-x_b)^3 x_b x_c \operatorname{Ei}(-x_c)
+ (x_a-x_b)^3 x_c^3 \operatorname{Ei}(-x_c)
\Bigg]\label{b70}.
\ee

\subsection{Limit in $T\to 0$ of $\Omega$}
\label{sec:limitT0}

In this subsection, we compute the zero-tachyon limit of the generic TCS action \eqref{eq:Omegagen}. We shall argue that it reduces to the standard CS action \eqref{336} up to gauge-invariant boundary terms. For this purpose, we rescale the tachyon field by a small parameter $\varepsilon$,
\be
T  \to \varepsilon T.\label{5c1}
\ee

We use a result from appendix \ref{sec:F} stating that the TCS forms obtained from two different paths through equation \eqref{335} differ by an exact form. To compute the limit $T\to 0$ of the TCS form, we shall choose a convenient path. Then, the previous result guarantees that the TCS form computed from the canonical path, will differ by an exact form from our result.

It is convenient to use the path $\gamma_{\mathrm{alt}}$, that appears as the first item in the itemized list in  \ref{path:galt}. $\gamma_{\mathrm{alt}}$ was  introduced in section \ref{sec:TCSQuillen}, in figure \ref{plane}, and studied in appendix \ref{sec:gammaalt}. The TCS form $\Omega_5^{\gamma_{\rm alt}}$ is the sum of the standard CS form $\omega_5$ and the gauge-invariant term $Z_5^0$, \eqref{b18-ii}, which is integrated on the path $II_{\rm alt}$ in figure \ref{plane}. Moreover, $Z_5^0$ vanishes in the $T\to 0$ limit, due to the following relations
\be
Z_5^0 (A_L, A_R, \varepsilon T) = \int_0^1 \intd s \;\str(i \varepsilon \mathcal{T} e^{i \mathcal{F}(1, \varepsilon s)}) = \int_0^\varepsilon \intd \tilde s\; \str(i \mathcal{T} e^{i \mathcal{F}(1, \tilde s)})\;, \label{5c2}
\ee
 where we changed variables $\tilde{s} = \varepsilon s$ in \eqref{z0}. The integrand in the last integral in \eqref{5c2} is independent of $\varepsilon$. In the small $\varepsilon$ limit, it is manifest that $Z_5^0$ vanishes, since the integrant is regular at any point on the path.
 This justifies the claim below equation \eqref{z0} that $Z_5^0 (T=0) =0$.

We deduce that the TCS form, obtained from the path $\gamma_{\mathrm{alt}}$ manifestly asymptotes to the standard CS form in the small $T$ limit. We conclude, using the result of appendix \ref{sec:F}, that the TCS form obtained from the path $\gamma$,  also asymptotes to the standard CS form, in the small $T$ limit up to boundary terms\footnote{Note that because the standard CS form does not satisfy criterion \ref{cr:noIR}, this type of boundary term is expected to appear in the IR.}. These boundary terms are shown to be gauge-invariant in section \ref{properties}.

\subsection{Variations of $\Omega$}
We now compute the variations of $\Omega$, defined in \eqref{eq:Omegagen} with respect to the gauge fields $A_L$, $A_R$ and the tachyon field $T$. Since $\Omega^b$ is a boundary contribution and $\Omega^c$ is closed they do not contribute to the bulk equations of motion. In the following we compute the contribution of $\Omega^0$. The purpose of this section is not only to derive the bulk equations of motion, but also the contribution of the TCS action to the conjugate momenta.

Indeed, if one is interested only in the equations of motion, it is more convenient to rely on the condensed formula derived in appendix \ref{sec:F}, which we recall here \eqref{12F},
\be
\frac{\delta \Omega_5}{\delta \mathcal{A}} = i \exp(i \mathcal{F})\label{eomTCS}.
\ee
In the rest of this section, we compute the variation of $\Omega_5^0$ while keeping track of boundary contributions.

For $\Omega_3^0$ we obtain
\be
\delta_{A_L} \Omega^0_3 = -i \tr_{\mathcal{L}}\big(T^\dag T \delta A_L, F_L, \mathbb{I} \big) + \intd \tr_{\mathcal{L}}\big(T^\dag DT, \delta A_L, \mathbb{I} \big) - \tr_{\mathcal{L}}\big(\intd(T^\dag DT), \delta A_L, \mathbb{I} \big)\nonumber
\ee
\be
 +i \tr_{\mathcal{L}}\big(T^\dag DT, \delta A_L A_L, \mathbb{I} \big) +i \tr_{\mathcal{L}}\big(T^\dag DT, A_L \delta  A_L, \mathbb{I} \big) + \intd \tr_{\mathcal{L}}\big(\delta A_L , DT^\dag T, \mathbb{I} \big)\nonumber
\ee
\be
 + \tr_{\mathcal{L}}\big(\delta A_L, \intd(DT^\dag T), \mathbb{I} \big) -i  \tr_{\mathcal{L}}\big(A_L \delta A_L, DT^\dag T, \mathbb{I} \big)  -i  \tr_{\mathcal{L}}\big(\delta A_L A_L,DT^\dag T, \mathbb{I} \big) \nonumber
\ee
\be
 +i  \tr_{\mathcal{L}}\big(F_L, \delta A_L T^\dag T, \mathbb{I} \big) + i  \tr_{\mathcal{L}}\big(T^\dag T \delta A_L,DT^\dag, DT,  \mathbb{I} \big)  +i \tr_{\mathcal{L}}\big(T^\dag DT,\delta A_L T^\dag, DT,  \mathbb{I} \big)\nonumber
\ee
\be
  +i \tr_{\mathcal{L}}\big(T^\dag DT, DT^\dag , T\delta A_L,  \mathbb{I} \big) + i \tr_{\mathcal{L}}\big(T^\dag T \delta A_L,F_R,  \mathbb{I} \big) - i \tr_{\mathcal{L}}\big(F_R,T  \delta A_L T^\dag,  \mathbb{I} \big)\nonumber
\ee
\be
  + i \tr_{\mathcal{L}}\big(T \delta A_L T^\dag,DT,DT^\dag,  \mathbb{I} \big) - i \tr_{\mathcal{L}}\big(T DT^\dag,T\delta A_L, DT^\dag,  \mathbb{I} \big) \nonumber
\ee
\be
 +i \tr_{\mathcal{L}}\big(T DT^\dag ,DT,\delta A_L T^\dag,  \mathbb{I} \big)\label{b71}.
\ee
To simplify \eqref{b71} we shall now show that $\tr_{\mathcal{L}}$ satisfies a graded cyclicity property following from the definition (\ref{440}). Consider the following sum
\be
\sum_{a_1, ..., a_{n+1}} (\mathcal{L})_{a_1 ... a_n a_{n+1}} (X_2)_{a_1 a_2} ...(X_n)_{a_{n-1}a_n} (\mathbb{I})_{a_{n} a_{n+1}} (X_1)_{a_{n+1} a_1}.\label{b72}
\ee
On the one hand, we apply the transformation
\be
a_i \to a_{i+1}, \qquad  a_{n+1} \to a_{1}\label{b73},
\ee
to \eqref{b72}, to obtain the next equation
\be
\sum_{a_1, ..., a_{n+1}} (\mathcal{L})_{a_2 ... a_{n+1} a_{1}} (X_2)_{a_2 a_3} ...(X_n)_{a_{n}a_{n+1}} (\mathbb{I})_{a_{n+1} a_{1}} (X_1)_{a_{1} a_2}.\label{b74}
\ee
We commute $(X_1)_{a_1 a_2}$ with all the other forms to place it on the left. Since the total form degree of \eqref{b74} is odd, this operation does not change the sign. Then, we use that $\mathcal{L}$ is symmetric under permutations of its indices. We obtain the exact definition of $\tr_{\mathcal{L}}(X_1, ..., X_n, \mathbb{I})$, \eqref{440}. On the other hand, equation \eqref{b72} can then be rewritten
\be
\sum_{a_1, ..., a_{n+1}} (\mathcal{L})_{a_1 ... a_n a_{n+1}} (X_2)_{a_1 a_2} ...(X_n)_{a_{n-1}a_n} (\mathbb{I})_{a_{n} a_{n+1}} (X_1)_{a_{n+1} a_1} = \tr_{\mathcal{L}}(X_2, ..., X_n, \mathbb{I}, X_1) \label{b75}.
\ee
We conclude that the weighted traces are graded\footnote{If the differential form degree of the weighted trace is even, one needs to keep track of the signs that arise from commuting the forms.} cyclic,
\be
\tr_{\mathcal{L}}(X_1, ..., X_n, \mathbb{I}) = \tr_{\mathcal{L}}(X_2, ..., X_n, \mathbb{I}, X_1)\label{b76}.
\ee
We use \eqref{b76} to rewrite \eqref{b71},
\be
\delta_{A_L} \Omega^0_3 =  -i \tr_{\mathcal{L}}\big(T^\dag T \delta A_L, F_L, \mathbb{I} \big) - \intd \tr_{\mathcal{L}}\big(\delta A_L , \mathbb{I} , T^\dag DT \big) - \tr_{\mathcal{L}}\big(\delta A_L, \mathbb{I}, \intd(T^\dag DT) \big)\nonumber
\ee
\be
 +i \tr_{\mathcal{L}}\big(A_L \delta A_L  + \delta A_L A_L, \mathbb{I}, T^\dag DT \big) + \intd \tr_{\mathcal{L}}\big(\delta A_L , DT^\dag T, \mathbb{I} \big)+ \tr_{\mathcal{L}}\big(\delta A_L , \intd(DT^\dag T), \mathbb{I} \big)\nonumber
\ee
\be
 -i  \tr_{\mathcal{L}}\big(A_L \delta A_L  + \delta A_L A_L, DT^\dag T, \mathbb{I} \big) +i  \tr_{\mathcal{L}}\big(\delta A_L  T^\dag T, \mathbb{I},F_L \big) + i  \tr_{\mathcal{L}}\big(T^\dag T \delta A_L ,DT^\dag, DT,  \mathbb{I} \big) \nonumber
\ee
\be
  +i \tr_{\mathcal{L}}\big(\delta A_L  T^\dag, DT,  \mathbb{I}, T^\dag DT \big)
 +i \tr_{\mathcal{L}}\big(T\delta A_L ,  \mathbb{I}, T^\dag DT, DT^\dag \big) + i \tr_{\mathcal{L}}\big(T^\dag T\delta A_L  ,F_R,  \mathbb{I} \big) \nonumber
\ee
\be
 - i \tr_{\mathcal{L}}\big(T \delta A_L  T^\dag,  \mathbb{I}, F_R \big)
+ i \tr_{\mathcal{L}}\big(T \delta A_L  T^\dag,DT,DT^\dag,  \mathbb{I} \big) - i \tr_{\mathcal{L}}\big(T\delta A_L , DT^\dag,  \mathbb{I}, T DT^\dag \big) \nonumber
\ee
\be
 +i \tr_{\mathcal{L}}\big(\delta A_L  T^\dag,  \mathbb{I} , T DT^\dag ,DT\big)\label{b77}.
\ee

We now specify the variation \eqref{b77} with respect to a given flavor matrix component of $A_L$. We write
\be
(M A_L N)_{a_1 a_2} = (M)_{a_1 p} (A_L)_{pq} (N)_{q a_2},\label{b78}
\ee
then we express the variation with respect to the component $pq$ as:
\be
\frac{\delta (M A_L N)_{a_1 a_2}}{\delta (A_L)_{pq}} =M_{a_1 p} N_{q a_2} \equiv (M_{. p} N_{q .})_{a_1 a_2}\label{b79}.
\ee
In \eqref{b79}, $M_{. p} N_{q .}$ is treated as a flavor space matrix. We can re-express the variation of $\Omega^0_3$ as
\be
\frac{\delta \Omega^0_3}{\delta (A_L)_{pq}} =  -i \tr_{\mathcal{L}}\big((T^\dag T)_{.p}  \mathbb{I}_{q.}, F_L, \mathbb{I} \big) - \intd \tr_{\mathcal{L}}\big( \mathbb{I}_{.p}  \mathbb{I}_{q.} , \mathbb{I} , T^\dag DT \big)\nonumber
\ee
\be
 - \tr_{\mathcal{L}}\big(\mathbb{I}_{.p}  \mathbb{I}_{q.}, \mathbb{I}, \intd(T^\dag DT) \big) +i \tr_{\mathcal{L}}\big((A_L)_{.p}  \mathbb{I}_{q.} \nonumber
\ee
\be
 + \mathbb{I}_{.p}  (A_L)_{q.}, \mathbb{I}, T^\dag DT \big) + \intd \tr_{\mathcal{L}}\big(\mathbb{I}_{.p}  \mathbb{I}_{q.} , DT^\dag T, \mathbb{I} \big)\nonumber
\ee
\be
 + \tr_{\mathcal{L}}\big(\mathbb{I}_{.p}  \mathbb{I}_{q.}, \intd(DT^\dag T), \mathbb{I} \big) -i  \tr_{\mathcal{L}}\big((A_L)_{.p}  \mathbb{I}_{q.} \nonumber
\ee
\be
  + \mathbb{I}_{.p}  (A_L)_{q.}, DT^\dag T, \mathbb{I} \big) +i  \tr_{\mathcal{L}}\big(\mathbb{I}_{.p}  \mathbb{I}_{q.} T^\dag T, \mathbb{I},F_L \big) \nonumber
\ee
\be
  + i  \tr_{\mathcal{L}}\big((T^\dag T)_{.p}  \mathbb{I}_{q.} ,DT^\dag, DT,  \mathbb{I} \big)  +i \tr_{\mathcal{L}}\big(\mathbb{I}_{.p}  T^\dag_{q.}, DT,  \mathbb{I}, T^\dag DT \big)\nonumber
\ee
\be
 +i \tr_{\mathcal{L}}\big(T_{.p}  \mathbb{I}_{q.},  \mathbb{I}, T^\dag DT, DT^\dag \big) + i \tr_{\mathcal{L}}\big((T^\dag T)_{.p}  \mathbb{I}_{q.} ,F_R,  \mathbb{I} \big) \nonumber
\ee
\be
 - i \tr_{\mathcal{L}}\big(T_{.p}  T^\dag_{q.},  \mathbb{I}, F_R \big) + i \tr_{\mathcal{L}}\big(T_{.p}  T^\dag_{q.} ,DT,DT^\dag,  \mathbb{I} \big)\nonumber
\ee
\be
  - i \tr_{\mathcal{L}}\big(T_{.p}  \mathbb{I}_{q.} , DT^\dag,  \mathbb{I}, T DT^\dag \big) +i \tr_{\mathcal{L}}\big(\mathbb{I}_{.p}  T^\dag_{q.},  \mathbb{I} , T DT^\dag ,DT\big)\label{b80}.
\ee
The variation with respect to $A_R$ can be obtained from \eqref{b80} by exchanging $L \leftrightarrow R$ and $T \leftrightarrow T^\dag$ in this formula and changing the overall sign.

We can compute likewise the variations of $\Omega^0_5$.
\be
\delta_{A_L} \Omega^0_5 =  - i \tr_{\mathcal{L}}\big( T^\dag T \delta A_L, DT^\dag , DT, DT^\dag, DT, \mathbb{I} \big) \nonumber
\ee
\be
-i \tr_{\mathcal{L}}\big(T\delta A_L T^\dag,DT,DT^\dag,DT,DT^\dag,\mathbb{I}\big) \nonumber
\ee
\be
 +i \tr_{\mathcal{L}}\big(T^\dag DT,\delta A_L T^\dag,DT,DT^\dag,DT,\mathbb{I} \big)
+i \tr_{\mathcal{L}}\big(TDT^\dag,T\delta A_L,DT,DT^\dag,DT,\mathbb{I}\big)\nonumber
\ee
\be
 -i \tr_{\mathcal{L}}\big(T^\dag DT,DT^\dag,T\delta A_L,DT^\dag,DT,\mathbb{I} \big)
-i \tr_{\mathcal{L}}\big(TDT^\dag,DT,\delta A_L T^\dag,DT,DT^\dag,\mathbb{I} \big)\nonumber
\ee
\be
 +i \tr_{\mathcal{L}}\big(T^\dag DT,DT^\dag,DT,\delta A_L T^\dag,DT, \mathbb{I} \big)
+i \tr_{\mathcal{L}}\big(TDT^\dag,DT,DT^\dag,T\delta A_L,DT^\dag,\mathbb{I} \big)\nonumber
\ee
\be
 -i \tr_{\mathcal{L}}\big(T^\dag DT,DT^\dag,DT,DT^\dag,T\delta A_L,\mathbb{I} \big)
-i \tr_{\mathcal{L}}\big(TDT^\dag,DT,DT^\dag,DT,\delta A_L T^\dag, \mathbb{I} \big)\nonumber
\ee
\be
 +i\tr_{\mathcal{L}}\big(T^\dag T\delta A_L,DT^\dag,DT,F_L,\mathbb{I} \big)
+i\tr_{\mathcal{L}}\big(T\delta A_L T^\dag,DT,DT^\dag,F_R,\mathbb{I} \big)\nonumber
\ee
\be
 -i\tr_{\mathcal{L}}\big(T^\dag DT,\delta A_LT^\dag,DT,F_L,\mathbb{I} \big)
-i\tr_{\mathcal{L}}\big(TDT^\dag,T\delta A_L,DT^\dag,F_R,\mathbb{I} \big)\nonumber
\ee
\be
 +i\tr_{\mathcal{L}}\big(T^\dag DT,DT^\dag,T\delta A_L,F_L,\mathbb{I} \big)
+i\tr_{\mathcal{L}}\big(TDT^\dag,DT,\delta A_L T^\dag,F_R,\mathbb{I} \big)\nonumber
\ee
\be
 -i \tr_{\mathcal{L}}\big(T^\dag DT,DT^\dag,DT,\delta A_L A_L + A_L \delta A_L,\mathbb{I} \big)
-\intd  \tr_{\mathcal{L}}\big(T^\dag DT,DT^\dag,DT,\delta A_L,\mathbb{I} \big)\nonumber
\ee
\be
 +\tr_{\mathcal{L}}\big(\intd(T^\dag DT),DT^\dag,DT,\delta A_L,\mathbb{I} \big)
-\tr_{\mathcal{L}}\big(T^\dag DT,\intd(DT^\dag),DT,\delta A_L,\mathbb{I} \big)\nonumber
\ee
\be
 +\tr_{\mathcal{L}}\big(T^\dag DT,DT^\dag,\intd(DT),\delta A_L,\mathbb{I} \big)
+i \tr_{\mathcal{L}}\big(T^\dag T\delta A_L,DT^\dag,F_R,DT,\mathbb{I}\big)\nonumber
\ee
\be
 +i \tr_{\mathcal{L}}\big(T \delta A_L T^\dag,DT,F_L,DT^\dag,\mathbb{I}\big)
-i \tr_{\mathcal{L}}\big(T^\dag DT,\delta A_LT^\dag,F_R,DT,\mathbb{I}\big)\nonumber
\ee
\be
 -i \tr_{\mathcal{L}}\big(T DT^\dag,T\delta A_L,F_L,DT^\dag,\mathbb{I}\big)
+i \tr_{\mathcal{L}}\big(T^\dag DT,DT^\dag,F_R,T\delta A_L,\mathbb{I}\big)\nonumber
\ee
\be
 +i \tr_{\mathcal{L}}\big(T DT^\dag,DT,F_L,\delta A_L T^\dag,\mathbb{I}\big)
+i \tr_{\mathcal{L}}\big(T DT^\dag,DT,\delta A_L A_L + A_L \delta A_L,DT^\dag,\mathbb{I}\big)\nonumber
\ee
\be
 -\intd \tr_{\mathcal{L}}\big(T DT^\dag,DT,\delta A_L,DT^\dag,\mathbb{I}\big)
+ \tr_{\mathcal{L}}\big(\intd(T DT^\dag),DT,\delta A_L,DT^\dag,\mathbb{I}\big)\nonumber
\ee
\be
 - \tr_{\mathcal{L}}\big(T DT^\dag,\intd(DT),\delta A_L,DT^\dag,\mathbb{I}\big)
- \tr_{\mathcal{L}}\big(T DT^\dag,DT,\delta A_L,\intd(DT^\dag),\mathbb{I}\big) \nonumber
\ee
\be
 +i \tr_{\mathcal{L}}\big(T^\dag T\delta A_L,F_L,DT^\dag,DT,\mathbb{I}\big)
+i \tr_{\mathcal{L}}\big(T \delta A_LT^\dag,F_R,DT,DT^\dag,\mathbb{I}\big)\nonumber
\ee
\be
 -i \tr_{\mathcal{L}}\big(T^\dag DT,F_L,\delta A_L T^\dag,DT,\mathbb{I}\big)
-i \tr_{\mathcal{L}}\big(T DT^\dag,F_R,T\delta A_L,DT^\dag,\mathbb{I}\big)\nonumber
\ee
\be
 +i \tr_{\mathcal{L}}\big(T^\dag DT,F_L,DT^\dag,T\delta A_L,\mathbb{I}\big)
+i \tr_{\mathcal{L}}\big(T DT^\dag,F_R,DT,\delta A_L T^\dag,\mathbb{I}\big)\nonumber
\ee
\be
 -i \tr_{\mathcal{L}}\big(T^\dag DT,\delta A_L A_L+ A_L \delta A_L,DT^\dag,DT,\mathbb{I}\big)
-\intd \tr_{\mathcal{L}}\big(T^\dag DT,\delta A_L,DT^\dag,DT,\mathbb{I}\big)\nonumber
\ee
\be
 + \tr_{\mathcal{L}}\big(\intd(T^\dag DT),\delta A_L,DT^\dag,DT,\mathbb{I}\big)
+ \tr_{\mathcal{L}}\big(T^\dag DT,\delta A_L,\intd(DT^\dag),DT,\mathbb{I}\big)\nonumber
\ee
\be
 - \tr_{\mathcal{L}}\big(T^\dag DT,\delta A_L,DT^\dag,\intd(DT),\mathbb{I}\big)
+i \tr_{\mathcal{L}}\big(F_L,\delta A_L T^\dag,DT,DT^\dag T,\mathbb{I}\big)\nonumber
\ee
\be
 +i \tr_{\mathcal{L}}\big(F_R,T \delta A_L,DT^\dag,DT T^\dag,\mathbb{I}\big)
-i \tr_{\mathcal{L}}\big(F_L,DT^\dag,T \delta A_L,DT^\dag T,\mathbb{I}\big)\nonumber
\ee
\be
 -i \tr_{\mathcal{L}}\big(F_R,DT,\delta A_L T^\dag,DT T^\dag,\mathbb{I}\big)
+i \tr_{\mathcal{L}}\big(F_L,DT^\dag,DT,\delta A_L T^\dag T,\mathbb{I}\big)\nonumber
\ee
\be
 +i \tr_{\mathcal{L}}\big(F_R,DT,DT^\dag,T\delta A_L T^\dag,\mathbb{I}\big)
+i \tr_{\mathcal{L}}\big(\delta A_L A_L + A_L \delta A_L,DT^\dag,DT,DT^\dag T,\mathbb{I}\big)\nonumber
\ee
\be
 -\intd \tr_{\mathcal{L}}\big(\delta A_L,DT^\dag,DT,DT^\dag T,\mathbb{I}\big)
- \tr_{\mathcal{L}}\big(\delta A_L,\intd(DT^\dag),DT,DT^\dag T,\mathbb{I}\big)\nonumber
\ee
\be
 + \tr_{\mathcal{L}}\big(\delta A_L,DT^\dag,\intd(DT),DT^\dag T,\mathbb{I}\big)
- \tr_{\mathcal{L}}\big(\delta A_L,DT^\dag,DT,\intd(DT^\dag T),\mathbb{I}\big)\nonumber
\ee
\be
 -i \tr_{\mathcal{L}}\big(T^\dag T\delta A_L,F_L,F_L,\mathbb{I}\big)
-i \tr_{\mathcal{L}}\big(T \delta A_LT^\dag,F_R,F_R,\mathbb{I}\big)\nonumber
\ee
\be
 +i \tr_{\mathcal{L}}\big(T^\dag DT,\delta A_L A_L + A_L \delta A_L,F_L,\mathbb{I}\big)
+i \tr_{\mathcal{L}}\big(T^\dag DT,F_L,\delta A_L A_L + A_L \delta A_L,\mathbb{I}\big)\nonumber
\ee
\be
 +\intd \tr_{\mathcal{L}}\big(T^\dag DT,\delta A_L,F_L,\mathbb{I}\big)
+\intd \tr_{\mathcal{L}}\big(T^\dag DT,F_L,\delta A_L,\mathbb{I}\big)\nonumber
\ee
\be
 - \tr_{\mathcal{L}}\big(\intd(T^\dag DT),\delta A_L,F_L,\mathbb{I}\big)
- \tr_{\mathcal{L}}\big(T^\dag DT,\delta A_L,\intd F_L,\mathbb{I}\big)\nonumber
\ee
\be
 - \tr_{\mathcal{L}}\big(\intd(T^\dag DT),F_L,\delta A_L,\mathbb{I}\big)
+ \tr_{\mathcal{L}}\big(T^\dag DT,\intd F_L,\delta A_L,\mathbb{I}\big) \nonumber
\ee
\be
 +i \tr_{\mathcal{L}}\big(T^\dag F_R,T\delta A_L,F_L,\mathbb{I}\big)
+i \tr_{\mathcal{L}}\big(T F_L,\delta A_L T^\dag,F_R,\mathbb{I}\big)\nonumber
\ee
\be
 -i \tr_{\mathcal{L}}\big(T^\dag F_R,DT,\delta A_L A_L + A_L \delta A_L,\mathbb{I}\big)
+i \tr_{\mathcal{L}}\big(T (\delta A_L A_L + A_L \delta A_L),DT^\dag,F_R,\mathbb{I}\big)\nonumber
\ee
\be
 -\intd \tr_{\mathcal{L}}\big(T^\dag F_R,DT,\delta A_L,\mathbb{I}\big)
-\intd \tr_{\mathcal{L}}\big(T \delta A_L,DT^\dag,F_R,\mathbb{I}\big)\nonumber
\ee
\be
 +\tr_{\mathcal{L}}\big(\intd(T^\dag F_R),DT,\delta A_L,\mathbb{I}\big)
+\tr_{\mathcal{L}}\big(T^\dag F_R,\intd(DT),\delta A_L,\mathbb{I}\big)\nonumber
\ee
\be
 +\tr_{\mathcal{L}}\big((\intd T) \delta A_L,DT^\dag,F_R,\mathbb{I}\big)
-\tr_{\mathcal{L}}\big(T \delta A_L,\intd(DT^\dag),F_R,\mathbb{I}\big)\nonumber
\ee
\be
 +\tr_{\mathcal{L}}\big(T \delta A_L,DT^\dag,\intd F_R,\mathbb{I}\big)
-i \tr_{\mathcal{L}}\big(F_L,F_L,\delta A_L T^\dag T,\mathbb{I}\big)\nonumber
\ee
\be
 -i \tr_{\mathcal{L}}\big(F_R,F_R,T\delta A_L T^\dag,\mathbb{I}\big)
-i \tr_{\mathcal{L}}\big(\delta A_L A_L + A_L \delta A_L,F_L,DT^\dag T,\mathbb{I}\big)\nonumber
\ee
\be
 -i \tr_{\mathcal{L}}\big(F_L,\delta A_L A_L + A_L \delta A_L,DT^\dag T,\mathbb{I}\big)
+\intd \tr_{\mathcal{L}}\big(\delta A_L,F_L,DT^\dag T,\mathbb{I}\big)\nonumber
\ee
\be
 +\intd \tr_{\mathcal{L}}\big(F_L,\delta A_L,DT^\dag T,\mathbb{I}\big)
+ \tr_{\mathcal{L}}\big(\delta A_L,\intd F_L,DT^\dag T,\mathbb{I}\big)\nonumber
\ee
\be
 + \tr_{\mathcal{L}}\big(\delta A_L,F_L,\intd (DT^\dag T),\mathbb{I}\big)
- \tr_{\mathcal{L}}\big(\intd F_L,\delta A_L,DT^\dag T,\mathbb{I}\big)
+ \tr_{\mathcal{L}}\big(F_L,\delta A_L,\intd (DT^\dag T),\mathbb{I}\big)
\label{b81}.
\ee

The equations of motion can be obtained from \eqref{b81} by replacing $\delta A_L$ by $\mathbb{I}_{.p}\mathbb{I}_{q.}$ as before (with the notation defined in \eqref{b79}):

\be
\frac{\delta \Omega^0_5}{\delta (A_L)_{pq}} =  - i \tr_{\mathcal{L}}\big( T^\dag T \mathbb{I}_{.p}\mathbb{I}_{q.}, DT^\dag , DT, DT^\dag, DT, \mathbb{I} \big) \nonumber
\ee
\be
-i \tr_{\mathcal{L}}\big(T\mathbb{I}_{.p}\mathbb{I}_{q.} T^\dag,DT,DT^\dag,DT,DT^\dag,\mathbb{I}\big) \nonumber
\ee
\be
 +i \tr_{\mathcal{L}}\big(T^\dag DT,\mathbb{I}_{.p}\mathbb{I}_{q.} T^\dag,DT,DT^\dag,DT,\mathbb{I} \big)
+i \tr_{\mathcal{L}}\big(TDT^\dag,T\mathbb{I}_{.p}\mathbb{I}_{q.},DT,DT^\dag,DT,\mathbb{I}\big)\nonumber
\ee
\be
 -i \tr_{\mathcal{L}}\big(T^\dag DT,DT^\dag,T\mathbb{I}_{.p}\mathbb{I}_{q.},DT^\dag,DT,\mathbb{I} \big)
-i \tr_{\mathcal{L}}\big(TDT^\dag,DT,\mathbb{I}_{.p}\mathbb{I}_{q.} T^\dag,DT,DT^\dag,\mathbb{I} \big)\nonumber
\ee
\be
 +i \tr_{\mathcal{L}}\big(T^\dag DT,DT^\dag,DT,\mathbb{I}_{.p}\mathbb{I}_{q.} T^\dag,DT, \mathbb{I} \big)
+i \tr_{\mathcal{L}}\big(TDT^\dag,DT,DT^\dag,T\mathbb{I}_{.p}\mathbb{I}_{q.},DT^\dag,\mathbb{I} \big)\nonumber
\ee
\be
 -i \tr_{\mathcal{L}}\big(T^\dag DT,DT^\dag,DT,DT^\dag,T\mathbb{I}_{.p}\mathbb{I}_{q.},\mathbb{I} \big)
-i \tr_{\mathcal{L}}\big(TDT^\dag,DT,DT^\dag,DT,\mathbb{I}_{.p}\mathbb{I}_{q.} T^\dag, \mathbb{I} \big)\nonumber
\ee
\be
 +i\tr_{\mathcal{L}}\big(T^\dag T\mathbb{I}_{.p}\mathbb{I}_{q.},DT^\dag,DT,F_L,\mathbb{I} \big)
+i\tr_{\mathcal{L}}\big(T\mathbb{I}_{.p}\mathbb{I}_{q.} T^\dag,DT,DT^\dag,F_R,\mathbb{I} \big)\nonumber
\ee
\be
 -i\tr_{\mathcal{L}}\big(T^\dag DT,\mathbb{I}_{.p}\mathbb{I}_{q.}T^\dag,DT,F_L,\mathbb{I} \big)
-i\tr_{\mathcal{L}}\big(TDT^\dag,T\mathbb{I}_{.p}\mathbb{I}_{q.},DT^\dag,F_R,\mathbb{I} \big)\nonumber
\ee
\be
 +i\tr_{\mathcal{L}}\big(T^\dag DT,DT^\dag,T\mathbb{I}_{.p}\mathbb{I}_{q.},F_L,\mathbb{I} \big)
+i\tr_{\mathcal{L}}\big(TDT^\dag,DT,\mathbb{I}_{.p}\mathbb{I}_{q.} T^\dag,F_R,\mathbb{I} \big)\nonumber
\ee
\be
 -i \tr_{\mathcal{L}}\big(T^\dag DT,DT^\dag,DT,\mathbb{I}_{.p}\mathbb{I}_{q.} A_L + A_L \mathbb{I}_{.p}\mathbb{I}_{q.},\mathbb{I} \big)
+\intd  \tr_{\mathcal{L}}\big(T^\dag DT,DT^\dag,DT,\mathbb{I}_{.p}\mathbb{I}_{q.},\mathbb{I} \big)\nonumber
\ee
\be
 +\tr_{\mathcal{L}}\big(\intd(T^\dag DT),DT^\dag,DT,\mathbb{I}_{.p}\mathbb{I}_{q.},\mathbb{I} \big)
-\tr_{\mathcal{L}}\big(T^\dag DT,\intd(DT^\dag),DT,\mathbb{I}_{.p}\mathbb{I}_{q.},\mathbb{I} \big)\nonumber
\ee
\be
 +\tr_{\mathcal{L}}\big(T^\dag DT,DT^\dag,\intd(DT),\mathbb{I}_{.p}\mathbb{I}_{q.},\mathbb{I} \big)
+i \tr_{\mathcal{L}}\big(T^\dag T\mathbb{I}_{.p}\mathbb{I}_{q.},DT^\dag,F_R,DT,\mathbb{I}\big)\nonumber
\ee
\be
 +i \tr_{\mathcal{L}}\big(T \mathbb{I}_{.p}\mathbb{I}_{q.} T^\dag,DT,F_L,DT^\dag,\mathbb{I}\big)
-i \tr_{\mathcal{L}}\big(T^\dag DT,\mathbb{I}_{.p}\mathbb{I}_{q.}T^\dag,F_R,DT,\mathbb{I}\big)\nonumber
\ee
\be
 -i \tr_{\mathcal{L}}\big(T DT^\dag,T\mathbb{I}_{.p}\mathbb{I}_{q.},F_L,DT^\dag,\mathbb{I}\big)
+i \tr_{\mathcal{L}}\big(T^\dag DT,DT^\dag,F_R,T\mathbb{I}_{.p}\mathbb{I}_{q.},\mathbb{I}\big)\nonumber
\ee
\be
 +i \tr_{\mathcal{L}}\big(T DT^\dag,DT,F_L,\mathbb{I}_{.p}\mathbb{I}_{q.} T^\dag,\mathbb{I}\big)
+i \tr_{\mathcal{L}}\big(T DT^\dag,DT,\mathbb{I}_{.p}\mathbb{I}_{q.} A_L + A_L \mathbb{I}_{.p}\mathbb{I}_{q.},DT^\dag,\mathbb{I}\big)\nonumber
\ee
\be
 -\intd \tr_{\mathcal{L}}\big(T DT^\dag,DT,\mathbb{I}_{.p}\mathbb{I}_{q.},DT^\dag,\mathbb{I}\big)
+ \tr_{\mathcal{L}}\big(\intd(T DT^\dag),DT,\mathbb{I}_{.p}\mathbb{I}_{q.},DT^\dag,\mathbb{I}\big)\nonumber
\ee
\be
 - \tr_{\mathcal{L}}\big(T DT^\dag,\intd(DT),\mathbb{I}_{.p}\mathbb{I}_{q.},DT^\dag,\mathbb{I}\big)
- \tr_{\mathcal{L}}\big(T DT^\dag,DT,\mathbb{I}_{.p}\mathbb{I}_{q.},\intd(DT^\dag),\mathbb{I}\big) \nonumber
\ee
\be
 +i \tr_{\mathcal{L}}\big(T^\dag T\mathbb{I}_{.p}\mathbb{I}_{q.},F_L,DT^\dag,DT,\mathbb{I}\big)
+i \tr_{\mathcal{L}}\big(T \mathbb{I}_{.p}\mathbb{I}_{q.}T^\dag,F_R,DT,DT^\dag,\mathbb{I}\big)\nonumber
\ee
\be
 -i \tr_{\mathcal{L}}\big(T^\dag DT,F_L,\mathbb{I}_{.p}\mathbb{I}_{q.} T^\dag,DT,\mathbb{I}\big)
-i \tr_{\mathcal{L}}\big(T DT^\dag,F_R,T\mathbb{I}_{.p}\mathbb{I}_{q.},DT^\dag,\mathbb{I}\big)\nonumber
\ee
\be
 +i \tr_{\mathcal{L}}\big(T^\dag DT,F_L,DT^\dag,T\mathbb{I}_{.p}\mathbb{I}_{q.},\mathbb{I}\big)
+i \tr_{\mathcal{L}}\big(T DT^\dag,F_R,DT,\mathbb{I}_{.p}\mathbb{I}_{q.} T^\dag,\mathbb{I}\big)\nonumber
\ee
\be
 -i \tr_{\mathcal{L}}\big(T^\dag DT,\mathbb{I}_{.p}\mathbb{I}_{q.} A_L+ A_L \mathbb{I}_{.p}\mathbb{I}_{q.},DT^\dag,DT,\mathbb{I}\big)
+\intd \tr_{\mathcal{L}}\big(T^\dag DT,\mathbb{I}_{.p}\mathbb{I}_{q.},DT^\dag,DT,\mathbb{I}\big)\nonumber
\ee
\be
 + \tr_{\mathcal{L}}\big(\intd(T^\dag DT),\mathbb{I}_{.p}\mathbb{I}_{q.},DT^\dag,DT,\mathbb{I}\big)
+ \tr_{\mathcal{L}}\big(T^\dag DT,\mathbb{I}_{.p}\mathbb{I}_{q.},\intd(DT^\dag),DT,\mathbb{I}\big)\nonumber
\ee
\be
 - \tr_{\mathcal{L}}\big(T^\dag DT,\mathbb{I}_{.p}\mathbb{I}_{q.},DT^\dag,\intd(DT),\mathbb{I}\big)
+i \tr_{\mathcal{L}}\big(F_L,\mathbb{I}_{.p}\mathbb{I}_{q.} T^\dag,DT,DT^\dag T,\mathbb{I}\big)\nonumber
\ee
\be
 +i \tr_{\mathcal{L}}\big(F_R,T \mathbb{I}_{.p}\mathbb{I}_{q.},DT^\dag,DT T^\dag,\mathbb{I}\big)
-i \tr_{\mathcal{L}}\big(F_L,DT^\dag,T \mathbb{I}_{.p}\mathbb{I}_{q.},DT^\dag T,\mathbb{I}\big)\nonumber
\ee
\be
 -i \tr_{\mathcal{L}}\big(F_R,DT,\mathbb{I}_{.p}\mathbb{I}_{q.} T^\dag,DT T^\dag,\mathbb{I}\big)
+i \tr_{\mathcal{L}}\big(F_L,DT^\dag,DT,\mathbb{I}_{.p}\mathbb{I}_{q.} T^\dag T,\mathbb{I}\big)\nonumber
\ee
\be
 +i \tr_{\mathcal{L}}\big(F_R,DT,DT^\dag,T\mathbb{I}_{.p}\mathbb{I}_{q.} T^\dag,\mathbb{I}\big)
+i \tr_{\mathcal{L}}\big(\mathbb{I}_{.p}\mathbb{I}_{q.} A_L + A_L \mathbb{I}_{.p}\mathbb{I}_{q.},DT^\dag,DT,DT^\dag T,\mathbb{I}\big)\nonumber
\ee
\be
 -\intd \tr_{\mathcal{L}}\big(\mathbb{I}_{.p}\mathbb{I}_{q.},DT^\dag,DT,DT^\dag T,\mathbb{I}\big)
- \tr_{\mathcal{L}}\big(\mathbb{I}_{.p}\mathbb{I}_{q.},\intd(DT^\dag),DT,DT^\dag T,\mathbb{I}\big)\nonumber
\ee
\be
 + \tr_{\mathcal{L}}\big(\mathbb{I}_{.p}\mathbb{I}_{q.},DT^\dag,\intd(DT),DT^\dag T,\mathbb{I}\big)
- \tr_{\mathcal{L}}\big(\mathbb{I}_{.p}\mathbb{I}_{q.},DT^\dag,DT,\intd(DT^\dag T),\mathbb{I}\big)\nonumber
\ee
\be
 -i \tr_{\mathcal{L}}\big(T^\dag T\mathbb{I}_{.p}\mathbb{I}_{q.},F_L,F_L,\mathbb{I}\big)
-i \tr_{\mathcal{L}}\big(T \mathbb{I}_{.p}\mathbb{I}_{q.}T^\dag,F_R,F_R,\mathbb{I}\big)\nonumber
\ee
\be
 +i \tr_{\mathcal{L}}\big(T^\dag DT,\mathbb{I}_{.p}\mathbb{I}_{q.} A_L + A_L \mathbb{I}_{.p}\mathbb{I}_{q.},F_L,\mathbb{I}\big)
+i \tr_{\mathcal{L}}\big(T^\dag DT,F_L,\mathbb{I}_{.p}\mathbb{I}_{q.} A_L + A_L \mathbb{I}_{.p}\mathbb{I}_{q.},\mathbb{I}\big)\nonumber
\ee
\be
 -\intd \tr_{\mathcal{L}}\big(T^\dag DT,\mathbb{I}_{.p}\mathbb{I}_{q.},F_L,\mathbb{I}\big)
-\intd \tr_{\mathcal{L}}\big(T^\dag DT,F_L,\mathbb{I}_{.p}\mathbb{I}_{q.},\mathbb{I}\big)\nonumber
\ee
\be
 - \tr_{\mathcal{L}}\big(\intd(T^\dag DT),\mathbb{I}_{.p}\mathbb{I}_{q.},F_L,\mathbb{I}\big)
- \tr_{\mathcal{L}}\big(T^\dag DT,\mathbb{I}_{.p}\mathbb{I}_{q.},\intd F_L,\mathbb{I}\big)\nonumber
\ee
\be
 - \tr_{\mathcal{L}}\big(\intd(T^\dag DT),F_L,\mathbb{I}_{.p}\mathbb{I}_{q.},\mathbb{I}\big)
+ \tr_{\mathcal{L}}\big(T^\dag DT,\intd F_L,\mathbb{I}_{.p}\mathbb{I}_{q.},\mathbb{I}\big) \nonumber
\ee
\be
 -i \tr_{\mathcal{L}}\big(T^\dag F_R,T\mathbb{I}_{.p}\mathbb{I}_{q.},F_L,\mathbb{I}\big)
-i \tr_{\mathcal{L}}\big(T F_L,\mathbb{I}_{.p}\mathbb{I}_{q.} T^\dag,F_R,\mathbb{I}\big)\nonumber
\ee
\be
 +i \tr_{\mathcal{L}}\big(T^\dag F_R,DT,\mathbb{I}_{.p}\mathbb{I}_{q.} A_L + A_L \mathbb{I}_{.p}\mathbb{I}_{q.},\mathbb{I}\big)
-i \tr_{\mathcal{L}}\big(T (\mathbb{I}_{.p}\mathbb{I}_{q.} A_L + A_L \mathbb{I}_{.p}\mathbb{I}_{q.}),DT^\dag,F_R,\mathbb{I}\big)\nonumber
\ee
\be
 -\intd \tr_{\mathcal{L}}\big(T^\dag F_R,DT,\mathbb{I}_{.p}\mathbb{I}_{q.},\mathbb{I}\big)
+\intd \tr_{\mathcal{L}}\big(T \mathbb{I}_{.p}\mathbb{I}_{q.},DT^\dag,F_R,\mathbb{I}\big)\nonumber
\ee
\be
 -\tr_{\mathcal{L}}\big(\intd(T^\dag F_R),DT,\mathbb{I}_{.p}\mathbb{I}_{q.},\mathbb{I}\big)
-\tr_{\mathcal{L}}\big(T^\dag F_R,\intd(DT),\mathbb{I}_{.p}\mathbb{I}_{q.},\mathbb{I}\big)\nonumber
\ee
\be
 -\tr_{\mathcal{L}}\big((\intd T) \mathbb{I}_{.p}\mathbb{I}_{q.},DT^\dag,F_R,\mathbb{I}\big)
+\tr_{\mathcal{L}}\big(T \mathbb{I}_{.p}\mathbb{I}_{q.},\intd(DT^\dag),F_R,\mathbb{I}\big)\nonumber
\ee
\be
 -\tr_{\mathcal{L}}\big(T \mathbb{I}_{.p}\mathbb{I}_{q.},DT^\dag,\intd F_R,\mathbb{I}\big)
-i \tr_{\mathcal{L}}\big(F_L,F_L,\mathbb{I}_{.p}\mathbb{I}_{q.} T^\dag T,\mathbb{I}\big)\nonumber
\ee
\be
 -i \tr_{\mathcal{L}}\big(F_R,F_R,T\mathbb{I}_{.p}\mathbb{I}_{q.} T^\dag,\mathbb{I}\big)
-i \tr_{\mathcal{L}}\big(\mathbb{I}_{.p}\mathbb{I}_{q.} A_L + A_L \mathbb{I}_{.p}\mathbb{I}_{q.},F_L,DT^\dag T,\mathbb{I}\big)\nonumber
\ee
\be
 -i \tr_{\mathcal{L}}\big(F_L,\mathbb{I}_{.p}\mathbb{I}_{q.} A_L + A_L \mathbb{I}_{.p}\mathbb{I}_{q.},DT^\dag T,\mathbb{I}\big)
+\intd \tr_{\mathcal{L}}\big(\mathbb{I}_{.p}\mathbb{I}_{q.},F_L,DT^\dag T,\mathbb{I}\big)\nonumber
\ee
\be
 +\intd \tr_{\mathcal{L}}\big(F_L,\mathbb{I}_{.p}\mathbb{I}_{q.},DT^\dag T,\mathbb{I}\big)
+ \tr_{\mathcal{L}}\big(\mathbb{I}_{.p}\mathbb{I}_{q.},\intd F_L,DT^\dag T,\mathbb{I}\big)\nonumber
\ee
\be
 + \tr_{\mathcal{L}}\big(\mathbb{I}_{.p}\mathbb{I}_{q.},F_L,\intd (DT^\dag T),\mathbb{I}\big)
- \tr_{\mathcal{L}}\big(\intd F_L,\mathbb{I}_{.p}\mathbb{I}_{q.},DT^\dag T,\mathbb{I}\big)
+ \tr_{\mathcal{L}}\big(F_L,\mathbb{I}_{.p}\mathbb{I}_{q.},\intd (DT^\dag T),\mathbb{I}\big)\nonumber
\ee
\be
\label{eq:variationOmega5}
\ee

Again, the variation with respect to $A_R$ can be deduced by exchanging $L \leftrightarrow R$ and $T \leftrightarrow T^\dag$ in \eqref{eq:variationOmega5} and changing the overall sign.

To compute the variation with respect to the tachyon field, one has to determine the variation of the $\mathcal{L}$ operators with respect to the tachyon. To do this, assuming $N_f \leq 4$ it is possible to use the explicit expression for the eigenvalues as described in section \ref{sec:explicit}. This method yields
\be
\frac{\partial \tr(TT^\dag)}{\partial T_{ij}}=  T^\dag_{ji},\label{b83}
\ee
with $T^*$ the complex conjugate of $T$. Then, the variation of the determinant is
\be
\frac{\partial \det(TT^\dag)}{\partial T_{ij}}= \det(TT^\dag) (T^{-1})_{ji}.\label{b84}
\ee
In the case of $SU(2)$, we obtain that the derivatives of the two eigenvalues of $TT^\dag$, $x_{\pm}$ defined in \eqref{442}, are given by
\be
\frac{\partial x_{\pm}}{\partial T_{ij}}= \frac{1}{2}  T^\dag_{ji} \pm \frac{1}{2} \frac{  \tr(TT^\dag) T^\dag_{ji}- 2 \det(TT^\dag) (T^{-1})_{ji}}{\sqrt{ \tr(TT^\dag)^2 - 4 \det(TT^\dag)}}
\label{eq:varsu2}.
\ee
In the case of $SU(3)$ it is easier to define intermediary objects before expressing the derivative of (\ref{eq:cardan}). In particular, one has
\be
\frac{\partial \tr((TT^\dag)^2)}{\partial T_{ij}} = (T^\dag T T^\dag)_{ji}\;.\label{b86}
\ee
We define
\be
I \equiv \frac{\frac{\tr(TT^\dag)^3}{18} - \frac{\tr(TT^\dag) \tr((TT^\dag)^2)}{2} - 3\det(TT^\dag)}{\frac{\tr(TT^\dag)^2}{3} - \tr((TT^\dag)^2) } \sqrt{\frac{3}{\frac{\tr((TT^\dag)^2)}{2} - \frac{\tr(TT^\dag)^2}{6}}},\label{b87}
\ee
such that we obtain
\be
x_a =   \frac{\tr(TT^\dag)}{3} + 2\sqrt{\frac{\tr((TT^\dag)^2)}{6} - \frac{\tr(TT^\dag)^2}{18}} \cos\bigg(\frac{2a\pi}{3} + \frac{1}{3}\arccos(I) \bigg).\label{b88}
\ee
Then we can write
\be
\frac{\partial x_a}{\partial T_{ij}} =   \frac{T^\dag_{ji}}{3} + \frac{\frac{(T^\dag T T^\dag)_{ji}}{3} - \frac{\tr(TT^\dag) T^\dag_{ji}}{9}}{\sqrt{\frac{\tr((TT^\dag)^2)}{6} - \frac{\tr(TT^\dag)^2}{18}}} \cos\bigg(\frac{2a\pi}{3} + \frac{1}{3}\arccos(I) \bigg) \nonumber
\ee
\be
+ \frac{2}{3}\sqrt{\frac{\tr((TT^\dag)^2)}{6} - \frac{\tr(TT^\dag)^2}{18}} \frac{\frac{\partial I}{\partial T_{ij}}}{\sqrt{1-I^2}}\sin\bigg( \frac{2a\pi}{3} + \frac{1}{3}\arccos(I)  \bigg)\label{b89},
\ee
and
\be
\frac{\partial I}{\partial T_{ij}} = \bigg[ \bigg( \frac{(\tr TT^\dag)^2 T^\dag_{ji}}{6}  - \frac{(\tr (TT^\dag)^2) T^\dag_{ji}}{2} - \tr(TT^\dag) (T^\dag T T^\dag)_{ji} - 3\det(TT^\dag)(T^{-1})_{ji} \bigg)\nonumber
 \ee
\be
\times \bigg( \frac{(\tr TT^\dag)^2  }{3} - \tr(( TT^\dag)^2)\bigg) - \bigg( \frac{\tr(TT^\dag)^3}{18} - \frac{\tr(TT^\dag) \tr((TT^\dag)^2)}{2} - 3\det(TT^\dag)\bigg)\nonumber
 \ee
\be
\times \bigg( \frac{2}{3} \tr (TT^\dag) T^\dag_{ji} - 2(T^\dag T T^\dag)_{ji} \bigg)\bigg] \times \frac{ \sqrt{\frac{3}{\frac{\tr((TT^\dag)^2)}{2} - \frac{\tr(TT^\dag)^2}{6}}} }{\big(\frac{\tr(TT^\dag)^2}{3} - \tr((TT^\dag)^2) \big)^2}
\nonumber
 \ee
\be
 +  \frac{\frac{\tr(TT^\dag)^3}{18} - \frac{\tr(TT^\dag) \tr((TT^\dag)^2)}{2} - 3\det(TT^\dag)}{\frac{\tr(TT^\dag)^2}{3} - \tr((TT^\dag)^2) } \times \bigg[ \frac{-\sqrt{3}}{2}  \frac{(T^\dag T T^\dag)_{ji} - \frac{1}{3} \tr(TT^\dag) T^\dag_{ji} }{\bigg(\frac{\tr((TT^\dag)^2)}{2} - \frac{\tr(TT^\dag)^2}{6} \bigg)^{3/2}} \bigg]\label{b90}.
\ee
We obtain $\frac{\partial \mathcal{L}}{\partial T}$ by using
\be
\frac{\partial \mathcal{L}}{\partial T_{pq}} = \sum_{i=1}^{N_f} \frac{\partial \mathcal{L}}{\partial x_i}  \frac{\partial x_i}{\partial T_{pq}}\label{b91}.
\ee

We already know $\frac{\partial x_i}{\partial T_{pq}}$ from \eqref{eq:varsu2} for $SU(2)$ or \eqref{b89}-\eqref{b90} for $SU(3)$. Next, we compute $\frac{\partial \mathcal{L}}{\partial x_i}$. For completion, we now report the values of $\frac{\partial \mathcal{L}}{\partial x_a}$ and $\frac{\partial \mathcal{L}}{\partial x_b}$. As before we begin with operators with at most two different indices. For the term $XXXXX$ we obtain
\be
\frac{\partial \mathcal{L}_{aaaaaa}}{\partial x_a} = \frac{e^{-x_a}\big(6 + x_a(6 + x_a(3 + x_a))\big)}{240 x_a^{4}}\label{b92},
\ee
\be
\frac{\partial \mathcal{L}_{aaaaba}}{\partial x_a} =
\frac{1}{48 x_a^{3}(x_a-x_b)^{6}} \Bigg[
-60 e^{-x_b} x_a^{3}(-1+x_b)\nonumber
\ee
\be
+ e^{-x_a}\Big(
- x_a^{3}\big(60 + x_a(4+x_a)(15 + x_a(3+x_a))\big)\nonumber
\ee
\be
+5 x_a^{3}\big(24 + x_a(18 + x_a(6 + x_a))\big)x_b
-10 x_a^{3}\big(11 + x_a(5 + x_a)\big)x_b^{2}
+10 x_a^{2}\big(6 + x_a(4 + x_a)\big)x_b^{3}\nonumber
\ee
\be
-5 x_a\big(3 + x_a(3 + x_a)\big)x_b^{4}
+(2 + x_a(2 + x_a))x_b^{5}
\Big)
-60 x_a^{3} x_b^{2}\big(\operatorname{Ei}(-x_a)-\operatorname{Ei}(-x_b)\big)
\Bigg]\label{b93},
\ee
\be
\frac{\partial \mathcal{L}_{aaaaba}}{\partial x_b} = \frac{1}{48 x_a^{2}(x_a-x_b)^{6}}\Big[
12 e^{-x_b} x_a^{2}(-5 + 2 x_a + 3 x_b)
\nonumber
\ee
\be+ e^{-x_a}\big(
x_a^{2}\big(60 + x_a(36 + x_a(9 + x_a))\big) \nonumber
\ee
\be
-4 x_a^{2}\big(24 + x_a(7 + x_a)\big)x_b
+6 x_a^{2}(5 + x_a)x_b^{2}
-4 x_a(3 + x_a)x_b^{3}
+(1 + x_a)x_b^{4}
\big)
\nonumber
\ee
\be
-12 x_a^{2} x_b(2 x_a + 3 x_b)\big(\operatorname{Ei}(-x_a)-\operatorname{Ei}(-x_b)\big)
\Big]\label{b94},
\ee
\be
\frac{\partial \mathcal{L}_{aaabba}}{\partial x_a} = \frac{1}{12 (x_a-x_b)^{6}}
\Big[
12 e^{-x_b}(-5 + 2 x_a + 3 x_b)
\nonumber
\ee
\be+ \frac{e^{-x_a}}{x_a^{2}}\big(
x_a^{2}\big(60 + x_a(36 + x_a(9 + x_a))\big)
-4 x_a^{2}\big(24 + x_a(7 + x_a)\big)x_b +6 x_a^{2}(5 + x_a)x_b^{2}
\nonumber
\ee
\be-4 x_a(3 + x_a)x_b^{3}
+(1 + x_a)x_b^{4}
\big)+12 x_b(2 x_a + 3 x_b)\big(\operatorname{Ei}(-x_b)-\operatorname{Ei}(-x_a)\big)
\Big]\label{b95},
\ee
\be
\frac{\partial \mathcal{L}_{aaabba}}{\partial x_b} =
\frac{1}{6 x_a (x_a-x_b)^{6}}
\Big[
-6 e^{-x_b} x_a(-5 + 4 x_a + x_b)
+ e^{-x_a}\big(
 -x_a(30 + x_a(6 + x_a))
\nonumber
\ee
\be
 +3 x_a(12 + x_a)x_b
-3 x_a x_b^{2} + x_b^{3}
\big)
+3 x_a (x_a^{2} + 6 x_a x_b + 3 x_b^{2})\big(\operatorname{Ei}(-x_a)-\operatorname{Ei}(-x_b)\big)
\Big]\label{b96},
\ee
\be
\frac{\partial \mathcal{L}_{aabbba}}{\partial x_a} =
\frac{1}{4 (x_a-x_b)^{6} x_b}
\Big[
e^{-x_b}\big(x_a^{3} - 3 x_a^{2} x_b + 3 x_a x_b(12 + x_b) - x_b(30 + x_b(6 + x_b))\big)
\nonumber
\ee
\be+3 e^{-x_a} x_b\big(-2(-5 + x_a + 4 x_b)\big)
-3 (3 x_a^{2} + 6 x_a x_b + x_b^{2})\big(\operatorname{Ei}(-x_a)-\operatorname{Ei}(-x_b)\big)
\Big]\label{b97}.
\ee
For the terms $XXXY$, $XXYX$, $XYXX$ and $YXXX$ we obtain
\be
\frac{\partial \mathcal{L}_{aaaaa}}{\partial x_a} =\frac{i\,e^{-x_a}\big(2 + x_a(2+x_a)\big)}{48 x_a^{3}}\label{b98},
\ee
\be
\frac{\partial \mathcal{L}_{aaaba}}{\partial x_a} =
-\frac{i}{12 x_a^{2}(x_a-x_b)^{5}}
\Bigg[
12 e^{-x_b} x_a^{2}(-1+x_b)
+ e^{-x_a}\Big(\!
x_a^{2}(2+x_a)(6+x_a(3+x_a))
\nonumber
\ee
\be-4 x_a^{2}(6+x_a(4+x_a))x_b
+6 x_a^{2}(3+x_a)x_b^{2}
-4 x_a(2+x_a)x_b^{3}
+(1+x_a)x_b^{4}\Big)
\nonumber
\ee
\be-12 x_a^{2} x_b^{2}\big(\operatorname{Ei}(-x_a)-\operatorname{Ei}(-x_b)\big)
\Bigg]\label{b99},
\ee
\be
\frac{\partial \mathcal{L}_{aaaba}}{\partial x_b} =
\frac{i\,e^{-x_a - x_b}}{12 x_a (x_a-x_b)^{5}}
\Big[
6 e^{x_a} x_a(-2 + x_a + x_b)
+ e^{x_b}\big(
x_a^{3}-3 x_a^{2}(-2+x_b)-x_b^{3}
\nonumber
\ee
\be+3 x_a(4+(-6+x_b)x_b)
\big)
-6 x_a x_b (x_a+x_b)\big(\operatorname{Ei}(-x_a)-\operatorname{Ei}(-x_b)\big)
\Big]\label{b100},
\ee
\be
\frac{\partial \mathcal{L}_{aabba}}{\partial x_a} =
\frac{i\,e^{-x_a - x_b}}{4 x_a (x_a-x_b)^{5}}
\Big[
6 e^{x_a} x_a(-2 + x_a + x_b)
+ e^{x_b}\big(
x_a^{3}-3 x_a^{2}(-2+x_b)-x_b^{3}
\nonumber
\ee
\be+3 x_a(4+(-6+x_b)x_b)
\big)
-6 x_a x_b (x_a+x_b)\big(\operatorname{Ei}(-x_a)-\operatorname{Ei}(-x_b)\big)
\Big]\label{b101},
\ee
\be
\frac{\partial \mathcal{L}_{aabba}}{\partial x_b} = i
\frac{1}{2 (x_a-x_b)^{5}}\Big[
-6 e^{-x_b}(-1+x_a)
+6 e^{-x_a}(-1+x_b)
\nonumber
\ee
\be
+ \big(x_a^{2}+4 x_a x_b+x_b^{2}\big)\big(\operatorname{Ei}(-x_a)-\operatorname{Ei}(-x_b)\big)
\Big]\label{b102}.
\ee
For the terms $XYY$, $YXY$, $YYX$ we obtain
\be
\frac{\partial \mathcal{L}^{(k=2)}_{aaaa}}{\partial x_a} =
-\frac{e^{-x_a}(1+x_a)}{12 x_a^{2}}\label{b103},
\ee
\be
\frac{\partial \mathcal{L}^{(k=2)}_{aaba}}{\partial x_a} =  \frac{1}{4 x_a (x_a-x_b)^{4}}\Big[
3 e^{-x_b} x_a(-1+x_b)
+ e^{-x_a}\big(x_a^{3}-3 x_a^{2}(-1+x_b)
\nonumber
\ee
\be+3 x_a(-1+x_b)^{2}-x_b^{3}\big)
-3 x_a x_b^{2}\big(\operatorname{Ei}(-x_a)-\operatorname{Ei}(-x_b)\big)
\Big]\label{b104},
\ee
\be
\frac{\partial \mathcal{L}^{(k=2)}_{aaba}}{\partial x_b} =  \frac{1}{4 (x_a-x_b)^{4}}\Big[
- e^{-x_b}\big(-3+2 x_a+x_b\big)
+ e^{-x_a}\big(-3 - x_a + 4 x_b\big)
\nonumber
\ee
\be+ x_b(2 x_a + x_b)\big(\operatorname{Ei}(-x_a)-\operatorname{Ei}(-x_b)\big)
\Big]\label{b105},
\ee
\be
\frac{\partial \mathcal{L}^{(k=2)}_{abba}}{\partial x_a} = \frac{1}{2 (x_a-x_b)^{4}}\Big[
- e^{-x_b}\big(-3+2 x_a+x_b\big)
+ e^{-x_a}\big(-3 - x_a + 4 x_b\big)
\nonumber
\ee
\be+ x_b(2 x_a + x_b)\big(\operatorname{Ei}(-x_a)-\operatorname{Ei}(-x_b)\big)
\Big]\label{b106}.
\ee
For operators with three different indices we obtain
\be
\frac{\partial \mathcal{L}^{(k=0)}_{aaabca}}{\partial x_a} =
\frac{1}{12 (x_a - x_b)^5 (x_b - x_c)} \Biggl[
12 e^{-x_b} (-1 + x_b)
+ \frac{e^{-x_a}}{x_a^2 (x_a - x_c)^5} \Bigl(\nonumber
\ee
\be
x_a^2 x_b \bigl(
x_a^4 (60 + x_a (36 + x_a (9 + x_a)))
- 4 x_a^3 (30 + x_a (30 + x_a (8 + x_a))) x_b\nonumber
\ee
\be
+ 6 x_a^2 (2 + x_a) (10 + x_a (5 + x_a)) x_b^2
- 4 x_a (15 + x_a (15 + x_a (6 + x_a))) x_b^3 \nonumber
\ee
\be
+ (2 + x_a) (6 + x_a (3 + x_a)) x_b^4
\bigr)
- x_a^2 \bigl(
x_a^4 (60 + x_a (36 + x_a (9 + x_a))) \nonumber
\ee
\be
- 10 x_a^3 (24 + x_a (7 + x_a)) x_b^2
+ 20 x_a^2 (12 + x_a (6 + x_a)) x_b^3 \nonumber
\ee
\be
- 15 x_a (8 + x_a (5 + x_a)) x_b^4
+ 4 (6 + x_a (4 + x_a)) x_b^5
\bigr) x_c\nonumber
\ee
\be
+ 2 x_a^2 \bigl(
2 x_a^3 (30 + x_a (30 + x_a (8 + x_a)))
- 5 x_a^3 (24 + x_a (7 + x_a)) x_b
+ 10 x_a^2 (5 + x_a) x_b^3 \nonumber
\ee
\be
- 10 x_a (4 + x_a) x_b^4
+ 3 (3 + x_a) x_b^5
\bigr) x_c^2 \nonumber
\ee
\be
- 2 x_a \bigl(
3 x_a^3 (2 + x_a) (10 + x_a (5 + x_a))
- 10 x_a^3 (12 + x_a (6 + x_a)) x_b
+ 10 x_a^3 (5 + x_a) x_b^2 \nonumber
\ee
\be
- 5 x_a (3 + x_a) x_b^4
+ 2 (2 + x_a) x_b^5
\bigr) x_c^3 \nonumber
\ee
\be
+ \bigl(
4 x_a^3 (15 + x_a (15 + x_a (6 + x_a)))
- 15 x_a^3 (8 + x_a (5 + x_a)) x_b
+ 20 x_a^3 (4 + x_a) x_b^2 \nonumber
\ee
\be
- 10 x_a^2 (3 + x_a) x_b^3
+ (1 + x_a) x_b^5
\bigr) x_c^4 \nonumber
\ee
\be
- \bigl(
x_a^2 (2 + x_a) (6 + x_a (3 + x_a))
- 4 x_a^2 (6 + x_a (4 + x_a)) x_b
+ 6 x_a^2 (3 + x_a) x_b^2 \nonumber
\ee
\be
- 4 x_a (2 + x_a) x_b^3
+ (1 + x_a) x_b^4
\bigr) x_c^5
\Bigr) \nonumber
\ee
\be
- e^{-(x_a - x_b)} x_b^2 (x_a - x_c)^5 \text{Ei}(-x_b)
+ e^{-x_c} (x_a - x_b)^5 (-1 + x_c) \nonumber
\ee
\be
+ (x_b - x_c)
\bigl(
-5 x_a^4 x_b x_c + 10 x_a^2 x_b^2 x_c^2
+ x_a^5 (x_b + x_c)
- 5 x_a x_b^2 x_c^2 (x_b + x_c) \nonumber
\ee
\be
+ x_b^2 x_c^2 (x_b^2 + x_b x_c + x_c^2)
\bigr) \text{Ei}(-x_a)
- (x_a - x_b)^5 x_c^2 \text{Ei}(-x_c)
\Biggr]\label{b107},
\ee
\be
\frac{\partial \mathcal{L}^{(k=0)}_{aaabca}}{\partial x_b} =
\frac{1}{24\,x_a\,(x_a-x_b)^5\,(x_a-x_c)^4\,(x_b-x_c)^2}
\Bigg[
-6 e^{-x_c} x_a (x_a-x_b)^5 (x_c-1) \nonumber
\ee
\be
-6 e^{-x_b} x_a (x_a-x_c)^4
\Bigl(x_a-5 x_b+x_a x_b+3 x_b^2-2(-2+x_a+x_b)x_c\Bigr)\nonumber
\ee
\be
-2 e^{-x_a}(x_b-x_c)^2
\Bigl(
x_a\bigl(
x_a^3(30+x_a(6+x_a))
-3 x_a^2(10+x_a(10+x_a))x_b \nonumber
\ee
\be
+3 x_a(5+x_a(5+x_a))x_b^2
-(3+x_a(3+x_a))x_b^3
\bigr) \nonumber
\ee
\be
-3 x_a\bigl(
x_a^2(20+x_a(8+x_a))
-x_a(10+3x_a(10+x_a))x_b \nonumber
\ee
\be
+(2+3x_a(4+x_a))x_b^2-(2+x_a)x_b^3
\bigr)x_c\nonumber
\ee
\be
+3 x_a\bigl(
x_a^3+x_a^2(7-3x_b)-(-3+x_b)(-1+x_b)x_b
+x_a(15+x_b(-23+3x_b))
\bigr)x_c^2 \nonumber
\ee
\be
+\bigl(
-x_a(12+x_a(6+x_a))
+3x_a(6+x_a)x_b
-3x_a x_b^2+x_b^3
\bigr)x_c^3
\Bigr) \nonumber
\ee
\be
+6 x_a (x_b-x_c)^2
\Bigl(
x_a^4(x_a+3x_b)-12x_a^3x_bx_c
+x_b(8x_a^2+5x_a x_b-x_b^2)x_c^2 \nonumber
\ee
\be
-2x_b(x_a+x_b)x_c^3
\Bigr)\,\mathrm{Ei}(-x_a)\nonumber
\ee
\be
+6 x_a x_b (x_a-x_c)^4
\Bigl(-x_b(x_a+3x_b)+2(x_a+x_b)x_c\Bigr)\,\mathrm{Ei}(-x_b) \nonumber
\ee
\be
-6 x_a (x_a-x_b)^5 x_c^2\,\mathrm{Ei}(-x_c)
\Bigg]\label{b108},
\ee
\be
\frac{\partial \mathcal{L}^{(k=0)}_{aabbca}}{\partial x_a} =
\frac{1}{4\,x_a\,(x_a-x_b)^5\,(x_a-x_c)^4\,(x_b-x_c)^2}
\Bigg[
-3 e^{-x_c} x_a (x_a-x_b)^5 (x_c-1) \nonumber
\ee
\be
- e^{-x_a-x_b}
\Bigl(
3 x_a^6 -15 x_a^5 x_b +3 x_a^6 x_b +9 x_a^5 x_b^2
-6 x_a^6 x_c +60 x_a^4 x_b x_c -18 x_a^5 x_b x_c\nonumber
\ee
\be
-36 x_a^4 x_b^2 x_c -30 x_a^4 x_c^2 +24 x_a^5 x_c^2
-90 x_a^3 x_b x_c^2 +42 x_a^4 x_b x_c^2\nonumber
\ee
\be
+54 x_a^3 x_b^2 x_c^2 +60 x_a^3 x_c^3 -36 x_a^4 x_c^3
+60 x_a^2 x_b x_c^3 -48 x_a^3 x_b x_c^3 \nonumber
\ee
\be
-36 x_a^2 x_b^2 x_c^3 -45 x_a^2 x_c^4 +24 x_a^3 x_c^4
-15 x_a x_b x_c^4 +27 x_a^2 x_b x_c^4\nonumber
\ee
\be
+9 x_a x_b^2 x_c^4 +12 x_a x_c^5 -6 x_a^2 x_c^5
-6 x_a x_b x_c^5
\Bigr)\nonumber
\ee
\be
- e^{-x_a}
(x_b-x_c)^2
\Bigl(
x_a (x_a^3(30+x_a(6+x_a))
-3 x_a^2(10+x_a(10+x_a))x_b \nonumber
\ee
\be
+3 x_a(5+x_a(5+x_a))x_b^2
-(3+x_a(3+x_a))x_b^3) \nonumber
\ee
\be
-3 x_a (x_a^2(20+x_a(8+x_a))
-x_a(10+3x_a(10+x_a))x_b\nonumber
\ee
\be
+(2+3x_a(4+x_a))x_b^2-(2+x_a)x_b^3)x_c \nonumber
\ee
\be
+3 x_a (x_a^3+x_a^2(7-3x_b)-(-3+x_b)(-1+x_b)x_b
+x_a(15+x_b(-23+3x_b)))x_c^2 \nonumber
\ee
\be
+(-x_a(12+x_a(6+x_a))+3x_a(6+x_a)x_b
-3x_a x_b^2+x_b^3)x_c^3
\Bigr)\nonumber
\ee
\be
-3 x_a (x_b-x_c)^2
\Bigl(
x_a^4(x_a+3x_b)-12x_a^3x_bx_c
+x_b(8x_a^2+5x_a x_b-x_b^2)x_c^2\nonumber
\ee
\be
-2x_b(x_a+x_b)x_c^3
\Bigr)\,\mathrm{Ei}(-x_a)\nonumber
\ee
\be
+3 x_a x_b (x_a-x_c)^4
\Bigl(x_b(x_a+3x_b)-2(x_a+x_b)x_c\Bigr)\,\mathrm{Ei}(-x_b)\nonumber
\ee
\be
+3 x_a (x_a-x_b)^5 x_c^2\,\mathrm{Ei}(-x_c)
\Bigg]\label{b109},
\ee
\be
\frac{\partial \mathcal{L}^{(k=0)}_{aabbca}}{\partial x_b} =
\frac{1}{2\,(x_a-x_b)^5\,(x_a-x_c)^3\,(x_b-x_c)^3}
\Bigg[
- e^{-x_c}(x_a-x_b)^5 (x_c-1) \nonumber
\ee
\be
+ e^{-x_a-x_b}
\Bigl(
-(x_a-x_c)^3\!\left[
-2(-5+x_b)x_b^2
+x_a^2(1+x_b-2x_c)\right. \nonumber
\ee
\be
-x_a(5x_b-3x_c)(1+x_b-2x_c)\left.
+(-15+x_b)x_b x_c+6x_c^2
\right] \nonumber
\ee
\be
+(x_b-x_c)^3\!\left[
-2x_a^3+x_a(-5+x_b)x_b+x_b^2
+(3-2x_b)x_b x_c \right.\nonumber
\ee
\be\left.
+x_a(-15+13x_b)x_c
-6(-1+x_b)x_c^2
+x_a^2(10-5x_b+x_c)
\right]
\Bigr) \nonumber
\ee
\be
-(x_b-x_c)^3
\Bigl(
3x_a^3(x_a+x_b)
-3x_a^2(x_a+3x_b)x_c
+(x_a^2+4x_a x_b+x_b^2)x_c^2
\Bigr)\,\mathrm{Ei}(-x_a) \nonumber
\ee
\be
+(x_a-x_c)^3
\Bigl(
3x_b^3(x_a+x_b)
-3x_b^2(3x_a+x_b)x_c
+(x_a^2+4x_a x_b+x_b^2)x_c^2
\Bigr)\,\mathrm{Ei}(-x_b)\nonumber
\ee
\be
-(x_a-x_b)^5 x_c^2\,\mathrm{Ei}(-x_c)
\Bigg]\label{b110},
\ee
\be
\frac{\partial \mathcal{L}^{(k=0)}_{aabbca}}{\partial x_c} =
\frac{1}{4\,(x_a-x_b)^4\,(x_a-x_c)^4\,(x_b-x_c)^3}
\Bigg[\nonumber
\ee
\be
e^{-x_c}(x_a-x_b)^4
\Bigl(
2 x_a (x_b-1)+x_b(x_c-3)+(5-3x_c)x_c
\Bigr) \nonumber
\ee
\be
- e^{-x_b}(x_a-x_c)^4
\Bigl(
2 x_a (x_c-1)-3x_c+x_b(5-3x_b+x_c)
\Bigr) \nonumber
\ee
\be
- e^{-x_a}(x_b-x_c)^3
\Bigl(
2x_a^3+4x_b(x_c-1)x_c-3x_c^2
+2x_a^2(-5+x_b+x_c) \nonumber
\ee
\be
+x_b^2(-3+4x_c)
-x_a\!\left(x_b^2+(-10+x_c)x_c+2x_b(-5+6x_c)\right)
\Bigr) \nonumber
\ee
\be
-(x_b-x_c)^3
\Bigl(
3x_a^4+2x_a x_b(-4x_a+x_b)x_c
+x_b(2x_a+x_b)x_c^2
\Bigr)\,\mathrm{Ei}(-x_a) \nonumber
\ee
\be
+ x_b (x_a-x_c)^4
\Bigl(
3x_b^2-(2x_a+x_b)x_c
\Bigr)\,\mathrm{Ei}(-x_b) \nonumber
\ee
\be
+ (x_a-x_b)^4 x_c
\Bigl(
2x_a x_b+(x_b-3x_c)x_c
\Bigr)\,\mathrm{Ei}(-x_c)
\Bigg]\label{b111-ii},
\ee
\be
\frac{\partial \mathcal{L}^{(k=0)}_{abbcca}}{\partial x_a} =
\frac{1}{2 (x_a - x_b)^4 (x_a - x_c)^4 (x_b - x_c)^3} \Biggl[
6 e^{-x_a} x_a (x_a - x_b) (x_b - x_c)^3
\nonumber
\ee
\be
- 3 e^{-x_a} x_a^2 (x_a - x_b) (x_b - x_c)^3
- 3 e^{-x_a} (x_a - x_b) x_b (x_b - x_c)^3 \nonumber
\ee
\be
+ 6 e^{-x_a} x_a (x_a - x_c) (x_b - x_c)^3
- 2 e^{-x_a} (x_a - x_b) (x_a - x_c) (x_b - x_c)^3\nonumber
\ee
\be
+ 4 e^{-x_a} x_a (x_a - x_b) (x_a - x_c) (x_b - x_c)^3
- 3 e^{-x_a} x_b (x_a - x_c) (x_b - x_c)^3
\nonumber
\ee
\be
+ e^{-x_a} x_b (-x_a + x_b) (x_a - x_c) (x_b - x_c)^3
+ e^{-x_b} (x_a - x_b) (x_a - x_c)^4 (-1 + x_c)\nonumber
\ee
\be
- 3 e^{-x_a} (x_a - x_b) (x_b - x_c)^3 x_c
+ 3 e^{-x_a} (x_a - x_b) x_b (x_b - x_c)^3 x_c\nonumber
\ee
\be
- e^{-x_a} (x_a - x_b) (x_a - x_c) (x_b - x_c)^3 x_c
+ 3 e^{-x_a} x_b (x_a - x_c) (x_b - x_c)^3 x_c \nonumber
\ee
\be
+ 3 e^{-x_a} x_a^2 (x_a - x_c) (-x_b + x_c)^3
+ 3 e^{-x_a} (x_a - x_c) x_c (-x_b + x_c)^3\nonumber
\ee
\be
+ 3 e^{-x_b} (x_a - x_c)^4 (x_a + (-2 + x_b) x_b + x_c - x_a x_c)
+ e^{-x_c} (x_a - x_b)^4 \bigl( 2 x_a (-1 + x_b)\nonumber
\ee
\be + x_b (-3 + x_c) + (5 - 3 x_c) x_c \bigr)
- 3 x_a^3 (x_a - x_b) (x_b - x_c)^3 \text{Ei}(-x_a)\nonumber
\ee
\be
+ 3 x_a^2 (x_a - x_b) (x_a - x_c) (x_b - x_c)^3 \text{Ei}(-x_a)
+ 3 x_a (x_a - x_b) x_b (x_b - x_c)^3 x_c \text{Ei}(-x_a) \nonumber
\ee
\be
+ 3 x_a x_b (x_a - x_c) (x_b - x_c)^3 x_c \text{Ei}(-x_a)
- (x_a - x_b) x_b (x_a - x_c) (x_b - x_c)^3 x_c \text{Ei}(-x_a)\nonumber
\ee
\be
+ 3 x_a^3 (x_a - x_c) (-x_b + x_c)^3 \text{Ei}(-x_a)
+ 3 x_b^3 (x_a - x_c)^4 \text{Ei}(-x_b) \nonumber
\ee
\be
- 3 x_a x_b (x_a - x_c)^4 x_c \text{Ei}(-x_b)
+ (x_a - x_b) x_b (x_a - x_c)^4 x_c \text{Ei}(-x_b)\nonumber
\ee
\be
+ 3 x_a (x_a - x_b)^4 x_b x_c \text{Ei}(-x_c)
- (x_a - x_b)^4 x_b (x_a - x_c) x_c \text{Ei}(-x_c)\nonumber
\ee
\be
- 3 (x_a - x_b)^4 x_c^3 \text{Ei}(-x_c)
\Biggr]\label{b112},
\ee

\be
\frac{\partial \mathcal{L}^{(k=1)}_{aabca}}{\partial x_a} =
-\frac{i}{4 (x_a - x_b)^4 (x_a - x_c)^4 (x_b - x_c)} \Biggl[
- 3 e^{-x_b} (-1 + x_b) (x_a - x_c)^4 \nonumber
\ee
\be+ 3 e^{-x_c} (x_a - x_b)^4 (-1 + x_c)
+ \frac{1}{x_a} e^{-x_a} \Bigl(
    x_a x_b \Bigl(
        -x_a^3 (12 + x_a (6 + x_a)) + 3 x_a^2 (6 + x_a (6 + x_a)) x_b \nonumber
\ee
\be
        - 3 x_a (2 + x_a)^2 x_b^2 + (3 + x_a (3 + x_a)) x_b^3
    \Bigr) \nonumber
\ee
\be
    + x_a \Bigl(
        x_a^3 (12 + x_a (6 + x_a)) - 6 x_a^2 (6 + x_a) x_b^2 + 8 x_a (3 + x_a) x_b^3 - 3 (2 + x_a) x_b^4
    \Bigr) x_c\nonumber
\ee
\be
    - 3 x_a \Bigl( x_a^2 (6 + x_a (6 + x_a)) - 2 x_a^2 (6 + x_a) x_b + 2 x_a x_b^3 - x_b^4 \Bigr) x_c^2\nonumber
\ee
\be
    + \Bigl( 3 x_a^2 (2 + x_a)^2 - 8 x_a^2 (3 + x_a) x_b + 6 x_a^2 x_b^2 - x_b^4 \Bigr) x_c^3 \nonumber
\ee
\be
    + \Bigl( -x_a^3 + 3 x_a^2 (-1 + x_b) - 3 x_a (-1 + x_b)^2 + x_b^3 \Bigr) x_c^4 \nonumber
\ee
\be
    + 3 e^{x_a} x_a (x_b - x_c) (x_a^2 - x_b x_c) (-4 x_a x_b x_c + x_a^2 (x_b + x_c) + x_b x_c (x_b + x_c)) \text{Ei}(-x_a) \nonumber
\ee
\be
    - 3 e^{x_a} x_a x_b^2 (x_a - x_c)^4 \text{Ei}(-x_b) \nonumber
\ee
\be
    + 3 e^{x_a} x_a \Bigl( (x_a - x_b)^4 x_c^2 \text{Ei}(-x_c) + (x_b - x_c) (-x_a^2 + x_b x_c) (-4 x_a x_b x_c + x_a^2 (x_b + x_c) \nonumber
\ee
\be+ x_b x_c (x_b + x_c)) \text{Ei}(-x_a)
    + x_b^2 (x_a - x_c)^4 \text{Ei}(-x_b) - (x_a - x_b)^4 x_c^2 \text{Ei}(-x_c) \Bigr)
\Bigr)
\Biggr]\label{b113},
\ee
\be
\frac{\partial \mathcal{L}^{(k=1)}_{aabca}}{\partial x_b} =
\frac{i}{4 (x_a - x_b)^4 (x_a - x_c)^3 (x_b - x_c)^2} \Biggl[\nonumber
\ee
\be
- e^{-x_c} (x_a - x_b)^4 (-1 + x_c) - e^{-x_b} (x_a - x_c)^3 \bigl( 2 x_b^2 + x_a (1 + x_b - 2 x_c) + 3 x_c - x_b (4 + x_c) \bigr)\nonumber
\ee
\be
- e^{-x_a} (x_b - x_c)^2 \bigl( x_b^2 - 2 (-1 + x_b) x_b x_c + (3 - 4 x_b) x_c^2 - 2 x_a^2 (-3 + 2 x_b + x_c) \nonumber
\ee
\be
+ x_a (x_b^2 + (-8 + x_c) x_c + 2 x_b (-2 + 5 x_c)) \bigr)
+ (x_b - x_c)^2 \bigl( x_a^3 (x_a + 2 x_b)
\nonumber
\ee
\be
- 6 x_a^2 x_b x_c + x_b (2 x_a + x_b) x_c^2 \bigr) \text{Ei}(-x_a)
- x_b (x_a - x_c)^3 \bigl( x_b (x_a + 2 x_b) - (2 x_a + x_b) x_c \bigr) \text{Ei}(-x_b)\nonumber
\ee
\be
- x_a^4 x_b^2 \text{Ei}(-x_a) - 2 x_a^3 x_b^3 \text{Ei}(-x_a) + 2 x_a^4 x_b x_c \text{Ei}(-x_a) + 4 x_a^3 x_b^2 x_c \text{Ei}(-x_a) + 6 x_a^2 x_b^3 x_c \text{Ei}(-x_a)\nonumber
\ee
\be
- x_a^4 x_c^2 \text{Ei}(-x_a) - 2 x_a^3 x_b x_c^2 \text{Ei}(-x_a) - 12 x_a^2 x_b^2 x_c^2 \text{Ei}(-x_a) - 2 x_a x_b^3 x_c^2 \text{Ei}(-x_a) - x_b^4 x_c^2 \text{Ei}(-x_a)\nonumber
\ee
\be
+ 6 x_a^2 x_b x_c^3 \text{Ei}(-x_a) + 4 x_a x_b^2 x_c^3 \text{Ei}(-x_a) + 2 x_b^3 x_c^3 \text{Ei}(-x_a) - 2 x_a x_b x_c^4 \text{Ei}(-x_a) - x_b^2 x_c^4 \text{Ei}(-x_a)\nonumber
\ee
\be
+ x_a^4 x_b^2 \text{Ei}(-x_b) + 2 x_a^3 x_b^3 \text{Ei}(-x_b) - 2 x_a^4 x_b x_c \text{Ei}(-x_b) - 4 x_a^3 x_b^2 x_c \text{Ei}(-x_b) - 6 x_a^2 x_b^3 x_c \text{Ei}(-x_b)\nonumber
\ee
\be
+ 6 x_a^3 x_b x_c^2 \text{Ei}(-x_b) + 6 x_a^2 x_b^2 x_c^2 \text{Ei}(-x_b) + 6 x_a x_b^3 x_c^2 \text{Ei}(-x_b) - 6 x_a^2 x_b x_c^3 \text{Ei}(-x_b)\nonumber
\ee
\be - 4 x_a x_b^2 x_c^3 \text{Ei}(-x_b)
- 2 x_b^3 x_c^3 \text{Ei}(-x_b) + 2 x_a x_b x_c^4 \text{Ei}(-x_b) + x_b^2 x_c^4 \text{Ei}(-x_b)\nonumber
\ee
\be + (x_a - x_b)^4 x_c^2 \text{Ei}(-x_c) - (x_a - x_b)^4 x_c^2 \text{Ei}(-x_c)
\Biggr]\label{b114},
\ee
\be
\frac{\partial \mathcal{L}^{(k=1)}_{abbca}}{\partial x_a} =
\frac{e^{-(x_a + x_b + x_c)}}{2 (x_a - x_b)^4 (x_a - x_c)^3 (x_b - x_c)^2} \Biggl[\nonumber
\ee
\be
- i \, e^{x_a + x_b} (x_a - x_b)^4 (-1 + x_c)
+ i \, e^{x_c} \Bigl(
    - e^{x_a} (x_a - x_c)^3 \bigl( 2 x_b^2 + x_a (1 + x_b - 2 x_c) \nonumber
\ee
\be+ 3 x_c - x_b (4 + x_c) \bigr)
    - e^{x_b} (x_b - x_c)^2 \bigl( x_b^2 - 2 (-1 + x_b) x_b x_c + (3 - 4 x_b) x_c^2 - 2 x_a^2 (-3 + 2 x_b + x_c)
	\nonumber
\ee
\be
 + x_a (x_b^2 + (-8 + x_c) x_c + 2 x_b (-2 + 5 x_c)) \bigr)
    + e^{x_a + x_b} (x_b - x_c)^2 \bigl( x_a^3 (x_a + 2 x_b) - 6 x_a^2 x_b x_c \nonumber
\ee
\be
+ x_b (2 x_a + x_b) x_c^2 \bigr) \text{Ei}(-x_a)
    - e^{x_a + x_b} \Bigl(
        x_b (x_a - x_c)^3 \bigl( x_b (x_a + 2 x_b) - (2 x_a + x_b) x_c \bigr) \text{Ei}(-x_b)\nonumber
\ee
\be
        + x_a^4 x_b^2 \text{Ei}(-x_a) + 2 x_a^3 x_b^3 \text{Ei}(-x_a) - 2 x_a^4 x_b x_c \text{Ei}(-x_a) - 4 x_a^3 x_b^2 x_c \text{Ei}(-x_a) - 6 x_a^2 x_b^3 x_c \text{Ei}(-x_a)\nonumber
\ee
\be
        + x_a^4 x_c^2 \text{Ei}(-x_a) + 2 x_a^3 x_b x_c^2 \text{Ei}(-x_a) + 12 x_a^2 x_b^2 x_c^2 \text{Ei}(-x_a) + 2 x_a x_b^3 x_c^2 \text{Ei}(-x_a) + x_b^4 x_c^2 \text{Ei}(-x_a)\nonumber
\ee
\be
        - 6 x_a^2 x_b x_c^3 \text{Ei}(-x_a) - 4 x_a x_b^2 x_c^3 \text{Ei}(-x_a) - 2 x_b^3 x_c^3 \text{Ei}(-x_a) + 2 x_a x_b x_c^4 \text{Ei}(-x_a) + x_b^2 x_c^4 \text{Ei}(-x_a)\nonumber
\ee
\be
        - x_a^4 x_b^2 \text{Ei}(-x_b) - 2 x_a^3 x_b^3 \text{Ei}(-x_b) + 2 x_a^4 x_b x_c \text{Ei}(-x_b) + 4 x_a^3 x_b^2 x_c \text{Ei}(-x_b) + 6 x_a^2 x_b^3 x_c \text{Ei}(-x_b)\nonumber
\ee
\be
        - 6 x_a^3 x_b x_c^2 \text{Ei}(-x_b) - 6 x_a^2 x_b^2 x_c^2 \text{Ei}(-x_b) - 6 x_a x_b^3 x_c^2 \text{Ei}(-x_b) + 6 x_a^2 x_b x_c^3 \text{Ei}(-x_b) \nonumber
\ee
\be  + (x_a - x_b)^4 x_c^2 \text{Ei}(-x_c) + 4 x_a x_b^2 x_c^3 \text{Ei}(-x_b)
        + 2 x_b^3 x_c^3 \text{Ei}(-x_b) - 2 x_a x_b x_c^4 \text{Ei}(-x_b) \nonumber
\ee
\be - x_b^2 x_c^4 \text{Ei}(-x_b) - (x_a - x_b)^4 x_c^2 \text{Ei}(-x_c)
    \Bigr)
\Bigr)
\Biggr]\label{b115},
\ee
\be
\frac{\partial \mathcal{L}^{(k=1)}_{abbca}}{\partial x_c} =
-\frac{i \, e^{-(x_a + x_b + x_c)}}{2 (x_a - x_b)^3 (x_a - x_c)^3 (x_b - x_c)^3} \Biggl[ \nonumber
\ee
\be
e^{x_a + x_c} (x_a - x_c)^3 \bigl( -((-2 + x_b) x_b) + x_a (-1 + x_c) - x_c \bigr) \nonumber
\ee
\be
- e^{x_a + x_b} (x_a - x_b)^3 \bigl( x_a (-1 + x_b) - x_b - (-2 + x_c) x_c \bigr) \nonumber
\ee
\be
+ e^{x_b + x_c} \Bigl(
    (x_b - x_c)^3 \bigl( (-2 + x_a) x_a + x_b + x_c - x_b x_c \bigr) \nonumber
\ee
\be
    + e^{x_a} \Bigl(
        x_a (x_b - x_c)^3 (x_a^2 - x_b x_c) \text{Ei}(-x_a) \nonumber
\ee
\be
        + x_b (x_a - x_c)^3 (-x_b^2 + x_a x_c) \text{Ei}(-x_b)
        - x_a^4 x_b x_c \text{Ei}(-x_c)
		\nonumber
\ee
\be
+ 3 x_a^3 x_b^2 x_c \text{Ei}(-x_c) - 3 x_a^2 x_b^3 x_c \text{Ei}(-x_c) + x_a x_b^4 x_c \text{Ei}(-x_c) \nonumber
\ee
\be
        + x_a^3 x_c^3 \text{Ei}(-x_c) - 3 x_a^2 x_b x_c^3 \text{Ei}(-x_c) + 3 x_a x_b^2 x_c^3 \text{Ei}(-x_c) - x_b^3 x_c^3 \text{Ei}(-x_c) \nonumber
\ee
\be
        - x_a^3 x_b^3 \text{Ei}(-x_a) + 3 x_a^3 x_b^2 x_c \text{Ei}(-x_a) + x_a x_b^4 x_c \text{Ei}(-x_a) - 3 x_a^3 x_b x_c^2 \text{Ei}(-x_a) \nonumber
\ee
\be
        - 3 x_a x_b^3 x_c^2 \text{Ei}(-x_a) + x_a^3 x_c^3 \text{Ei}(-x_a) + 3 x_a x_b^2 x_c^3 \text{Ei}(-x_a) - x_a x_b x_c^4 \text{Ei}(-x_a) \nonumber
\ee
\be
        + x_a^3 x_b^3 \text{Ei}(-x_b) - x_a^4 x_b x_c \text{Ei}(-x_b) - 3 x_a^2 x_b^3 x_c \text{Ei}(-x_b) + 3 x_a^3 x_b x_c^2 \text{Ei}(-x_b) \nonumber
\ee
\be
        + 3 x_a x_b^3 x_c^2 \text{Ei}(-x_b) - 3 x_a^2 x_b x_c^3 \text{Ei}(-x_b) - x_b^3 x_c^3 \text{Ei}(-x_b) + x_a x_b x_c^4 \text{Ei}(-x_b) \nonumber
\ee
\be
        + (x_a - x_b)^3 x_c (x_a x_b - x_c^2) \text{Ei}(-x_c)
    \Bigr)
\Bigr)
\Biggr]\label{b116},
\ee
\be
\frac{\partial \mathcal{L}^{(k=2)}_{abca}}{\partial x_a} =
\frac{1}{2 (x_a - x_b)^3 (x_a - x_c)^3 (x_b - x_c)} \Biggl[
e^{x_a + x_b} (x_a - x_b)^3 (-1 + x_c)\nonumber
\ee
\be
+ e^{x_c} \Bigl(
    e^{x_a} (-1 + x_b) (-x_a + x_c)^3
    + e^{x_b} (-x_b + x_c) \bigl(x_a^3 + x_b^2 (1 - 2 x_c) + x_b (1 - 2 x_c) x_c \nonumber
\ee
\be
    + x_c^2 - 3 x_a^2 (-1 + x_b + x_c) + x_a (x_b^2 + (-3 + x_c) x_c + x_b (-3 + 7 x_c))\bigr)
\Bigr) \nonumber
\ee
\be
- e^{x_a + x_b} \Bigl(
    (x_b - x_c) (-3 x_a^2 x_b x_c + x_b^2 x_c^2 + x_a^3 (x_b + x_c)) \text{Ei}(-x_a) \nonumber
\ee
\be
    + x_b^2 (-x_a + x_c)^3 \text{Ei}(-x_b)\nonumber
\ee
\be
    + x_c^2 (x_a^3 - 3 x_a^2 x_b x_c + 3 x_a x_b^2 x_c - x_b^3 x_c) \text{Ei}(-x_c)
    - (-x_a + x_b)^3 x_c^2 \text{Ei}(-x_c)
\Bigr)
\Biggr]\label{b117},
\ee
\be
\frac{\partial \mathcal{L}^{(k=2)}_{abca}}{\partial x_b} = \frac{1}{4 (x_a - x_b)^3 (x_a - x_c)^2 (x_b - x_c)^2} \Biggl[
e^{x_a + x_b + x_c} \Bigl(\nonumber
\ee
\be
- e^{x_a + x_c} \bigl( -((-3 + x_b) x_b) - x_a (1 + x_b - 2 x_c) - 2 x_c \bigr) (x_a - x_c)^2 \\
+ e^{x_a + x_b} (x_a - x_b)^3 (-1 + x_c) \nonumber
\ee
\be
- e^{x_b + x_c} \Bigl(
    (x_b - x_c)^2 (x_b + x_a (-3 + x_a + x_b) + 2 x_c - 2 x_b x_c) \nonumber
\ee
\be
    + e^{x_a} \bigl(
        x_a (x_b - x_c)^2 (x_a (x_a + x_b) - 2 x_b x_c) \text{Ei}(-x_a)
		\nonumber
\ee
\be
        - x_b (x_a - x_c)^2 (x_b (x_a + x_b) - 2 x_a x_c) \text{Ei}(-x_b)\nonumber
\ee
\be
        - x_a^3 x_c^2 \text{Ei}(-x_c) + 3 x_a^2 x_b x_c^2 \text{Ei}(-x_c) - 3 x_a x_b^2 x_c^2 \text{Ei}(-x_c) + x_b^3 x_c^2 \text{Ei}(-x_c) \nonumber
\ee
\be
        - x_a^3 x_b^2 \text{Ei}(-x_a) - x_a^2 x_b^3 \text{Ei}(-x_a) + 2 x_a^3 x_b x_c \text{Ei}(-x_a) + 2 x_a^2 x_b^2 x_c \text{Ei}(-x_a) + 2 x_a x_b^3 x_c \text{Ei}(-x_a)\nonumber
\ee
\be
        - x_a^3 x_c^2 \text{Ei}(-x_a) - x_a^2 x_b x_c^2 \text{Ei}(-x_a) - 4 x_a x_b^2 x_c^2 \text{Ei}(-x_a) + 2 x_a x_b x_c^3 \text{Ei}(-x_a) \nonumber
\ee
\be
        + x_a^3 x_b^2 \text{Ei}(-x_b) + x_a^2 x_b^3 \text{Ei}(-x_b) - 2 x_a^3 x_b x_c \text{Ei}(-x_b) - 2 x_a^2 x_b^2 x_c \text{Ei}(-x_b) - 2 x_a x_b^3 x_c \text{Ei}(-x_b) \nonumber
\ee
\be
        + 4 x_a^2 x_b x_c^2 \text{Ei}(-x_b) + x_a x_b^2 x_c^2 \text{Ei}(-x_b) + x_b^3 x_c^2 \text{Ei}(-x_b) - 2 x_a x_b x_c^3 \text{Ei}(-x_b)  \nonumber
\ee
\be
        + (x_a - x_b)^3 x_c^2 \text{Ei}(-x_c)
    \bigr)
\Bigr)
\Bigr)
\Biggr]\label{b118}.
\ee

We can then compute the variation of $\frac{\delta \Omega^0_5}{\delta T_{pq}}$. We integrate by parts covariant derivatives of the tachyon. For instance, we rewrite
\be
\tr_\mathcal{L}(T^\dag F_R, DT, F_L, \mathbb{I}) = \intd \tr_\mathcal{L}(T^\dag F_R, T,  F_L, \mathbb{I})- \tr_\mathcal{L}(\intd(T^\dag F_R), T,  F_L, \mathbb{I}) \nonumber
\ee
\be
 +\tr_\mathcal{L}(T^\dag F_R, T,  \intd(F_L), \mathbb{I}) + i \tr_\mathcal{L}(T^\dag F_R, T A_L,  F_L, \mathbb{I}) - i \tr_\mathcal{L}(T^\dag F_R, A_R T,  F_L, \mathbb{I})\label{b119}.
\ee
After some tedious but straightforward algebra we obtain
\be
\frac{\delta \Omega^0_5}{\delta T_{pq}} =   -\tr_{\frac{\partial \mathcal{\mathcal{L}}}{\partial T_{pq}}}\big(T^\dagger DT, DT^\dagger, DT, DT^\dagger, DT, \mathbb{I} \big) +  \tr_{\frac{\partial \mathcal{\mathcal{L}}}{\partial T_{pq}}}\big(T^\dagger DT, DT^\dagger, DT, F_L, \mathbb{I} \big) \nonumber
\ee
\be
 + \tr_{\frac{\partial \mathcal{\mathcal{L}}}{\partial T_{pq}}}\big(T^\dag DT, DT^\dag, F_R, DT, \mathbb{I} \big)   +  \tr_{\frac{\partial \mathcal{\mathcal{L}}}{\partial T_{pq}}}\big( T^\dag DT, F_L ,DT^\dag, DT , \mathbb{I} \big) \nonumber
\ee
\be
  - \tr_{\frac{\partial \mathcal{\mathcal{L}}}{\partial T_{pq}}}\big(F_L, DT^\dag, DT, DT^\dag T , \mathbb{I} \big)
-\tr_{\frac{\partial \mathcal{\mathcal{L}}}{\partial T_{pq}}}\big(T^\dagger DT, F_L, F_L , \mathbb{I} \big) \nonumber
\ee
\be
  -\tr_{\frac{\partial \mathcal{\mathcal{L}}}{\partial T_{pq}}}\big(T^\dag F_R, DT, F_L, \mathbb{I} \big)  + \tr_{\frac{\partial \mathcal{\mathcal{L}}}{\partial T_{pq}}}\big(F_L, F_L , DT^\dag T, \mathbb{I} \big) \nonumber
\ee
\be
- \intd \tr_\mathcal{L}(T^\dag \mathbb{I}_{.p}\mathbb{I}_{q.}, DT^\dag, DT, DT^\dag, DT, \mathbb{I})+ \tr_\mathcal{L}((\intd T^\dag ) \mathbb{I}_{.p}\mathbb{I}_{q.}, DT^\dag, DT, DT^\dag, DT, \mathbb{I}) \nonumber
\ee
\be
-\tr_\mathcal{L}(T^\dag  \mathbb{I}_{.p}\mathbb{I}_{q.},  \intd DT^\dag, DT, DT^\dag, DT, \mathbb{I})
+\tr_\mathcal{L}(T^\dag  \mathbb{I}_{.p}\mathbb{I}_{q.}, DT^\dag,  \intd DT, DT^\dag, DT, \mathbb{I})  \nonumber
\ee
\be
-\tr_\mathcal{L}(T^\dag  \mathbb{I}_{.p}\mathbb{I}_{q.}, DT^\dag, DT,  \intd DT^\dag, DT, \mathbb{I})
+\tr_\mathcal{L}(T^\dag  \mathbb{I}_{.p}\mathbb{I}_{q.}, DT^\dag, DT, DT^\dag, \intd DT, \mathbb{I})  \nonumber
\ee
\be
-i\tr_\mathcal{L}(T^\dag ( \mathbb{I}_{.p}\mathbb{I}_{q.}A_L - A_R  \mathbb{I}_{.p}\mathbb{I}_{q.}), DT^\dag, DT, DT^\dag, DT, \mathbb{I})  \nonumber
\ee
\be
- \intd \tr_\mathcal{L}(T^\dag DT, DT^\dag,  \mathbb{I}_{.p}\mathbb{I}_{q.}, DT^\dag, DT, \mathbb{I})  \nonumber
\ee
\be
+\tr_\mathcal{L}(\intd (T^\dag DT), DT^\dag,  \mathbb{I}_{.p}\mathbb{I}_{q.}, DT^\dag, DT, \mathbb{I})
-\tr_\mathcal{L}(T^\dag DT, \intd DT^\dag,  \mathbb{I}_{.p}\mathbb{I}_{q.}, DT^\dag, DT, \mathbb{I})  \nonumber
\ee
\be
-\tr_\mathcal{L}(T^\dag DT, DT^\dag,  \mathbb{I}_{.p}\mathbb{I}_{q.}, \intd DT^\dag, DT, \mathbb{I})
+\tr_\mathcal{L}(T^\dag DT, DT^\dag,  \mathbb{I}_{.p}\mathbb{I}_{q.}, DT^\dag,\intd DT, \mathbb{I})  \nonumber
\ee
\be
-i \tr_\mathcal{L}(T^\dag DT, DT^\dag,  \mathbb{I}_{.p}\mathbb{I}_{q.}A_L - A_R  \mathbb{I}_{.p}\mathbb{I}_{q.}, DT^\dag, DT, \mathbb{I})   \nonumber
\ee
\be
- \intd \tr_\mathcal{L}(T^\dag DT, DT^\dag, DT, DT^\dag, \mathbb{I}_{.p}\mathbb{I}_{q.}, \mathbb{I}) \nonumber
\ee
\be
+\tr_\mathcal{L}(\intd (T^\dag DT), DT^\dag, DT, DT^\dag, \mathbb{I}_{.p}\mathbb{I}_{q.}, \mathbb{I})
-\tr_\mathcal{L}(T^\dag DT, \intd DT^\dag, DT, DT^\dag, \mathbb{I}_{.p}\mathbb{I}_{q.}, \mathbb{I})   \nonumber
\ee
\be
+\tr_\mathcal{L}(T^\dag DT, DT^\dag, \intd DT, DT^\dag, \mathbb{I}_{.p}\mathbb{I}_{q.}, \mathbb{I})
-\tr_\mathcal{L}(T^\dag DT, DT^\dag, DT, \intd DT^\dag, \mathbb{I}_{.p}\mathbb{I}_{q.}, \mathbb{I})   \nonumber
\ee
\be
\tr_\mathcal{L}(\mathbb{I}_{.p}\mathbb{I}_{q.} DT^\dag, DT, DT^\dag,DT, DT^\dag, \mathbb{I})
- \intd\tr_\mathcal{L}(T DT^\dag, \mathbb{I}_{.p}\mathbb{I}_{q.}, DT^\dag,DT, DT^\dag, \mathbb{I}) \nonumber
\ee
\be
+\tr_\mathcal{L}( \intd(T DT^\dag), \mathbb{I}_{.p}\mathbb{I}_{q.}, DT^\dag,DT, DT^\dag, \mathbb{I})
+\tr_\mathcal{L}(T DT^\dag,  \mathbb{I}_{.p}\mathbb{I}_{q.}, \intd DT^\dag,DT, DT^\dag, \mathbb{I}) \nonumber
\ee
\be
-\tr_\mathcal{L}(T DT^\dag, \mathbb{I}_{.p}\mathbb{I}_{q.}, DT^\dag, \intd DT, DT^\dag, \mathbb{I})
+\tr_\mathcal{L}(T DT^\dag, \mathbb{I}_{.p}\mathbb{I}_{q.}, DT^\dag, DT, \intd DT^\dag, \mathbb{I}) \nonumber
\ee
\be
- \intd \tr_\mathcal{L}(T DT^\dag, DT, DT^\dag, \mathbb{I}_{.p}\mathbb{I}_{q.}, DT^\dag, \mathbb{I})
+\tr_\mathcal{L}( \intd(T DT^\dag), DT, DT^\dag,\mathbb{I}_{.p}\mathbb{I}_{q.}, DT^\dag, \mathbb{I}) \nonumber
\ee
\be
-\tr_\mathcal{L}(T DT^\dag, \intd DT, DT^\dag,\mathbb{I}_{.p}\mathbb{I}_{q.}, DT^\dag,\mathbb{I})
+\tr_\mathcal{L}(T DT^\dag, DT, \intd DT^\dag,\mathbb{I}_{.p}\mathbb{I}_{q.}, DT^\dag, \mathbb{I}) \nonumber
\ee
\be
+\tr_\mathcal{L}(T DT^\dag, DT, DT^\dag, \mathbb{I}_{.p}\mathbb{I}_{q.}, \intd DT^\dag, \mathbb{I})   \nonumber
\ee
\be
+i\tr_\mathcal{L}(T DT^\dag, \mathbb{I}_{.p}\mathbb{I}_{q.}A_L - A_R \mathbb{I}_{.p}\mathbb{I}_{q.}, DT^\dag,DT, DT^\dag, \mathbb{I}) \nonumber
\ee
\be
+i\tr_\mathcal{L}(T DT^\dag, DT, DT^\dag, \mathbb{I}_{.p}\mathbb{I}_{q.} A_L - A_R \mathbb{I}_{.p}\mathbb{I}_{q.}, DT^\dag, \mathbb{I})\nonumber
\ee
\be
+\intd \tr_{\mathcal{L}}\big(T^\dagger  \mathbb{I}_{.p}\mathbb{I}_{q.}, DT^\dagger, DT, F_L, \mathbb{I} \big)
-\tr_{\mathcal{L}}\big(\intd T^\dagger  \mathbb{I}_{.p}\mathbb{I}_{q.}, DT^\dagger, DT, F_L, \mathbb{I} \big) \nonumber
\ee
\be
+\tr_{\mathcal{L}}\big(T^\dagger  \mathbb{I}_{.p}\mathbb{I}_{q.},\intd DT^\dagger, DT, F_L, \mathbb{I} \big)
-\tr_{\mathcal{L}}\big(T^\dagger  \mathbb{I}_{.p}\mathbb{I}_{q.}, DT^\dagger, \intd DT, F_L, \mathbb{I} \big) \nonumber
\ee
\be
+\tr_{\mathcal{L}}\big(T^\dagger  \mathbb{I}_{.p}\mathbb{I}_{q.}, DT^\dagger, DT, \intd F_L, \mathbb{I} \big)
+i\tr_{\mathcal{L}}\big(T^\dagger ( \mathbb{I}_{.p}\mathbb{I}_{q.} A_L -  \mathbb{I}_{.p}\mathbb{I}_{q.} A_R), DT^\dagger, DT, F_L, \mathbb{I} \big) \nonumber
\ee
\be
+\intd\tr_{\mathcal{L}}\big(T^\dagger DT, DT^\dagger,  \mathbb{I}_{.p}\mathbb{I}_{q.}, F_L, \mathbb{I} \big)
-\tr_{\mathcal{L}}\big(\intd(T^\dagger DT), DT^\dagger,  \mathbb{I}_{.p}\mathbb{I}_{q.}, F_L, \mathbb{I} \big) \nonumber
\ee
\be
+\tr_{\mathcal{L}}\big(T^\dagger DT,\intd DT^\dagger,  \mathbb{I}_{.p}\mathbb{I}_{q.}, F_L, \mathbb{I} \big)
+\tr_{\mathcal{L}}\big(T^\dagger DT, DT^\dagger, \mathbb{I}_{.p}\mathbb{I}_{q.},\intd F_L, \mathbb{I} \big) \nonumber
\ee
\be
+i \tr_{\mathcal{L}}\big(T^\dagger DT, DT^\dagger,  \mathbb{I}_{.p}\mathbb{I}_{q.} A_L - A_R  \mathbb{I}_{.p}\mathbb{I}_{q.}, F_L, \mathbb{I} \big)\nonumber
\ee
\be
-\tr_{\mathcal{L}}\big(\mathbb{I}_{.p}\mathbb{I}_{q.} DT^\dag, DT, DT\dag, F_R, \mathbb{I} \big)
+\intd \tr_{\mathcal{L}}\big(T DT^\dag, \mathbb{I}_{.p}\mathbb{I}_{q.}, DT\dag, F_R, \mathbb{I} \big)\nonumber
\ee
\be
-\tr_{\mathcal{L}}\big(\intd(T DT^\dag), \mathbb{I}_{.p}\mathbb{I}_{q.}, DT\dag, F_R, \mathbb{I} \big)
-\tr_{\mathcal{L}}\big(T DT^\dag, \mathbb{I}_{.p}\mathbb{I}_{q.}, \intd DT\dag, F_R, \mathbb{I} \big)\nonumber
\ee
\be
+\tr_{\mathcal{L}}\big(T DT^\dag, \mathbb{I}_{.p}\mathbb{I}_{q.}, DT\dag, \intd F_R, \mathbb{I} \big)
-i\tr_{\mathcal{L}}\big(T DT^\dag, \mathbb{I}_{.p}\mathbb{I}_{q.}A_L - A_R \mathbb{I}_{.p}\mathbb{I}_{q.}, DT\dag, F_R, \mathbb{I} \big)\nonumber
\ee
\be
+\intd \tr_{\mathcal{L}}\big(T^\dag \mathbb{I}_{.p}\mathbb{I}_{q.}, DT^\dag, F_R, DT, \mathbb{I} \big)
-\tr_{\mathcal{L}}\big(\intd (T^\dag )\mathbb{I}_{.p}\mathbb{I}_{q.}, DT^\dag, F_R, DT, \mathbb{I} \big) \nonumber
\ee
\be
+\tr_{\mathcal{L}}\big(T^\dag \mathbb{I}_{.p}\mathbb{I}_{q.},\intd  DT^\dag, F_R, DT, \mathbb{I} \big)
-\tr_{\mathcal{L}}\big(T^\dag \mathbb{I}_{.p}\mathbb{I}_{q.}, DT^\dag, \intd  F_R, DT, \mathbb{I} \big) \nonumber
\ee
\be
-\tr_{\mathcal{L}}\big(T^\dag \mathbb{I}_{.p}\mathbb{I}_{q.}, DT^\dag, F_R, \intd  DT, \mathbb{I} \big)
+i\tr_{\mathcal{L}}\big(T^\dag (\mathbb{I}_{.p}\mathbb{I}_{q.}A_L - A_R \mathbb{I}_{.p}\mathbb{I}_{q.}), DT^\dag, F_R, DT, \mathbb{I} \big) \nonumber
\ee
\be
+\intd \tr_{\mathcal{L}}\big(T^\dag DT, DT^\dag, F_R, \mathbb{I}_{.p}\mathbb{I}_{q.}, \mathbb{I} \big)
-\tr_{\mathcal{L}}\big(\intd (T^\dag DT), DT^\dag, F_R, \mathbb{I}_{.p}\mathbb{I}_{q.}, \mathbb{I} \big) \nonumber
\ee
\be
+\tr_{\mathcal{L}}\big(T^\dag DT, \intd DT^\dag, F_R, \mathbb{I}_{.p}\mathbb{I}_{q.}, \mathbb{I} \big)
-\tr_{\mathcal{L}}\big(T^\dag DT, DT^\dag, \intd F_R, \mathbb{I}_{.p}\mathbb{I}_{q.}, \mathbb{I} \big) \nonumber
\ee
\be
+i \tr_{\mathcal{L}}\big(T^\dag DT, DT^\dag, F_R, (\mathbb{I}_{.p}\mathbb{I}_{q.}A_L - A_R \mathbb{I}_{.p}\mathbb{I}_{q.}), \mathbb{I} \big)\nonumber
\ee
\be
-\tr_{\mathcal{L}}\big(\mathbb{I}_{.p}\mathbb{I}_{q.} DT^\dag, DT, F_L, DT^\dag, \mathbb{I} \big)
+\intd \tr_{\mathcal{L}}\big(T DT^\dag, \mathbb{I}_{.p}\mathbb{I}_{q.}, F_L, DT^\dag, \mathbb{I} \big) \nonumber
\ee
\be
-\tr_{\mathcal{L}}\big(\intd(T DT^\dag),\mathbb{I}_{.p}\mathbb{I}_{q.}, F_L, DT^\dag, \mathbb{I} \big)
-\tr_{\mathcal{L}}\big(T DT^\dag, \mathbb{I}_{.p}\mathbb{I}_{q.}, \intd F_L, DT^\dag, \mathbb{I} \big) \nonumber
\ee
\be
-\tr_{\mathcal{L}}\big(T DT^\dag, \mathbb{I}_{.p}\mathbb{I}_{q.}, F_L, \intd DT^\dag, \mathbb{I} \big)
-i\tr_{\mathcal{L}}\big(T DT^\dag, (\mathbb{I}_{.p}\mathbb{I}_{q.}A_L - A_R \mathbb{I}_{.p}\mathbb{I}_{q.}), F_L, DT^\dag, \mathbb{I} \big) \nonumber
\ee
\be
+ \intd \tr_{\mathcal{L}}\big( T^\dag \mathbb{I}_{.p}\mathbb{I}_{q.}, F_L ,DT^\dag, DT , \mathbb{I} \big)
-\tr_{\mathcal{L}}\big(\intd T^\dag \mathbb{I}_{.p}\mathbb{I}_{q.}, F_L ,DT^\dag, DT , \mathbb{I} \big)\nonumber
\ee
\be
+\tr_{\mathcal{L}}\big( T^\dag \mathbb{I}_{.p}\mathbb{I}_{q.},\intd F_L ,DT^\dag, DT , \mathbb{I} \big)
+\tr_{\mathcal{L}}\big( T^\dag \mathbb{I}_{.p}\mathbb{I}_{q.}, F_L ,\intd DT^\dag, DT , \mathbb{I} \big)\nonumber
\ee
\be
-\tr_{\mathcal{L}}\big( T^\dag \mathbb{I}_{.p}\mathbb{I}_{q.}, F_L ,DT^\dag, \intd DT , \mathbb{I} \big)
+i\tr_{\mathcal{L}}\big( T^\dag (\mathbb{I}_{.p}\mathbb{I}_{q.}A_L - A_R \mathbb{I}_{.p}\mathbb{I}_{q.}), F_L ,DT^\dag, DT , \mathbb{I} \big)\nonumber
\ee
\be
+\intd \tr_{\mathcal{L}}\big( T^\dag DT, F_L ,DT^\dag,\mathbb{I}_{.p}\mathbb{I}_{q.}, \mathbb{I} \big)
-\tr_{\mathcal{L}}\big( \intd(T^\dag DT), F_L ,DT^\dag,\mathbb{I}_{.p}\mathbb{I}_{q.}, \mathbb{I} \big)\nonumber
\ee
\be
+\tr_{\mathcal{L}}\big( T^\dag DT,\intd F_L ,DT^\dag, \mathbb{I}_{.p}\mathbb{I}_{q.}, \mathbb{I} \big)
+\tr_{\mathcal{L}}\big( T^\dag DT, F_L ,\intd DT^\dag,\mathbb{I}_{.p}\mathbb{I}_{q.}, \mathbb{I} \big)\nonumber
\ee
\be
+i \tr_{\mathcal{L}}\big( T^\dag DT, F_L ,DT^\dag, (\mathbb{I}_{.p}\mathbb{I}_{q.} A_L - A_R \mathbb{I}_{.p}\mathbb{I}_{q.}) , \mathbb{I} \big)\nonumber
\ee
\be
-\tr_{\mathcal{L}}\big(\mathbb{I}_{.p}\mathbb{I}_{q.} DT^\dag, F_R ,DT, DT^\dag , \mathbb{I} \big)
+\intd\tr_{\mathcal{L}}\big( T DT^\dag, F_R ,\mathbb{I}_{.p}\mathbb{I}_{q.}, DT^\dag , \mathbb{I} \big)\nonumber
\ee
\be
-\tr_{\mathcal{L}}\big(\intd(T DT^\dag), F_R ,\mathbb{I}_{.p}\mathbb{I}_{q.}, DT^\dag , \mathbb{I} \big)
+\tr_{\mathcal{L}}\big( T DT^\dag,\intd F_R ,\mathbb{I}_{.p}\mathbb{I}_{q.}, DT^\dag , \mathbb{I} \big)\nonumber
\ee
\be
-\tr_{\mathcal{L}}\big( T DT^\dag, F_R ,\mathbb{I}_{.p}\mathbb{I}_{q.}, \intd DT^\dag , \mathbb{I} \big)
-i\tr_{\mathcal{L}}\big( T DT^\dag, F_R ,(\mathbb{I}_{.p}\mathbb{I}_{q.}A_L - A_R \mathbb{I}_{.p}\mathbb{I}_{q.}), DT^\dag , \mathbb{I} \big)\nonumber
\ee
\be
+\intd \tr_{\mathcal{L}}\big(F_L, DT^\dag, \mathbb{I}_{.p}\mathbb{I}_{q.}, DT^\dag T , \mathbb{I} \big)
-\tr_{\mathcal{L}}\big(\intd F_L, DT^\dag, \mathbb{I}_{.p}\mathbb{I}_{q.}, DT^\dag T , \mathbb{I} \big) \nonumber
\ee
\be
+\tr_{\mathcal{L}}\big(F_L,\intd  DT^\dag, \mathbb{I}_{.p}\mathbb{I}_{q.}, DT^\dag T , \mathbb{I} \big)
-\tr_{\mathcal{L}}\big(F_L, DT^\dag, \mathbb{I}_{.p}\mathbb{I}_{q.},\intd ( DT^\dag T) , \mathbb{I} \big) \nonumber
\ee
\be
-i\tr_{\mathcal{L}}\big(F_L, DT^\dag,  \mathbb{I}_{.p}\mathbb{I}_{q.}A_L  - A_R  \mathbb{I}_{.p}\mathbb{I}_{q.}, DT^\dag T , \mathbb{I} \big)
-\tr_{\mathcal{L}}\big(F_L, DT^\dag, DT, DT^\dag  \mathbb{I}_{.p}\mathbb{I}_{q.}, \mathbb{I} \big) \nonumber
\ee
\be
+\intd \tr_{\mathcal{L}}\big(F_R,  \mathbb{I}_{.p}\mathbb{I}_{q.}, DT^\dag, DT T^\dag , \mathbb{I} \big)
-\tr_{\mathcal{L}}\big(\intd F_R,  \mathbb{I}_{.p}\mathbb{I}_{q.}, DT^\dag, DT T^\dag , \mathbb{I} \big)\nonumber
\ee
\be
+\tr_{\mathcal{L}}\big(F_R,  \mathbb{I}_{.p}\mathbb{I}_{q.}, \intd DT^\dag, DT T^\dag , \mathbb{I} \big)
-\tr_{\mathcal{L}}\big(F_R,  \mathbb{I}_{.p}\mathbb{I}_{q.}, DT^\dag,\intd ( DT T^\dag) , \mathbb{I} \big)\nonumber
\ee
\be
+i\tr_{\mathcal{L}}\big(F_R,  \mathbb{I}_{.p}\mathbb{I}_{q.}A_L  - A_R  \mathbb{I}_{.p}\mathbb{I}_{q.}, DT^\dag, DT T^\dag , \mathbb{I} \big)
+\intd \tr_{\mathcal{L}}\big(F_R, DT, DT^\dag,  \mathbb{I}_{.p}\mathbb{I}_{q.} T^\dag , \mathbb{I} \big)\nonumber
\ee
\be
-\tr_{\mathcal{L}}\big(\intd F_R, DT, DT^\dag,  \mathbb{I}_{.p}\mathbb{I}_{q.} T^\dag , \mathbb{I} \big)
-\tr_{\mathcal{L}}\big(F_R,\intd DT, DT^\dag,  \mathbb{I}_{.p}\mathbb{I}_{q.} T^\dag , \mathbb{I} \big)\nonumber
\ee
\be
+\tr_{\mathcal{L}}\big(F_R, DT, \intd DT^\dag,  \mathbb{I}_{.p}\mathbb{I}_{q.} T^\dag , \mathbb{I} \big)
+\tr_{\mathcal{L}}\big(F_R, DT, DT^\dag,  \mathbb{I}_{.p}\mathbb{I}_{q.} \intd T^\dag , \mathbb{I} \big)\nonumber
\ee
\be
+i\tr_{\mathcal{L}}\big(F_R, DT, DT^\dag, ( \mathbb{I}_{.p}\mathbb{I}_{q.}A_L - A_R \mathbb{I}_{.p}\mathbb{I}_{q.}) T^\dag , \mathbb{I} \big)\nonumber
\ee
\be
-\intd \tr_{\mathcal{L}}\big(T^\dagger \mathbb{I}_{.p}\mathbb{I}_{q.}, F_L, F_L , \mathbb{I} \big)
+\tr_{\mathcal{L}}\big((\intd T^\dagger) \mathbb{I}_{.p}\mathbb{I}_{q.}, F_L, F_L , \mathbb{I} \big)
-\tr_{\mathcal{L}}\big(T^\dagger \mathbb{I}_{.p}\mathbb{I}_{q.},\intd F_L, F_L , \mathbb{I} \big)\nonumber
\ee
\be
-\tr_{\mathcal{L}}\big(T^\dagger \mathbb{I}_{.p}\mathbb{I}_{q.}, F_L,\intd F_L , \mathbb{I} \big)
-i \tr_{\mathcal{L}}\big(T^\dagger (\mathbb{I}_{.p}\mathbb{I}_{q.}A_L - A_R \mathbb{I}_{.p}\mathbb{I}_{q.}), F_L, F_L , \mathbb{I} \big) \nonumber
\ee
\be
+ \tr_{\mathcal{L}}\big( \mathbb{I}_{.p}\mathbb{I}_{q.} DT^\dagger, F_R, F_R , \mathbb{I} \big)
-\intd \tr_\mathcal{L}(T^\dag F_R, \mathbb{I}_{.p}\mathbb{I}_{q.},  F_L, \mathbb{I})
+\tr_\mathcal{L}(\intd(T^\dag F_R), \mathbb{I}_{.p}\mathbb{I}_{q.},  F_L, \mathbb{I}) \nonumber
\ee
\be
-\tr_\mathcal{L}(T^\dag F_R, \mathbb{I}_{.p}\mathbb{I}_{q.},  \intd(F_L), \mathbb{I})
- i \tr_\mathcal{L}(T^\dag F_R, \mathbb{I}_{.p}\mathbb{I}_{q.} A_L - A_R \mathbb{I}_{.p}\mathbb{I}_{q.},  F_L, \mathbb{I}) \nonumber
\ee
\be
 + \tr_{\mathcal{L}}\big(\mathbb{I}_{.p}\mathbb{I}_{q.} F_L, DT^\dag, F_R, \mathbb{I} \big)
 +  \tr_{\mathcal{L}}\big(F_L, F_L , DT^\dag \mathbb{I}_{.p}\mathbb{I}_{q.}, \mathbb{I} \big)
-\intd \tr_{\mathcal{L}}\big(F_R, F_R, \mathbb{I}_{.p}\mathbb{I}_{q.} T^\dag , \mathbb{I} \big)\nonumber
\ee
\be
+\tr_{\mathcal{L}}\big(\intd F_R, F_R, \mathbb{I}_{.p}\mathbb{I}_{q.} T^\dag , \mathbb{I} \big)
+\tr_{\mathcal{L}}\big(F_R, \intd F_R, \mathbb{I}_{.p}\mathbb{I}_{q.} T^\dag , \mathbb{I} \big)
-\tr_{\mathcal{L}}\big(F_R, F_R, \mathbb{I}_{.p}\mathbb{I}_{q.} \intd T^\dag , \mathbb{I} \big) \nonumber
\ee
\be
-i\tr_{\mathcal{L}}\big(F_R, F_R, (\mathbb{I}_{.p}\mathbb{I}_{q.}A_L - A_R\mathbb{I}_{.p}\mathbb{I}_{q.}) T^\dag , \mathbb{I} \big)
  -  \big( L \leftrightarrow R,T \leftrightarrow T^\dag \big)
\label{eq:variationT}
\ee
The first four lines in this formula are the terms obtained when differentiating $\mathcal{L}$. The following $62$ lines are the differentiation of the words.

\section{The expansion near the defect}
\label{sec:instanton}

In this appendix, we compute the integrals around the baryon defect that appear in the computations of the baryon number, \eqref{NB2} and the anomaly, \eqref{320}. The defect is the point where $T=0$ in the bulk defined in section \ref{sec:bar}. As is shown in section \ref{sec:bar}, the defect is also the position where the tachyon phase winds. It can be viewed as a $D_0$-brane. We derive an expansion for the tachyon and gauge fields near the defect to study the integrals in \eqref{NB2} and \eqref{320}. To analyze the behavior of the fields near the defect, we consider a simplified model that retains the ingredients relevant to the local structure of the instanton while neglecting effects that are subleading in the $\xi \to 0$ limit.

First, we assume that $m_q$ and $\sigma_q$, appearing in the near-boundary expansion of the tachyon \eqref{51} are {proportional to the identity matrix} and therefore we use the $T= \tau U$ ansatz.

Second, to compute the integrals \eqref{NB2}, \eqref{320} around the defect, we rely on expansions of the fields in the (small) distance to the defect center. We denote this distance by $\xi$, so that the defect is at $\xi=0$. In the small $\xi$ expansion, the space is locally flat and therefore the effects of curvature only appear at subleading order in $\xi$, so to leading order we can use a flat space metric. An analogous model has been considered in curved space in AdS in \cite{Gorsky:2012eg, Gorsky2015}, and it was found that the geometry produces a linear term in the holographic coordinate $r$ in the energy, which pushes the soliton to the infrared, but did not affect the leading order expansions of the fields near the soliton's center.

We use the quadratic approximation to the DBI action, which reduces to the Yang-Mills action and a standard kinetic term and potential for the tachyon. This simplification can be shown in retrospect to be justified.

We consider separately the effect of the CS action on the instanton solution and of the tachyon field. In section \ref{sec:c1}, we study the effect of a CS term on the BPST instanton solution in the absence of a tachyon field, and show that in this case the finiteness of the energy implies that the zero-size instanton solution is forbidden. In section \ref{sec:c2} we consider a finite size BPST instanton and study how the gauge field scales near the center of the instanton with no tachyon. Finally, in section \ref{sec:c3} we add the tachyon and check how it modifies the analysis.

\subsection{Effect of the Chern-Simons term}
\label{sec:c1}
In this subsection, we study the Yang-Mills-CS action in order to see how the presence of the CS action affects the BPST instanton solution. The main result we show is that the instanton density generates an electric field through the CS action, making the energy of the zero-size instanton infinite. The action is
\be
S = S_{YM} + S_{CS}, \label{c1}
\ee
for a single $\text{U}(N_f)$ gauge field, $A$. We decompose the gauge field into a $\text{U}(1)$ field $\hat{A}$ and $SU(N)$ fields $A^a$, and denote by $\hat{F}$ and $F^a$ the associated field strengths. The flat space Yang-Mills action can then be written as
\be
S_{YM} = - \int \intd^5x \left[ \frac{1}{4} \hat{F}_{mn}^2  + \frac{1}{2} \tr(F_{mn}^{2})\right].\label{c2}
\ee
We write with an index $0$ the time component, $r$ the holographic coordinate, $m,n = 0, 1, 2, 3, r$ and $i,j = 1,2,3,r$. The CS action in the absence of the tachyon is given\footnote{Here we consider the CS theory at level $1$. The overall constant multiplying the CS action does not affect the qualitative result of this section.} by
\be
S_{CS} = \frac{1}{24\pi^2} \int \intd^5x \varepsilon^{mnpqs} \tr\left(A_m \partial_n A_p \partial_q A_s - \frac{3}{2}i A_m A_n A_p \partial_q A_s - \frac{3}{5} A_m A_n A_p A_q A_s \right) \label{c3},
\ee
where $ \varepsilon$ is the Levi-Civita symbol. We now consider the $N_f=2$ case, then generalize to generic $N_f$. The $N_f=2$ case is analogous\footnote{Up to neglecting the same terms in the large $\lambda$ expansion, and considering a similar metric.} to the case studied in the Witten-Sakai-Sugimoto model, \cite{Hata:2007mb}.

We now assume that the fields are time-independent. The equations of motion for the $SU(2)$ gauge fields then read
\be
- D_i^2  A_0 + \frac{1}{64\pi^2} \varepsilon_{ijkl} F_{ij} \hat{F}_{kl} = 0 \label{c4},
\ee
\be
D_i F_{ji} = 0\label{c5}.
\ee
For the $\text{U}(1)$ gauge fields, the equations of motion are
\be
- \partial_i^2 \hat{A}_{0} + \frac{1}{64\pi^2}  \varepsilon_{ijkl} \big(\tr(F_{ij} F_{kl}) +  \frac{1}{2} \hat{F}_{ij} \hat{F}_{kl} \big) = 0\label{c6-ii},
\ee
\be
\partial_i \hat{F}_{ji} = 0\label{c7}.
\ee
In \eqref{c4}-\eqref{c7} we considered static solutions. We observe that, due to the CS term, the instanton density appearing in \eqref{c6-ii} sources the $\hat{A}_0$ component. We can repeat the analysis in \cite{Hata:2007mb}: we solve \eqref{c7} by setting
\be
\hat{A}_i = 0\label{c8}.
\ee
Then, we solve \eqref{c4} by
\be
D_i^2 A_0 = 0 \to A_0 = 0\label{c9}.
\ee
To find an instanton solution to \eqref{c6-ii}, we start from the BPST instanton solution for $F_{IJ}$, which reads
\be
F_{IJ} = \frac{4 \rho^2}{(\rho^2 + \xi^2)^2}\eta_{IJa} \lambda^a, \quad F_{Ir}  = \frac{4 \rho^2}{(\rho^2 + \xi^2)^2} \lambda^I,\label{c11}
\ee
where we noted the distance to the center of the instanton $\xi$ and $\eta^a_{ i j }$ is the 't Hooft symbol. Moreover, $\rho$ is the size of the instanton, $\lambda^a$ the $SU(2)$ generators, and $I = 1,2,3$.
We inject \eqref{c11} into \eqref{c6-ii} and integrate twice, imposing regularity at $\xi=0$, to obtain
\be
\hat{A}_0 = \frac{1}{8\pi^2 \xi^2} \left(1  -\frac{\rho^4}{(\rho^2 + \xi^2)^2} \right)\label{c10}.
\ee

In the pure Yang-Mills case in flat space, $\rho$ is a modulus, and the energy does not depend on $\rho$. The presence of the CS term generates an electric field contribution through $\hat{A}_0$. The CS term then contributes to the energy, and its contribution scales like $1/\rho^2$, therefore $\rho$ is no longer a modulus \cite{Hata:2007mb}. Moreover, the presence of the CS term makes the energy of the zero size $U(2)$ instanton infinite.

These results can be generalized to the generic $N_f$ case. While in the $N_f = 2$ case the symmetric structure constant $d_{abc} \equiv 2 \tr({T^a, T^b}T^c)$ is zero, for general $N_f$ it is in general nonzero. The equations of motion of the non-abelian fields are then given by
\be
D_m( F_{mn}^{a} ) + \frac{1}{32 \pi^2} d^{abc} \varepsilon^{npqrs} F_{pq}^b F_{rs}^c = 0.\label{c12}
\ee
Then the equation of motion for $A_0$ becomes
\be
D_i^2 A_{0}^{a} + \frac{1}{16 \pi^2} d^{abc}  F_{ri}^b \tilde{F}^{ri, c} = 0,
\label{eomA0}
\ee
where we wrote
\be
\tilde{F}^{ri, c} \equiv \frac{1}{2} \varepsilon^{rijk} F_{jk}^{c}.\label{c14}
\ee
The CS term can now also source the non-abelian part of $A^0$. The conclusion that the zero-size instanton has infinite energy also holds in this case.

\subsection{The expansion of the BPST instanton}
\label{sec:c2}
We now show that, for any finite energy instanton with a single gauge field $A$, the gauge field asymptotes to pure gauge around the defect. This means that the field strengths vanish and, together with the vanishing of the angular components of the tachyon kinetic term, we  establish in the next subsection that this justifies expanding the DBI action to quadratic order in the fields.

In this section we show this result for the usual BPST instanton. For this, we consider a static solution and start from the BPST instanton solution written in $\mathbb{R}^4$. We assume that the size of the instanton is nonzero so that the energy is finite. We write the BPST solution for $F$ in an orthonormal coframe $e^i$,
\be
F^a = \frac{4\rho^2}{(\xi^2 + \rho^2)^2} \eta^a_{ i j } e^i \wedge e^j.\label{f14f}
\ee
In spherical coordinates, the coframe is given by
\be
e^i = \{ \intd\xi, \xi \intd\theta, \xi  \sin \theta \intd\f, \xi \sin \theta \sin \f \intd\psi \}\label{c16},
\ee
where $\xi$ is the distance to the center of the defect.
In \eqref{c16}, all angular $e^i$'s are proportional to $\xi$, so we obtain from \eqref{f14f} that in the $\xi \to 0$ regime, the \textit{form components} scale as
\be
F_{\theta \phi}, F_{\theta \psi}, F_{\phi \psi} = \mathcal{O}\left(\frac{\xi^2}{\rho^2}\right).\label{c17}
\ee
Recall now that the integrals appearing in section \ref{sec:bar} are performed over a sphere $S^3(r_*)$ of infinitesimal radius around the defect. The integrals are then performed only on angular coordinates. If we shrink the radius of the sphere around the singularity to zero, from \eqref{c17} we obtain $F\to 0$, therefore the pullback of the gauge fields to $S^3(r_*)$ asymptotes to pure gauge in the small radius limit of the $S^3$,
\be
F_{S^3(r_*)} \to 0, \qquad A_{S^3(r_*)} \to i g^\dag \intd g,
\ee
where $g$ is a group element.

To address the near-defect integrals relevant for the baryon number, \eqref{NB2}, and for the anomaly, \eqref{320}, we shall study the behavior of the gauge field $A$, which depends on the gauge choice. We consider two examples. In the regular Landau gauge, where the gauge field $A$ winds at spatial infinity,
\be
A^a = \frac{2 \eta_{i j}^a \xi^i}{\xi^2 + \rho^2} e^j,
\ee
the near-defect expansion for the gauge field is of order
\be
A_{\theta}, A_\phi, A_\psi =  \mathcal{O}\left(\frac{\xi^2}{\rho^2} \right),\label{c18}
\ee
such that in this gauge we can assume the gauge fields to go to zero as well around the singularity.

However, in the singular Landau gauge, where $A$ winds around the defect,
\be
A^a = \frac{2 \rho^2 \eta_{i j}^a \xi^i}{\xi^2(\xi^2 + \rho^2)} e^j,
\ee
the near-defect behavior of $A$ is
\be
A_{\theta}, A_\phi, A_\psi  =  \mathcal{O}(1).\label{c19}
\ee

We now return to the integrals appearing in \eqref{NB2}. Note that in both cases, one has
\be
\int_{S^3(r_*)}\tr( A\wedge F)= 0\label{c20},
\ee
but
\be
\int_{S^3(r_*)} \tr(A^3) \label{c20-ii}
\ee
depends on the gauge; more precisely it is sensitive to large gauge transformations. Importantly, \eqref{c20-ii} is quantized by topology\footnote{Indeed, as $A$ is pure gauge, it can be written as a Maurer-Cartan form $g^\dag \intd g$, with group element $g$, therefore the integral of $\tr(A^3)$ lies in the third homotopy group $\pi_3(G)$, which is $\mathbb{Z}$ for the unitary groups we are considering here.}, therefore it can change only under large gauge transformations\footnote{By large gauge transformations, we mean gauge transformations that are not homotopic to the trivial gauge transformation.}, like the transformation that relates the regular gauge \eqref{c18} to the singular gauge \eqref{c19}. Indeed, the gauge \eqref{c18} yields zero for \eqref{c20-ii}, while \eqref{c19} produces a finite result. We shall see in the next section, that in the singular gauge there can be a nonzero contribution to the anomaly from the defect, so that this gauge should be excluded.

\subsection{Field expansion near the defect with a tachyon field}
\label{sec:c3}

We  now study a more complicated model that is adapted to our setup, which includes a tachyon field and two gauge fields $A_L$ and $A_R$, which the full TCS action.

It is known that in holographic flavor models with a tachyon, the presence of the tachyon field tends to make the soliton shrink \cite{Gorsky:2012eg, Gorsky2015}. However, since the TCS term prevents collapse to zero size, we continue to assume a finite instanton size $\rho>0$.

For simplicity, in the rest of this appendix we assume that the fields are spherically symmetric around the defect. Even though the spherical ansatz is not general, most of the results derived here rely only on the fact that the field strength is well defined at the defect. This regularity condition enforces that the gauge fields in angular directions around the defect asymptote to pure gauge at the defect, which is the result used in section \ref{sec:bar}. This result also holds in a more general ansatz like the cylindrically symmetric ansatz of \cite{jknp}.

\subsubsection{Spherical expansion}

We still take the quadratic approximation to the DBI action. The simplified tachyon-DBI action is composed of two Yang-Mills actions for the left and right gauge fields, with a standard kinetic term and potential for the tachyon.

The total action also contains the TCS action, defined in \eqref{eq:Omegagen}, such that the full action is
\be
S =  \int \intd^5x \bigg(\frac{1}{2}\tr (F_L^2 + F_R^2) +  \tr( DT DT^\dag )+ V_T(TT^\dag)\bigg) + \frac{i N_c}{4\pi^2} \int \Omega_5\label{c21}.
\ee
The Euler-Lagrange equations are
\be
D_m F_L^{mn} = J_{T,L}^n +  J_{TCS,L}^n,  \qquad J_{T,L}^n \equiv i (T D^n T^\dag - D^n T T^\dag) ,\label{c22}
\ee
where $DT$ is defined in \eqref{a9}, and similarly for the right fields. For the tachyon it is
\be
\frac{1}{\sqrt{g}}D_m (\sqrt{g} D^m T)  = \frac{\partial V_T}{\partial T^\dag } + J_{TCS, T} \label{g22},
\ee
where $J_{TCS,L}$ and $J_{TCS,T}$ are the terms obtained from the variation of the CS action with respect to $A_L$ and $T$. In \eqref{g22} $D_m$ is the covariant derivative in flat space. We ignore the currents and the potential term in the equations for now as they are subleading, which we shall verify a posteriori.

We look for separable solutions $T = \tau(\xi) U(\theta, \phi, \psi)$ to \eqref{g22}. We obtain in spherical coordinates
\be
\frac{1}{\sqrt{g}}D_m (\sqrt{g} D^m T) =  \frac{\partial V_T}{\partial T^\dag } + J_{TCS, T}\label{c24}.
\ee
Requiring that the winding numbers of $A_L$ and $A_R$ are finite, and that the spherical symmetry is restored around the defect, the angular components of the gauge field $A_{L/R,\theta}, A_{L/R,\varphi}, A_{L/R,\phi}$ are proportional to their winding numbers $n_L$, $n_R$ in the small $\xi$ limit,
\be
A_{L,R} \propto n_{L,R} .\label{c25}
\ee
It was shown in \cite{jknp} that $n_L = -n_R$ is required by finiteness of the energy. We also fix the gauge to require that $A_\xi= 0$. Because $n_L = -n_R$, to leading order the contributions from gauge fields to \eqref{g22} cancel each other. With all the gauge field contributions vanishing, \eqref{c24} becomes a Poisson equation,
\be
 \frac{\partial^2 T}{\partial \xi^2 } + \frac{3}{\xi} \frac{\partial T}{\partial \xi} - \frac{\nabla_{S^3}^2 T}{\xi^2} = \frac{\partial V_T}{\partial T^\dag } + J_{TCS,T}.\label{c26}
\ee
To make use of the spherical symmetry we decompose the tachyon in spherical harmonics $Y$,
\be
T = \sum_{\ell=0}^\infty \sum_{m_1, m_2} T_{\ell m_1 m_2} (\xi) Y^{(\ell)}_{m_1 m_2}(\theta, \f, \psi)\label{c27}.
\ee
This decomposition diagonalizes the Laplacian on $S^3$ and reduces the PDE to ordinary differential equations. Indeed, fitting this into \eqref{c26} we obtain the following ODE for $T_{\ell m_1 m_2}$,
\be
 \frac{\intd^2 T_{\ell m_1 m_2}}{\intd \xi^2 } + \frac{3}{\xi} \frac{\intd T_{\ell m_1 m_2}}{\intd \xi} - \frac{\ell(\ell+2)}{\xi^2} T_{\ell m_1 m_2}=\frac{\partial V_T}{\partial T^\dag } +  J_{TCS, T}\label{c28}.
\ee
We shall first solve the homogeneous equation. We shall then check a posteriori that to leading order it is consistent to consider the right hand side as a subleading perturbation. Equation \eqref{c28} is an ODE that is homogeneous in $\xi$, therefore it has power-law solutions. We substitute in \eqref{c28} a power law ansatz and we obtain
\be
T_{\ell m_1 m_2} \propto \xi^\ell, \xi^{-\ell  -2}\label{c29},
\ee
where $\ell$ is a positive integer. Then, because $T$ goes to zero in $r_*$, both the solution $T\propto \xi^{-\ell -2}$ and the zero mode $T \propto \xi^0$ are forbidden. We conclude that
\be
T \propto \xi^\ell, \qquad \ell \geq 1. \label{c29-ii}
\ee
The ansatz $T= \tau(\xi) U(\theta, \phi, \psi)$ then corresponds to truncating \eqref{c27} to the smallest $\ell$ term, and enforcing that $T_{lm_1 m_2}$ does not depend on $m_1$ or $m_2$. Then, the derivative of $\tau$ is at most constant,
\be
T= \tau(\xi) U(\theta, \phi, \psi), \qquad \tau = T_1 \xi + \mathcal{O}(\xi^2), \quad  \tau' = T_1 + \mathcal{O}(\xi)\label{c30},
\ee
where $T_1$ is a non-negative constant.

We now choose a BPST-like ansatz for the gauge fields, adapted to $U(2)$. As we showed previously, with the standard CS action one can obtain a solution to the equations of motion by turning on only the abelian component of $A_{L, 0}$ and $A_{R, 0}$. $F_L$ and $F_R$ asymptote to zero, \eqref{c17}, so that the $SU(2)$ part is pure gauge. We therefore make the following ansatz for $A_L$ and $A_R$,
\be
A_{L/R} = i f(\xi) g_{L/R}^\dag \intd g_{L/R}(\theta, \phi, \psi) + h(\xi) I \intd t,\label{c31}
\ee
where $g_{L}, g_R$ are group elements that are assumed to depend only on angular coordinates and have winding $n_L$, $n_R$ around $S^3(r_*)$ respectively. $h(\xi)$ encodes the electric potential $\hat{A}_0$ sourced by the instanton density. $h(\xi)$ is a function of $\xi$ to be determined, and $I$ is the flavor space identity matrix. Then, we obtain
\be
F_{L/R} = if'(\xi) \intd \xi \wedge g_{L/R}^\dag \intd g_{L/R} + if(\xi)(f(\xi)-1)(g_{L/R}^\dag \intd g_{L/R})^2 + h'(\xi) I \intd \xi \wedge \intd t.
\label{Fansatz}
\ee
Using the gauge $A_{\xi (L/R)} = 0$, time independence and the ansatz \eqref{c31}, the equations of motion for the gauge fields from \eqref{c21} become
\be
\frac{1}{\sqrt{-g}}\partial_\xi (\sqrt{-g}  \partial_\xi A_{0 (L/R)} )  = J_{TCS (L/R)}^{0} + J_{T (L/R)}^{0}\label{c34},
\ee
\be
\frac{1}{\sqrt{-g}}D_\alpha(\sqrt{-g} g^{\alpha \beta} \partial_\xi A_{\beta (L/R)}) = - J_{TCS (L/R)}^{\xi}+ J_{T (L/R)}^{\xi}\label{c35},
\ee
\be
\frac{1}{\sqrt{-g}}\partial_\xi (\sqrt{-g} g^{\alpha \beta} \partial_\xi A_{\beta (L/R)}) + \frac{1}{\sqrt{-g}}D_\beta (\sqrt{-g} g^{\alpha \gamma} g^{\beta \delta} F_{\gamma \delta (L/R)}) = J_{TCS (L/R)}^{\alpha} + J_{T (L/R)}^{\alpha}
\label{g36}.
\ee
In these equations of motion and later, we use the indices $\alpha, \beta \dots = \theta, \phi, \psi$ to denote $S^3$ indices. In order to trust the BPST-like solution, we need to estimate the right hand side of these equations and check that they produce subleading contributions.

We shall start by determining the expansions of the tachyon current in \eqref{c22}, for which we need to compute the expansion of $U^\dag D U$. Using the ansatz $T=\tau(\xi) U(\theta, \phi, \psi)$, we obtain
\be
U^\dag \pa_0 U = 0, \qquad U^\dag \pa_\xi  U = 0.\label{c44}
\ee
Using $A_\xi = 0$ and the fact that the time component of $A_{L/R}$ is abelian, we obtain $U^\dag D_0 U = U^\dag D_\xi U = 0$. For the angular components, the result depends on if $A_{L/R}$ or $U$ wind around the defect. From now on, the discussion depends on if $A_L$ and $A_R$ wind, and we omit the indices $L,R$ when the expressions apply to both $A_L$ and $A_R$.

If $U$ winds, then one has
\be
U^\dag \pa_\alpha U  =\mathcal{O}(1).\label{c44-ii}
\ee
Instead, using \eqref{c25} and \eqref{c44-ii}, if $U$ or $A$ winds, then
\be
U^\dag D_\alpha U = U^\dag \partial_\alpha U + i A_{L,\a} - i U^\dag A_{R,\alpha} U =\mathcal{O}(1),\label{c45}
\ee
and if nothing winds, then
\be
U^\dag D_\alpha U = \mathcal{O}(\xi^a), \qquad a>0.\label{c46}
\ee
The most singular expansion possible for $U^\dag D_\alpha U$, if $U$ or $A$ wind, is then
\be
U^\dag D_\alpha U =  \mathcal{O}(1)\label{c47}.
\ee

Using the ansatz \eqref{c31} in the tachyon currents \eqref{c22}, we obtain
\be
J_{T ,0} = J_{T ,\xi} =0, \quad J_{T, \alpha} = -2i \tau^2 D_\alpha U U^\dag = \mathcal{O} (\xi^2)\label{c48}.
\ee
\smallbreak
Therefore, these currents can safely be ignored in the equations of motion if the scaling of the tachyon is leading with respect to the scalings of \eqref{c48}, which we shall verify later.

We now proceed and analyze the behavior of the TCS current $J_{TCS}$. For the TCS contribution, it was shown in \cite{jknp} that upon a small variation of the gauge field $ A_L \to  A_L  + \d  A_L $, $\Omega_5^0$ (see \eqref{416}) changes as:
\begin{align}
\nonumber \d_{ A_L }\Omega_5^0 &= \mathrm{Tr}\bigg\{ \delta A_L \wedge \bigg(\! -if_1(\tau)\Big( F_L \wedge  F_L  + U^\dagger F_R \wedge  F_R  U -  A_L \wedge  F_L \wedge U^\dagger DU -  \\
\nonumber &\hphantom{= \mathrm{Tr}\bigg\{ \delta A_L \wedge\Big( -if_1(\tau)}  \, -  F_L \wedge U^\dagger DU\wedge  A_L  -  A_L \wedge U^\dagger DU\wedge  F_L  -  \\
\nn &\hphantom{= \mathrm{Tr}\bigg\{ \delta A_L \wedge\Big( -if_1(\tau)} \, - U^\dagger DU\wedge  F_L \wedge  A_L  \Big) - \\
\nonumber &\hphantom{= \mathrm{Tr}\bigg\{ \delta A_L \wedge\Big(}  - \mathrm{d}\Big[f_1(\tau)(U^\dagger DU\wedge  F_L  +  F_L \wedge U^\dagger DU ) \Big] - \\
\nonumber &\hphantom{= \mathrm{Tr}\bigg\{ \delta A_L \wedge\Big(}  -if_2(\tau) \Big( F_L \wedge U^\dagger DU\wedge U^\dagger DU + U^\dagger DU\wedge U^\dagger DU\wedge  F_L  +\\
\nn &\hphantom{= \mathrm{Tr}\bigg\{ \delta A_L \wedge\Big( -if_2(\tau) \Big(} + U^\dagger DU\wedge  F_L \wedge U^\dagger DU +  \\
\nonumber &\hphantom{= \mathrm{Tr}\bigg\{ \delta A_L \wedge\Big( -if_2(\tau) \Big(} + F_R \wedge DUU^\dagger\wedge DUU^\dagger + DUU^\dagger\wedge DUU^\dagger\wedge  F_R  + \\
\nn &\hphantom{= \mathrm{Tr}\bigg\{ \delta A_L \wedge\Big( -if_2(\tau) \Big(} + DUU^\dagger\wedge  F_R \wedge DUU^\dagger \Big)- \\
\nonumber &\hphantom{= \mathrm{Tr}\bigg\{ \delta A_L \wedge\Big(}  - \mathrm{d}\Big[f_2(\tau)\, U^\dagger DU\wedge U^\dagger DU\wedge U^\dagger DU\Big]- \\
\nonumber &\hphantom{= \mathrm{Tr}\bigg\{ \delta A_L \wedge\Big(} -if_3(\tau)\Big( F_L \wedge U^\dagger   F_R  U \!+\! U^\dagger F_R U\wedge  F_L  \!-\!  A_L \wedge U^\dagger F_R \wedge DU - \\
\nn &\hphantom{= \mathrm{Tr}\bigg\{ \delta A_L \wedge\Big( -if_3(\tau) \Big(} - U^\dagger F_R \wedge DU\wedge  A_L  -  A_L \wedge U^\dagger DU\wedge U^\dagger F_R U - \\
\nonumber &\hphantom{= \mathrm{Tr}\bigg\{ \delta A_L \wedge\Big( -if_3(\tau) \Big(} - U^\dagger DU\wedge U^\dagger F_R U\wedge  A_L \Big)- \\
\nonumber &\hphantom{= \mathrm{Tr}\bigg\{ \delta A_L \wedge\Big(}  - \mathrm{d}\Big[f_3(\tau)(U^\dagger DU\wedge U^\dagger F_R U + U^\dagger F_R \wedge  DU) \Big] - \\
\label{dO50L}  &\hphantom{= \mathrm{Tr}\bigg\{ \delta A_L \wedge\Big(} -4if_4(\tau) U^\dagger DU\wedge U^\dagger DU\wedge U^\dagger DU\wedge U^\dagger DU \,\bigg) \bigg\} \,.
\end{align}
Where the functions $f_i(\tau)$ were defined in~\cite{jknp}. We only need their behavior at small values of the tachyon:
\be
f_1=  -\frac{1}{6} e^{-\tau^2} =  -\frac{1}{6}  + \mathcal{O}(\xi^2), \qquad f_2=  \frac{i}{12} (1+\tau^2) e^{-\tau^2} = \frac{i}{12} +  \mathcal{O}(\xi^2)\label{c38},
\ee
\be
f_3 = -\frac{1}{12}e^{-\tau^2} = -\frac{1}{12}  + \mathcal{O}(\xi^2), \qquad f_4  = \frac{1}{120}(2+2\tau^2+\tau^4)e^{-\tau^2} = \frac{1}{60} +  \mathcal{O}(\xi^2)\label{c39}.
\ee
where we inserted the tachyon expansion $\tau = \mathcal{O}(\xi)$ to obtain the final estimates.

In order to obtain the small $\xi$ leading expansion of all the terms in~\eqref{dO50L}, we need to discuss if $U$ or $g$(and therefore, $A$) wind at the center. For $A$ to wind, $f$ must go to one at the defect. Note that, since in \eqref{Fansatz}, there is a term proportional to $f-1$, we then need to consider the subleading behavior of $f$ if $A$ winds. Without assuming that $A$ winds, we denote the first two terms in the expansion of $f$ as
\be
f(\xi) = C_0 \xi^a + C_1 \xi^k +  \dots\label{c40}
\ee
with $k>a \geq 0$. In the case where $A$ winds, $f$ goes to one, such that one has $C_0 = 1$ and $a=0$. Since the leading behavior depends only on whether the angular derivatives are $\mathcal{O}(1)$ or suppressed, the asymptotics reduce to four possible winding sectors. With this expansion we obtain
\be
\text{A winds: } a=0, \; f(\xi) = 1+ \mathcal{O}(\xi^k) , \quad \text{A does not wind: } a>0,\; f(\xi) = \mathcal{O}(\xi^a)\label{c41}.
\ee
This implies, using \eqref{Fansatz}
\be
\text{A winds: } F_{\xi \alpha} = \mathcal{O}(\xi^{k -1}), \quad \text{A does not wind: } F_{\xi \alpha} =\mathcal{O}(\xi^{a-1})\label{c42},
\ee
\be
\text{A winds: } F_{\beta \gamma} =\mathcal{O}( \xi^{k}), \quad \text{A does not wind: } F_{\beta \gamma} =\mathcal{O}( \xi^{a})\label{c43}.
\ee
We shall now determine the expansion of terms appearing in \eqref{dO50L}. For the TCS current with only spatial indices, the following terms contribute:
\begin{itemize}
  \item If $A$ winds, $U$ winds:
\be
F_{\xi \alpha} F_{\beta \gamma} \sim \xi^{2k -1}, \ F_{\xi \alpha} A_\beta A_\gamma \sim \xi^{k -1}\label{c50}
\ee
\be
F_{\xi \alpha} A_\beta U^\dag \partial_\gamma U \sim \xi^{k -1}, \ F_{\xi \alpha}  U^\dag \partial_\beta U  U^\dag \partial_\gamma U  \sim \xi^{k -1}\label{c51}
\ee
  \item If $A$ winds, $U$ does not wind:
\be
F_{\xi \alpha} F_{\beta \gamma} \sim \xi^{2k -1}, \ F_{\xi \alpha} A_\beta A_\gamma \sim \xi^{k -1}\label{c52}
\ee
  \item If $A$ does not wind, $U$ winds:
\be
F_{\xi \alpha} F_{\beta \gamma} \sim \xi^{2a -1}, \ F_{\xi \alpha} A_\beta A_\gamma \sim \xi^{3a -1}\label{c53}
\ee
\be
F_{\xi \alpha} A_\beta U^\dag \partial_\gamma U \sim \xi^{2a -1}, \ F_{\xi \alpha}  U^\dag \partial_\beta U  U^\dag \partial_\gamma U  \sim \xi^{a -1}\label{c54}
\ee
  \item If $A$ does not wind, $U$ does not wind:
\be
F_{\xi \alpha} F_{\beta \gamma} \sim \xi^{2a -1}, \ F_{\xi \alpha} A_\beta A_\gamma \sim \xi^{3a -1}\label{c55}
\ee
\end{itemize}
We expect the equation of motion for $A_0$ to be corrected to leading order by the TCS current, as it happens without scalar. We substitute \eqref{c31} into the equation of motion for $A_0$, \eqref{c34},
\be
\partial_\xi (\xi^3 h'(\xi)) = \mathcal{O} (\xi^{c}),\label{c56}
\ee
where $c$ depends on the case considered. This implies
\be
h(\xi) \sim \xi^b, \quad b= c - 1.\label{c57}
\ee
\be
\text{$A$, $U$ wind: } b=k-2, \  \text{$A$ winds: } b=k-2, \ \nonumber \ee\be \text{$U$ winds: } b=a-2, \  \text{nothing winds: } b=2a-2. \ \label{c58}
\ee
Inserting the scalings \eqref{c46}, \eqref{c47}, \eqref{c41}, \eqref{c42}, \eqref{c43} and \eqref{c58} into the ansatz \eqref{c31}, and substituting it inside $J_{TCS,T}$, it can be checked that the $J_{TCS, T}$ contribution is subleading in the equation of motion for $T$ and so is every other component of the TCS currents.

To conclude the argument we study the potential contribution, $\pa V_T / \pa T^\dag$, using that $V_T$ is a function of $TT^\dag$, we deduce
\be
\frac{\pa V_T}{\pa T^\dag} = T V'(TT^\dag).\label{c59-ii}
\ee
Due to \eqref{c29-ii}, the right hand side of \eqref{c59-ii} scales at most like $\xi^\ell$. Therefore, as can be seen from \eqref{c28}, this term affects $T$ at order $\xi^{\ell + 2}$, which is always subleading, justifying the first approximation to neglect this contribution to leading order.

\subsubsection{Determination of admissible exponents from finite energy}

Now that we have established the leading order expansion of the gauge fields and of the tachyon and determined that to leading order, the only contribution from the TCS term is the one that sources $h$ in \eqref{c34}, we substitute the ansatz in the action to obtain the constraints on the expansions of $f$ and $h$ imposed by finiteness of the action
\be
S_{YM} = \frac{1}{2}\int \tr(F_L^2+ F_R^2) = 12 \pi^2 \int  \intd\xi \, \left( \xi f'(\xi)^2 + 4\frac{f^2(\xi) (f(\xi) - 1)^2}{\xi} + \xi^3 h'(\xi)^2 \right) \label{c49}.
\ee
The overall factor comes from the integration over $S^3$.

For the first term to be integrable in $\xi=0$ one must ensure that we have $a>0$. Therefore, the tachyon current term $J_{TL, \alpha}$ can be ignored if the left-hand side of \eqref{g36} scales like a power of $\xi$ bigger than $-2$, which is guaranteed for the BPST solution. This justifies using the BPST-like ansatz \eqref{c31} with $A_0$ turned on at leading order.

If $A$ does not wind the second term in the action is also integrable if $a>0$. If $A$ winds it enforces instead $k>0$.

Integrability of the third term implies $b>-1$, which gives a constraint on $k$ and $a$ from \eqref{c58},

\begin{table}[htb]
\centering
\begin{tabular}{|c|c|c|}
 \hline
 Field & $A_L, A_R$ wind & $A_L, A_R$ do not wind \\
 \hline
 $U$ winds & $k>1$  & $a>1$ \\
 \hline
 $U$ does not wind & $k>1$  & $a>\frac{1}{2}$ \\
 \hline
\end{tabular}
\caption{Constraint on the fields scaling implied by finiteness of the energy}
\label{table:scal}
\end{table}

\subsubsection{Explicit computation of the integrals}

Then, we shall use these constraints to compute the integrals of \eqref{320} and \eqref{NB2}. With the ansatz \eqref{c31}, the first integral in \eqref{NB2} becomes
\be
\int_{S^3(r_*)} \tr A\wedge F = -  \int_{S^3(r_*)} f(\xi)^2 (f(\xi)-1) \tr(( g^\dag \intd g )^3)\label{c60},
\ee
which scales like $f(\xi)^2 \sim \xi^{2a}$ with $a>0$ if $A$ does not wind, and like $\xi^k$ if it does, therefore this integral vanishes in the small $\xi$ limit at least linearly regardless of the winding of $A$. In the BPST case we recover the result of the previous subsection
\be
\int_{S^3(r_*)} \tr A\wedge F =  \int_{S^3(r_*)}  \frac{2\xi^4 \rho^2}{(\xi^2 + \rho^2)^3}\tr(( g^\dag \intd g )^3)\label{c61},
\ee
which is regular for $\rho>0$. In particular the BPST solution in regular gauge has
\be
\text{BPST: }\quad a=2, \quad b=0,\label{c62}
\ee
and $a=b=0$ in the singular gauge. In the singular gauge, we obtain $k = 2$, so the integral \eqref{c60} vanishes like $\xi^2$.

We shall now compute the contribution of the anomaly from $r_*$ in this model. The anomaly from $r_*$ can be deduced from \eqref{320},
\be
\frac{i}{24\pi^2} \int_{S^3(r_*) \times \mathbb{R}_t} \str(\Lambda (F^2 + \frac{i}{2} (AAF + AFA + FAA) - \frac{1}{2} A^4))\label{c63}.
\ee
The terms that possibly contribute to the anomaly are
\be
\int_{S^3(r_*) \times \mathbb{R}_t} \str(\Lambda F_{\gamma 0} F_{\alpha\beta}) = 0,\label{c64}
\ee
\be
\int_{S^3(r_*) \times \mathbb{R}_t} \str(\Lambda F_{\gamma 0} A_\alpha A_\beta) = 0.\label{c65}
\ee
Both vanish by symmetries of the solution near the instanton center, and two more terms for which we need to distinguish the different cases:
\begin{itemize}
\item If $A$ winds, $U$ winds:
\be
\int_{S^3(r_*) \times \mathbb{R}_t} \str(\Lambda A_0  A_\alpha F_{\beta \gamma}) \sim \xi^{2k-2} \to 0, \ \nonumber \ee\be
\int_{S^3(r_*) \times \mathbb{R}_t} \str(\Lambda A_0  A_\alpha A_\b A_\g) \sim \xi^{k-2} \to C\label{c66}
\ee
\item If $A$ winds, $U$ does not wind:
\be
\int_{S^3(r_*) \times \mathbb{R}_t}  \str(\Lambda A_0  A_\alpha F_{\beta \gamma}) \sim \xi^{2k-2} \to 0, \ \nonumber \ee\be
\int_{S^3(r_*) \times \mathbb{R}_t}  \str(\Lambda A_0  A_\alpha A_\b A_\g) \sim \xi^{k-2} \to C\label{c67}
\ee
  \item If $A$ does not wind, $U$ winds:
\be
\int_{S^3(r_*) \times \mathbb{R}_t}  \str(\Lambda A_0  A_\alpha F_{\beta \gamma}) \sim \xi^{3a-2} \to 0, \ \nonumber \ee\be
\int_{S^3(r_*) \times \mathbb{R}_t}  \str(\Lambda A_0  A_\alpha A_\b A_\g) \sim \xi^{4a-2} \to 0\label{c68}
\ee
  \item If $A$ does not wind, $U$ does not wind:
\be
\int_{S^3(r_*) \times \mathbb{R}_t}  \str(\Lambda A_0  A_\alpha F_{\beta \gamma}) \sim \xi^{4a-2} \to 0, \ \nonumber \ee\be
\int_{S^3(r_*) \times \mathbb{R}_t}  \str(\Lambda A_0   A_\alpha A_\b A_\g) \sim \xi^{5a-2} \to 0\label{c69}
\ee
\end{itemize}
where we have inserted the constraints of finite energy on $a$ and $k$, and required that $a$ and $b$ are integers for the nonzero limits. Note that this implies that if $A$ winds around $r_*$, there can in general be a nonzero anomaly coming from the defect. In all the remaining cases, which includes the radial gauge  $A_{L,r} = A_{R,r}=0$ case where the gauge fields cannot wind (see section \ref{sec:bar}), we conclude that the anomaly contribution of the singularity at $r_*$ is zero.

To conclude this appendix, we justify a posteriori our use of the Yang-Mills-Scalar action instead of the DBI action \eqref{210}. As we have already shown, the field strength asymptotes to pure gauge at the singularity. To fully justify using the simplified Yang-Mills-Scalar action, we need to study the tachyon kinetic term.

Substituting the ansatz \eqref{c30} and the gauge condition $A_{L,\xi}= A_{R,\xi} =0$ in the tachyon kinetic term we obtain
\be
\tr(DT DT^\dag) = \tr( \tau'^2 + \tau^2 DU DU^\dag)\label{c70}
\ee
From \eqref{c30}, it can be shown that the second term in \eqref{c70}, the kinetic term for $U$, vanishes at least quadratically. The field strengths and $\tau^2 DU DU^\dag$ terms in the DBI action vanish at the center of the instanton, therefore we can expand the DBI action near the center of the instanton in these fields. The term $\tau'^2$ in \eqref{c70} is however a priori a nonzero constant. As a consequence, we introduce an effective metric $\tilde{g}$ as in \eqref{f75}, defined as
\be
\tilde{g}^{mn} \equiv g_{mn} + \kappa \partial_m \tau \partial_m \tau, \label{c71}
\ee
where $g$ is the spherical coordinates metric centered in $r_*, x=0$. Note that near the center of the instanton $\tau$ is a function of $\xi$ only, so only the term $\tilde{g}^{\xi\xi}$ differs from the spherical metric. Expanding the DBI action in $DU$ and $F_{L,R}$ around the effective metric as in \cite{jknp}, we obtain \eqref{f74},
\be
S_{\mathrm{DBI}} =  - M^3 N_c \int \intd^5x V_f(\lambda, \tau)\sqrt{- \det \tilde{g}}\bigg(\frac{1}{2} + \frac{\kappa \tau^2}{2}\tilde{g}^{mn} \tr(D_m U^\dag D_n U) \nonumber \ee \be + \frac{w^2 }{8} \tilde{g}^{mp}\tilde{g}^{nq} \tr(F_{L, mn} F_{L, pq} + F_{R, mn} F_{R, pq})\bigg)\label{f74A}.
\ee
This justifies using the simplified DBI action expanded to quadratic order \eqref{f74}. The action \eqref{f74} and the simplified one \eqref{c21} differ only by multiplicative functions of the metric and the background, which are smooth near the center of the instanton and therefore only contribute to subleading corrections to the equations of motion obtained from \eqref{c21}. Therefore, to leading order in $\xi$ the result obtained with the DBI-TCS action and the simplified Yang-Mills-Scalar-TCS action \eqref{c21} are equal.

\section{Details of the boundary effective action}
\label{sec:effectiveappendix}
\subsection{Effective action in the U(1) case}
\label{sec:effectiveappendix1}
In this subsection, we compute the boundary effective action for an $N_f = 1$ simplified version of the V-QCD model considered in section \ref{sec:effective}. The holographic bulk theory contains a $\text{U}(1)$ gauge field $A$ and a complex scalar $\phi$. The gauge field plays the role of the axial gauge field in the V-QCD model, while $\phi$ plays the role of the tachyon field. The effect of the axial anomaly is ignored in this appendix.

To simplify the analysis, we consider this theory in AdS. It is expected that the difference between the AdS metric and the V-QCD model metric will only affect the spectrum of bulk modes quantitatively. V-QCD has also log asymptotics near the boundary but this difference is not essential here. We  indicate later on when this simplified model qualitatively differs from the V-QCD model.

We derive the quadratic action for the pions, and for this purpose it is enough to work with the linearized equations of motion. Consequently, we ignore the contribution of the TCS term which only appears at higher order in the fields. The action for this simplified model then reads
\be
S= \int \intd^5x \left( - \frac{\ell}{4r} F_{mn}^2 - \frac{\ell^3}{2 r^3}(D_m \phi D_m \phi^*) - \frac{\ell^5}{r^5} V(|\phi|)   \right) \label{f1}.
\ee
We use the following ansatz for the scalar field,
\be
\phi(x, r)= R(r) e^{i\theta(x, r)}.\label{f2}
\ee
This ansatz assumes that the modulus of $\phi$, $R(r)$, is homogeneous in $x$ because we expect that its fluctuations will be heavier mesons that we neglect here. We take $R(r)$ to be given by the solution to the background equations of motion in the absence of gauge fields. We consider $R$ to be fixed, and study perturbations of the fields around this vacuum solution.

With the ansatz \eqref{f2} the action \eqref{f1} becomes
\be
S = \int \intd^5x  \left(  - \frac{\ell}{4r} F_{\mu \nu}^2 - \frac{\ell}{2r}(F_{r\mu})^2 - \frac{\ell^3}{2 r^3}\Big( (\pa_r R)^2 + R^2(\partial_\mu \theta - A_\mu)^2  \right.+ \nonumber \ee
\be \left. + R^2(\partial_r \theta- A_r)^2\Big) - \frac{\ell^5}{ r^5}V(|\phi|)  \right)\label{f3}.
\ee
The background equation of motion for $R$ is
\be
\pa_r  \bigg(\frac{\ell^3}{r^3} \pa_r R \bigg) - \frac{\ell^5}{r^5} V'(R)= 0.\label{f4}
\ee
For simplicity, we consider a quadratic potential
\be
V(R) = \frac{m^2}{2} R^2 = - \frac{3 }{2 \ell^2} R^2.\label{f5}
\ee
We then obtain the background solution in AdS as
\be
R(r) = R_0 r + R_1 r^3.\label{f6}
\ee
This is an exact solution with two integration constants $R_0$ and $R_1$ which correspond respectively to the quark mass (source), and the absolute value of the vacuum expectation value of the chiral condensate (vev).

On this background the dynamical fluctuating fields are the tachyon phase $\theta$ in \eqref{f2} and the gauge field $(A_r,A_\mu)$. The linearized equations of motion for $A_\m$, $A_r$ and $\theta$ are
\be
\pa_r \bigg(\frac{\ell}{r}F_{r\m}\bigg) +\frac{\ell}{r} \eta^{\alpha \n} \pa_\a F_{\n \m} + \frac{\ell^3}{r^3}R^2(\pa_\m \theta - A_\m) = 0,\label{f7}
\ee
\be
\eta^{\alpha \n} \pa_\a \big(\frac{\ell}{r}F_{ \n r}\big) + \frac{\ell^3}{r^3}R^2(\pa_r \theta - A_r) = 0,\label{f8}
\ee
\be
\pa_r \bigg(\frac{\ell^3 R^2}{r^3} (\pa_r\theta - A_r)\bigg) + \frac{\ell^3 R^2}{r^3} \eta^{\alpha \n} \pa_\a( \pa_\n \theta - A_\n) = 0.\label{f9}
\ee
We impose the radial gauge $A_r=0$, for which the equations of motion become
\be
\pa_r \bigg(\frac{\ell}{r}\pa_r A_\mu\bigg) + \frac{\ell}{r} \eta^{\alpha \n} \pa_\a F_{\n \m} + \frac{\ell^3}{r^3}R^2(\pa_\m \theta - A_\m) = 0, \label{f10}
\ee
\be
\frac{\ell}{r}\pa_r \big(\eta^{\alpha \n} \pa_\a A_\nu\big) - \frac{\ell^3}{r^3}R^2 \pa_r \theta = 0,\label{f11}
\ee
\be
\pa_r \bigg(\frac{\ell^3 R^2}{r^3} \pa_r\theta \bigg) + \frac{\ell^3 R^2}{r^3} \eta^{\alpha \n} \pa_\a( \pa_\n \theta - A_\n) = 0.\label{f12}
\ee
Before considering the massive quark case, we shall investigate the massless quark case where $R_0=0$ in \eqref{f6}.

\subsubsection{The massless case}

For vanishing quark mass, $R_0=0$, it is expected that the lightest mode is massless, as in this case, the chiral symmetry is spontaneously broken. We shall study whether such a mode exists.  We  work in Fourier space and expand the vector $A_\mu(k)$ on a basis of momenta associated to massless particles\be
A_\mu(k,r) \equiv A(k,r) \hat{k}_\mu + \bar{A}(k,r) \hat{\bar{k}}_\mu + A^i(k,r) \hat{k}_{i\mu}\,, \label{f12b}
\ee
with
\be
\hat{k}_\mu \equiv (1,0,0,1), \quad \hat{\bar{k}}_\mu \equiv \frac{1}{2}(1,0,0,-1)\label{f13},
\ee
\be
\hat{k}_{1, \mu} \equiv (0,1,0,0), \quad \hat{k}_{2,\mu} \equiv (0,0,1,0)\label{f14}.
\ee
We consider a lightlike 4-momentum $k_\mu$, that we parametrize without loss of generality as
\be
k_\mu = k \hat{k}_\mu. \label{f15}
\ee

The vectors in \eqref{f13}-\eqref{f14} satisfy
\be
\eta^{\mu \nu} \hat{k}_\mu \hat{\bar{k}}_\nu = 1, \quad  \eta^{\mu \nu} \hat{k}_{i\mu} \hat{k}_{j \nu} = \delta_{ij}, \qquad \eta^{\mu \nu} \hat{k}_{i\mu} \hat{k}_{\nu} = \eta^{\mu \nu} \hat{k}_{i\mu} \hat{\bar{k}}_\nu = \eta^{\mu \nu} \hat{\bar{k}}_\mu \hat{\bar{k}}_\nu =  \eta^{\mu \nu}\hat{k}_\mu \hat{k}_\nu = 0\label{f14b}.
\ee

Decomposing the equations \eqref{f10}, \eqref{f11} and \eqref{f12} on the basis \eqref{f12b}, we find that the transverse sector, proportional to $\hat{k}_{1,2}$, decouples from the longitudinal sector, proportional to $\hat{k}_\m, \hat{\bar{k}}_\mu$. The former is expected not to contribute to the pion effective action, so we focus on the latter, whose equations of motion are given by
\be
\pa_r \left(\frac{\ell}{r} \pa_r A\right) + \frac{\ell}{r}k^2 \bar{A} - \frac{\ell^3 R^2}{r^3} (A + ik \theta) = 0,
\label{f18}
\ee
\be
\pa_r \left(\frac{\ell}{r} \pa_r \bar{A}\right) - \frac{\ell^3 R^2}{r^3} \bar{A} = 0,
\label{f19}
\ee
\be
-ik \frac{\ell}{r}\pa_r \bar{A} = \frac{\ell^3 R^2}{r^3}\pa_r \theta, \label{f20}
\ee
\be
\pa_r \left( \frac{\ell^3 R^2}{r^3}\pa_r \theta\right) + ik  \frac{\ell^3 R^2}{r^3}  \bar{A} = 0.\label{f21}
\ee
For massless quarks, the background scalar is given by
\be
R(r) = R_1 r^3.\label{f22}
\ee
The generic solution to \eqref{f19} is then
\be
\bar{A}(k_\mu, r) = C_{1, \bar{A}}(k_\mu) r I_{-1/3}(\ell R_1 r^3/3) + C_{2, \bar{A}}(k_\mu) r I_{1/3}(\ell R_1 r^3/3),\label{f23}
\ee
where $I_{1/3}$, $I_{-1/3}$ are modified Bessel functions.

In the UV, these two solutions have the following expansion,
\be
r I_{-1/3}(\ell R_1 r^3/3)  = c_0 + c_6 r^6 + \mathcal{O}(r^{12}),\label{f24}
\ee
\be
r I_{1/3}(\ell R_1 r^3/3)  = c_2 r^2 + \mathcal{O}(r^{8}).\label{f25}
\ee
To check if a combination of these solutions can give rise to a normalizable mode, we shall now make an ansatz for $\theta$ and $A$. We verify that a normalizable mode can be found with the following ansatz,
\be
\theta(k_\n, r) \equiv \eta'(k_\nu) \theta_0(r), \qquad A_\m(k_\n, r) = -i k_\m \eta'(k_\nu) \xi(r),  \label{f26-ii}
\ee
where $\eta'$ will be identified with the boundary $\eta'$ meson and $\theta_0(r)$ and $\xi(r)$ are radial wavefunctions to be determined. We then substitute \eqref{f26-ii} in the action \eqref{f3} to deduce the normalizability condition on $\theta$ and $A$. We find
\be
S = - \frac{1}{2} \!\int\! \intd^4x\! \left[\! \bigg( \!\int\! \intd r \left(  \frac{\ell}{r} (\pa_r\xi)^2  - \frac{\ell^3}{r^3} R^2 ( \xi \!-\! \theta_0)^2  \right) \bigg)  (\pa_\mu \eta')^2  \right. \nonumber \ee \be  \left. +  \bigg( \!\int\!\intd r   \frac{\ell^3}{r^3} R^2 (\pa_r\theta_0)^2 \bigg)  (\eta')^2     \right] \label{f26-iv}\!.
\ee
The mass and kinetic term of the $\eta'$ can be identified in \eqref{f26-iv}, with coefficients that take the form of integrals over the holographic coordinate $r$. Both of these have to be finite for the mode to be normalizable. We now return to equation \eqref{f23} and see if these conditions can be satisfied. To compute $\pa_r \theta_0$ we use \eqref{f20},
\be
\pa_r \theta_0 = -\frac{ip}{\ell^2 R_1^2 r^4} \pa_r \bar{A},\label{f27}
\ee
where we can substitute \eqref{f23}.
Computing the near boundary expansion of \eqref{f27} we obtain
\be
\pa_r \theta_0 = \alpha \frac{C_{2, \bar{A}} }{r^3} + \beta C_{1, \bar{A}} r + \mathcal{O}(r^3),\label{f29}
\ee
where $\alpha$ and $\beta$ are unimportant nonzero numerical factors obtained from the series expansion of the Bessel functions. For the radial integral in the mass term in \eqref{f26-iv} to converge in the UV one must then enforce $C_{2, \bar{A}} =0 $.

In the IR, $\pa_r \theta_0$ diverges as
\be
\pa_r \theta_0 \sim \beta_{IR} C_{1, \bar{A}} e^{\frac{r^3 \ell R_1}{3} + \mathcal{O}(\log(r))},\label{f30}
\ee
where $\beta_{IR}$ is a nonzero numerical constant. So normalizability also imposes that $C_{1, \bar{A}}$ should be zero. We conclude that in the sector we are considering, normalizability requires
\be
\theta_0(r) = \theta_0 = \text{constant}, \quad \bar{A}=0\label{f31}.
\ee
Crucially, this implies that the mass term in \eqref{f26-iv} vanishes.

We now reinsert \eqref{f31} into the equation for $A$, \eqref{f18}, and solve it. We introduce $\tilde{\xi}\equiv A + i k \theta_0$, the Fourier space equivalent of the gauge-invariant combination $A_\mu - \pa_\mu \theta_0$. We find that $\tilde{\xi}$ satisfies
\be
\pa_r \left(\frac{\ell}{r} \pa_r \tilde{\xi}\right) - \ell^3 R_1^2 r^3 \tilde{\xi}= 0.
\label{f32}
\ee
Observe that this is the same equation \eqref{f19} as the one satisfied by $\bar{A}$, so that the solution for $\tilde{\xi}$ is also of the form \eqref{f23}. Inserting \eqref{f31} and $\tilde{\xi} \equiv A + ik \theta_0$ in \eqref{f3}, the effective action for the boundary mode $\eta'$ becomes
\be
S = - \int_0^\infty  \intd r\, \bigg( \frac{\ell}{2 r} (\pa_r \tilde{\xi})^2  + \frac{\ell^3 R_1^2}{2r^3} (\tilde{\xi})^2 \bigg) \int \intd^4x (\pa_\mu \eta')^2 .\label{f33}
\ee
We can now analyze the convergence of the radial integral in \eqref{f33} in terms ot the solutions of \eqref{f32}. From the general solution \eqref{f23},
we now show that the following linear combination is normalizable both at $r=0$ and $r\to +\infty$:
\be
 \tilde{\xi}(r) =  C_{\tilde{\xi}} r K_{1/3}(\ell \sigma r^3/3),\label{f34}
\ee
where $K$ is the Bessel K function, and we have denoted $C_{\tilde{\xi}}$ a multiplicative constant. Observe that substituting \eqref{f34} in the radial integral in \eqref{f33} provides a finite result, proportional to $C_{\tilde{\xi}}^2$. We choose the value of $C_{\tilde{\xi}}$ such that the kinetic term is canonically normalized
\be
\int\intd r\, \bigg( \frac{\ell}{ r} (\pa_r \tilde{\xi})^2 +  \frac{\ell^3 R^2}{ r^3} (\tilde{\xi})^2 \bigg)= - \bigg(\frac{\ell}{r} \tilde{\xi} \pa_r \tilde{\xi}\bigg)(r \to 0) =  1,\label{f37}
\ee
where we have used that $\tilde{\xi}$ in \eqref{f34} goes to zero exponentially in the IR as can be shown from the analytical solution above. Inserting \eqref{f37} into \eqref{f33} we find
\be
S_{\eta'} =  \int_{r=0} \intd^4x \bigg(\frac{\ell}{2 r} \tilde{\xi} \pa_r \tilde{\xi} \bigg)(\pa_\mu \eta')^2 =  - \frac{1}{2} \int \intd^4x (\pa_\mu \eta')^2.\label{f39}
\ee
Note that since the mass term obtained from substituting \eqref{f31} in \eqref{f26-iv} vanishes, the normalizable $\eta'$ mode we derived is massless.

\subsubsection{The massive quark case}
We shall now return to the massive quark case. It is expected that the lightest mode is massive, as in this case the chiral symmetry is explicitly broken. We can decompose the gauge field into transverse and longitudinal massive modes
\be
A_\mu \equiv A_\mu^{\perp} + \pa_\mu \varphi, \quad \pa^\mu A_\mu^{\perp} =0 \label{f40},
\ee
The equations of motion in the radial gauge \eqref{f10}, \eqref{f12} and constraint equation \eqref{f11} are now
\be
\partial_r \bigg(\frac{\ell^3}{ r^3}  R^2 \pa_r \theta \bigg) + \frac{\ell^3}{ r^3}  R^2 \eta^{\a\m} \pa_\a \pa_\m (\theta - \varphi)=0,\label{f42}
\ee
\be
\partial_r \bigg(\frac{\ell}{ r} \partial_r \pa_\mu \varphi \bigg) + \frac{\ell^3}{ r^3}  R^2  \partial_\mu (  \theta - \varphi)=0,\label{f42-ii}
\ee
\be
\frac{\ell}{r} \pa_r(\Box \varphi) = \frac{\ell^3}{r^3} R^2 \pa_r \theta.\label{f41}
\ee
We  show later that the longitudinal gauge fields and phase of the scalar can both be expanded in separable modes as
\be
\varphi = \sum_n \varphi_n(r) \alpha_n(x), \quad \theta = \sum_n \theta_n(r)\alpha_n(x).\label{f43}
\ee
Substituting the mode expansion \eqref{f43} in \eqref{f42-ii}, isolating $\theta_n$, taking a derivative with respect to $r$ and inserting the result in \eqref{f41}, we obtain
\be
\partial_r \bigg(\frac{r^3 }{R^2 \ell^3} \pa_r f_n \bigg) \alpha_n - \frac{r}{\ell} f_n \alpha_n = -  \frac{r^3}{\ell^3 R^2} f_n \Box \alpha_n,\label{f48}
\ee
where we defined $f_n = \frac{\ell}{r} \pa_r \varphi_n$. In order to solve \eqref{f48}, we introduce a Sturm-Liouville equation for $f_n$,
\be
-\frac{\ell^3 R^2}{r^3} \left( \partial_r \bigg(\frac{r^3 }{R^2 \ell^3} \pa_r f_n \bigg) - \frac{r}{\ell} f_n \right) = m_n^2  f_n,\label{f48b}
\ee
where we denoted $ m_n^2$ the eigenvalue of the Sturm-Liouville operator.

Substituting $f_n = \frac{\ell}{r} \pa_r \varphi_n$ into \eqref{f48b} and using the equations of motion we obtain
\be
\partial_r \bigg(\frac{\ell^3}{ r^3}  R^2 \pa_r \theta_n \bigg) + \frac{\ell^3}{ r^3}  R^2 m_n^2  (\theta_n - \varphi_n)=0,\label{f45}
\ee
\be
\partial_r \bigg(\frac{\ell}{ r} \partial_r \pa_\m \varphi_n \bigg) + \frac{\ell^3}{ r^3}  R^2 \pa_\m (  \theta_n - \varphi_n)=0,\label{f46}
\ee
\be
m_n^2 \frac{r^2}{\ell^2}\partial_r\varphi_n = R^2 \partial_r\theta_n \label{f47}.
\ee

The Sturm-Liouville equation \eqref{f48b} can be recast into a Schr\"odinger problem. The behavior of the potential qualitatively differs in AdS and in the V-QCD background. In the V-QCD model the potential goes to infinity in the IR and therefore, there exists a discrete tower of orthogonalizable modes. Instead, in AdS there is a continuum of modes solution to \eqref{f48}. In order to reproduce this feature of V-QCD, we use an IR cutoff $r_{IR}$, which discretizes the modes. These modes can then be taken orthogonal to each other with respect to the following inner product,
\be
(f_n, f_m) = \int \intd r \frac{r}{\ell R^2} \pa_r \varphi_n \pa_r \varphi_m = C_{n} C_{m} \delta_{mn},\label{f49}
\ee
where $C_{n,m}$ are normalization constants that we determine by enforcing that the kinetic term in the effective action is canonically normalized. Substituting \eqref{f47} into \eqref{f49}, we obtain
\be
\int \intd r \frac{\ell^3}{r^3} R^2 \pa_r \theta_n \pa_r \theta_m = C_n C_m m_n^2 m_m^2 \delta_{mn}.\label{f50}
\ee
Equation \eqref{f50} is the radial integral appearing in the mass term in the effective action. To compute the effective action, we insert \eqref{f43} in \eqref{f3} and obtain
\be
S  = -\frac{1}{2} \sum_{m,n} \int \intd^4x \left[ \pa_\m \a_m \pa^\m \a_n \int  \intd r \bigg( \frac{\ell}{r}  \pa_r \varphi_n \pa_r \varphi_m+  \frac{\ell^3}{r^3} R^2( \theta_n - \varphi_n)( \theta_m - \varphi_m)  \bigg) \right. \nonumber \ee\be + \left.  \a_m \a_n \int \intd r \bigg(  \frac{\ell^3 R^2}{r^3} \pa_r \theta_m \pa_r \theta_n \bigg)   \right] .\label{f52}
\ee
Requiring that the kinetic term is canonically normalized yields
\be
\int  \intd r \bigg( \frac{\ell}{r} \pa_r \varphi_n \pa_r \varphi_m+  \frac{\ell^3}{r^3} R^2( \theta_n - \varphi_n)( \theta_m - \varphi_m)  \bigg) =  \delta_{mn} .\label{f56}
\ee
We integrate by parts the first term in the radial integral in \eqref{f56} and use the equations of motion to obtain
\be
\int \intd r \frac{\ell}{2r} \pa_r \varphi_n \pa_r \varphi_m = \int \intd r \frac{\ell^3 R^2}{2r^3} \varphi_n ( \theta_m - \varphi_m).\label{f53}
\ee
This leads to
\be
\int  \intd r \bigg( \frac{\ell}{2 r} \pa_r \varphi_n \pa_r \varphi_m+  \frac{\ell^3}{2 r^3} R^2( \theta_n - \varphi_n)( \theta_m - \varphi_m)  \bigg) = \int  \intd r \frac{\ell^3}{2 r^3} R^2 \theta_n ( \theta_m - \varphi_m).\label{f54}
\ee
We use once again the equations of motion, integrate by parts and use \eqref{f56} to obtain
\be
\int  \intd r \frac{\ell^3}{2 r^3} R^2 \theta_n ( \theta_m - \varphi_m) = \int  \intd r \frac{\ell^3}{2 m_n^2 r^3} R^2 \pa_r \theta_n \pa_r \theta_m =  \frac{1}{2}\delta_{mn},\label{f55}
\ee
Comparing \eqref{f55} and \eqref{f50} fixes the value of $C_n$ to
\be
C_n = m_n^{-1}.\label{f51}
\ee
Using \eqref{f56} and \eqref{f55} in \eqref{f52}, we find the following effective action
\be
S = - \sum_{n=0}^{+\infty} \frac{1}{2}\int \intd^4x \bigg( (\partial_\mu \alpha_n)^2 + m_n^2 \alpha_n^2\bigg) .\label{f57}
\ee

\subsection{Details of the non-abelian case}
\label{sec:effectiveappendix2}
In this subsection we give more details on some of the derivations in section \ref{sec:effective}. The purpose of this appendix is to justify the normalization conditions used in section \ref{sec:effective}, and to derive the UV and IR behavior of the pion mode in order to determine the WZW term coefficient in \eqref{f177}. Specifically:

\begin{itemize}
\item In \ref{sec:d21} we give the asymptotic expansions of the $B_i$ functions appearing in \eqref{f76}
\item In \ref{sec:d22} we provide more details on the computation of $\theta_0$ (see \eqref{f116}, \eqref{f162})
\item In \ref{sec:d23} we derive the separable mode expansion \eqref{f125}, \eqref{f126} and the result \eqref{f137}
\end{itemize}

\subsubsection{Asymptotics of the functions $B_i(r)$}
\label{sec:d21}
Recall that the action from section \ref{sec:effective}, \eqref{f76}, is
\be
S_{\mathrm{DBI}}  =  -\frac{1}{2}  \int \intd^5x \bigg[ 2 B_1(r) \tr(D_r U^\dag D_r U)  +2 B_2(r) \eta^{\m\n}\tr(D_\mu U^\dag D_\nu U)  \nonumber \ee\be + B_3(r) [(F_{\mu r}^{A,a})^2 -  ([A_r, A_\m]^a)^2] + B_4(r)[ (F_{\mu \nu}^{A,a})^2 - ([A_\m, A_\n]^a)^2]\bigg],
\label{f761}
\ee
where the $B_i$ functions are given by
\be
B_1 \equiv M^3 N_c V_f \sqrt{- \tilde{g}} \frac{\kappa \tau^2}{2} \tilde{g}^{rr}, \quad
B_2 \equiv M^3 N_c V_f \sqrt{- \tilde{g}} \frac{\kappa \tau^2}{2} g^{xx},
\label{f78A}\ee
\be
B_3 \equiv M^3 N_c V_f \sqrt{- \tilde{g}} \frac{w^2}{2} \tilde{g}^{rr} g^{xx},\quad
B_4 \equiv  M^3 N_c V_f \sqrt{- \tilde{g}} \frac{w^2}{4} (g^{xx})^2.
\label{f80A}\ee
We shall compute the asymptotic scalings of $B_i$, $i=1,2,3,4$. The UV asymptotics depend on whether the quark mass is zero or nonzero.

In the massless quark case, the UV asymptotics of the tachyon field can be deduced from \eqref{51}
\be
\tau = (\sigma r^3  (-\log(r \Lambda_{UV}))^{\rho}) \left(1 + \mathcal{O}\left(\frac{1}{\log(r \Lambda_{UV})}\right)\right),\label{f98A}
\ee
We deduce the UV scaling of the $B_i$ functions,
\be B_1 \sim B_{1, UV} r^3\log(r \Lambda_{UV})^{2\rho}, \quad B_2 \sim B_{2, UV} r^3\log(r \Lambda_{UV})^{2\rho}, \nonumber \ee\be B_3 \sim B_{3, UV} r^{-1}, \quad B_4 \sim B_{4, UV} r^{-1} \label{f99A}
\ee
In the massive quark case, \eqref{51} becomes
\be
\label{f100A}
\tau =  (m_q r (-\log(r \Lambda_{UV}))^{-\rho} + \sigma_q r^3 (-\log(r \Lambda_{UV}))^{\rho})\left(1 + \mathcal{O}\left(\frac{1}{\log(r \Lambda_{UV})}\right)\right).
\ee
In this case, the UV scaling of the $B_i$ functions is
\be B_1 \sim B_{1, UV} r^{-1}\log(r \Lambda_{UV})^{-2\rho}, \quad B_2 \sim B_{2, UV} r^{-1}\log(r \Lambda_{UV})^{-2\rho},\nonumber \ee\be  B_3 \sim B_{3, UV} r^{-1}, \quad B_4 \sim B_{4, UV} r^{-1} \label{f101A}
\ee
Note that the behavior of $B_1$ and $B_2$ is substantially different in the massive case, where they diverge as $r\to 0$, and the massless case, where they remain finite in the UV.

In order to compute the IR scaling of the $B_i(r)$, we use the IR scalings of the VQCD model in \cite{spectrum}. The asymptotic scalings used in \cite{spectrum} are specifically adjusted to produce exactly linear meson trajectories. The flavor potential $V_f$ scales as
\be
V_f \sim V_{IR} e^{-a_{IR} \tau^2}.\label{f81A-ii}
\ee
The IR scalings of the V-QCD background solution are
\be
\lambda \sim e^{\frac{3r^2}{2\ell^2} + \lambda_c}, \quad g^{xx} \sim \frac{\ell}{r} e^{\frac{2r^2}{\ell^2} - 2A_c}, \quad \tau \sim \tau_0 \frac{r^{C_\tau}}{\ell^{C_\tau}},\label{f81A}
\ee
where $C_\tau >1$, $V_{IR}$, $\tau_0>0$ and $a_{IR}$ are constants of the model. From \eqref{f81A-ii} and \eqref{f81A}, we deduce the behavior of $w$ and $\kappa$,
\be
w \sim w_{IR} \lambda^{-4/3} \log(\lambda) \sim w_{IR} r^2  e^{\frac{-2r^2}{\ell^2}}, \quad \kappa \sim \kappa_{IR} \lambda^{-4/3} \log(\lambda)^{1/2} \sim \kappa_{IR} r e^{\frac{-2r^2}{\ell^2}},\label{f82A}
\ee
with $\kappa_{IR}$ and $w_{IR}$ two constants. The IR expansions (in the $r\to \infty$ limit) of the $B$ functions are then
\be
B_1 \sim B_{1, IR}  r^{\frac{7}{2} + C_\tau} e^{- 5\frac{r^2}{ \ell^2}}e^{-a_{IR}\tau_0^2 \frac{r^{2C_\tau}}{\ell^{2C_\tau}}}  , \quad B_2 \sim B_{2, IR}  r^{\frac{3}{2} + 3 C_\tau}  e^{- 5 \frac{r^2}{\ell^2}}e^{-a_{IR}\tau_0^2 \frac{r^{2C_\tau}}{\ell^{2C_\tau}}}  \label{f83A},
\ee
\be B_{3} \sim B_{3, IR} r^{\frac{11}{2}- C_\tau}  e^{-5\frac{r^2}{\ell^2}} e^{-a_{IR}\tau_0^2 \frac{r^{2C_\tau}}{\ell^{2C_\tau}}}, \quad B_{4} \sim B_{4, IR} r^{\frac{7}{2} + C_\tau}  e^{-5\frac{r^2}{\ell^2}} e^{-a_{IR}\tau_0^2 \frac{r^{2C_\tau}}{\ell^{2C_\tau}}}  \label{f84A}.
\ee

\subsubsection{Details on the computation of $\theta_0$}
\label{sec:d22}
In this section we provide more details on the computation of the wavefunction of the phase of the tachyon $\theta_0$, defined in \eqref{f94} in the massless quark case and in \eqref{f126} in the massive quark case. The purpose of this computation is to determine the IR and UV values of $\theta_0$, which fix the coefficient of the WZW term (see \eqref{f177}).

We begin with the massless case, where we showed in \eqref{f103} that $\theta_0$ is a constant for $r>0$. A subtlety arises, because if $\theta_0$ was continuous in $r=0$, as can be seen from \eqref{f177} the total derivative $\pa_r (\theta_0^5)$ would vanish and lead to a vanishing WZW term. We show that the physically relevant solution is discontinuous at $r=0$, i.e.
\be
\theta_0(0) = 0, \qquad \theta_0(r>0)=\frac{1}{f_\pi}.
\ee

Recall that we use the definition of $f_\pi$ established in \cite{casero} from the transverse modes of the gauge field at zero momentum, \eqref{f110}
\be
f_\pi^2 = - B_3 (\upsilon \pa_r \upsilon)(r=0, p^2 = 0), \label{f110A}
\ee
where we have defined $\upsilon$ as the wave-function of a transverse mode, for instance $A^{\perp}_1$. This wave-function satisfies the equation of motion
\be
\pa_r (B_3 \pa_r \upsilon) - 4 B_2  \upsilon =0.\label{f111A}
\ee
Equation \eqref{f111A} admits a normalizable mode, with normalization
\be
\upsilon(0) = 1.\label{f112A}
\ee
$\upsilon$ and $\tilde{\xi}$, defined in \eqref{f106} satisfy the same linear differential equation \eqref{f106} and \eqref{f111A}. Moreover, it can be shown from the UV asymptotics of the solutions that there is a single UV-normalizable solution. The differential equation \eqref{f111A} is linear, therefore $\upsilon$ and $\tilde{\xi}$ differ by an overall constant factor that we call $c$,
\be
\tilde{\xi} = c\upsilon\label{f113A}.
\ee
To determine this factor we match \eqref{f109},
\be
\int_0^\infty \intd r(B_3 (\pa_r \tilde{\xi})^2 + 4B_2 \tilde{\xi}^2)   = - (B_3 \tilde{\xi} \pa_r \tilde{\xi})(r=0) = 1,
\ee
and \eqref{f110A}, to obtain
\be
 1= - B_3 \tilde{\xi} \pa_r \tilde{\xi} (\epsilon) = c^2 \big( - B_3 \pa_r \upsilon (\epsilon)\big)  = c^2 f_\pi^2\label{f114A}.
\ee
We use then \eqref{f113A} and \eqref{f114A} to obtain
\be
\upsilon = \tilde{\xi} f_\pi.\label{f115A}
\ee
Because the boundary condition on $\upsilon$ is given in the UV, $\upsilon(0)=1$, we evaluate \eqref{f115A} in the UV. We then use the definition of $\tilde{\xi}$, \eqref{f108-ii}, and the absence of chemical potential boundary condition in the UV, $A = 0$, so that $\tilde{\xi} = \theta_0$ as $r\to 0$. After some algebra, using $\upsilon(0)=1$ we conclude that equation \eqref{f116} holds,
\be
\theta_0 =  \frac{1}{f_\pi}.\label{f116A}
\ee

Equation \eqref{f116A} holds for $r>0$. Indeed, equation \eqref{f97} is automatically satisfied in $r=0$, because \eqref{f78A}, \eqref{f80A} imply $B_1(0) = B_3(0) = 0$. Therefore, the {\em function} $\theta_0(r)$ which solves \eqref{f97} can admit solutions which are discontinuous in $r=0$, i.e. \eqref{f116A} is valid for $r>0$, but $\theta_0(0)$ can take any value. In general $\theta_0$ is then
\be
\theta_0 (r) = \theta_0(0)+ \Theta(r) \left(\frac{1}{f_\pi} - \theta_0(0)\label{f117A}\right),
\ee
where we defined the Heaviside function
\be
 \Theta(r) = \left\{ \begin{matrix}
 0 \text{ if } r\leq 0\\
 1 \text{ if } r> 0
 \end{matrix} \right.
\ee

To fix $\theta_0(0)$, we require that $\theta_0$ in the massless quark case coincides pointwise with the small mass limit of the finite quark mass case. We turn on a small quark mass and call $\theta_0$ the lightest mode in the spectrum, which becomes massless in the massless quark limit. We solve the equations of motion \eqref{f127}, \eqref{f129} for $\theta_0$,
\be
\partial_r(B_3 \partial_r \varphi_0) + 4 B_2 (\theta_0 - \varphi_0) = 0,
\label{feom1}
\ee
\be
B_3 m_\pi^2 \partial_r \varphi_0 = 4  B_1 \partial_r \theta_0. \label{feom2}
\ee
Here we identified the zero mode with the pions, $m_0 = m_\pi$. From the normalizability condition, i.e. the UV convergence of the radial integral \eqref{f137}, we deduce \eqref{f144}, $\theta_0(0) = \varphi_0(0)$. The boundary condition of the gauge fields is imposed by the absence of chemical potential, $A_\mu(0)=0$, so $\varphi_0(0)=0$. Therefore, we obtain
\be
\theta_0(0)=0. \label{f118A}
\ee
The correct solution in the massless quark case should be the limit of the solution for finite quark mass as $m_q\to 0$\footnote{This is expected from the boundary theory picture as the chiral effective field theory is continuous in the $m_q\to 0$ limit.}. In this limit a discontinuity develops at the UV boundary. Combining \eqref{f117A} and \eqref{f118A}, we obtain the value of $\theta_0(r)$ in the massless case,
\be
\theta_0 (r) = \Theta(r) \left(\frac{1}{f_\pi}\label{f118A-ii}\right).
\ee
Note that \eqref{f118A-ii} is consistent with the result obtained in \eqref{f162}.

We now provide an approximation of the function $\theta_0(r)$ (which is {\em not} a constant) in the small $m_q$ regime. We begin with \eqref{f139},
\be
\theta_0(r) = \frac{m_\pi^2}{4} \int_0^r \intd r' \frac{\pa_r \varphi_0 B_3}{B_1}. \label{f119A}
\ee
As was motivated in section \ref{sec:effective} in the discussion following \eqref{f146b}, the integrand in \eqref{f119A} is only significant in the UV and quickly vanishes as $r$ increases\footnote{The crucial observation leading to this fact is that $B_3/B_1$ is sharply localized near the UV boundary in the small $m_q$ limit, as can be guessed from the UV expansion \eqref{f99A}. It can be numerically checked that in this limit the ratio also goes to zero everywhere else.}. To approximate $\theta_0(r)$, we replace the integrand in \eqref{f119A} by its UV expansion. The UV expansion of $\varphi_0$ and $\theta_0$ were given in \eqref{f142}, \eqref{f143}
\be
\theta_0 = \theta_0(0)  + c_{\theta, 1}r^2 + \OO(r^4) \label{f142A},
\ee
\be
\varphi_0 =  \varphi_0(0)  +  c_{\varphi, 1}r^2 + \OO(r^4) \label{f143A},
\ee
Substituting \eqref{f99A} (with $\rho=0$, i.e. we do not consider logarithms in the expansion) into \eqref{f119A}, we obtain
\be
\theta_0(r) \cong \frac{m_\pi^2}{4} \int_0^r \intd v \frac{2 c_{\varphi,1} v^3}{2 \kappa(0) \ell^2 (m_q v + \sigma v^3)^2}, \label{f170A-ii}
\ee

We shall now compute the value of $c_{\varphi,1}$. For this purpose we compute $c_{\theta,1}$ and relate it to $c_{\varphi,1}$ using \eqref{feom2}. This can be done by using that the wave-function of the pions goes to the massless quark wave-function \eqref{f118A-ii} in the $m_q\to 0$ limit. We insert in \eqref{a26} the expansions that we use for $\theta_0$ and use $U = \exp(2 i\theta_0 \pi)$, then we obtain
\be
U = I + 2 i c_{\theta, 1} r^2 \pi(x) + \dots,\label{f154A}
\ee
which after matching with $\mathcal{A}$ as defined in \eqref{a26} gives
\be
\mathcal{A} = 2 i c_{\theta_1} \pi(x) + \mathcal{O}(\pi^2),\label{f155A}
\ee
and then using the ansatz $\mathcal{H} = \sigma I$, \eqref{a27} and \eqref{a28} we obtain
\be
\mathcal{U} = I + \frac{2 i m_q  c_{\theta_1} }{\sigma} \pi(x) + \mathcal{O}(\pi^2).\label{f156A}
\ee
We can also compute the boundary pion matrix $\mathcal{U}$ from the massless case result \eqref{f118A-ii},
\be
\mathcal{U} = \exp\bigg(2 i \frac{\pi}{f_\pi}\bigg) \label{f156A-ii}
\ee

Matching \eqref{f156A} with \eqref{f156A-ii} implies
\be
c_{\theta_1}  = \frac{\sigma}{m_q f_\pi}.\label{f157A}
\ee
Inserting the near boundary expansion of $\varphi_0$, $\theta_0$, \eqref{f142}, \eqref{f143A} and $c_{\varphi,0} = c_{\theta,0}=0$ in the constraint equation \eqref{feom2} we then obtain
\be
c_{\varphi, 1} = \lim_{r\to 0} \frac{4B_1 c_{\theta,1}}{m_\pi^2 B_3} = \frac{8m_q^2 \kappa(0) c_{\theta, 1}\ell^2}{m_\pi^2 w(0)^2 },\label{f158A}
\ee
Finally, reinjecting \eqref{f157A} into \eqref{f158A} we obtain
\be
c_{\varphi_1} =  \frac{8 m_q \sigma \kappa(0) \ell^2}{m_\pi^2 w^2(0) f_\pi}.\label{f159A}
\ee
From this, we can deduce
\be
\frac{m_\pi^2 \varphi_0'}{m_q r} = \frac{2m_\pi^2  c_{\varphi, 1}}{m_q} + \mathcal{O}(r^2) = \frac{16 \sigma \kappa(0) \ell^2}{w^2(0) f_\pi} + \mathcal{O}(m_q).\label{f160A}
\ee
We substitute \eqref{f159A} into \eqref{f119A} to obtain finally
\be
\theta_0(r) =  \frac{2 m_q \sigma}{f_\pi} \int_0^r \intd v \frac{v^3 }{(m_q v + \sigma v^3)^2} \nonumber \ee \be = \frac{1}{f_\pi} \frac{\sigma r^2}{m_q + \sigma r^2},\label{f160A-ii}
\ee
which interpolates smoothly between zero and $1/f_\pi$. Importantly, \eqref{f160A-ii} gives exactly \eqref{f118A-ii} pointwise in the massless limit. We obtained that the difference between the value of $\theta_0$ in the UV and in the IR is independent of $m_q$. This statement guarantees that the coefficient of the WZW term is indeed the one appearing in \eqref{f179}.

\subsubsection{Details about the modes $\theta_n$, $\varphi_n$}
\label{sec:d23}
In this section we derive the mode expansion \eqref{f125}, \eqref{f126} and show that the normalization condition of the mass term \eqref{f134} is equivalent on shell to \eqref{f137}. We assume here that $m_q \neq 0$.

We start from the Sturm-Liouville equation for $f_n=  B_3\pa_r \varphi_n$, \eqref{f132}, which we recall here:
\be
\partial_r \left(\frac{1}{B_2} \pa_r f_n\right) - \frac{4}{B_3} f_n = - \frac{1}{B_1} m_n^2 f_n.\label{f132A}
\ee
This Sturm-Liouville equation can be rewritten as a Schr\"odinger equation. Due to the asymptotics of the tachyon in V-QCD, the potential of the associated Schr\"odinger equation diverges to infinity in the IR and therefore there exists a discrete orthonormal basis of solutions to the Schr\"odinger equation. This implies the existence of a discrete basis of eigenfunctions with eigenvalue $m_n^2$ that are solutions to \eqref{f132A}, which are orthogonal to each other with respect to the inner product
\be
(f_n, f_m) = \int \intd r \frac{1}{B_1} f_n f_m \label{f133A}.
\ee
We may summarize the normalization freedom with generic constants $C_n$, that appear in the orthogonalization condition as
\be
(f_n, f_m)= C_n C_m \delta_{mn}\label{f133A-ii}
\ee
We can then replace $f_n=  B_3 \varphi_n'$ and use \eqref{f129} twice to deduce the norm for $\pa_r \theta$
\be
\int_0^{\infty} \intd r \left[ B_1 (\partial_r \theta_n ) (\partial_r \theta_m) \right] = \frac{C_n C_m m_n^2 m_m^2}{16} \delta_{mn}.\label{f134A}
\ee

We now fix the value of $C_n$ by relating the normalization of the mass and the kinetic term, and imposing that the kinetic term in the effective action is canonically normalized. Recall now that the bulk action evaluated on the ansatz \eqref{f125}, \eqref{f126} is given by equation \eqref{f131}
\be
S = -\frac{1}{2} \sum_{m,n} \int \intd^4x \left[ \bigg( \int \intd r 4 B_1 (\pa_r \theta_n)(\pa_r \theta_m) \bigg) \a_n \a_m+  \right. \nonumber \ee \be \left. + \bigg(\int \intd r B_3 \pa_r \varphi_n \pa_r \varphi_m + 4B_2 (\theta_n - \varphi_n)(\theta_m - \varphi_m)   \bigg) \pa_\mu \a_n \pa_\mu \a_m\right]\label{f134A-ii}.
\ee
Note that the integral in \eqref{f134A} is the radial integral multiplying the mass term in \eqref{f134A-ii}. We impose that the kinetic term is canonically normalized,
\be
\int \intd r B_3 \pa_r \varphi_n \pa_r \varphi_m + 4B_2 (\theta_n - \varphi_n)(\theta_m - \varphi_m)  = \delta_{mn}.  \label{f137A}
\ee
We integrate by parts in the first term of \eqref{f137A} and use \eqref{f129}. Then we add both terms in \eqref{f137A} and use \eqref{f127}. We obtain
\be
\int \intd r 4B_2 \theta_n (\theta_m - \varphi_m) = \delta_{mn}\label{f137A-ii}
\ee
Then we use \eqref{f129}, \eqref{f127} and integrate by part to obtain
\be
\int \intd r 4 B_1 \pa_r \theta_n \pa_r \theta_m = m_n^2 \delta_{mn}\label{f137A-iii}
\ee
Comparing \eqref{f134A} and \eqref{f137A-iii}, we see that \eqref{f137A} and \eqref{f137A-iii} hold if $C_n$ is set to
\be
C_n = \frac{2}{m_n}.\label{f135A}
\ee
Using \eqref{f137A} and \eqref{f137A-iii}, we finally find the following effective action,
\be
S = -\frac{1}{2} \sum_{n} \int \intd^4x \left[ m_n^2 \a_n^2 + (\pa_\mu \a_n)^2 \right]\label{f136A}.
\ee

\section{Details on the anomaly}
\label{sec:F}
This section contains additional material related to the descent equations, the definition of the TCS form and the anomaly. We recall that the descent equations do not fully fix the TCS form or the anomaly \cite{Bilal:2008qx}, and we then relate this ambiguity to the ambiguity in the choice of path in section \ref{sec:TCSQuillen}. We then review the difference between the consistent and covariant anomalies and derive how they appear in the bulk equations of motion.
\subsection{Ambiguity in the descent equations}
We start from the descent equations \eqref{312}, \eqref{471}:
\be
\chi_6 = \intd \Omega_5, \label{1f0}
\ee
\be
\delta \Omega_5 = \intd I_4^1. \label{1f}
\ee
We indicate with a subscript the differential form degree and with a superscript $1$ terms which contain a single instance of the gauge parameter $\Lambda$ defined in \eqref{471-b}.

We shall study the possible modifications of $I_4^1$ or $\Omega_5$ leaving \eqref{1f0}, \eqref{1f} invariant. We distinguish several different types of ambiguity in the definition of $\Omega_5$ and in the definition of $I_4^1$.

Firstly, note that it is possible to add an exact form to $I_4$, i.e. the modification
\be
I_4^1 \to I_4^1 + \intd \beta_3^1 \label{3f}
\ee
leaves both \eqref{1f0} and \eqref{1f} invariant. The modification \eqref{3f} then adds a boundary term to the variation of the effective action $\Gamma$,
\be
\delta \Gamma = \int_M I_4^1 \to \int_M I_4^1  + \int_{\pa M} \beta_3^1.
\ee
In the literature it is common \cite{Bilal:2008qx} to enforce that $I_4^1$ contains $\Lambda$ and not $\intd \Lambda$. This procedure requires an integration by part which leaves a boundary term in $\delta \Gamma$. Then the ambiguity \eqref{3f} can be lifted by choosing $\beta_3^1$ to cancel the boundary term coming from the integration by part used to remove $\intd \Lambda$. Note that the modification \eqref{3f} does {\em not} affect the conservation equation of anomalous currents.

Secondly, \eqref{1f0} is also ambiguous, in the sense that it is possible to add to the TCS form a closed form. Here we consider adding an exact form
\be
\Omega_5 \to \Omega_5 + \intd \alpha_4. \label{5f}
\ee
This modification is different from \eqref{3f}, as the left-hand side in the descent equation \eqref{1f} is modified, and requires a modification of $I_4^1$ for the descent equation \eqref{1f} to hold. The associated modification of $I_4^1$ is given by
\be
\delta \Omega_5 \to  \delta \Omega_5 + \delta \intd \alpha_4 = \intd I_4^1 - \intd \delta \alpha_4. \label{5fb}
\ee
\be
I_4^1 \to I_4^1 - \delta \alpha_4.\label{6f}
\ee
Under the transformation \eqref{6f} the divergence of the anomalous current also transforms. We see in \eqref{5f}, \eqref{6f} that shifting the TCS action by an exact form shifts the anomaly by a related BRST-exact form. We now show that the difference between two choices of path corresponds to a transformation \eqref{5f}, \eqref{6f}.

\subsection{Loops in the plane $(a,b)$ provide  exact forms}

We now prove that forms obtained from loops $\bar{\gamma}$ in the plane $(a,b)$ according to (\ref{335}) (illustrated in figure \ref{plane}), differ by an exact form.

Recall first that for a loop, $\mathcal{F}_0 = \mathcal{F}_1$ in \eqref{329}. Therefore, the left-hand side of \eqref{329} is zero and consequently, the differential form $\Omega^{(\bar{\gamma})}$, defined in \eqref{335},
\be
\Omega^{(\bar{\gamma})} = \oint_{\bar{\gamma}} \intd s \str(\exp(i\mathcal{F}_s) \dot{D}_s) \label{7f},
\ee
is closed. We shall show a stronger statement, that $\Omega^{(\bar{\gamma})}$ is globally exact under some hypothesis on the loop we shall specify. This will prove that two forms corresponding to two different paths with same end-points differ by an exact form.

To prove the above, we generalize the arguments leading to \eqref{322} to a two-parameter family of superconnections. The first parameter, $s$, corresponds to the coordinate along the path, while the second parameter, $t$, corresponds to the parameter varying between the two paths. We lift the connection to $\mathcal{M} \times[ 0,1]^2$, and define the total differential,
\be
\tilde{\intd} = \intd +  \intd s\pa_s +  \intd t \pa_t.\label{8f}
\ee
Equation \eqref{311} becomes
\be
\tilde{\intd}  \tilde{\chi}  = 0,\label{9f}
\ee
where $\tilde{\chi}$ is the lifted Chern character associated to the lifted superconnection $\tilde{D}$,
\be
\tilde{D} \equiv \tilde{\intd} - i a(s,t) A - ib(s,t) \mathcal{T}.
\ee
We expand $\tilde{\chi}$ into $0$, $1$ and $2$-forms on $[0,1]^2$,
\be
\tilde{\chi} = \chi_{(0)} + \intd s \, \chi_s   +\intd t \, \chi_t   + \intd s \wedge \intd t \, \chi_{st} .\label{10f}
\ee
Note that $\chi_{(0)}$, $\chi_{s}$, $\chi_t$ and $\chi_{st}$ are sums of forms of all orders on $\mathcal{M}$. Substituting \eqref{10f} into \eqref{9f} we find
\be
\intd \chi_{(0)} + \intd s (\pa_s \chi_{(0)} - \intd  \chi_s)  + \intd t  (\pa_t \chi_{(0)} - \intd  \chi_t) + \intd s \wedge \intd t (\pa_s \chi_t - \pa_t \chi_s + \intd  \chi_{st})  =0.\label{11f}
\ee
The vanishing of the zero-form term states the known fact that the Chern character on $\mathcal{M}$ is closed regardless of the value of $a,b$. The vanishing of the $\intd s$ term recovers the transgression formula \eqref{326}. The vanishing ot the $\intd t$ term states how the Chern character varies between the two paths. The interesting term for our purposes is the last term in (\ref{11f}), or double transgression,
\be
\pa_s \chi_t - \pa_t \chi_s + \intd  \chi_{st} = 0.\label{12f}
\ee
The first two terms in \eqref{12f} are the exterior derivative $\intd_{s,t}$ on $[0,1]^2$ of $\chi_{(1)}$, the 1-form component of \eqref{10f} in the plane $(s,t)$. Therefore \eqref{12f} can be multiplied by $\intd s \wedge \intd t$ and rewritten
\be
\intd_{s,t} \chi_{(1)} = - \intd \chi_{(2)}, \label{13f}
\ee
\begin{equation}
\chi_{(1)} \equiv  \chi_s \intd s  + \chi_t \intd t, \qquad \chi_{(2)} \equiv \chi_{st} \intd s \wedge \intd t. \label{13fb}
\end{equation}
We proceed and integrate \eqref{13f} over the square $[0,1]^2$. Using Stokes theorem in the $(s,t)$ plane, we obtain that the left-hand side of (\ref{13f}) is
\be
\int_{[0,1]^2} \intd_{s,t} \chi_{(1)} = \oint_{\bar{\gamma}} \chi_{(1)} = \Omega^{(\bar{\gamma})},\label{14f}
\ee
i.e. the difference between the TCS actions obtained from both paths. The integral of \eqref{13f} therefore reads
\be
\label{17f} \Omega^{(\bar{\gamma})} = - \intd \int_{\Sigma} \chi_{(2)},
\ee
where $\Sigma$ is the surface enclosed by the loop $\bar{\gamma}$. (\ref{17f}) implies what we wanted to show: integrals of the TCS density over closed loops yield exact forms.

Equation \eqref{17f} gives an expression for the exact form generated by a given loop. After some algebra similar to what was derived in section \ref{sec:TCSQuillen}, the double transgression $\int_{\Sigma} \chi_{(2)}$ can indeed be expressed as
\be
\int_{\Sigma} \chi_{(2)} = \int_{0}^{1}\int_{0}^{1} \intd s\, \intd t\, \str\left(i \frac{\pa D_{s,t}}{\pa s} i \frac{\pa D_{s,t}}{\pa t} \exp(i \mathcal{F}_{s,t})\right),
\ee
where $D_{s,t}$ is the connection at fixed $s,t$. In virtue of \eqref{5f} and \eqref{6f}, changing path therefore yields a counterterm $\delta \alpha_4$ to the anomaly,
\be
I_4^1 \to I_4^1 + \delta  \int_{0}^{1}\int_{0}^{1} \intd s\, \intd t\, \str\left(\frac{\pa D_{s,t}}{\pa s} \frac{\pa D_{s,t}}{\pa t} \exp(i \mathcal{F}_{s,t})\right)\label{diffpath}.
\ee

\subsection{Bardeen term}
This section recalls known facts about the link between the consistent and covariant anomalies, and gives the expression of the 4d term relating the consistent anomaly $I_4^1$ to the covariant anomaly $I_4^{1,cov}$, known as the Bardeen term \cite{Bardeen:1984pm}
\be
\alpha_4^1 \equiv \int_{0}^{1}\intd s a(s)\str\left(\Lambda \wedge \intd \left(i \frac{\partial\mathcal{A}_s}{\partial s} \wedge e^{i\mathcal{F}(a,b)}\right)\right).\label{1F}
\ee
Note that this term cannot be obtained by a adding a boundary counterterm in the action. Instead, adding this term to the action corresponds to a redefinition of the current.

We compute the change of the consistent anomaly form $I_4^1$ in presence of this term, and show that the newly obtained anomaly is the covariant anomaly $I_4^{1, cov}$. In the absence of $\alpha_4$, $I_4^1$ is given by equation \eqref{eq:anompath} that we recall here,
\be
 I_4^1  =  \int_{0}^{1}\intd s (1-a(s))\str\left(\Lambda \wedge \intd \left(i \frac{\partial\mathcal{A}_s}{\partial s} \wedge e^{i\mathcal{F}(a,b)}\right)\right).\label{2F}
\ee
Note that the anomaly is dependent on the renormalization scheme in field theory \cite{Bilal:2008qx,Sugimoto}, and path-dependent in our description as discussed in the previous subsection. Adding $\alpha_4^1$ yields,
\be
 I_4^1 + \alpha_4^1  =  \int_{0}^{1}\intd s  \str\left(\Lambda \wedge \intd \left(i \frac{\partial\mathcal{A}_s}{\partial s} \wedge e^{i\mathcal{F}(a,b)}\right)\right).\label{3F}
\ee
We compute the integral in $s$ and obtain
\be
 I_4^1 + \alpha_4^1  =   \str(\Lambda \wedge \exp(i\mathcal{F})_4),\label{4F}
\ee
where the exponential is the quantity that appears in the second Chern number density $\chi_4$, defined in \eqref{542}. It is manifestly gauge invariant and takes the same form as the covariant anomaly found in \cite{Sugimoto}. Indeed, using (B.31) from appendix B of \cite{Sugimoto}, we obtain
\be
D_\m J_a^{\m, cov}= - \frac{\delta ( I + \alpha_4)}{\delta \Lambda} = \chi_{4L} + \chi_{4R}.\label{5F}
\ee

\subsection{Contribution of the TCS action to the anomalous conservation equation}
In this last subsection we compute the contribution of the TCS action to the divergence of the anomalous axial current using the holographic dictionary. For this purpose, we need to compute the near-boundary behavior of the fields using the full equations of motion. The action is composed of the TCS action and the simplified DBI action of section \ref{sec:effective}. We begin by computing the variation of the TCS action.

For this purpose we recall the descent equation \eqref{312} for $\Omega_5$,
\be
\intd \Omega_5 = \chi_6,\label{6F}
\ee
where $\chi_6$ is the 6-form component of the Chern character defined in \eqref{39}. We now perform a generic transformation of the superconnection
\be
\mathcal{A} \to \mathcal{A} + \delta \mathcal{A}\label{6F2},
\ee
and compute the exterior derivative of the variation $\delta \Omega_5$,
\be
\intd \delta \Omega_5  = \delta \intd \Omega_5 = \delta \str(\exp(i \mathcal{F})) = \str(\delta \exp(i \mathcal{F})) \label{7F}.
\ee
To compute the last term in \eqref{7F} we expand the exponential in series,
\be
\intd \delta \Omega_5 = \sum_{n=0}^{\infty} \frac{1}{n!}\str(\delta  (i \mathcal{F})^n),\label{8F}
\ee
and we use the cyclicity of the supertrace and resum the exponential series,
\be
\intd \delta \Omega_5 = \sum_{n=0}^{\infty} \frac{1}{(n-1)!}\str((i \delta \mathcal{F})  (i \mathcal{F})^{n-1}) =  \str((i\delta \mathcal{F})\exp(i \mathcal{F})).\label{9F}
\ee
We then use $\delta \mathcal{F} = [D, \delta\mathcal{A}]$, and the Bianchi identity to obtain
\be
\intd \delta \Omega_5 =  \str([D, i\delta \mathcal{A} \exp(i \mathcal{F})]).\label{10F}
\ee
Using property \eqref{a32} from appendix \ref{sec:convention2} we finally obtain
\be
\intd \delta \Omega_5 = \intd  \str( i\delta \mathcal{A} \exp(i \mathcal{F})).\label{11F}
\ee
Locally, we then obtain
\be
\delta \Omega_5 = \str( i\delta \mathcal{A} \exp(i \mathcal{F})) + \intd(\dots).\label{12F}
\ee
The exact piece in \eqref{12F} becomes a boundary variation. It does not contribute to the equations of motion, but it contains the contribution of the TCS action to the conjugate momenta, which can be read in  \eqref{eq:variationOmega5}, \eqref{eq:variationT}.

Importantly, equation \eqref{12F} is manifestly path-independent, while the contribution to the conjugate momentum depends on the path. Indeed, it was shown in the previous subsections that loops in the plane $(a,b)$ are exact, such that they contribute to the dots in \eqref{12F}. The contribution of the TCS action to the equations of motion is then path-independent, while the contribution to the conjugate momenta depends on the path.

We now return to the computation of the near-boundary expansion of the fields. For superconnection variations that only affect gauge fields (i.e. with $\delta T = 0$), equation \eqref{12F} leads to
\be
\frac{\delta \Omega_5}{\delta A_{L/R}}  = i \chi_{4,L/R},\label{13F}
\ee
where $\chi_{4,L}$ and $\chi_{4,R}$ are defined as
\begin{equation}
\label{c4LR} \exp{(i\mathcal{F})}|_{4} =
\begin{pmatrix}
\chi_{4,L} && \dots \\
\dots && \chi_{4,R}	
\end{pmatrix} \, .
\end{equation}
Note that, in terms of $\chi_{4,L/R}$, the quantity $\chi_4$ in \eqref{eq:instdensity} is expressed as
\be
 \chi_4 = \str\exp{(i\mathcal{F})}|_{4} = \mathrm{Tr}(\chi_{4,L}) - \mathrm{Tr}(\chi_{4,R})\,.
\ee
Equation \eqref{13F} is manifestly covariant, as is expected for a contribution to the bulk equations of motion.

It can be shown using the near-boundary equations of motion and the holographic dictionary that the right hand side of equation \eqref{13F} is what appears in the conservation of the anomalous current. After some algebra it can be schematically rewritten
\be
\eta^{\m\n} D_{V,\m}  J_\n^{cov}  =  c_A \epsilon^{\m\n\rho\sigma} (\chi_{L, \m\n\rho\sigma} +  \chi_{R, \m\n\rho\sigma}) + c_B m_q \sigma ( \mathcal{U} - \mathcal{U}^\dag),\label{17F}
\ee
where $ J_\n^{cov}$ is the covariant current, and $c_A, c_B$ are numerical factors. This is the holographic covariant anomalous conservation equation \eqref{5F}, see for instance \cite{Landsteiner:2012kd}. The mass term in \eqref{17F} comes from the DBI action. It can be shown by computing $c_A$ that equation \eqref{17F} is then consistent with the field theory results, as can be checked from \eqref{5F} (at $m_q = 0$) \cite{Sugimoto}.

In order to obtain the consistent anomaly, one needs to relate $J_\n^{cov}$ to the current associated with the gauge variation of the boundary effective action, $J_\n$,
\be
\delta_\Lambda S_{TCS} = \int \intd^4x  D^\n \Lambda J_\n,
\ee
where $S_{TCS} = \frac{iN_c}{4\pi^2} \int \Omega_5$. $J_\n$ can be read in \eqref{320}. This current is related to $J_\n^{cov}$ by the Bardeen term \eqref{1F}, as $J^{cov}_\nu = J_\nu + P_\nu$, with $P_\nu$ defined as
\be
\int \alpha_4^1  \equiv \int \intd^4x D^\n \Lambda P_\n,
\ee
and $P_\nu$ can be read off from \eqref{1F}. Recall that $J_\n$ also depends on boundary terms appearing in $\Omega_5$, and therefore on the path taken to compute $\Omega_5$, as is expected of the consistent anomaly.

\end{appendix}
	

\end{document}